\documentclass{aa}        
\usepackage{txfonts}
\usepackage{graphicx,longtable,lscape,natbib,psfig,textcomp,fontenc,amssymb}  
\bibpunct{(}{)}{;}{a}{}{,} % to follow the A&A style
\newcommand{\ltsima} {$\; \buildrel < \over \sim \;$}  
\newcommand{\gtsima} {$\; \buildrel > \over \sim \;$}  
\newcommand{\lta} {\lower.5ex\hbox{\ltsima}}  
\newcommand{\gta} {\lower.5ex\hbox{\gtsima}}

\newcommand{\ergs}{\>{\rm erg}\,{\rm s}^{-1}}

\newcommand{\ergsHz}{\ensuremath{{\rm erg}\,{\rm s}^{-1}\,{\rm Hz}^{-1}}}
\newcommand{\mum}{$\mu$m}

\begin{document}

\title{The radio-loud AGN population at $z\gtrsim1$  in the COSMOS
  field. I. Selection and Spectral Energy Distributions}
  
\titlerunning{The radio-loud AGN population at $z\gtrsim1$ in the COSMOS field} 
\authorrunning{Baldi, R.D. et al.}
  
\author{Ranieri D. Baldi \inst{1,2,3} \and Alessandro Capetti \inst{4} \and Marco
  Chiaberge \inst{5,6,7} \and Annalisa Celotti \inst{1,8,9}}
   
%\author{Ranieri D. Baldi  \inst{1}  \and
%  Alessandro Capetti \inst{2} et al }

%\altaffiltext{1}{SISSA, via Bonomea 265, 34136 Trieste, Italy}
%\email{baldi@ph.technion.ac.il} 
\institute{SISSA-ISAS, via Bonomea 265, I-34136 Trieste, Italy \and Physics
  Department, The Technion, Haifa 32000, Israel, baldi@ph.technion.ac.il \and  Physics Department,
  Faculty of Natural Sciences, University of Haifa, Israel \and INAF -
 Osservatorio Astrofisico di Torino, Strada Osservatorio 20, I-10025 Pino
 Torinese, Italy, \and Space Telescope Science
 Institute, 3700 San Martin Drive, Baltimore, MD 21218, USA, \and
 INAF-Istituto di Radio Astronomia, via P. Gobetti 101, I-40129 Bologna,
 Italy, \and Center for Astrophysical Sciences, Johns Hopkins University,
 3400 N. Charles Street Baltimore, MD 21218, USA \and INAF-Osservatorio
  Astronomico di Brera, via E. Bianchi 46, 23807 Merate, Italy, \and INFN-
  Sezione di Trieste, via Valerio 2, 34127, Trieste, Italy}

\date{}

\abstract{We select a sample of radio galaxies at high redshifts ($z \gtrsim
  1$) in the COSMOS field, by cross-matching optical and infrared (IR) images
  with the FIRST radio data. The aim of this study is to explore the high-$z$
  radio-loud (RL) AGN population at much lower luminosities than the classical
  samples of distant radio sources and similar to those of the local
  population of radio galaxies. Precisely, we extended a previous analysis
  focused on low-luminosity radio galaxies by \citet{chiaberge09} and
  \citet{baldi13}.  The wide multiwavelength coverage provided by the COSMOS
  survey allows us to derive their Spectral Energy Distributions (SEDs). We
  model them with our own developed technique {\it 2SPD} that includes old and
  young stellar populations and dust emission. When added to those previously
  selected we obtain a sample of 74 RL AGN. The SED modeling returns several
  important quantities associated with the AGN and host properties. The
  resulting photometric redshifts range from z$\sim$0.7 to 3. The sample
  mostly includes compact radio sources, but also 21 FR~IIs sources; the radio
  power distribution of the sample covers $\sim$10$^{31.5} - 10^{34.3}$
  $\ergsHz$, thus straddling the local FR~I/FR~II break. The inferred range of
  stellar mass of the hosts is $\sim$10$^{10} - 10^{11.5}$ M$_{\odot}$.  The
  SEDs are dominated by the contribution from an old stellar population with
  an age of $\sim 1 - 3$ Gyr for most of the sources. However, UV and mid-IR
  (MIR) excesses are observed for half of the sample. The dust luminosities
  inferred from the MIR excesses are in the range $L_{\rm dust} \sim10^{43}
  -10^{45.5}$ erg s$^{-1}$, associated with temperatures approximately of
  350-1200 K.  Estimates of the UV component yield values of $\sim
  10^{41.5}-10^{45.5}$ erg s$^{-1}$ at 2000 \AA. UV emission is significantly
  correlated with both IR and radio luminosities, the former being the
  stronger link. However, the origin of UV and dust emission, whether it is
  produced by the AGN of by star formation, is still unclear. Our results show
  that this RL AGN population at high redshifts displays a wide variety of
  properties. Low-power radio galaxies, associated with UV- and IR-faint hosts
  are generally similar to red massive galaxies of the local FR~Is. At the
  opposite side of the radio luminosity distribution, large MIR and UV
  excesses are observed in objects consistent with quasar-like AGN, as also
  proved by their high dust temperatures, more similar to local FR~IIs.

 }

\maketitle

\section{Introduction}

The advent of multiband dataset from large area surveys marked the starting
point of a new scientific approach based on large samples of sources through a
multiwavelength analysis.  The immense number of sources as well as the
completeness of the sample is fundamental for obtaining results with high
statistical foundations. Clear examples are represented by large surveys, such
as Sloan Digital Sky Survey (SDSS) \citep{york00} and COSMOS
\citep{scoville07} which provide wide multi-wavelength coverage. The
association of targets at different wavelengths helps us to determine the
properties of the sources and, especially to derive their SEDs.

Among the most energetic phenomena in the Universe, radio galaxies occupy an
important position in the study of the fundamental issues of modern
astrophysics, such as accretion onto black holes (BH), the co-evolution
between the host galaxy and BH, the feedback process of the AGN on the
interstellar and intercluster medium (e.g. \citealt{hopkins06a,fabian06}). In
this context, the cross-match of radio and optical surveys specifically
provides a unique tool in the analysis of the RL AGN. It consists in
identifying optically large numbers of radio sources, to obtain
spectroscopic/photometric redshifts, and finally to investigate the links
between the radio structures, associated with the central engine, and the host
galaxies. For example, \citet{best05b,best05a} selected a sample of radio
galaxies by cross-correlating optical SDSS, and radio NVSS \citep{NVSS} and
FIRST \citep{FIRST} catalogs. This sample constitutes a very good
representation of radio galaxies in the local Universe. Recently, the advent
of the COSMOS survey which give X-ray, UV, optical, infrared, and radio data
all at once, facilitates the community to select large sample of sources on a
wider range of wavelengths even then SDSS, but on a 2 deg$^{2}$ area of the
sky.

Since the COSMOS catalogs are available for the community, several studies
have already been carried out on radio
sources. \citet{schinnerer04,schinnerer07} selected $\sim$3600 radio-emitting
galaxies (starburst and AGN) in the COSMOS field based on VLA radio maps at
1.4 GHz. \citet{smolcic08} explore the properties of the sub-mJy radio
population, using the VLA-COSMOS dataset, separating star-forming galaxies
from AGN, reaching L$_{\rm 1.4 GHz}$ $\sim$ 10$^{33}$ $\ergsHz$.  Their sample
is a mixture of objects with z $\lesssim$ 1.2 where AGN dominate over
star-bust galaxies for L$_{\rm 1.4 GHz} > 10^{31} \ergsHz$. \citet{bardelli10}
investigate the properties and the environment of radio sources, at $z < 1$,
by combining the VLA-COSMOS dataset and the redshift-survey zCOSMOS \citep{lilly07}. This
sample includes low-luminosity radio-emitting objects (L$_{\rm 1.4 GHz} <
10^{32} \ergsHz$) associated with red massive ($\sim$ 7 $\times$10$^{10}$
M$_{\odot}$) hosts in over-dense regions.

The basic idea of this study is to select a sample representative of the RL
AGN population at $z\gtrsim1$, in order to investigate the properties of the
radio galaxies in a cosmic era, where the AGN activity plays a fundamental
role in the galaxy formation.  Such a sample can be used to answer to
different astrophysical questions, such as, cosmological evolution of radio
galaxies, study of their luminosity function, comparison with local active and
quiescent galaxies, all in light of the symbiotic relation between AGN and
host.  For this purpose we choose the COSMOS field because the large
multi-wavelength dataset provided by the survey consists in a unique tool to
perform a multi-band selection procedure.

The main problem of the selection of sources at high redshifts is represented
by the observational bias which makes flux-limited samples more abundant of
powerful sources due to the tight redshift-luminosity dependence.  In fact
available samples of RL AGN at $z\gtrsim1$ (e.g., 3CR and its deeper
successors) mostly include powerful 'edge-brightened' (FR~II,
\citealt{fanaroff74}) objects. This implies that our knowledge about the
high-z RL AGN population misses a fundamental piece, consisting of the weak
'edge-darkened' radio galaxies (FR~I). Unlike previous studies, in this work
we pay much attention to include the low-power radio sources in order to
satisfy the sample requirements of completeness and homogeneity, as needed.
The first steps in that direction were done by \citet{chiaberge09} (hereafter
C09). They selected in the COSMOS field the first seizable sample of FR~I
candidates at z$\gtrsim$1. Such a research is motivated by the peculiarity of
this class of sources. In fact in the local Universe, FR~Is typically live in
massive early-type galaxy in clusters.  This behavior, if shared by their
high-z counterparts, will help the community to address a number of other
unsolved problems in current astrophysics, such as the evolution of the
elliptical galaxies, assessing the relationship between giant elliptical and
their central BH at low nuclear luminosities, searching for cluster to study
their formation and evolution, and studying the possible FR~I-QSO association,
not common in the local Universe.

In order to study the overall high-z RL AGN population, we deem necessary to
relax the selection criteria and include radio sources regardless of their
morphology and radio flux. In particular we also include FR~II radio sources
as well as those with a flux higher than 13 mJy, the high flux threshold used
by C09. This resulting sample will include, by definition, the low-power radio
galaxies selected by C09, despite of the different goals of selection. As we
will show later, the selection procedure yields a sample of radio sources of
much lower luminosity than classical samples of distant radio sources
(approximatively 2.5 orders of magnitude fainter, e.g., \citealt{willott01}
for z$\gtrsim$0.7) and similar to those of the local population of radio
galaxies.

Once the sample is selected, we will study the multi-wavelength properties
(from radio to IR) of the high-z RL AGN population .  We will address
different studies of such a population in forthcoming papers. Precisely, in
this work, we will also perform an analysis analogous to what \citet{baldi13}
(hereafter B13) operated on the sample selected by C09.  B13, taking advantage
of the large multi-wavelength coverage provided by the COSMOS survey,
carefully identified the correct counterparts of the selected radio sources at
different wavelengths to derive their genuine emission. They thus constructed
their SEDs from the FUV to the MIR wavelengths.  The SED modeling with
stellar templates returned their photo-z and their AGN and host properties.
Their main results are the following.  The resulting photometric redshifts
range from $\sim$ 0.7 to 3. The radio power distribution of their sample,
$\sim10^{31.5} - 10^{33.3} \ergsHz$ at rest-frame 1.4 GHz, indicates that
their radio power is indeed mostly consistent with local FR~Is. Yet, a small
contribution of sources show larger radio power, above the local FR~I/FR~II
break.\footnote{FR~I galaxies typically have a radio power lower than that of
  FR~II sources, with the FR~I/FR~II break set at $L_{\rm 178 MHz} \sim 2
  \times 10^{33} \ergsHz$ \citep{fanaroff74} . The transition is rather smooth
  and both radio morphologies are present in the population of sources around
  the break.}  Most of the hosts of these high-z low-luminosity radio sources
are massive galaxies ($\sim$7$\times$10$^{10}$ M$_{\odot}$) dominated by an old stellar
population (a few Gyr) but significant excesses in either the UV or in the MIR
band are often present.

In this paper we select a sample of RL AGN in the COSMOS field with a
multiwavelength identification of their hosts described in
Section~\ref{sample} and Section~\ref{identification}. Since the entire sample
includes the FR~I candidates selected by C09 and their properties are already
studied by B13, we initially study the 'newly selected' radio sources
following B13 and then we consider the entire sample.  Using the FUV-MIR data
provided by the COSMOS survey, we derive their SED
(Section~\ref{fittingsed}) and we model them using 2SPD ({\it 2 Stellar
  Population and Dust component}) technique for the radio sources selected in
this work. The results returned from the SED modeling are presented in
Section~\ref{results}: the photometric redshifts, and the host properties,
such as stellar ages, masses and dust, and UV components. In
Section~\ref{totalsample} we gather the entire sample of distant RL AGN in the
COSMOS field to study their radio power distribution (Section~\ref{radio}) and
global properties (Section~\ref{global}). In Section~\ref{multi} we look for
relations between radio, IR, and UV luminosities to investigate the origin of
their emission at these wavelengths.  In Section~\ref{discussion}, we
summarize the results and we discuss our findings.

We adopt a Hubble constant of H$_{0}$ = 71 km s$^{-1}$ Mpc$^{-1}$,
$\Omega_{m}$ = 0.27 and $\Omega_{vac}$ = 0.73, as given by the {\it WMAP}
cosmology (e.g., \citealt{spergel03,jarosik11}). All the magnitudes are in AB
mag system, if not otherwise specified.

\section{Dataset}

The photometric data used to derive the SEDs of our sources are taken from the
COSMOS survey \citep{scoville07}. This survey comprises ground based as well
as imaging and spectroscopic observations from radio to X-rays wavelengths,
covering a 2 deg$^{2}$ area in the sky.

Optical and IR observations and data reduction are presented in
\citet{capak07}, \citet{capak08} and \citet{taniguchi07}. A multiwavelength
photometric catalog was generated using SExtractor \citep{bertin96} and
includes objects with total magnitude {\it I}$<$ 25.\footnote{The COSMOS
  catalog is derived from a combination of the CFHT $i^{*}$ and Subaru $i^{+}$
  images, to which the authors refer as {\it I}-band images} The survey
includes HST/ACS data \citep{koekemoer07} which provides the highest angular
resolution ($\sim0\farcs09$) among the COSMOS images. Furthermore, the survey
gathers data from GALEX, Subaru, CFHT, UKIRT, NOAO, and Spitzer.

The COSMOS collaboration provides different catalogs. For this optical
identification we use the COSMOS Intermediate and Broad Band Photometry
Catalog 2008
\citep{capak08}\footnote{http://irsa.ipac.caltech.edu/cgi-bin/Gator/nph-dd}
from which we take the broad-band magnitudes of our sources from the FUV to the K
bands. At IR wavelengths, we also used the S-COSMOS IRAC 4-channel Photometry
Catalog June 2007 and S-COSMOS MIPS 24 Photometry Catalog October 2008 (or
S-COSMOS MIPS 24um DEEP Photometry Catalog June 2007) to search for the
Spitzer/IRAC and MIPS counterparts \citep{sanders07}.

The selection of the sample is also based on radio data from the FIRST survey
\citep{FIRST}. The data are obtained with the VLA in B configuration with an
angular resolution of $\sim$5\arcsec and reach a flux limit of $\sim$1 mJy. We
also used data from the NVSS survey \citep{NVSS} (VLA in D configuration). The
NVSS has a lower spatial resolution ($\sim$45 \arcsec) and with a higher flux
density limit ($\sim$2.5 mJy) than the FIRST survey, but these data are useful
since they are more sensitive to diffuse low surface brightness radio emission
that the FIRST data.  We will also use data from the VLA-COSMOS Large and Deep
Projects \citep{schinnerer04,schinnerer07}, i.e. VLA observations (in A-C array)
of the COSMOS field at a resolution of 1\farcs5 and with a mean rms noise of
$\sim$ 10 $\mu$Jy/beam.

The COSMOS survey also provides spectroscopic data from the Very Large
Telescope (VLT) (zCOSMOS, \citealt{lilly07}) and from the Magellan (Baade)
telescope \citep{trump07}.  In addition, the COSMOS collaboration performed
their own photo-z derivation mostly for sources with $i^{+}< 25.5$ with a
relative redshift accuracy of 0.007 at $i^{+}<$ 22.5 \citep{ilbert09} and,
also, for optically identified sources detected with XMM, achieving a relative
accuracy of 0.014 for $i^{+}<$ 22.5 \citep{salvato09}. The COSMOS Photometric
Redshift Catalog Fall 2008 (mag {\it I} = 25 limited) gathers the photometric
redshifts measured from those authors.

\section{The sample} 
\label{sample}

\begin{table*}
\addtolength{\tabcolsep}{-2pt}
  \begin{center}
  \caption{Identified radio galaxies}
  \label{1table}
  \begin{tabular}{l|lcccccc|cccc}
    \hline \hline
n & radio ID                                                       &                RA  & DE  &  N$_{c}$  & F$_{FIRST}$ &  F$_{NVSS}$  & radio morph & host ID & mag$_{i}$  & z$_{photo}$ & z$_{spec}$  \\
\hline
  1  & J100046.91+020726.5             & 10 00 46.944 & +02 07 26.02  & 1 &   1.79 &       2.6 & compact & 766333  &  22.28  & 1.210$_{1.19}^{1.50}$  &   1.1577$^{a}$  \\   %Less secure BLAGN  92\%
  2  & J100109.28+021721.7$^{(2)}$   & 10 01 09.280 & +02 17 21.49  & 1 &   3.21 &       3.7 & FR~II      &  $-$       & $<$26.72     &  $-$                        &      $-$             \\
  3  & J100101.26+020118.0             & 10 01 01.258 & +02 01 16.98  & 1 &   1.68 & $<$2.5& compact & 756907  & 24.97  & 1.876$_{1.45}^{2.35}$  &      $-$       \\
  4  & J100016.57+022638.4             & 10 00 16.575 & +02 26 38.28  & 1 &   5.18 &       5.1 & compact &  $-$      &  25.90     &      $-$                      &  $-$   \\
  5  & J100114.85+020208.8$^{(4)}$  & 10 01 14.942 & +02 02 21.48  & 2 &   4.78 &       6.2 & FR~II      & 754529  &   21.21 &  1.120$_{1.10}^{1.15}$  &  0.9707$^{a}$    \\ %Less secure BLAGN  92\%  
  6  & J100114.12+015444.3$^{(6)}$  & 10 01 14.542 & +01 54 51.72  & 2 &   4.99 &       6.2 & FR~II      & 526188  &  23.30  & 1.426$_{1.35}^{1.48}$   &   $-$              \\
  7  & J100058.05+015129.0$^{(4)}$  & 10 00 58.107 & +01 51 41.15  & 2 & 10.10 &     12.4 & FR~II      & 534525  &  25.24  & 1.455$_{1.26}^{2.29}$  &  $-$       \\
  8  & J100201.17+021327.1$^{(3)}$  & 10 02 01.235 & +02 13 26.67  & 1 &   4.89 &       6.3 & FR~II      & 952745  &  21.45  & 0.832$_{0.81}^{0.85}$  &  0.8357$^{a}$  \\     %92\%
  9  & J095959.16+014837.8$^{(2)}$  & 09 59 59.127 & +01 48 37.79  & 1 &   7.70 &       8.4 & FR~II      & 591011  &  25.80  &    $-$                         &  $-$                \\
 10 & J100120.06+023443.7            & 10 01 20.090 & +02 34 43.62  & 1 &   9.07 &       9.6 & compact& 1425414 & 26.82  &   $-$                         &    $-$       \\
 11 & J100140.12+015129.9$^{(4)}$  & 10 01 39.193 & +01 51 39.49  & 1 &   7.85 &      11.1& FR~II      &  509607  & 23.22 & 0.959$_{0.93}^{1.01}$    &    $-$     \\
 12 & J100006.17+024000.5            & 10 00 06.137 & +02 40 00.13  & 1 &   3.47 &        *   & compact & $-$        &  25.95    &      $-$                       &  $-$   \\
 13 & J100007.29+024049.8$^{(2)}$  &10 00 07.294  & +02 40 49.79  & 1 &   3.47 &        *   & FR~II       & 1703047&  23.99 &  1.238$_{0.93}^{1.43}$  &   $-$   \\    
 14 & J095835.44+020543.7            & 09 58 35.473 & +02 05 43.81  & 1 & 11.73 &     13.1 & compact & 845386 &  24.58 &  1.259$_{1.10}^{1.91}$   &   $-$ \\ 
 15 & J095927.25+023729.2$^{(3)}$ & 09 59 27.221 & +02 37 37.32  & 1 &   1.94 &       5.9 & FR~II       & 1490892& 22.66 & 1.006$_{0.98}^{1.04}$   &  $-$ \\
 16 & J100137.77+014811.7           & 10 01 37.793 & +01 48 11.38  & 1 &   2.60 & $<$2.5 & compact & 517639  & 21.87 & 0.836$_{0.81}^{0.86}$   &  0.8442$^{b}$ \\   %nla >90\%
 17 & J100230.11+020912.4$^{(5)}$ & 10 02 29.915 & +02 09 10.72  & 2 &   6.40 &       6.3 & FR~II       & 939988  &  24.08 & 1.458$_{1.32}^{1.58}$  &  $-$  \\
 18 & J100007.90+024315.4$^{(4)}$ & 10 00 07.835 & +02 43 10.52  & 2 &   9.99 &     12.0 & FR~II       & 1695922 & 22.38 & 1.483$_{1.44}^{1.63}$  &  $-$  \\
 19 & J100218.03+015555.7           & 10 02 18.083 & +01 55 56.85  & 1 &   1.04 & $<$2.5 & compact &  $-$       &  21.52   &      $-$                     &  $-$ \\
 20 & J100212.06+023134.8$^{(4)}$ & 10 02 11.867 & +02 31 34.40  & 2 & 16.28 &     17.4 & FR~II       & 1408636 & 23.81 & 1.014$_{0.91}^{1.21}$     &   $-$ \\
 21 & J100159.82+023904.8           & 10 01 59.861 & +02 39 04.53  & 1 &   2.67 &       4.3 & compact & 1633838 & 22.40 & 0.813$_{0.79}^{0.84}$     &  $-$ \\ 
 22 & J095837.11+023549.0           & 09 58 37.168 & +02 35 49.37  & 1 &   1.05 & $<$2.5 &compact  & 1519463 & 24.91  & 2.604$_{2.25}^{2.86}$     &  $-$   \\
 23 & J100028.31+013507.8$^{(5)}$ & 10 00 26.611 & +01 35 27.67  & 2 & 14.65 &     26.6 & FR~II       & 115652  & 22.42  &  0.835$_{0.82}^{0.86}$    &  0.83933$^{b}$  \\ %nla >90\%
 24 & J095826.95+023711.7           & 09 58 26.966 & +02 37 11.55  & 1 &   2.24 &       3.0 & compact & 1516040 & 25.63 &       $-$                       &  $-$ \\
 25 & J100124.09+024936.3$^{(4)}$ & 10 01 24.122 & +02 49 36.58  & 1 &   3.39 &       3.9 & FRI/FRII & 1874867 & 21.21 & 0.825$_{0.81}^{0.84}$   &  0.82510$^{b}$ \\ %a >90\%
 26 & J095756.52+022717.3           & 09 57 56.541 & +02 27 17.25  & 1 &   2.57 &       2.3 & compact & 1319327 & 21.14 & 0.731$_{0.72}^{0.75}$     &   $-$         \\
 27 & J095908.87+013606.6           & 09 59 08.861 & +01 36 06.64  & 1 &   5.46 &       6.5 & compact & 159456   & 25.22 &  $-$                            &    $-$    \\ 
 28 & J095839.24+013557.8$^{(4)}$ & 09 58 39.742 & +01 35 56.85  & 1 &   2.56 &       4.9 & FR~II       & 182240   & 24.27 &  1.676$_{1.40}^{2.06}$   &    $-$   \\
 29 & J095821.65+024628.1 	        & 09 58 21.700 & +02 46 28.07  & 1 &   4.62 &     10.3 & compact-QSO & 1738294 & 19.35 &  0.781$_{0.77}^{0.79}$ &  1.4050$^{c}$  \\ 
 30 & J095838.01+013217.1 	        & 09 58 37.991 & +01 32 17.12  & 1 &   4.21 &       4.3 &  extended &   $-$     & $<$26.10    &      $-$                      &  $-$  \\
 31 & J095835.71+025328.9 	        & 09 58 35.719 & +02 53 28.67  & 1 &   3.61 &       3.5 & extended  & 1957693 & 25.84 &     $-$                      &  $-$   \\
 32 & J095738.38+023837.7$^{*}$  & 09 57 38.375 & +02 38 37.62  & 1 &   2.78 &       3.2 & extended  & 1780946 & 24.35 &     $-$                       &  $-$  \\  
 33 & J100331.82+014901.4$^{*}$  & 10 03 31.849 & +01 49 01.77  & 1 &   1.08 &       2.7 & extended  &  $-$       &  $<$25.40   &      $-$                       &  $-$ \\  
 34 & NVSS J100250+013017          & 10 02 50.681 & +01 30 19.26  & 1 &   8.61 &       7.0 & extended  &  61719   & 24.07 &   1.388$_{1.26}^{1.46}$   &   $-$  \\     
 35  & J100217.97+015836.4         & 10 02 17.988 & +01 58 36.13  & 1 &  26.83 & 26.3 & compact  & 714756   & 21.63  & 0.902$_{0.89}^{0.91}$ &   $-$   \\   
 36 & J095803.21+021357.7         & 09 58 03.223 & +02 13 57.58  & 1 &  24.71 & 25.2 & compact & 1103009 & 24.89 & 2.218$_{1.91}^{2.77}$ &    $-$  \\
 37 & J095908.32+024309.6$^{(4)}$ & 09 59 07.629 & +02 43 02.59  & 3 &  55.92 & 59.4 & FR~II-QSO        & 1721832  & 19.20 & 0.787$_{0.78}^{0.80}$  &  1.3197$^{c}$ \\  
 38 & NVSS J095758+015832$^{*}$  & 09 58 00.807 & +01 58 56.75  & 3 &  43.91 & 52.2 & FR~II      & 887322    & 23.89 & 2.368$_{2.19}^{2.76}$ &  $-$  \\
 39 & J100252.88+015549.7        & 10 02 52.887 & +01 55 49.66  & 1 & 16.65  & 16.7 & compact & 477930   & 25.72 &          $-$              &   $-$\\
 40 & J095908.95+024813.4$^{(3)}$ & 09 59 09.143 & +02 48 16.45  & 2 & 20.66  & 28.8 & FR~II       & 1947189 & 23.06 & 1.111$_{1.08}^{1.15}$ &  $-$ \\
 41 & J095742.30+020426.0         & 09 57 42.313 & +02 04 25.97  & 1 & 18.63  & 18.5 & extended & 873336   & 22.39 & 0.858$_{0.85}^{0.87}$ & $-$\\
 42 & J100153.77+024954.0        & 10 01 53.822 & +02 49 53.94  & 1 & 16.77  & 16.0 & compact & 1852665 & 23.29 & 1.112$_{1.06}^{1.24}$  & $-$ \\
 43 & J100303.66+014736.0$^{(6)}$ & 10 03 04.903 & +01 47 24.21  & 2 & 23.88  & 32.6 & FR~II       & 454341    & 22.97 & 1.192$_{0.89}^{1.37}$  & $-$ \\ 
 44 & J100251.11+024248.5$^{(4)}$ & 10 02 50.858 & +02 42 50.14  & 2 &176.05 &174.3& FR~II      & 1599142  & 22.95 & 1.185$_{1.16}^{1.21}$   & $-$ \\
 45 & J095741.10+015122.509$^{(6)}$ & 09 57 39.795 & +01 51 41.87  & 3 & 31.17  & 43.2 & FR~II     & 657685     & 22.59 & 0.820$_{0.79}^{1.01}$  & $-$ \\
 46 & J095822.30+024721.3$^{(5)}$ & 09 58 22.881 & +02 47 28.14  & 2 & 17.36  & 22.7 & FR~II     & 1736088   & 22.57 & 0.907$_{0.85}^{0.94}$  &  0.8784$^{a}$ \\   %95\% 
\hline
%     &  COSMOS-FR~I 7          &  10 01 04.527 & +02 02 03.14 & 1 &   1.14 &  $<$2.5& compact &  $-$      &  $<$27.20     &      $-$                      &  $-$   \\  % ogg 7 di marco      
%COSMOS-FR~I 7  & J100104.51+020203.6             &  10 01 04.527 & +02 02 03.14 & 1 &   1.14 &  $<$2.5& compact &  $-$      &   $-$     &      $-$                      &  $-$   \\  % ogg 7 di marco      
\hline
\end{tabular}
\end{center}
Column description: (1) identification number; (2) COSMOS-VLA
(Large Project) ID number of the object \citep{schinnerer07}. In case of multiple objects, the
number of components is shown on the superscript. The obejcts marked with
$*$ are identified in the COSMOS-VLA Deep Project \citep{schinnerer04}. The object 34 has the
NVSS ID; (3)-(4) right ascension and declination of (one of the components
of) FIRST radio source; (5) number of matches found in the FIRST catalog
associated with the same radio galaxy; (6) (total) FIRST radio flux (mJy) of
the entire radio source; (7) NVSS radio flux (mJy). The two objects, 12 and
13, are included in the same NVSS radio sources with a flux of 7.6 mJy. (8)
COSMOS-VLA radio morphology; (9) ID number of the host associated with the radio
galaxy from the COSMOS Intermediate and Broad Band Photometry Catalog 2008 ;
(10) Subaru $i^{+}$, CFHT $i^{*}$ or ACS/HST F814W magnitude of the host
galaxy from the the COSMOS catalog or measured on the image; (11)
photometric redshift with 99\% confidence-level errors; (12) spectroscopic
redshift from the zCOSMOS catalog \citep{lilly07} marked with $a$, the
Magellan catalog \citep{trump07} marked with $b$, and SDSS QSO spectra
\citep{hewett10} marked with $c$.
\end{table*}

% nla = nla': Type 2 AGN and red galaxy hybrid  red galaxy continuum with absorption and strong emission lines    
%a =no emission lines, red galaxy-type spectrum (obscured AGN) 

The aim of the project is to select the RL AGN population in the COSMOS field
at high redshifts. With respect to the objects considered in C09 and studied
by B13, we relax the selection criteria in order to include not just high
redshift FR~I candidates, but all radio sources likely to be associated with
galaxies at z $\gtrsim$ 1. Differently from C09 we did not set a high radio
flux limit (they only included objects with a FIRST flux ranging between 1 and
13 mJy) and we also decided not to exclude u-band dropout galaxies.
 
We then simply search for sources with FIRST radio emission larger than 1 mJy
over the COSMOS field. The total number of FIRST radio detection sources in a
circular area of radius 5100 arcsec\footnote{We consider a radius of 5100
  arcsec because it corresponds to the radius of a circle which circumscribes
  the 'squared' COSMOS field. This allows us to include all the radio sources
  which satisfy our selection criteria but are located at the edges of the
  COSMOS field.} is 515.\footnote{This number corresponds to each FIRST radio
  detection. Therefore, it does not correspond to the total number of radio
  sources, since double or even triple FIRST radio sources are present.} This
number also includes the radio sources selected by C09.  Given that the radio
morphology is fundamental to identify the host galaxy, we only consider FIRST
sources which have VLA-COSMOS counterpart. This strongly facilitates the host
identification because of its higher spatial resolution than FIRST images. In
10 objects the VLA image does not show any radio emission in correspondence of
the FIRST source. These are probably diffuse objects not visible in the
VLA-COSMOS image due to its higher spatial resolution. Such radio sources have
FIRST radio flux less than 3.74 mJy (typically $\sim$1.2 mJy).

Similarly to the selection procedure performed by C09, we use the assumption
that the properties of the host galaxies of the RL AGN population at high
redshifts are similar to those of distant FR~IIs.  Typical FR~II radio galaxy
at $1<z<2$ has a K-band magnitude fainter than 17 \citep{willott03}. Since the
typical $I-K$ color for FR~II host is not smaller than 4, this assumption sets
a lower limit on the {\it I}-band magnitude of the host of 21 (in the Subaru
or CFHT images). We use this optical limit to select the optical counterparts
of the FIRST radio sources. In such a process, we perform a rough
identification of the I-band counterpart. In fact a more detailed host
identification is aim of the next step of the selection.  Furthermore, we
also include in the sample radio sources which satisfy the radio selection but
are either spectroscopically classified as QSO or candidates QSO based on the
point-like appearance in the ACS images. Such objects might have {\it I}$<$
21, since the AGN emission out-shines the galaxy emission in optical
band. Excluding the sources of the C09 sample, the new selected objects in the
COSMOS field based on the radio emission and on the {\it I}-band magnitude are
56\footnote{To identify the sources, we use for simplicity a number ranging
  from 1 to 56. This number does not correspond to the nomenclature used by
  C09. In fact, the sources selected by C09 are named as 'COSMOS-FR~I'.}

We identify the optical/infrared counterpart to the radio emission, by
checking the multi-wavelength images of each radio source. Details of the host
identification are in Section~\ref{identification}. For 46 objects this
procedure is successful (see Table~\ref{1table}), while for 10 FIRST radio
sources we do not trust much in the host identification because of the
complexity of the radio morphology or the ambiguity of the correct optical
counterpart. Therefore we prefer to leave these 10 sources out. Those sources
are presented in Table~\ref{2table}.

The sample of 46 members includes 2 spectroscopically confirmed QSOs from the
SDSS catalog \citep{hewett10}. From the point of view of the radio morphology
classification, the sample includes 18 compact (or marginally resolved at the
resolution provided by the COSMOS-VLA images) radio structures, 6 slightly
extended sources, one intermediate FR~I/FR~II, and 21 FR~IIs (see
Sect~\ref{radio}). Nine of the sources we selected as part of this work
  have properties that satisfy the selection criteria of C09. These objects
  were not included in that sample most likely because their radio flux in the
  version of the FIRST catalog those authors used was not listed as $>$ 1 mJy.
  This is due to the continuously ongoing updating of the FIRST catalog over
  the years.\footnote{http://sundog.stsci.edu/first/catalogs/}

  The COSMOS collaboration performed a similar radio selection in the COSMOS
  field, called the VLA-COSMOS Large Project \citep{schinnerer07}. It consists
  of identification of radio-emitting sources at 1.4 GHz observed with the VLA
  at higher radio resolution (A and C arrays) than NVSS and FIRST we
  used. However, our selection procedure is more sensitive to low brightness
  radio sources than that used by the VLA-COSMOS Large Project.  In fact we
  detect radio sources in FIRST which do not show emission in the VLA-COSMOS
  images.  Most of our sources are included in the VLA-COSMOS catalog
  (Table~\ref{1table}). However, those not included are still visible in the
  VLA-COSMOS radio maps.  Therefore our sample does not totally overlap with
  the catalog created by \citet{schinnerer07}.  This suggests that our careful
  visual inspection is fundamental to identify weak sources. Furthermore,
  differently from the VLA-COSMOS catalog, we perform a multi-band
  cross-matching to isolate the distant radio sources and exclude the possible
  star-forming galaxies.

\section{Multi-band counterparts identification}
\label{identification}

The radio morphology of the source plays a fundamental role in this
procedure. In order to inspect that of our sources, we mainly use the
  radio maps (180 arcsec, corresponding to a size of $\sim$1.5 Mp at z=1)
  from the VLA-COSMOS Large and Deep Projects in comparison with the
  low-resolution FIRST and NVSS images. The VLA-COSMOS images provide
  sufficient angular resolution to recognize the presence of radio cores,
  useful (but not necessary) to identify the corresponding host galaxies.

The VLA-COSMOS radio morphologies of the selected sample are variegated
  (see Figure~\ref{radio1} and \ref{hard} and Table~\ref{1table}). 21 FR~IIs with their
  classical double-lobed structures clearly stand out. One source, namely 25,
  shows an intermediate FR~I/FR~II radio morphology, since its two-sided jets show
  surface brightness peaks approximately at half of its radio extended
  size. Conversely, no 'bona-fide' FR~I structures are present. Half of the
  sample appears as compact sources (unresolved, or slightly resolved at the
  resolution of the VLA-COSMOS survey, 1\farcs5). 6 objects show only slightly
  elongated radio structures. One QSO is associated with a FR~II, while the
  other has a compact radio morphology. Unfortunately, we leave out 10 sources
  because we fail in the host identification. These sources show compact,
  complex and FR~II radio morphologies (see Figure~\ref{noident} and
  Table~\ref{2table}). 

A further tool of identification is given by the high-resolution HST/ACS
  images, which allow us to locate the host associated with the radio source
  and distinguish it from possible mis-identified companions. All the sources,
  but 8 objects, have HST/ACS images.  Further five objects are not bright
  enough to be detected in the HST/ACS maps. For these objects $i^+$-band Subaru
  images represent the best alternatives for their larger sensitivity despite
  lower resolution than HST/ACS. For one source (namely 33), HST/ACS and
  $i^+$-band Subaru images are not present, but $i^{*}$-band CFHT
  observation.

Let us focus on the identifications.  21 targets show a compact
  (unresolved, or slightly resolved at the resolution of the VLA-COSMOS
  survey, 1\farcs5), and another 6 sources show extended structures on a
scale of $\sim$3-4\arcsec. For such objects, we start the multiband
identification process looking for a {\it I}-band counterpart to the radio
source within a 0\farcs3\ radius from the radio position in the COSMOS
catalog. Then we check the co-spatiality of the position of the optical host,
the VLA radio source and the IRAC infrared emission. If this occurs with a
separation of less than 0\farcs5\ (the astrometry accuracy of IRAC maps), the
three counterparts are considered as associated with the same source. The
majority of the radio sources are optically identified at distances smaller
than 0\farcs1\ from the radio source.  We also use the Spitzer data to search
for the infrared counterparts.  In such case we use a larger search radius
(2\arcsec) due to their coarser resolutions. For 2 compact radio sources the
host identification failed because of the presence of multiple optical sources
falling within the radio structure. In Fig.~\ref{radio1} we show the
successful identification, while the two 'failed' objects are shown in
Fig.~\ref{noident}.

Considering the sources which show extended radio emission, the sample
selected in this work also includes sources with a FR~II morphology, contrary
to the C09 sample. The optical identification for those extended radio objects
is clearly more difficult than for the compact ones.  In order to strengthen
the host identification, we visually check the counterparts in each image
available from the UV to the IR band.

In 11 cases, the radio source has a triple structure with a central unresolved
radio component (most likely the radio core) associated with an optical/IR
counterpart (see Fig.~\ref{radio1}). In these cases we proceed as above, using
the location of the central component as reference for the optical/IR
counterpart (Fig.~\ref{hard}).

Conversely, in the remaining 18 radio sources the association is less
obvious. This is the case of 2 objects with a complex radio morphology (object 49 and
50), of 14 double radio sources (without radio core), or of 2 sources with a
triple morphology, but where the central component is extended. For the 2
triple sources (objects 7 and 17), the central radio component indicates the
approximative location of the radio core and it is probably elongated due to
the contribution of a radio jet. We look for an optical/IR host in this area
and indeed a single bright galaxy is found in both cases, which we identify as
the counterpart.

For the remaining 14 FR~IIs we look for the host galaxy along a straight line
connecting the hot spots in the VLA-COSMOS images. In 8 cases we identify as
host galaxy the brightest optical/IR counterpart which also corresponds to the
closest to the center of the radio structure as shown in Fig.~\ref{hard}.
Conversely, in objects 47, 48, and 52 there is no visible optical/IR source
between the lobes; finally, in objects 54, 55, and 56, the association is not
univocal since several optical/IR sources are present. These 6 objects are
shown in Fig.~\ref{noident} together with the 2 complex radio sources, which
we exclude from the sample.

Clearly, while the identifications of the hosts of 10 (2 triples and 8
doubles) FR~II are plausible, having excluded the less convincing ones, we
cannot exclude that there might still be spurious associations. This is most
likely the case of radio sources of large angular size and where the host is
far from the center of the radio structure, such as object 46. However, in
most cases the proximity of the radio lobes reduces significantly such a risk
since the host search area is limited to only a few squared arcseconds.

Summarizing, the identification is successful for 46 objects. All but 6 are
present in the COSMOS broad band Photometry catalog. Furthermore, each radio
source has a Spitzer infrared counterpart. For the identified objects, we take
from these catalogs the 3\arcsec-aperture photometric magnitudes from $\sim$
0.15 $\mu$m to 24 $\mu$m. The careful visual inspection of the images of the
identified sources enables us to recognize the presence of three sources
associated with a stellar-like optical counterpart (see Sect.~\ref{host} for
details). This might indicate a compact nucleus out-shining the host galaxy,
sign of the presence a quasar in the center. In fact two of them are
spectroscopically confirmed QSOs (namely, 29 and 37). Unfortunately, the third
object (namely, 35) does not have any available spectroscopic information. It
is associated with a compact radio source. All the sources are detected in
optical band, but three objects (namely, 2, 30, and 34) which are only
detected at longer wavelengths.

\begin{table*}
  \begin{center}
  \caption{Radio galaxies with no clear host identification}
  \label{2table}
  \begin{tabular}{l|lcccccc}
    \hline \hline
n     & radio ID & RA  & DEC  &  N$_{c}$ & F$_{FIRST}$  &    F$_{NVSS}$ & radio morph \\
\hline
  47  & J100102.38+020529.1$^{(3)}$     & 10 01 02.402 & +02 05 26.77  & 1 &  2.68  &       2.3  &  complex \\
 48 & J095949.80+015650.7$^{(2)}$        & 09 59 49.787 & +01 56 49.97  & 1 &  1.77  & $<$2.5 & FR~II  \\
 49 & J100049.58+014923.7$^{(4)}$ & 10 00 48.705 & +01 49 22.29  & 2 &  5.48   &    12.4  & composite \\
 50 & J100129.35+014027.1$^{(2)}$        & 10 01 29.328 & +01 40 27.01  & 1 &1.69   &      7.2 & complex \\
 51 & J095856.19+024127.9 	            & 09 58 56.220 & +02 41 27.39  & 1 &  3.67   &      4.2 & compact \\
 52 & J095901.52+024740.6$^{(4)}$ & 09 59 01.632 & +02 47 39.82  & 2 &  4.69   &      6.7 & FR~II \\
 53 & J100320.60+021608.9$^{*}$               & 10 03 20.613 & +02 16 08.99  & 1 &  3.38   &      3.9 & compact \\  
 54 & J100245.39+024519.8$^{(2)}$        & 10 02 44.947 & +02 45 00.99  & 2 & 20.00  &     28.2 & FR~II \\
 55 & J095822.93+022619.8$^{(6)}$ & 09 58 24.989 & +02 26 49.36 & 2 &  84.73  & 103.2 & FR~II\\ 
 56 & J100243.20+015942.1$^{(3)}$          & 10 02 42.622 & +01 59 37.74 & 2 &  53.52  & 58.7   & FR~II \\
\hline
\end{tabular}
\end{center}
Column description: (1) identification number increasing with
  the distance from the center of the COSMOS field; (2) COSMOS-VLA (Large
  Project) ID number of the object. In case of multiple objects, the number of
  components is shown on the superscript. The obejcts marked with $*$ are identified in the COSMOS-VLA Deep
  Project; (3)-(4) right ascension and
  declination of (one of the components of) FIRST radio source; (5) number of matches found in the FIRST
  catalog associated with the same radio galaxy; (6) (total) FIRST radio flux
  (mJy) of the entire radio source; (7) NVSS radio flux (mJy); (8) COSMOS-VLA radio morphology.
\end{table*}

The COSMOS catalogs are affected by the limitations typical of
multiband surveys, such as misidentification of targets with close neighbor or
the contamination by nearby bright sources. We then check the multi-band
counterpart identification of each source by visually inspecting its multiband
images, rather than blindly using the data provided by the COSMOS catalog.
More specifically, we identify all objects where i) nearby sources are present
within the 3\arcsec\ radius used for the aperture photometry (thus
contaminating the genuine emission from the radio galaxies) or when ii)
the counterparts to the i-band object does not correspond to the same object
over the various bands. In these circumstances, 1) in case of contamination
from a nearby source(s), we subtract from the flux resulting from the
photometry centered on the radio source the emission from the neighbor(s), 2)
we perform a new 3\arcsec-aperture photometry properly centered on the position of the
radio source. When needed, we applied the required aperture corrections (see
B13).  

When an object cannot be separate from a close companion or when the
counterpart is not visible in a given band, we measure a 2-$\sigma$ flux upper
limit.  Six sources are visible in the optical band (Subaru or CFHT) but they
do not reach the flux threshold of the COSMOS broad band Photometry catalog, or are
contaminated by a companion, or are wrongly identified by the COSMOS
catalog. In only three cases (namely, 2, 30, and 33) instead, no
i-band identification is possible, but the identification with the radio
source is nonetheless straightforward in the infrared images.

As already outlined in B13, in some cases, the GALEX and MIPS catalog
photometry returns apparently wrong identification. However, we visually check
all the counterparts to confirm the UV and IR identifications.  If a source is
not detected in GALEX we prefer not to include upper limits in our analysis,
because the corresponding NUV or FUV flux is substantially higher than the
detections at larger wavelengths and are not useful to constrain the SED. For
six sources, we measure the 24$\mu$m flux when not detected by the catalog.
Furthermore, we compute the photometry correction to the COSMOS UKIRT J-band
magnitudes as observed by B13.

The corrected 3\arcsec-aperture photometric measurements of all 46 sources are
tabulated in Tables~A.1 and A.2 and includes all the
multi-band magnitudes associated with the optical/IR counterparts of the radio
galaxies.

\section{SED fitting}
\label{fittingsed}

The SEDs are derived by collecting multiband data from the FUV to the MIR
wavelengths.  Since not all of the objects are detected in the entire set of
available bands, the number of datapoints used to constrain the SED fitting
ranges from 19 (object 29) to 2 (object 33). However, the upper limits to the
magnitudes, especially in blue and IR band, can, at least, roughly constrain
the contribution of the YSP and dust component.

As discussed in B13, the residuals of the SED fitting becomes smaller when a
second stellar component and dust emission are included. Taking into account
the larger number of parameters in the fit than in the case of a single
stellar population (such as {\it Hyperz}, \citealt{bolzonella00}), the fitting
improvement is basically due to the fact that two different stellar
populations, typically one younger and one older (YSP and OSP, respectively)
can reproduce better the complex morphology of the SED, which might represent
the complex star formation history of the galaxy.  Furthermore, the dust
emission is necessary to account for the MIR component, not compatible with
stellar emission.  The simultaneous inclusion of such components is allowed by
our developed code {\it 2SPD} (see B13 for details on the code) which we
prefer to use in this work rather than {\it Hyperz}. Since we use stellar
templates, we left out the two spectroscopically confirmed QSOs from this
procedure.

The stellar synthetic models used are from \citet{bc09} (priv. comm.) and
\citet{ma05}, the two sets differing for their Initial Mass Function (IMF)
\citep{salpeter55,kroupa01,chabrier03}. We considered models of solar
metallicity, single stellar population with ages ranging from 1 Myr to 12.5
Gyr. We adopt a dust-screen model for the extinction normalized with the free
parameter $A_V$, and the \citet{calzetti00} law. On the other hand, we model
the dust component with a single (or, in some cases, two) temperature
black-body emission.

The code {\it 2SPD} searches for the best match between the sum of the
different components and the photometric points minimizing the appropriate
$\chi^{2}$ function. {\it 2SPD} returns the following free parameters: $z$,
$A_V$, the age of the two stellar populations, the temperature of the dust
component(s), and the normalization factors. For the 9 objects whose
spectroscopic redshifts are available, we prefer to keep fixed their redshifts
as the observed values, since in the case of free parameter, as performed in
B13, the photo-z obtained are always consistent with the spectroscopic
redshifts.  From the fit we measure the stellar mass content of the two
stellar populations at 4800 \AA\ rest frame. However, caution should be
exerted before associating these values to physical quantities because of
degeneracy in the parameter space, apart from the photometric redshifts
(Table~\ref{2spd}).  Furthermore, since the infrared excess is not often
evident, the dust emission is usually poorly constrained. Thence we prefer not
to give any particular physical meaning to each value of the dust parameters
(temperature and luminosity). We will return to the dust properties in more
detail in Sect.~\ref{dust}.

To estimate the errors on the photo-z and mass derivations, we measure the
99\%- confident solutions for these quantities. This is computed by varying
the value of the parameter of interest (photo-z or mass), until the $\chi^{2}$
value increases by $\Delta \chi^2$ = 6.63, corresponding to a confidence level
of 99\% for that parameter.

Figure~\ref{sed1} shows the plots of the SED fitting, while Table~\ref{2spd}
presents the resulting parameters of the fit.

\begin{table*}
\addtolength{\tabcolsep}{-2pt}
  \begin{center}
  \caption{{\it 2SPD} SED fitting}
  \label{2spd}
  \begin{tabular}{c|c|cccc|cc|c|cc|cc|c}
    \hline \hline
ID  & redshift        &\multicolumn{4}{c|}{YSP} & \multicolumn{2}{c|}{OSP}& log M$_{*}$ &   \multicolumn{2}{c}{Dust} &   \multicolumn{2}{|c|}{IR excess}   &  UV \\
\hline
    & z$_{phot}$          &  Age &  A$_V$  & $f_{YSP}$ & Log M$_{*}$ &  Age  &  A$_V$    &            &  T$_{dust}$ & L$_{dust}$                   &   L$_{IR \,\,exc.}$   &  $\alpha_{8-24}$   &  L$_{UV}$\\        
\hline
1  & 1.1577$^{s}$                 &    0.004   &  0.73   &  21.9\%   & 0.14\%  &  2.0  & 0.72    &  11.31$^{+0.04}_{-0.04}$      &  118-528    & 224.5-213.8  &  45.74   &   -0.87  & 43.51\\ 
2  & 1.767$^{+0.570}_{-0.236}$  &    0.006   &  1.24   &  23.8\%   & 0.14\%  &  3.0  & 1.21    &  10.67$^{+0.05}_{-0.05}$      & 181        &   17.0  &  $<$43.96    &     & $<$42.00  \\ 
3  &  1.983$^{+0.34}_{-0.37}$  &    0.001   &  1.07   &  21.8\%   & 0.51\%  &  2.0  & 0.81    &  10.79$^{+0.06}_{-0.05}$      & 166        & 29.4   &  44.15    &  $>$-0.42  & 42.98 \\ 
4  &  1.104$^{+0.76}_{ -0.28}$  &    0.003   &  1.24   &  6.6\%   & 0.04\%  &  3.0  & 0.94    &  10.32$^{+0.16}_{-0.18}$      & 179-468   &  24.8-2.5   &  44.41     &   0.62  &  $<$41.43 \\ 
5  &  0.9707$^{s}$                  &    0.002   &  0.76   &  27.3\%   & 0.49\%  &  4.0  & 0.10    &  11.17$^{+0.09}_{-0.10}$      & 134-911   & 3.8-6.8   &   44.22   &   -0.53   &  43.80\\ 
6  &  1.963$^{+0.24}_{-0.28}$  &    0.003   &  0.49   &  32.1\%   & 0.30\%  &  3.0  & 0.21    &  10.85$^{+0.05}_{-0.05}$      & 189    &   11.0   &   43.79   &  $>$-0.38   & 43.47 \\ 
7  & 1.384$^{+0.16}_{-0.06}$  &    0.004   &  1.31  &  28.1\%   & 0.09\%  &  4.0  & 1.30    &   10.89$^{+0.06}_{-0.06}$      & 123      &  4.8   &  $<$43.49    &  & 42.19\\ 
8  & 0.8357$^{s}$                   &    0.002   &  2.80  &  10.3\%   & 1.7\%    &  4.0  & 0.13   &   11.59$^{+0.07}_{-0.07}$      & 151-769        &   2.8-4.2 &  $<$44.10    &  & $<$42.20 \\ 
9  &  2.152$^{+0.17}_{-0.44}$  &    0.003   &  0.98  &  10.4\%   & 0.22\%  &  1.0  & 0.62   &   10.60$^{+0.05}_{-0.06}$      & 200        &  26.2  &   $<$44.15   &     & 42.83 \\ 
{\bf 10}  &  1.268$^{+0.30}_{-0.33}$ &   0.002 &  1.5  &  34.9\%  & 1.88\% &   1.0 &  1.19   &    9.42$^{+0.08}_{-0.08}$      & 221-720  &   3.8-1.8  &   44.11  & -0.24  &  41.65 \\ 
11  &  1.018$^{+0.15}_{-0.08}$  &    0.004   &  1.38  &  9.8\%   & 0.18\%  &  0.8  & 1.21   &   10.69$^{+0.06}_{-0.06}$      & 122       &  4.2  &    $<$43.44   &     & 42.23$^{m}$\\ 
12  &   1.860$^{+0.73}_{-0.56}$  &    0.009   &  1.47  &  48.8\%   & 2.09\%  &  2.0  & 1.19   &  10.49$^{+0.10}_{-0.09}$      & 119       &   10.2 &   $<$43.66   &     & 42.38 \\ 
13  &   1.018$^{+0.02}_{-0.05}$  &    0.004   &  1.61  &  13.0\%   & 0.44\%  &  0.7  & 0.99   &  10.16$^{+0.04}_{-0.04}$      & 128       &  2.3  &    $<$43.16   &     &   41.88$^{m}$ \\ 
14  &    1.377$^{+0.27}_{-0.103}$  &    0.001   &  1.09  &  3.3\%   & 0.09\%  &  2.0  & 0.54   & 10.92$^{+0.07}_{-0.08}$      & 131       &   10.3  &   43.87   &    $>$-0.12 &   42.34 \\ 
15  &   1.015$^{+0.03}_{-0.02}$  &    0.004   &  1.45  &  0.6\%   & 0.02\%  &  1.0  & 0.58   &  10.76$^{+0.03}_{-0.03}$      & 139       &  2.1  &   $<$43.15    &     &  $<$42.06 \\ 
16  &   0.8442$^{s}$                   &    0.003   &  1.10  &  1.6\%   & 0.03\%  &  1.0  & 0.71   &  10.78$^{+0.06}_{-0.06}$      & 124       & 3.1   &   43.26   &   $>$-0.59  &  $<$42.22 \\ 
17  &   1.391$^{+0.22}_{-0.11}$  &    0.002   &  0.84  &  7.4\%   & 0.14\%  &  2.0  & 0.46   &   10.89$^{+0.04}_{-0.04}$      & 185       & 9.3  &    $<$43.83   &     & 42.87 \\ 
18  &    1.193$^{+0.123}_{-0.16}$  &    0.008   &   1.36  &  59.0\%   & 3.15\%  &  2.0  & 0.64   &  10.94$^{+0.03}_{-0.04}$      & 198-643  &  26.6-3.3  &   44.71     &   0.43  & 43.30 \\ 
19  &    0.932$^{+0.03}_{-0.01}$  &    0.003   &   2.26  &  5.3\%   & 0.33\%  &  2.0  & 0.28   & 11.26$^{+0.03}_{-0.03}$      & 110       & 8.9   &   43.66   &    $>$0.91  &  $<$42.20 \\ 
20  &    1.016$^{+0.20}_{-0.09}$  &    0.001   &   0.88  &  4.6\%   & 0.16\%  &  0.9  & 0.86   &    10.33$^{+0.08}_{-0.08}$      & 210       &   4.5 &   $<$43.52    &     & 42.33$^{m}$   \\ 
21  &    0.894$^{+0.06}_{-0.062}$  &    0.004   &   0.91  &  3.8\%   & 0.04\%  &  2.0  & 0.27   &  10.71$^{+0.04}_{-0.04}$      &196       & 3.2   &  $<$43.42    &     &  42.20$^{m}$ \\ 
{\bf 22}  &    2.393$^{+1.3}_{-0.88}$  &    0.002   &   0.56  &  31.5\%   & 2.21\%  &  0.9  & 0.07  &    9.77$^{+0.15}_{-0.17}$      & 181-932       &  3.0-6.6  &   44.40   &   0.28   & 43.32 \\ 
23  &    0.8393$^{s}$                &    0.004   &   1.17  &  5.6\%   & 0.12\%  &  0.8  & 0.79  &   10.50$^{+0.02}_{-0.02}$      & 135       &  2.6  &   $<$43.30   &     & 42.25$^{m}$ \\ 
24  &    2.006$^{+0.30}_{-0.27}$  &    0.002   &   0.84  &  7.1\%   & 0.10\%  &  2.0  & 0.71  &   11.08$^{+0.04}_{-0.04}$      & 178   &    98.1 &  44.69    &   $>$0.66  & 42.91 \\ 
25  &    0.825$^{+0.03}_{-0.03}$  &    0.003   &   3.02  &  25.8\%   & 3.04\%  &  2.0  & 0.60  &   11.17$^{+0.02}_{-0.02}$      & 134   &  3.1    &   43.27   &   $>$2.50  &  $<$41.98 \\ 
26  &    0.830$^{+0.11}_{-0.11}$  &              &            &               &              &  2.0  & 0.10  &    11.04$^{+0.06}_{-0.06}$      & 125   &  2.9  &  $<$43.32    &     &  $<$42.07 \\ 
27  &    2.523$^{+0.68}_{-0.32}$  &    0.004   &   0.57  &  10.7\%   & 0.17\%  &  1.0  & 0.30  &    10.68$^{+0.07}_{-0.07}$      & 131  & 33.8   &  $<$44.17    &     & 43.22 \\ 
28 &     1.788$^{+0.43}_{-0.33}$  &    0.005   &   0.69  &  34.7\%   & 0.67\%  &  2.0  & 0.08  &    10.51$^{+0.09}_{-0.08}$      & 212-505      &  5.2-1.9  &  44.12    &  -0.23    & 43.30 \\ 
29 &    1.4050$^{s}$                  &   \multicolumn{4}{|c|}{QSO}                    &         &             &                                            &                   &                 &               &               &            \\               
30 &     2.268$^{+0.56}_{-0.21}$  &    0.005   &   1.53  &  49.3\%   & 0.80\%  &  2.0  & 1.22  &     10.74$^{+0.05}_{-0.05}$      &161       &  25.7  &  $<$44.16    &     & 42.53$^{m}$ \\ 
31 &    1.698$^{+0.49}_{-0.19}$  &    0.001   &   1.08  &  15.1\%   & 0.25\%  &  2.0  & 1.04  &    10.65$^{+0.05}_{-0.05}$      & 227   &   16.9 &   $<$44.06   &     & 42.57  \\ 
32 &    1.067$^{+0.32}_{-0.10}$  &    0.003  &   2.08  &  64.8\%   & 3.27\%  &  1.0  & 1.75  &   10.08$^{+0.07}_{-0.07}$      &206       &   5.2  &   $<$43.59   &     & 42.05$^{m}$ \\ 
{\bf 33} &     1.421$^{+0.67}_{-0.61}$  &    0.003  &   1.32  &  4.9\%   & 0.03\%  &  3.0  & 1.01  &   10.92$^{+0.07}_{-0.07}$      &179       &  19.1  &  $<$44.02    &     & $<$41.79 \\ 
34 &      1.402$^{+0.31}_{-0.32}$  &    0.002  &   0.67  &  0.8\%   & 0.03\%  &  0.9  & 0.30  &   10.65$^{+0.23}_{-0.29}$      &  204-731   & 14.4-18.0   &   44.61   &   -1.08   &  $<$42.63 \\ 
35 &  1.028$^{+0.27}_{ -0.17}$        &  0.006 & 0.45   &  59.2\%   & 0.96\%  &  3.0  & 0.09    &  10.71$^{+0.06}_{-0.06}$      &  206-625   & 5.1-7.4   &  44.34    &  -0.69   & 43.89 \\ 
36  &   2.255$^{+0.70}_{-0.35}$       &  0.04 & 0.31   &  34.3\%   & 1.50\%  &  2.0  & 0.71    &  10.91$^{+0.07}_{-0.07}$        & 155   &  25.1  &    $<$44.09   &      & 43.37   \\ 
37  & 1.3176$^{s}$                    &  \multicolumn{4}{|c|}{QSO}                 &        &                    &                                           &           &              &                    &       &     \\
38 &   2.355$^{+0.45}_{-0.16}$   &  0.009 & 0.47   &  22.5\%   & 0.48\%  &  2.0  & 0.31    &  11.09$^{+0.04}_{-0.04}$        & 132  &   67.0 &   44.42   &   $>$0.58  & 43.70 \\   
39 & 1.851$^{+0.36}_{-0.30}$      &  0.006 & 1.24   &  11.0\%   & 0.16\%  &  3.0  & 0.34    &  10.94$^{+0.07}_{-0.06}$     & 155    & 17.7   &   $<$43.94   &     & 42.34$^{m}$  \\   
40 & 1.110$^{+0.24}_{ -0.20}$       &            &           &               &              &  0.7  & 0.73    &  10.70$^{+0.12}_{-0.11}$        & 155    &  4.6  &   $<$43.51    &     &  $<$42.24   \\   
41 & 1.028$^{+0.07}_{-0.18}$       &  0.001 &  0.14   &  3.0\%   & 0.16\%  &  0.07  & 1.40    &  10.22$^{+0.19}_{-0.07}$        & 211-646   & 26.4- 1.9  &   44.69   &   0.15  & 43.27 \\   
42 & 1.137$^{+0.44}_{-0.30}$       &  0.003 &  1.10   &  4.4\%   & 0.10\%  &  0.9  & 0.68    &  10.75$^{+0.16}_{-016}$        & 156       &  4.9   &  $<$43.66     &     & 42.58$^{m}$ \\   
43 &  1.247$^{+0.46}_{-0.44}$       &  0.005 &  1.00   &  2.9\%   & 0.06\%  &  0.9  & 0.64   &  11.03$^{+0.19}_{-0.32}$        & 96       &  9.9  &  $<$43.63    &     & 42.68$^{m}$  \\   
44 & 1.062$^{+0.14}_{-0.05}$       &  0.003 &  0.15   &  4.0\%   & 0.04\%  &  1.0& 0.48   &  10.61$^{+0.03}_{-0.04}$        & 189-644   & 8.2-1.6    &   44.37   &   0.53   & 43.12 \\   
45 & 0.918$^{+0.07}_{-0.07}$       &  0.003 &  1.04   &  2.0\%   & 0.04\%  &  0.8  & 0.94   &   10.74$^{+0.05}_{-0.05}$        & 101   & 4.9   &    $<$43.35   &     & 42.28$^{m}$  \\   
46 & 0.8784$^{s}$                     &  0.007 &  1.35   &  10.7\%   & 0.49\%  &  0.8 & 0.78   &   10.47$^{+0.09}_{-0.13}$        & 100   &  4.4  &   $<$43.29   &     & 42.27$^{m}$  \\   
\hline
%{\bf COSMOS-FR~I 7}  & 4.51$^{+1.26}_{-0.13}$  &    0.003   &  0.51   &  2.7\%   & 0.09\%  & 0.9   & 0.10    &  10.83$^{+0.11}_{-0.06}$      & 119        &   206.9  &   44.70   &   $>$0.31  &  $<$43.45 \\ 
\hline
\end{tabular}
\end{center}
Results from the analysis of the SEDs with {\it 2SPD}. Column
  description: (1) ID number of the object; (2) photometric redshift measured
  with {\it 2SPD}; (3)-(4)-(5)-(6) age in Gyr, A$_V$, flux fraction and mass
  fraction of the young stellar population (YSP) at 4800 \AA rest frame;
  (7)-(8) age in Gyr and A$_V$ of the old stellar population (OSP); (9) the
  total stellar mass of the galaxy in M$_{\odot}$; (10)-(11) the effective
  temperature (in K) of the one or two dust components and their luminosities,
  L$_{dust}$ (in units of 10$^{9}$ L$_{\odot}$); (12)-(13) the infrared excess
  luminosity (in erg s$^{-1}$) defined in the text (Section~\ref{dust}) and
  the spectral index measured on the infrared excess at 8 and 24$\mu$m; (14)
  UV luminosity at 2000 \AA\, in the rest frame in erg s$^{-1}$. The marginal
  UV excesses are marked with a $^m$. The objects with the ID in bold
  charachter (namely, 10, 22, and 33) are excluded from the final sample for
  their ambiguous SED properties.
\end{table*}

\section{RESULTS}
\label{results}

The SED modeling process has been performed for all the 46 objects (except for
the spectroscopically-confirmed QSO, namely 29 and 37), by using the
template-fitting techniques, {\it 2SPD}. We also model the SED of object 35,
although it is optical appearance and its SED shape suggests an identification
as a QSO.

The SEDs for approximately half of the sample show a '{\it bell}' shape. Such a
behavior is ascribed to the dominance of the OSP over the YSP and dust
component(s). However, the other half of the sample show excesses in UV and/or
MIR wavelengths.

Generally, the number of photometric detections is crucial for the SED fitting
reliability. The case of 33 is a clear example. The object is detected only in
3.6 and and 5.8 $\mu$m IRAC bands, because the source is on the border of the
COSMOS field.  Furthermore, since the OSP is the component which fits the SED
on a larger range of wavelengths than YSP and dust, the dominance of the OSP
over the other two components determines the quality/reliability of the
modeling. In fact another aspect crucial for the fitting is the intrinsic
prominence of some spectral properties of the OSP: the 4000 \AA\ break and the
blue or infrared part of the spectrum. If these features are not evident in
the SED because the YSP and/or dust component dominate over the OSP, the
resulting fit is not unique and thus is not reliable.  This is the case of
the objects 10, 22 and 35. Therefore, for these four radio sources (including
object 33), we do not consider reliable the SED model we obtain with the
{\it 2SPD} technique.

\subsection{Photometric redshifts}
\label{photoz}

The photometric redshifts obtained with {\it 2SPD} (Table~\ref{2spd}) for the
46 objects selected in this work range between $\sim$0.8 and 2.4. 13 out of 46
sources are not present in the COSMOS photo-z catalog mainly because their
I-band magnitudes are beyond the $I=25$ limit of the COSMOS Photometric
Redshift Catalog and, marginally because of the mis-identification of the
photometric counterparts (see B13 for details).  For such objects we do not
have another photo-z measurement apart from our derivation.

\begin{figure}
\centerline{
\includegraphics[scale=0.40]{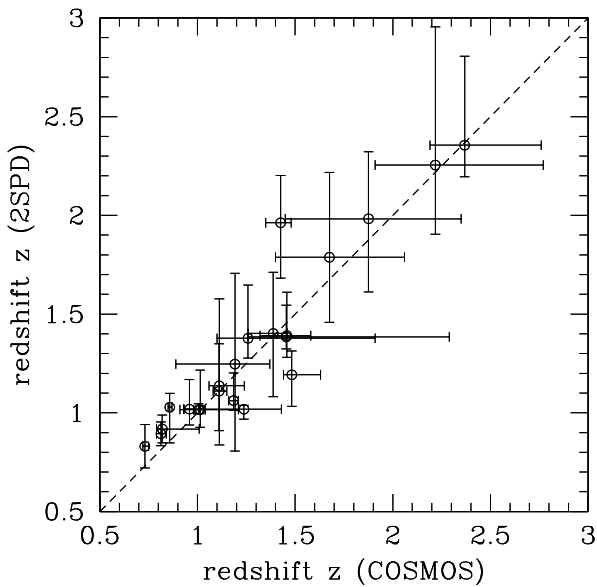}}
\caption{Comparison of the photometric-z measured with {\it 2SPD} with those
  obtained by the COSMOS collaboration \citep{ilbert09,salvato09}. The dashed
  line is the bisector of the plane.}
\label{myzilb}
\end{figure}

  In order to test the reliability of our photo-z derivation, we first compare
  the photometric redshifts measured with {\it 2SPD} with those obtained with
  the template-fitting technique performed by the COSMOS collaboration
  \citep{ilbert09,salvato09} (Fig.~\ref{myzilb}). Obviously, we do not
  consider for the comparison the objects 10, 22, 33, and 35 (for the reasons
  explained in Sect.~\ref{results}), the sources not included in the COSMOS
  photo-z catalog, and those whose spectroscopic redshifts are available.
  Generally, our photometric redshifts are consistent with the COSMOS photo-z
  within the errors. On average, our photo-z uncertainties are slightly
  smaller than those provided by the COSMOS collaboration. This indicates that
  the redshift does not depend much on the single changes in the
  datapoints. However, our SEDs are more realistic because of our careful
  visual inspection of the multi-band counterpart identifications. The
  normalized redshift differences ($\Delta z/(1+z)$, between our values and the
  COSMOS photo-z) are smaller than 0.08, similar to what B13 found, for all
  but 3 objects that reach $\Delta z/ z \sim 0.22-0.27$ (objects 6, 13, and
  18).

Since the sample includes spectroscopically-confirmed QSOs and other potential
QSOs whose SEDs appear power-law dominated, we use the method to derive the
photometric redshifts, introduced by \citet{richards01b,richards01a} for
quasars. They construct an empirical color-redshift relation based on the
median colors of quasars from the SDSS survey as a function of
redshift. Photometric redshifts are then determined by minimizing the
$\chi^{2}$ between all four observed colors and the median colors (obtained by
combining the five SDSS magnitudes, $u'$, $g'$, $r'$, $i'$, and $z'$) as a
function of redshift.

We apply this method to the sources which show a QSO-like SED and detected in
at least 2 SDSS bands, i.e. objects 22 and 35 (object 10 is excluded for this
reason). As a sanity check, we include in this analysis the two
spectroscopically-confirmed QSO, namely 29 and 37.  We also decided to
re-analyze objects with similar spectral shapes found by B13, namely
COSMOS-FR~I 32, 37, 226, because they are potentially QSO for their power-law
spectral behavior, and the spectroscopically-confirmed QSO, COSMOS-FR~I 236.

\begin{figure*}
\begin{center}$
\begin{array}{cccc}
\includegraphics[scale=0.25,angle=90]{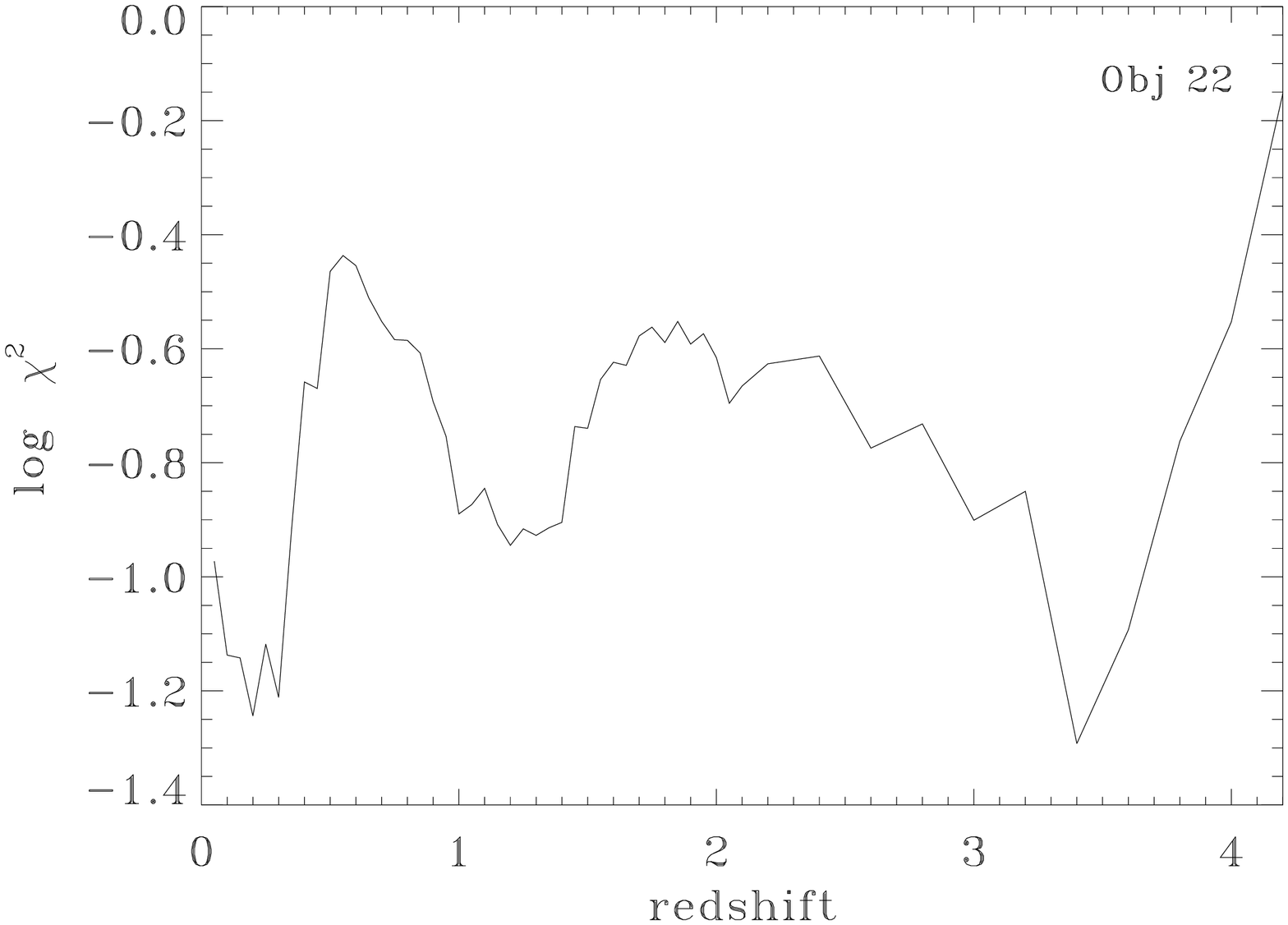} &
\includegraphics[scale=0.25,angle=90]{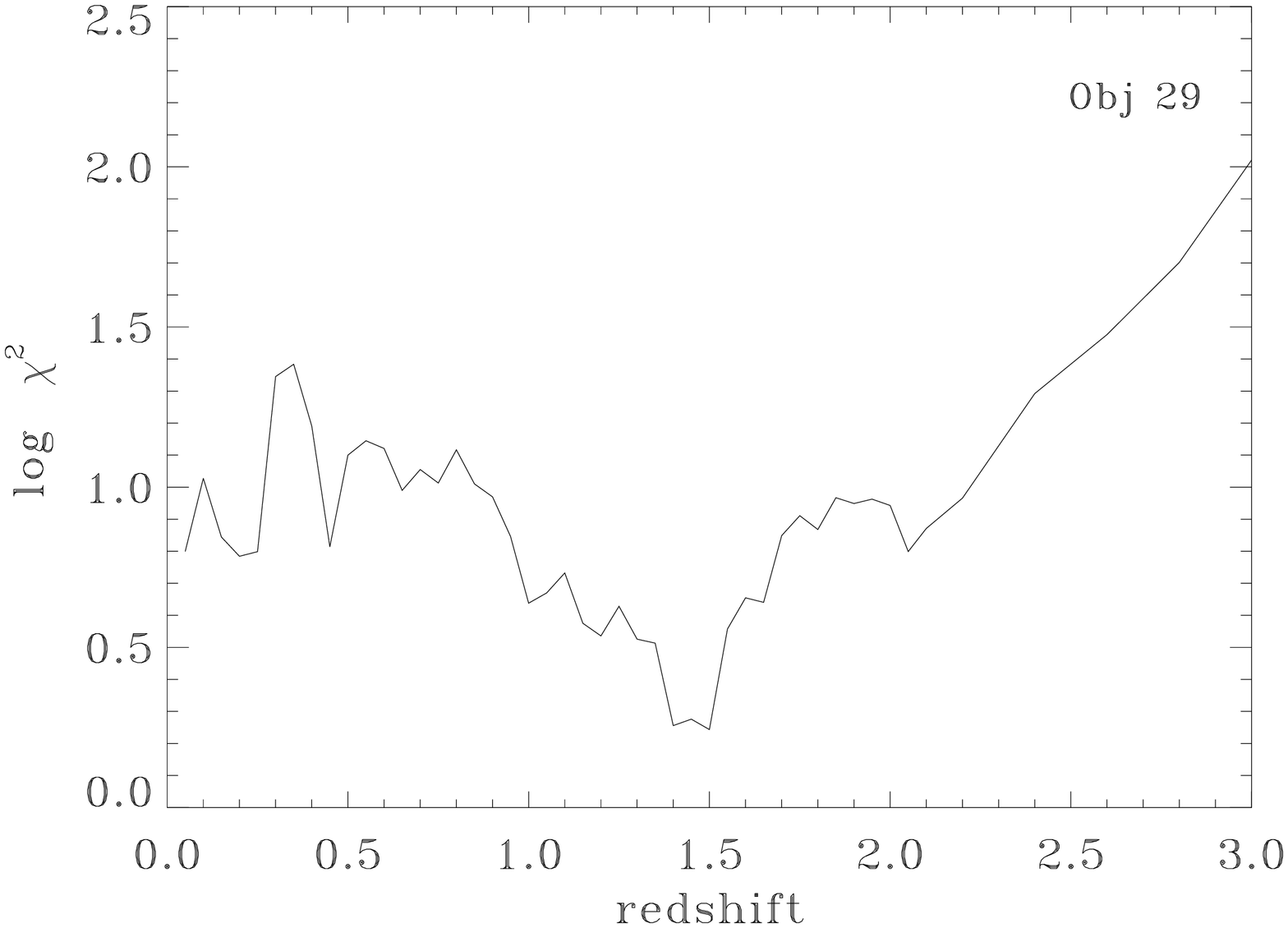}  &
\includegraphics[scale=0.25,angle=90]{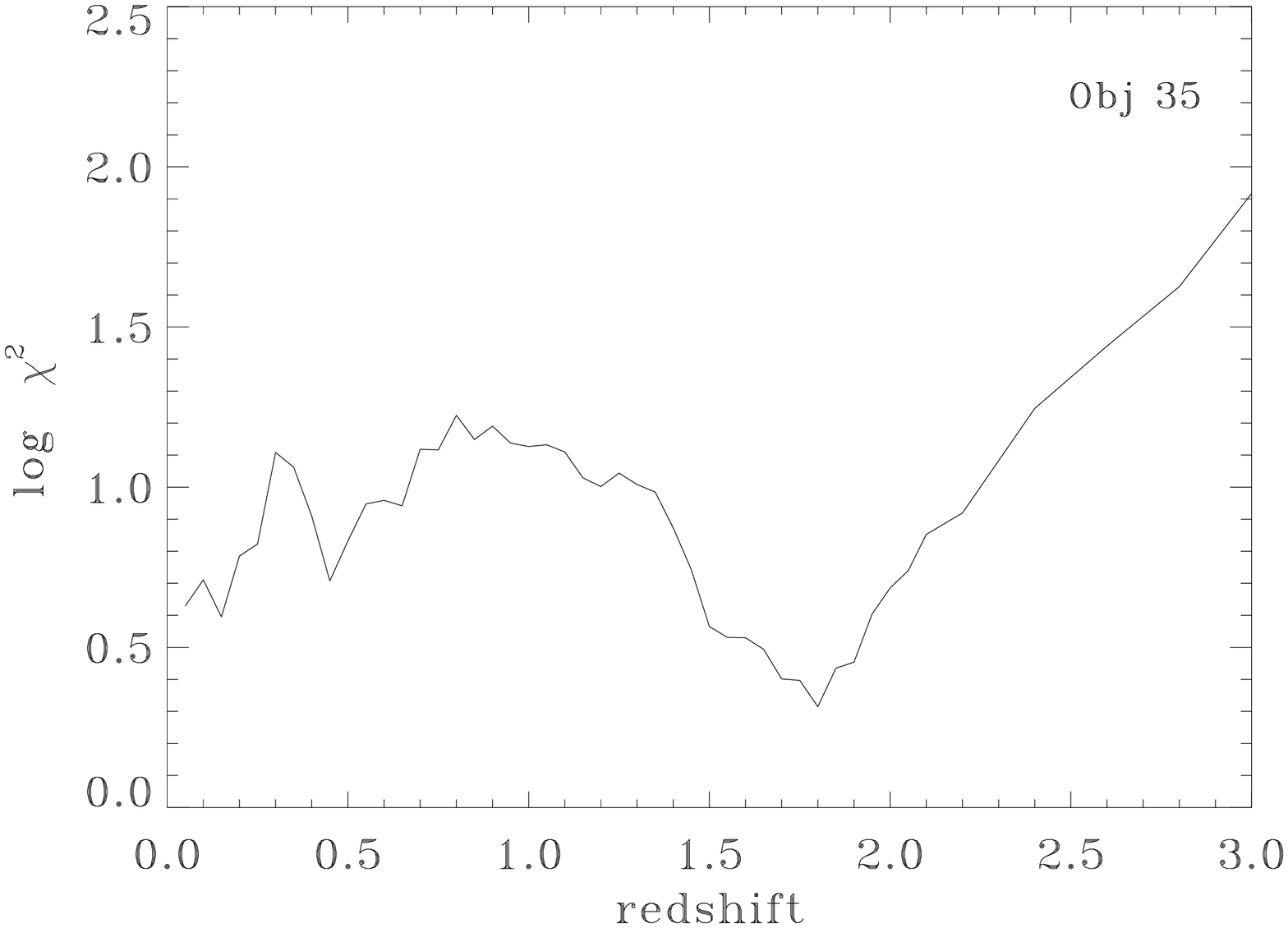} \\
\vspace{1em}
\includegraphics[scale=0.25,angle=90]{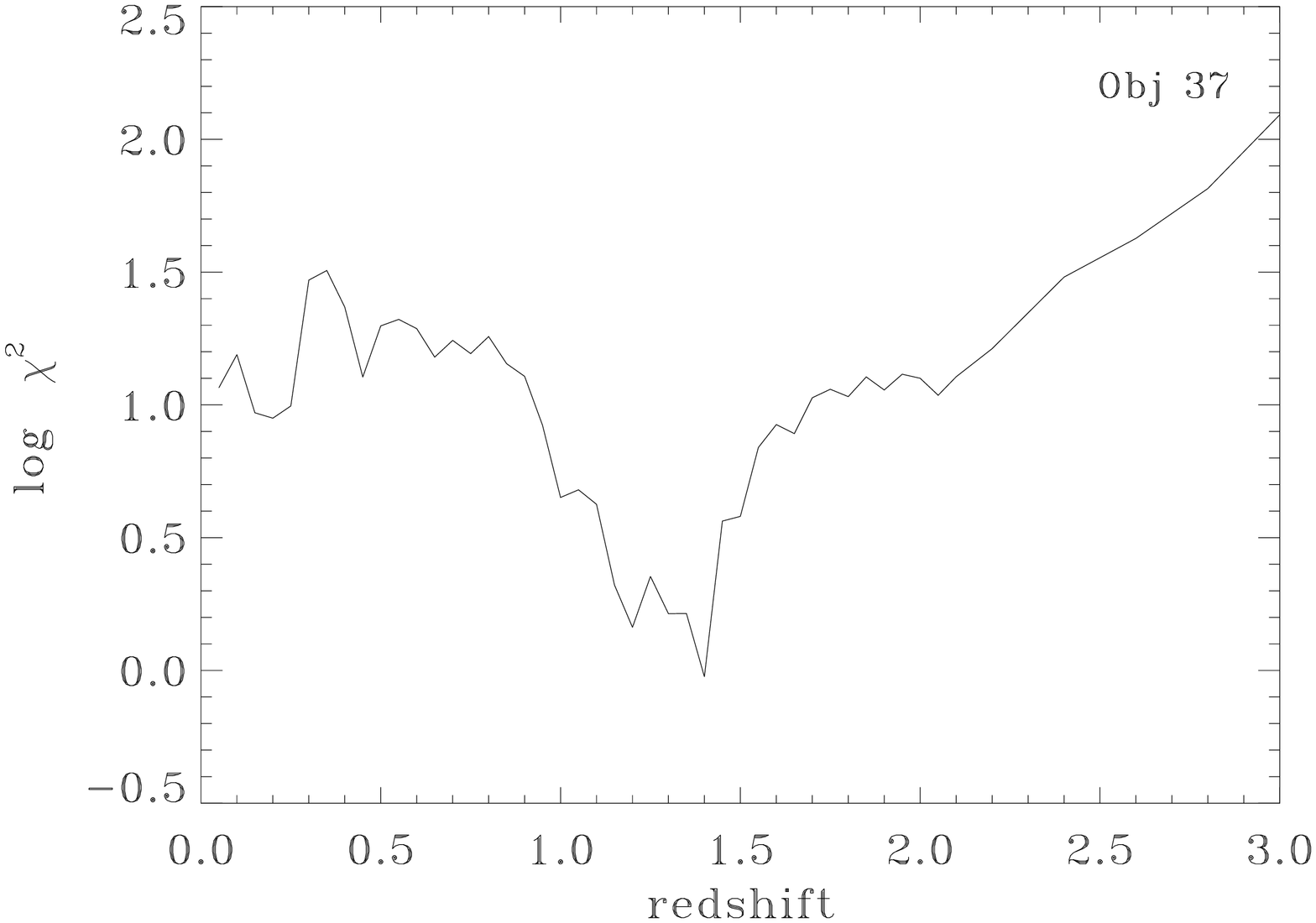}  &
\includegraphics[scale=0.25,angle=90]{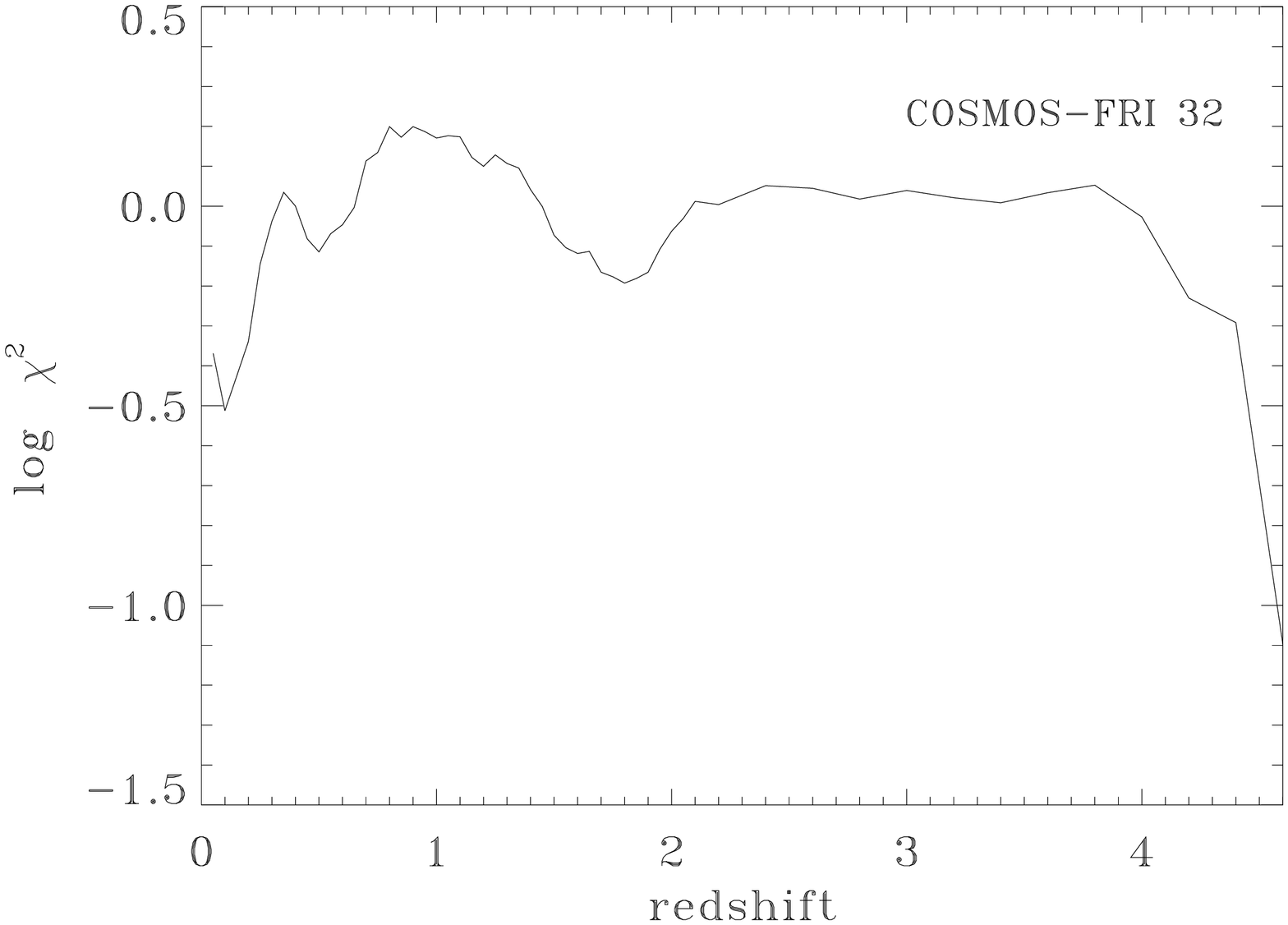} &
\includegraphics[scale=0.25,angle=90]{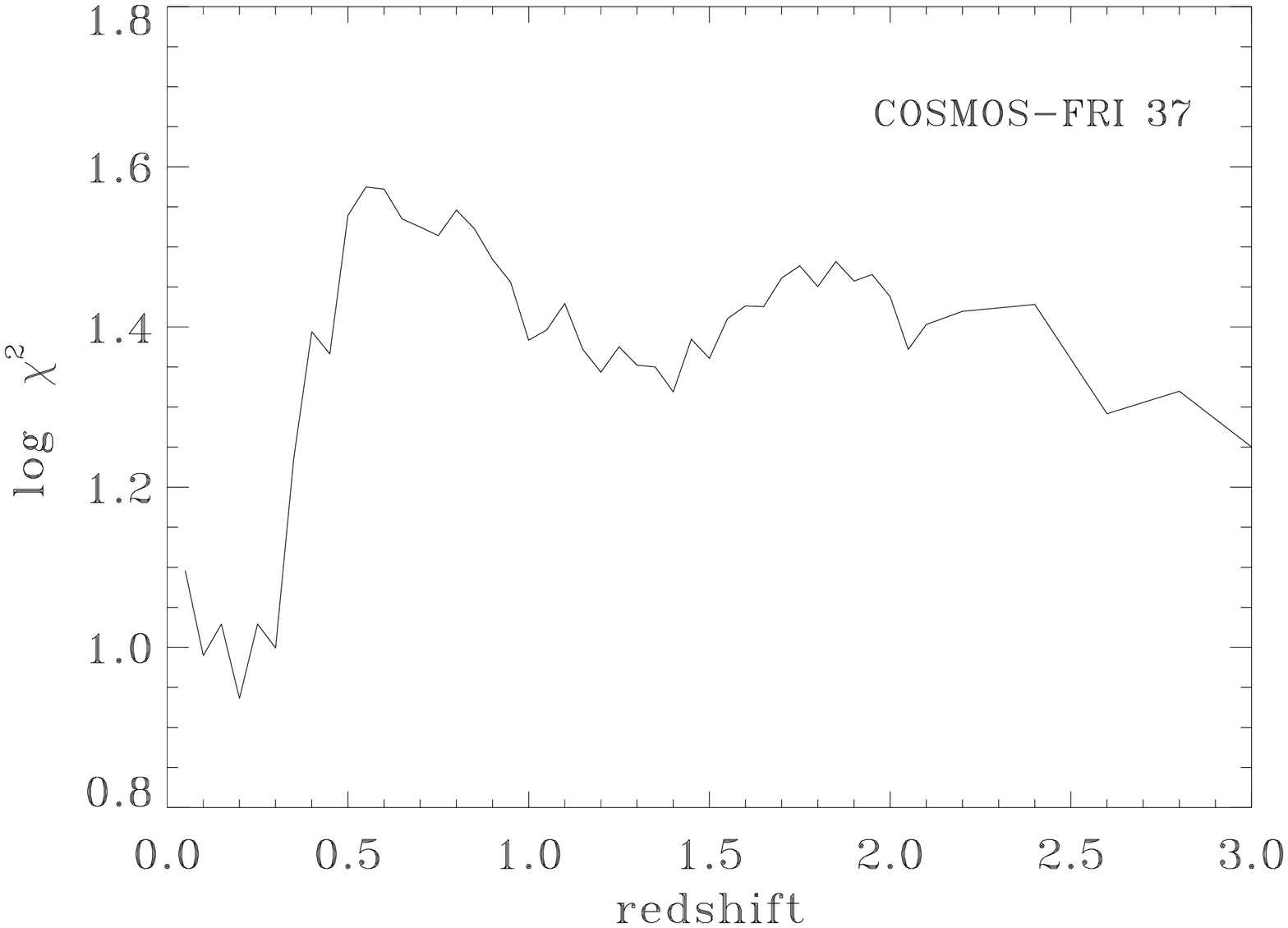}  \\
\vspace{1em}
\includegraphics[scale=0.25,angle=90]{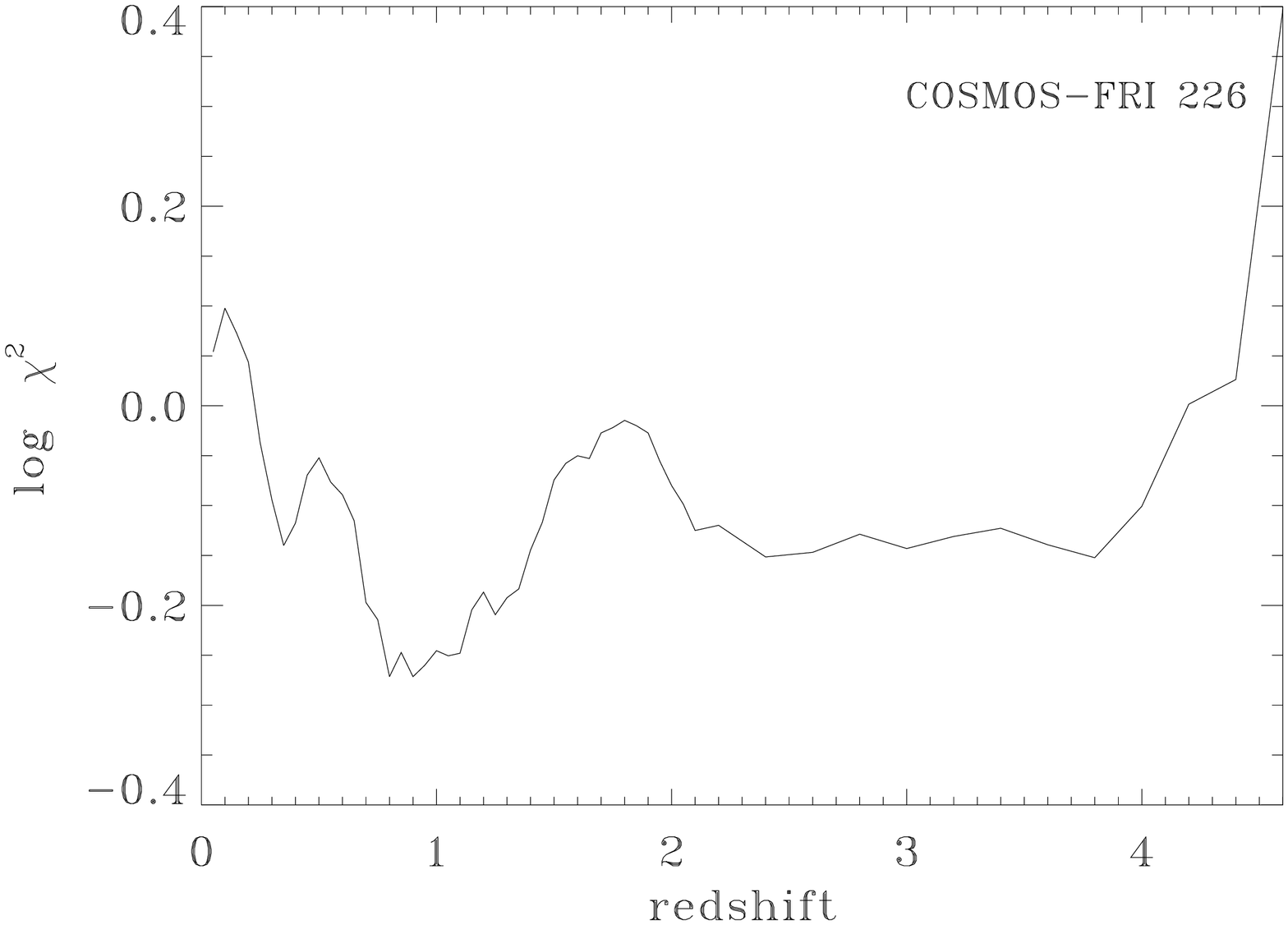} &
\includegraphics[scale=0.25,angle=90]{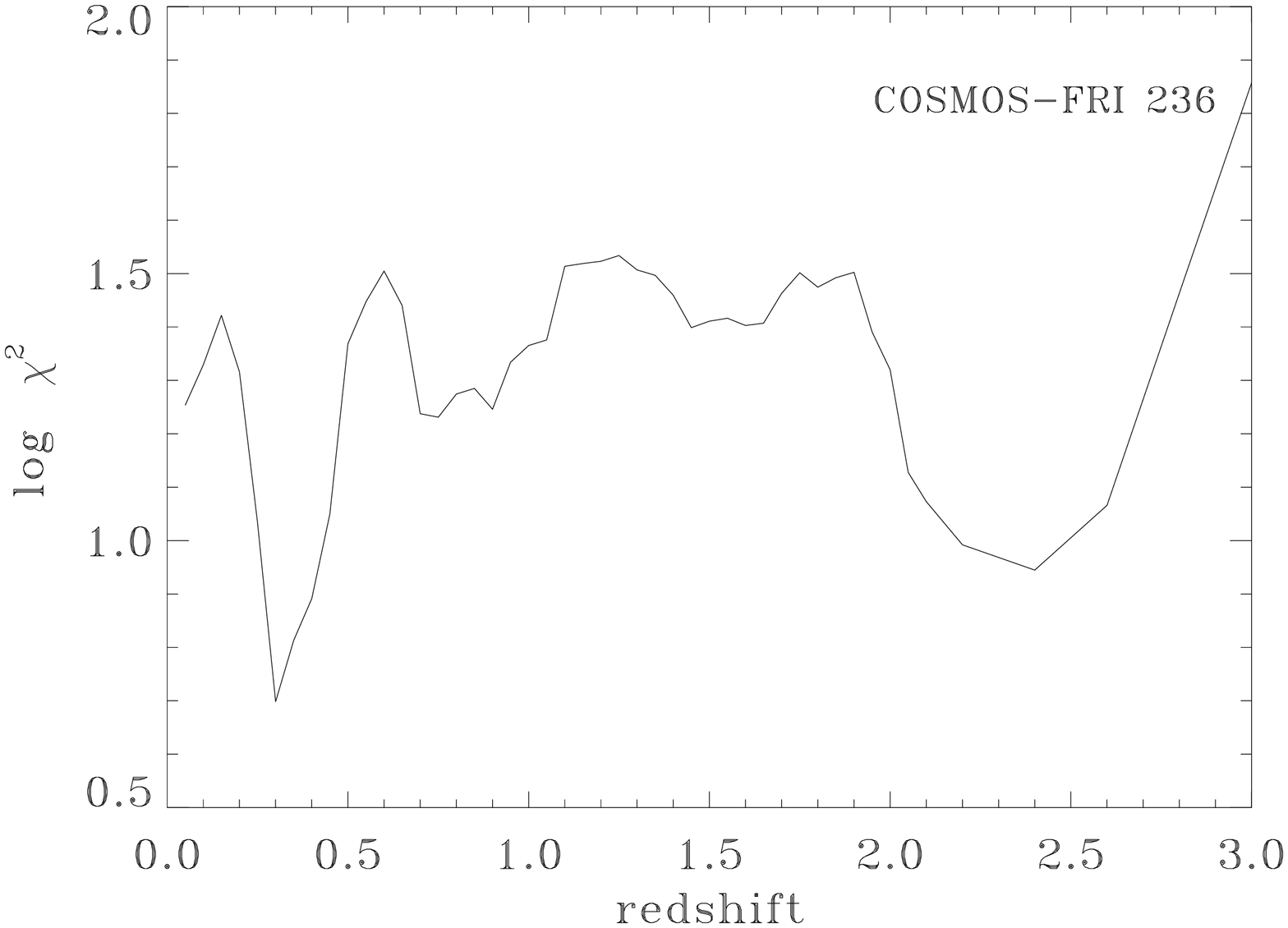}  
\end{array}$
\end{center}
\caption{Distribution of the $\chi^{2}$ function (see text) at varying
  redshift for objects 22, 29, 35, and 37 and for COSMOS-FR~I 32,
  37, 226, and 236 obtained from the color-redshift relation using SDSS colors
  \citep{richards01a}.}
\label{zphotsdss}
\end{figure*}

Table~\ref{sdsscolor} reports the SDSS colors for these 8 objects and
Figure~\ref{zphotsdss} shows their $\chi^{2}$ curves. The $\chi^{2}$
  minima smaller than the unity are not considered statistically significant.
For the three spectroscopic QSO (29, 37 and COSMOS-FR~I 236), the $\chi^{2}$
minima is consistent with the spectroscopic redshifts. For object 35 the
$\chi^{2}$ minimum indicates a redshift of 1.8, different from that we derive
from the SED modeling (1.03). We then finally change the classification of this
source to QSO because of its SED and optical appearance and use as redshift
$z=1.80^{+0.40}_{-0.40}$. Conversely, for the objects 22 and 33 and
COSMOS-FR~I 32, 37, and 226, the SDSS color fitting does not produce a
reliable evidence for a redshift within the range of our interest. Therefore,
we finally exclude these 6 objects (including object 10 not detected in SDSS
bands) from the entire sample since their photometric derivation is not
convincing.

\begin{figure}
\centerline{
\includegraphics[scale=0.40]{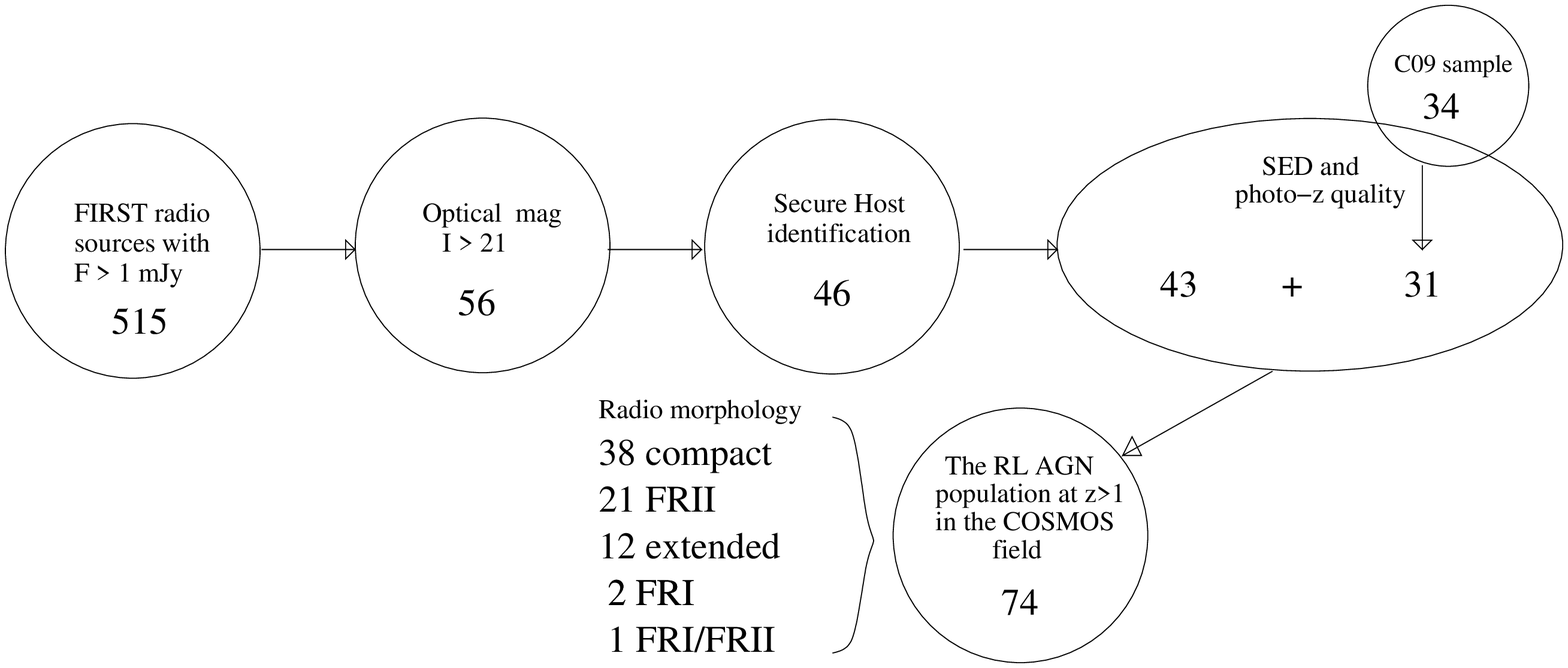}}
\caption{Flow-chart describing our selection procedure. The number of sources
  that survive each step is reported inside each circle. See text for more details.}
\label{diag}
\end{figure}

Summarizing, after the SED quality test the sample selected in this work is
reduced to 43 objects and we also exclude three sources from the sample studied by
B13. Figure~\ref{diag} summarizes our selection procedure, providing the
number of the sources selected in each step of the selection.

\begin{table*}
\begin{center}
\caption{COSMOS SDSS color}
\begin{tabular}{l|ccccc|c}
\hline
Object & $u'$    & $g'$ & $r'$ & $i'$ & $z'$  & z$_{zphot,SDSS}$  \\
\hline
%10       & $<$23.50              &    $<$23.90             &	$<$23.90               &   $<$23.00   &   $<$22.20 &  \\
22       & $<$23.50              &    25.41$\pm$1.02  &     24.70$\pm$0.82  &   $<$23.00    &    $<$22.20 &  \\
29       & 20.01$\pm$0.01  &    19.81$\pm$0.01  &     19.54$\pm$0.01   &   19.37$\pm$0.01  &   19.37$\pm$0.03 &   1.50$^{+0.30}_{-0.55}$ \\  
35       & 22.18$\pm$0.09 &      22.09$\pm$0.04 &    21.96$\pm$0.06    &   21.54$\pm$0.06  &   21.65$\pm$0.28 &   1.80$^{+0.40}_{-0.40}$\\  
37      & 19.31$\pm$0.01  &    19.21$\pm$0.01   &    18.88$\pm$0.01    &  18.75$\pm$0.01   &   18.80$\pm$0.01 &  1.40$^{+0.20}_{-0.40}$\\
\hline 
COSMOS-FR~I 32      & $<$23.50            &	$<$23.90               &  24.86$\pm$0.61      &    23.99$\pm$0.41 &    $<$22.20  &  \\
COSMOS-FR~I 37      & 23.70$\pm$0.48 &    22.78$\pm$0.08    &  22.06$\pm$0.06      &   21.68$\pm$0.08  &  21.22$\pm$0.26 &  \\     
COSMOS-FR~I 226    & 24.41$\pm$0.65 &    $<$23.90               &  23.84$\pm$0.28      &    24.90$\pm$1.31 &    $<$22.20 &  \\
COSMOS-FR~I 236    & 20.89$\pm$0.03 &   20.44$\pm$0.01     &  20.17$\pm$0.01      &  20.04$\pm$0.02   & 19.58$\pm$0.04 &  2.40$^{+0.30}_{-0.35}$ \\
\hline
\end{tabular}
\label{sdsscolor}
\label{log}
\end{center}
SDSS color, $u'$, $g'$,$r'$, $i'$, and $z'$, for the
  sources which show 'power-low' SEDs and the photometric redshifts,
  z$_{zphot,SDSS}$, derived using SDSS color (see Section~\ref{photoz}).
\end{table*}

\subsection{Host galaxy properties}
\label{host}

\begin{figure}
  \includegraphics[scale=0.45]{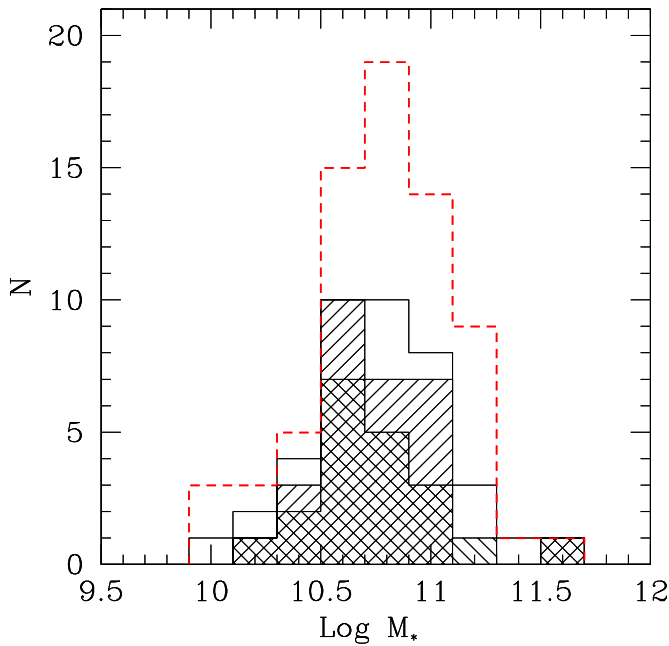}
  \caption{Distribution of stellar masses (in M$_{\odot}$) of our sample
    obtained with {\it 2SPD}. The solid line represents the sample of
      radio galaxies selected in this work. The forward-slash shaded
    histogram represents the HPs, while the back-slash shaded histogram the
    FR~IIs. The dashed histogram represents the entire COSMOS RL AGN sample at
    high redshifts.}
\label{Msisto}
\end{figure}

\begin{figure}
  \includegraphics[scale=0.40]{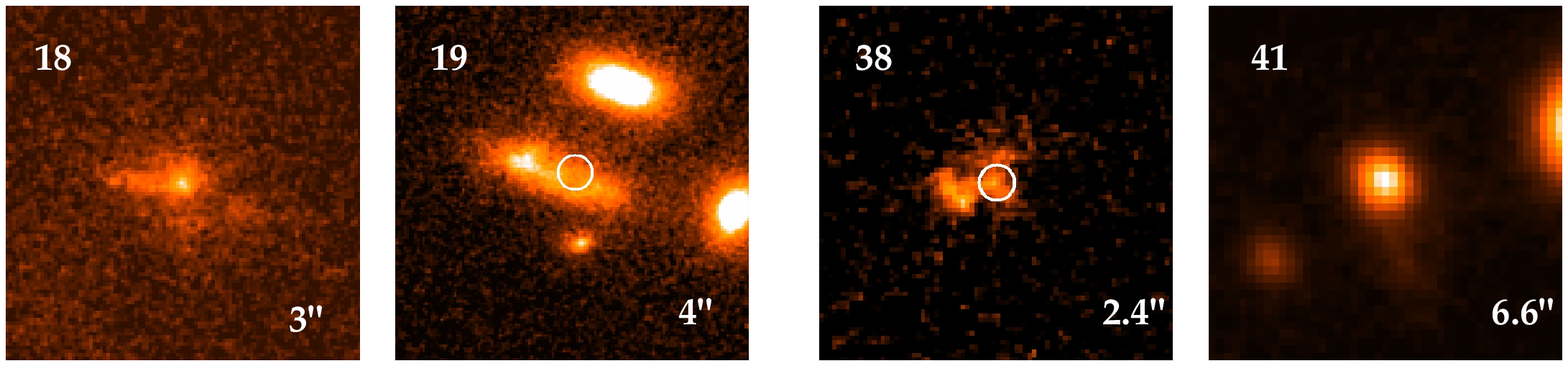}
  \caption{HST/ACS images of the four sources which show irregular optical
  morphologies. The image size is indicated in the panel. The white
  circle locates the position of the radio source.}
\label{irrhost}
\end{figure}

We now focus on the properties of the host galaxies inferred from the SED
modeling and in particular on their stellar populations for the 43 sources
selected in this work, similarly to what done by B13.

The stellar mass of the galaxy, M$_{*}$, is one of the most
robust result of the modeling.  However, as discussed in B13, the presence of
additional dust component to the OSP might affect the stellar mass estimate
more than the inclusion of a YSP. The inferred mass range is $\sim$10$^{10} -
10^{11.5}$ M$_{\odot}$ (Fig.~\ref{Msisto}), reported in Table~\ref{2spd}.

Although the other host parameters derived from the SED modeling are less
constrained than the estimate of the stellar content in the galaxy, we can
globally state that the hosts are dominated by an OSP with
an age of $\sim$ 1$-$3 $\times$ 10$^{9}$ years, similar to the results
obtained by B13. Assuming that the emission at short wavelengths is associated
with a YSP, the most significant UV component are reproduced by stellar
populations of a few Myr (Table~\ref{2spd}), with a contribution to the total
mass of the galaxy less than 1\%.

In order to qualitatively study the host type, the optical HST/ACS images
provides the highest resolution view of the galaxy, although the maps are
single orbit pointings.  For the remaining (Subaru, CHFT, and Spitzer),
optical and infrared images can only provide a tentative indication of the
host morphology for all the sources, apart from the four QSO which show a
point-like optical nucleus outshining the weaker galaxy. With a visual
inspection of the multi-band counterparts of the host, we can attempt to
recognize the presence of clear spiral/disk galaxies or galaxies which clearly
differ from smooth ellipticals as observed at z$\sim$1 (e.g.,
\citealt{huertas07}).  We do not see any evident late-type galaxy, but
generally we can tentatively classify them as bulge-dominated objects. Only
four sources (namely 18, 19, 38 and 41) show more irregular morphologies than
classical undisturbed ellipticals (Fig.~\ref{irrhost}), which might be
possible sign of an interaction with companion(s). In fact, the optical images
of several objects show rich environments in their surroundings, similarly to
what C09 found for the COSMOS-FR~I sources.

\subsection{Dust emission}
\label{dust}

Concerning the 43 objects selected in this work, dust emission is required to
adequately model the SEDs of 16 objects (not considering the three QSOs) due
to the detection of emission at 24 $\mu$m, and significant excesses above the
stellar emission are observed also at shorter infrared wavelengths in 8 of
these galaxies. However, as discussed in Sect.~\ref{fittingsed}, the results
of the SED fitting concerning the dust components should not be used to infer dust
properties. In order to explore the dust properties we estimated the residuals
between the best fitting stellar model and the data-points, looking for an
excess at the Spitzer wavelengths. We then integrated the residuals (by
assuming that the spectrum is represented by a multiple step function) to
obtain the infrared excess luminosity, $L_{\rm IR \,\ excess}$ in the range
covered by the Spitzer data, i.e. $\sim$ 3 - 26 \mum\ (see B13 for the
details). The estimated dust properties are reported in Table~\ref{2spd}.

The analysis misses the information on the dust content for the quasars,
because they have been treated differently from the rest of the sample. In
fact we do not model their SEDs with stellar templates. Therefore, for the
three QSO selected in this work, we estimate the dust luminosities with the
method mentioned above and by assuming that the emission at the Spitzer
wavelengths has only a dust origin. We also estimate the dust component for
the QSO selected by C09 (COSMOS-FR~I 236) with the same
method. Table~\ref{allsample} shows the inferred IR luminosities for the
quasars.

The dust luminosities of the sample selected in this work, expressed as
infrared excess luminosities, are in the range $ L_{\rm  IR \,\ excess}
\sim10^{43} -10^{45.5}$ erg s$^{-1}$, similar to that found for the low-power
radio galaxies studied by B13. FR~IIs cover the entire range of IR excess
luminosities, showing also non detections at 24 $\mu$m.

Similarly to what done by B13, we also measured the spectral index of the
infrared residuals over the OSP between 8 \mum\ and 24 \mum, $\alpha_{IR}$.
Taking into account only significant ($>$ 3 $\sigma$) excesses at 8 \mum, this
value can be estimated in 8 cases, with values spanning between $\alpha_{IR}
\sim$ 1 and -1. For other 8 objects with only a 24 \mum\ detection, the upper
limit to the 8 \mum\ flux translates into a lower limit of $\alpha_{IR}
\gtrsim -1$. For the four QSO (29, 35, 37, and COSMOS-FR~I 236), we derive
again the spectral index assuming that the emission at 8 \mum\ has only a dust
origin (Table~\ref{2spd} and \ref{allsample}).

Since the temperatures associated with the thermal component are poorly
constrained, we prefer to crudely estimate the overall dust temperature from
the spectral index $\alpha_{IR}$.  By assuming a single black-body dust
component, the values of $\alpha_{IR}$ translate into a temperature range of
750-1200 K and 350-600 K for $\alpha_{IR} = -1$ and $1$, respectively. The
derived temperature depends on redshift, with the lower (upper) values of T
being derived for $z=0.8$ ($z=3$).

 \subsection{UV excess}
\label{uv}

Inspection of the SED fits obtained with {\it 2SPD} indicates that the UV
excesses (above the contribution of the OSP) are usually poorly
constrained. Furthermore, the very stellar origin is not granted and the UV
excess might be due to an AGN contribution. It is necessary to introduce a
model-independent criterion to assess which sources really show an UV excess
and to estimate its luminosity. We visually check all SEDs, searching for
sources with a substantial flattening in the SED at short wavelengths or with
a change of the slope between the OSP and the emission in the UV band.  A
clear (marginal) UV excess in seen in 18 (12) sources, properly marked in
Table~\ref{2spd}. The remaining SEDs drop sharply in the UV and are well
reproduced by the emission from the OSPs.

In order to quantify the UV contribution, for the objects showing an UV excess
we measure the flux at 2000 \AA\ in the rest frame, L$_{\rm UV}$, from the
best fitting model, similarly to what done by B13. For the UV-faint sources, we estimate an
upper limit on the UV emission. For the four QSOs, we measure the flux at 2000
\AA\ by using a power-law fit on the photometric data-points in the UV. In
addition, we measure the upper limits to the UV excess also for the UV-faint
objects studied in B13. The new UV excess measurements, included in the
observed range $10^{41.5} \lesssim L_{\rm UV} \lesssim 10^{45.5}$ erg
s$^{-1}$, are reported in Table~\ref{allsample}.

\section{The RL AGN population at $z \gtrsim 1$ in the COSMOS field}
\label{totalsample}

The final sample of radio galaxies selected in the COSMOS field at $z\gtrsim1$
counts 74 objects (precisely, 43 selected in this work plus 31 from B13). It
includes 4 QSO, namely 29, 35, 37, and COSMOS-FR~I 236. Figure~\ref{diag}
  summarizes the selection procedure which drives to the final sample. The
redshifts of the entire sample range between $\sim$0.7 and $\sim$3 with a
median of 1.2. In the next sections we will study the properties of the entire
population, gathering the results obtained in this work and those obtained by
B13. Table~\ref{allsample} collects the information on the members of the
sample.

\begin{table*}
\tiny
\renewcommand{\arraystretch}{0.5}
\addtolength{\tabcolsep}{-2pt}
  \begin{center}
  \caption{The radio-loud AGN population in the COSMOS field at $z\gtrsim1$}
  \label{allsample}
  \begin{tabular}{c|c|cccc|c|cc|c}
    \hline \hline
ID  & redshift        &\multicolumn{4}{c|}{radio} &                                             log M$_{*}$ &   \multicolumn{2}{c}{Dust}    &  UV \\
\hline
    & z$_{phot}$          &  L$_{FIRST}$  & L$_{NVSS}$ & radio class & radio morph & M$_{\odot}$  &   L$_{IR \,\, exc.}$   &  $\alpha_{8-24}$   &  L$_{UV}$\\        
\hline
1  & 1.1577$^{s}$                   &  32.07  &    32.23     &  LP   &   compact  & 11.31$^{+0.04}_{-0.04}$    &  45.74      &   -0.87     & 43.51\\ 
2  & 1.767$^{+0.570}_{-0.236}$  &  32.76  &    32.82     &   HP &   FR~II        & 10.67$^{+0.05}_{-0.05}$    & $<$43.96 &                 & $<$42.00  \\ 
3  &  1.983$^{+0.34}_{-0.37}$    &  32.59  & $<$32.77  &  HP  & compact    & 10.79$^{+0.06}_{-0.05}$    &  44.15      & $>$-0.42 & 42.98 \\ 
4  &  1.104$^{+0.76}_{ -0.28}$   &  32.48  &      32.47    &  LP   & compact     & 10.32$^{+0.16}_{-0.18}$    &  44.41      &   0.62        &  $<$41.43 \\ 
5  &  0.9707$^{s}$                  &   32.13 &      32.24    & LP   &  FR~II          & 11.17$^{+0.09}_{-0.10}$    &   44.22     &   -0.53      &  43.80\\ 
6  &  1.963$^{+0.24}_{-0.28}$    &   32.75  &      32.84   & HP  &  FR~II          & 10.85$^{+0.05}_{-0.05}$     &   43.79     & $>$-0.38 & 43.47 \\ 
7  & 1.384$^{+0.16}_{-0.06}$    &   32.76   &   32.85      & HP  & FR~II           & 10.89$^{+0.06}_{-0.06}$     &  $<$43.49 &                & 42.19\\ 
8  & 0.8357$^{s}$                  &   32.50   &      32.61   & HP  & FR~II           & 11.59$^{+0.07}_{-0.07}$     &  $<$44.10 &                & $<$42.20 \\ 
9  &  2.152$^{+0.17}_{-0.44}$    &  33.34   & 33.37        &  HP & FR~II           & 10.60$^{+0.05}_{-0.06}$     &   $<$44.15 &               & 42.83 \\ 
11  &  1.018$^{+0.15}_{-0.08}$  &    32.58  &  32.73      &  HP & FR~II           & 10.69$^{+0.06}_{-0.06}$     & $<$43.44   &                & 42.23$^{m}$\\ 
12  &   1.860$^{+0.73}_{-0.56}$  & 32.84   &                  &  HP  & compact     & 10.49$^{+0.10}_{-0.09}$    &   $<$43.66 &                & 42.38 \\ 
13  &   1.018$^{+0.02}_{-0.05}$  & 32.22   &                  &   LP  & FR~II          & 10.16$^{+0.04}_{-0.04}$    &   $<$43.16  &                &   41.88$^{m}$ \\ 
14  &    1.377$^{+0.27}_{-0.103}$ &33.06   &  33.11       &  HP  & compact    & 10.92$^{+0.07}_{-0.08}$    &   43.87        & $>$-0.12 &   42.34 \\ 
15  &   1.015$^{+0.03}_{-0.02}$  &  31.97   & 32.45       &  LP   & FR~II          & 10.76$^{+0.03}_{-0.03}$    &   $<$43.15  &                  &  $<$42.06 \\ 
16  &   0.8442$^{s}$                &  31.90   & $<$31.88 &  LP  &  compact   & 10.78$^{+0.06}_{-0.06}$     &   43.26        & $>$-0.59 &  $<$42.22 \\ 
17  &   1.391$^{+0.22}_{-0.11}$  &  32.81   &  32.80      & HP  & FR~II           & 10.89$^{+0.04}_{-0.04}$    &    $<$43.83  &                 & 42.87     \\ 
18  &    1.193$^{+0.123}_{-0.16}$ & 32.84   &  32.92     &   HP & FR~II          & 10.94$^{+0.03}_{-0.04}$     &   44.71        &   0.43        & 43.30 \\ 
19  &    0.932$^{+0.03}_{-0.01}$  & 31.61   & $<$31.99 &  LP  & compact    & 11.26$^{+0.03}_{-0.03}$     & 43.66          &  $>$0.91  &  $<$42.20 \\ 
20  &    1.016$^{+0.20}_{-0.09}$  &  32.89   &  32.92     &   HP & FR~II         & 10.33$^{+0.08}_{-0.08}$     & $<$43.52    &                 & 42.33$^{m}$   \\ 
21  &    0.894$^{+0.06}_{-0.062}$  & 31.97   & 32.18     &   LP  & compact    & 10.71$^{+0.04}_{-0.04}$    &  $<$43.42    &                &  42.20$^{m}$ \\ 
23  &    0.8393$^{s}$                &   32.64  &  32.90    & HP   &  FR~II          & 10.50$^{+0.02}_{-0.02}$   &   $<$43.30   &                 & 42.25$^{m}$ \\ 
24  &    2.006$^{+0.30}_{-0.27}$  &  32.73   & 32.86     &  HP & compact      & 11.08$^{+0.04}_{-0.04}$     &  44.69         & $>$0.66  & 42.91 \\ 
25  &    0.825$^{+0.03}_{-0.03}$  &  31.99   &  32.05    &  LP &   FR~I/FR~II  & 11.17$^{+0.02}_{-0.02}$     &   43.27        &   $>$2.50 &  $<$41.98 \\ 
26  &    0.830$^{+0.11}_{-0.11}$  &  31.88   &  31.83    &  LP  & compact      & 11.04$^{+0.06}_{-0.06}$     &  $<$43.32   &                &  $<$42.07 \\ 
27  &    2.523$^{+0.68}_{-0.32}$  &  33.35   &  33.42    &  HP & compact      & 10.68$^{+0.07}_{-0.07}$      &  $<$44.17  &                & 43.22 \\ 
28 &     1.788$^{+0.43}_{-0.33}$  &  32.67   &   32.95    & HP &  FR~II          & 10.51$^{+0.09}_{-0.08}$      &  44.12         &  -0.23     & 43.30 \\ 
29 &    1.4050$^{s}$                 &   32.68   &  33.03   &   HP & compact     & QSO                                &    45.27      &  -0.92      &  45.69       \\               
30 &     2.268$^{+0.56}_{-0.21}$  &   33.13   & 33.14     &  HP   & extended   & 10.74$^{+0.05}_{-0.05}$    &  $<$44.16   &                & 42.53$^{m}$ \\ 
31 &    1.698$^{+0.49}_{-0.19}$  &    32.77   &  32.75    &  HP  &  extended   & 10.65$^{+0.05}_{-0.05}$    &   $<$44.06  &                & 42.57  \\ 
32 &    1.067$^{+0.32}_{-0.10}$  &    32.17   &   32.23    &  LP  & extended   & 10.08$^{+0.07}_{-0.07}$      &   $<$43.59 &               & 42.05$^{m}$ \\ 
34 &      1.402$^{+0.31}_{-0.32}$  &  32.95   &  32.86    &  HP  & extended   & 10.65$^{+0.23}_{-0.29}$     &   44.61         &   -1.08    &  $<$42.63 \\ 
35 &  1.80$^{+0.40}_{ -0.40}$      &   33.70   &   33.69    & HP  & compact     & QSO                               &  44.89          &  -0.78     & 44.66     \\
36  &   2.255$^{+0.70}_{-0.35}$    &   33.89   & 33.90     &  HP & compact      &  10.91$^{+0.07}_{-0.07}$   & $<$44.09    &               & 43.37   \\ 
37  & 1.3176$^{s}$                   &   33.70   &  33.72     & HP  &  FR~II          &  QSO                              &   45.47        & -0.95      & 45.55         \\
38 &   2.355$^{+0.45}_{-0.16}$   &    34.17   &   34.24    & HP &  FR~II           & 11.09$^{+0.04}_{-0.04}$      & 44.42         & $>$0.58  & 43.70 \\   
39 & 1.851$^{+0.36}_{-0.30}$     &     33.52   &  33.52    & HP & compact       & 10.94$^{+0.07}_{-0.06}$     &   $<$43.94 &                & 42.34$^{m}$  \\   
40 & 1.110$^{+0.24}_{ -0.20}$     &    33.08   &  33.23    & HP &  FR~II           &  10.70$^{+0.12}_{-0.11}$     &   $<$43.51  &               &  $<$42.24   \\   
41 & 1.028$^{+0.07}_{-0.18}$      &    32.96   &  32.96     & HP  & extended   & 10.22$^{+0.19}_{-0.07}$      &   44.69        &   0.15      & 43.27 \\   
42 & 1.137$^{+0.44}_{-0.30}$     &     33.02   &  33.00     & HP  & compact     & 10.75$^{+0.16}_{-016}$      &  $<$43.66  &                & 42.58$^{m}$ \\   
43 &  1.247$^{+0.46}_{-0.44}$    &     33.27   &  33.40      &HP &  FR~II           & 11.03$^{+0.19}_{-0.32}$     &  $<$43.63  &                & 42.68$^{m}$  \\   
44 & 1.062$^{+0.14}_{-0.05}$      &   33.97   &   33.96     & HP &  FR~II          & 10.61$^{+0.03}_{-0.04}$       &   44.37       &   0.53       & 43.12 \\   
45 & 0.918$^{+0.07}_{-0.07}$     &     33.07   &  33.21     & HP  &  FR~II         & 10.74$^{+0.05}_{-0.05}$      &  $<$43.35   &                & 42.28$^{m}$  \\   
46 & 0.8784$^{s}$                    &   32.77   &   32.88     & HP  &  FR~II         & 10.47$^{+0.09}_{-0.13}$     &  $<$43.29   &                 & 42.27$^{m}$  \\   
\hline
COSMOS-FR~I 1   & 0.8827$^{s}$              &   31.78  &  $<$31.93  & LP  & compact    &  10.08$^{+0.04}_{-0.04}$  & $<$43.31   &                & $<$41.92             \\ 
COSMOS-FR~I 2   & 1.33$^{+0.10}_{-0.09}$  &   32.02  &      32.40     & LP  & extended   &  11.00$^{+0.04}_{-0.04}$  & $<$43.65  &                &   42.53  \\ 
COSMOS-FR~I 3   & 2.20$^{+0.32}_{-0.44}$  &   33.10   &     33.19     & HP & compact    &  10.59$^{+0.08}_{-0.10}$   & 45.27        &  -0.01     &  43.21  \\
COSMOS-FR~I 4   & 1.37$^{+0.10}_{-0.06}$  &   32.77   &     32.86     & HP  & FR~I          &  11.16$^{+0.04}_{-0.03}$   & $<$43.55  &               &  42.71  \\
COSMOS-FR~I 5   & 2.01$^{+0.22}_{-0.35}$  &   32.47   &     32.89     & HP  & compact    &  11.49$^{+0.04}_{-0.03}$  & 44.76         & $>$0.49 &  $<$43.00    \\ 
COSMOS-FR~I 11   & 1.57$^{+0.14}_{-0.09}$ &  32.18   &  $<$32.52  & LP  & compact    &   10.98$^{+0.10}_{-0.05}$  &  $<$43.80  &               &  $<$42.11 \\ 
COSMOS-FR~I 13   & 1.19$^{+0.08}_{-0.11}$ &  32.02   &     32.22     & LP  & compact     &  10.72$^{+0.04}_{-0.03}$   & 44.23        & -0.41      &   42.65$^m$  \\  
COSMOS-FR~I 16   & 0.9687$^s$              & 32.38     &   32.27      & LP   & compact    &  10.74$^{+0.06}_{-0.06}$   & 43.60        & $>$-0.12 &  42.42$^m$  \\  
COSMOS-FR~I 18   & 0.92$^{+0.14}_{-0.11}$ &   32.22   &     32.28    & LP   & extended   &   10.02$^{+0.08}_{-0.08}$  & $<$43.96  &                 &   $<$42.23\\  
COSMOS-FR~I 20   & 0.88$^{+0.02}_{-0.02}$ &   31.66   & $<$31.93 & LP  & extended     &  11.03$^{+0.02}_{-0.03}$   & $<$43.02  &                &  42.25$^m$ \\  
COSMOS-FR~I 22   & 1.30$^{+0.05}_{-0.04}$ &   32.38   & $<$32.34  & LP  & compact     & 11.16$^{+0.02}_{-0.03}$    & 43.87         &  $>$0.66 &  $<$41.79    \\  
COSMOS-FR~I 25   & 1.33$^{+0.11}_{-0.13}$ &   32.29   &     32.39   & LP   & compact     &  10.75$^{+0.04}_{-0.05}$   & 43.75         & $>$0.26   &  42.87$^m$  \\  
COSMOS-FR~I 26   & 1.09$^{+0.12}_{-0.07}$ &   32.02   &     32.25   & LP   & extended     & 11.12$^{+0.04}_{-0.04}$  &  $<$43.45  &                  &  42.50$^m$  \\  
COSMOS-FR~I 28   & 2.90$^{+0.20}_{-0.26}$ &   32.99   &     33.13   & HP  & compact      &  11.38$^{+0.04}_{-0.04}$   & 44.28       & $>$0.22   &   $<$43.35 \\  
COSMOS-FR~I 29   & 1.32$^{+0.23}_{-0.24}$ &   32.28   &     32.31   & LP   & compact      &  10.03$^{+0.05}_{-0.05}$ &  $<$43.68  &                 &   42.83  \\    
COSMOS-FR~I 30   & 1.06$^{+0.11}_{-0.07}$ &   31.83   &     32.11   & LP   & compact      & 11.03$^{+0.05}_{-0.05}$  & $<$43.47   &                 &  $<$41.81  \\  
COSMOS-FR~I 31   & 0.9132$^s$               & 32.14     &   32.18    & LP   & compact      & 10.75$^{+0.03}_{-0.03}$  & $<$43.35   &                &   42.86  \\  
COSMOS-FR~I 34   & 1.55$^{+0.41}_{-0.19}$  &   32.84   &     32.77  & HP   & compact      &  10.99$^{+0.07}_{-0.07}$ & $<$44.04  &                 &   42.82  \\ 
COSMOS-FR~I 36   & 1.07$^{+0.10}_{-0.04}$  &   32.23   &     32.25  & LP  & compact       &  10.83$^{+0.02}_{-0.03}$  &  43.46       &  $>$0.20  &  $<$41.91 \\    
COSMOS-FR~I 38   & 1.30$^{+0.17}_{-0.28}$  &  32.94    &   33.00    & HP   & compact     &  10.65$^{+0.07}_{-0.07}$  & 43.77         & 0.45        &   43.03 \\    
COSMOS-FR~I 39   & 1.10$^{+0.05}_{-0.05}$  &   31.90  & $<$32.16 & LP  & compact      &  10.88$^{+0.03}_{-0.03}$  & $<$43.20  &                 &   $<$42.23  \\  
COSMOS-FR~I 52   & 0.7417$^s$               &  31.54    & $<$31.73 & LP  & compact      &  10.78$^{+0.10}_{-0.10}$  & 43.33        &  0.61        &   42.96   \\ 
COSMOS-FR~I 70   & 2.32$^{+0.53}_{-0.20}$   & 33.12   &     33.18    & HP  & compact     &  10.65$^{+0.07}_{-0.03}$  & 44.07         & $>$-0.50  &   43.54 \\  
COSMOS-FR~I 202   & 1.31$^{+0.09}_{-0.12}$ & 31.98   &     32.46    & LP  & extended     &   10.86$^{+0.03}_{-0.06}$ & $<$43.66   &                 &   42.21  \\    
COSMOS-FR~I 219   & 1.03$^{+0.02}_{-0.04}$ & 31.96   & $<$32.09 & LP  & compact      &  10.67$^{+0.03}_{-0.03}$  & $<$43.32  &                  &   42.69$^m$ \\  
COSMOS-FR~I 224   & 1.10$^{+0.10}_{-0.04}$ & 32.28   &     32.27   & LP   & extended     &  10.71$^{+0.03}_{-0.04}$  & $<$43.53   &                &    42.53$^m$  \\ 
COSMOS-FR~I 228   & 1.31$^{+0.05}_{-0.07}$ & 32.25   &     32.51   & LP   & compact      &  11.06$^{+0.03}_{-0.03}$ & $<$43.64    &                &   $<$41.91      \\  
COSMOS-FR~I 234   & 1.10$^{+0.14}_{-0.08}$  & 32.41   &     32.48  & LP  & FR~I              &  10.83$^{+0.04}_{-0.04}$ & $<$43.52    &                &   $<$42.02  \\  
COSMOS-FR~I 236   & 2.1318$^s$               & 33.29   &     33.29  & HP  & compact       &  QSO                           & 45.45           & -1.01     & 45.60  \\
COSMOS-FR~I 258   & 0.9009$^s$               & 31.90   &   32.12    & LP  & compact        &  10.64$^{+0.08}_{-0.09}$  & $<$43.41   &              &    43.27 \\ 
COSMOS-FR~I 285   & 1.10$^{+0.13}_{-0.08}$  & 32.23   &     32.31  & LP  & extended       &  10.43$^{+0.04}_{-0.04}$ & 43.73          &  -0.22  &    43.08  \\            
\hline
\end{tabular}
\end{center}
Column description: (1) ID number of the object; (2) photometric or spectroscopic (if available) redshift; (3)-(4) FIRST and NVSS radio luminosities; (5) classification based on the radio power: low or high power (LP or HP) radio sources;  (6) radio morphology based on the VLA-COSMSO images; 
  (7)  the total stellar mass of the galaxy in M$_{\odot}$; (8)-(9) the infrared excess
  luminosity (in erg s$^{-1}$)  and
  the spectral index measured on the infrared excess at 8 and 24$\mu$m; (10)
  UV luminosity at 2000 \AA\, in the rest frame in erg s$^{-1}$. The marginal
  UV excesses are marked with a $^m$. 
\end{table*}

\subsection{Radio properties}
\label{radio}

\begin{figure*}
\includegraphics[scale=0.45]{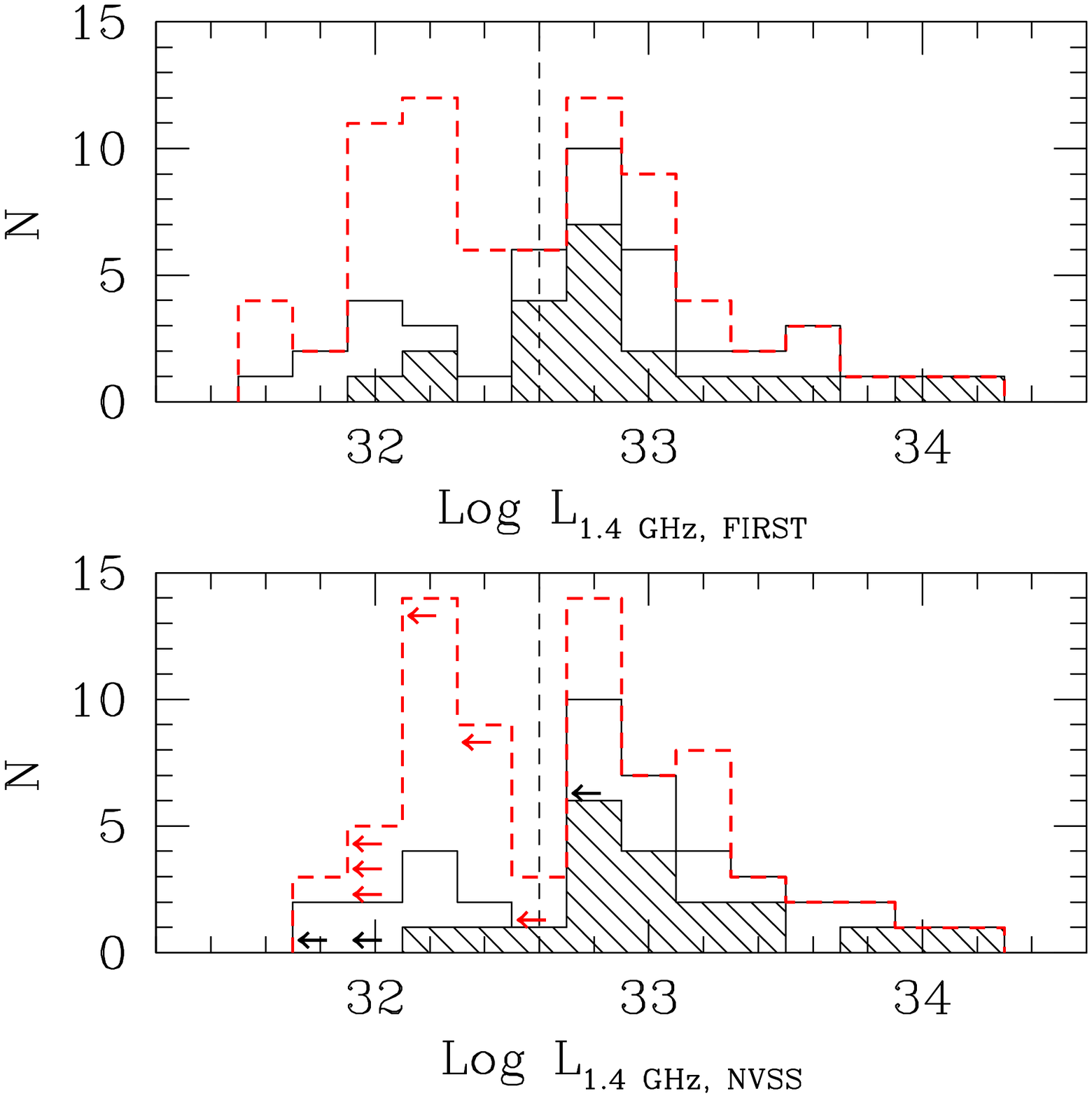}
\includegraphics[scale=0.45]{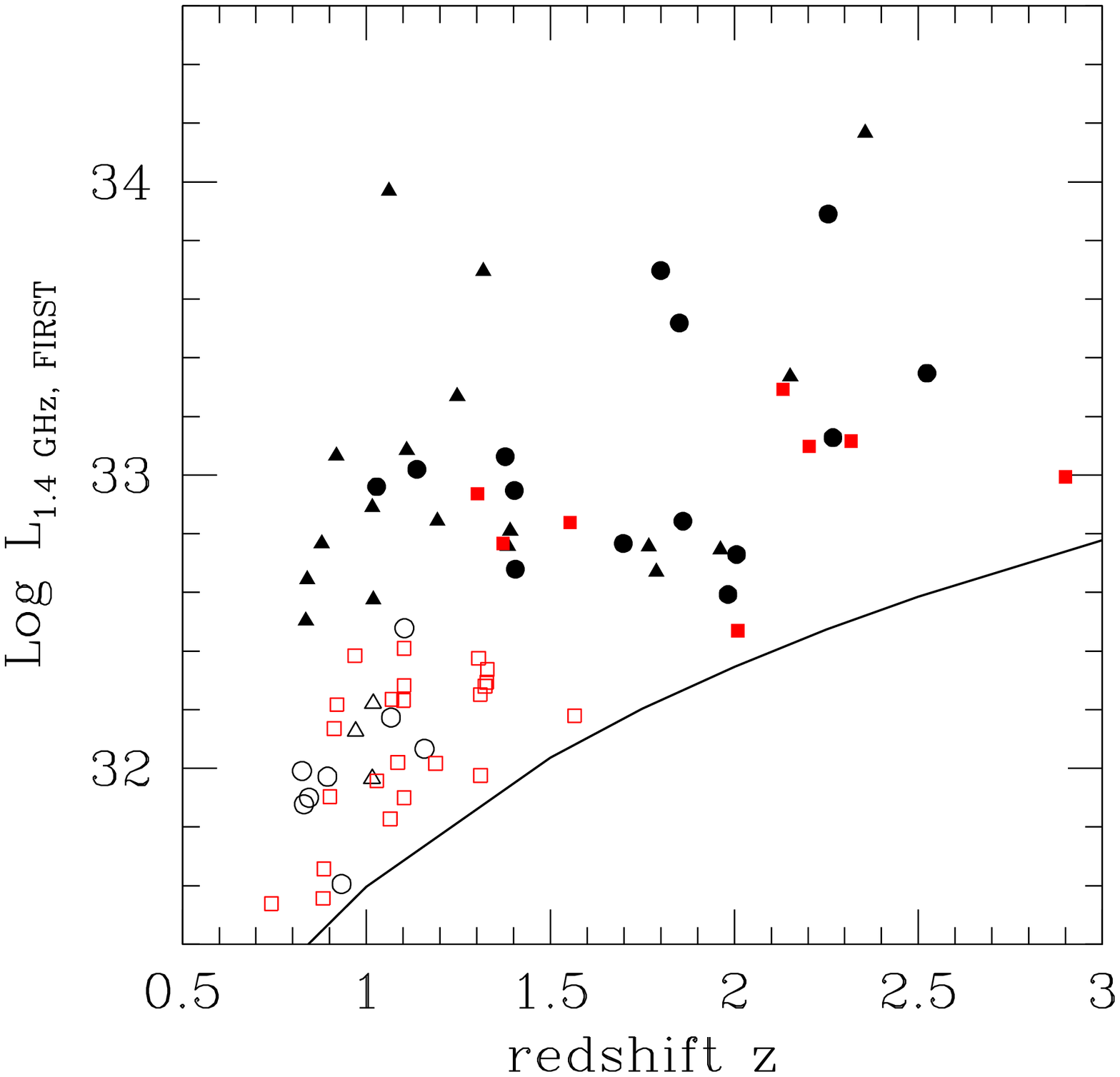}
\caption{Left panel: distribution of the rest-frame total radio luminosity (in
  erg s$^{-1}$ Hz$^{-1}$) at 1.4 GHz from the FIRST data (upper panel) and
  NVSS data (lower panel). The dashed lines correspond to the local FR~I/FR~II
  break used to separate the sample in HP and LP sources. The FR~IIs
  distribution is represented by the back-slash shaded histogram. The dashed
  histogram represents the entire COSMOS RL AGN sample at high
  redshifts. Right panel: redshifts versus the rest-frame FIRST radio
  luminosity at 1.4 GHz (in erg s$^{-1}$ Hz$^{-1}$). The empty points are the
  LPs and the full points are the HPs. The FR~IIs are the triangles. The red
  squared points represent the sample studied by B13: LPs are the empty
  points, and HPs the filled ones. The solid line represents the
    luminosity-redshift relation corresponding to the flux limit of the FIRST
    survey (between 1-0.6 mJy but for the plot we use 0.75 mJy.)}
\label{radioplot}
\end{figure*}

Considering all the radio sources selected in the COSMOS field (see
Table~\ref{allsample}), most of them appear compact or marginally extended.
Although it is possible that the non-detection of large-scale structures is a
result of the high radio frequency at which the VLA-COSMOS catalog is carried
out, it is likely that the compact sources are intrinsically small, with sizes
smaller than a few tens of kpc. The FR~IIs, which show structures of, at most,
some hundreds of kpc, are approximately one third of the entire sample, while
the bona-fide FR~Is are two (both in C09 sample).

In Fig.~\ref{radioplot} (left panel) we show the histograms of rest-frame
(using a radio spectral index $\alpha$ = 0.8) radio powers at 1.4 GHz measured
with FIRST and NVSS, reported in Table~\ref{allsample}.\footnote{For the
  objects 12 and 13 we cannot measure their NVSS luminosities because their
  NVSS emissions are incorporated in the same source.}  The resulting
luminosities are in the range 10$^{31.5}-10^{34.3}$ erg s$^{-1}$ Hz$^{-1}$,
straddling the local FR~I/FR~II break (L$_{1.4 \,\,\rm GHz} \sim 10^{32.6}$ erg
s$^{-1}$ Hz$^{-1}$ converting from 178 MHz to 1.4 GHz,
\citealt{fanaroff74}). For the entire population of high-z RL AGN in the
COSMOS field, the median FIRST luminosity is $10^{32.59}$ $\ergsHz$.

The radio luminosities of our sample are far larger than $\sim$10$^{30}$ erg
s$^{-1}$ Hz$^{-1}$, the radio luminosity below which starburst activity may
significantly contribute (e.g., \citealt{mauch07,wilman08}). Therefore, we can
be confident that our sources are associated with RL AGN, where radio emission
has a non-thermal synchrotron origin from the relativistic jets. Furthermore,
the radio-loudness parameter (here measured as the radio-to-UV flux ratio,
\citealt{white07}) is always far higher than the threshold that separates
radio-quiet from RL AGN.

Similarly to the approach used by B13, we prefer to separate the sample in
high and low power (HP and LP, respectively) using the local FR~I/FR~II break
as divide (see Table~\ref{allsample}).  This helps us to investigate the
properties of the sample and the role of the AGN, whose radio luminosity is a
good estimator of its power. 40 out of 74 objects are HPs. All FR~IIs are HPs,
with only 3 exceptions (namely 5, 13, and 15) although still with radio
luminosities larger than $10^{32}$ erg s$^{-1}$ Hz$^{-1}$. With respect to
B13, the inclusion of sources of higher fluxes corresponds to a broader range
of luminosity at each given redshift (Fig.~\ref{radioplot}, right panel), thus
improving our ability to explore the properties of high-z radio
galaxies. Furthermore, it is noteworthy the redshift coverage of the FR~II as
broad as the entire observed range.

\subsection{Global photometric properties of the sample}
\label{global}

\begin{figure*}
\includegraphics[scale=0.45]{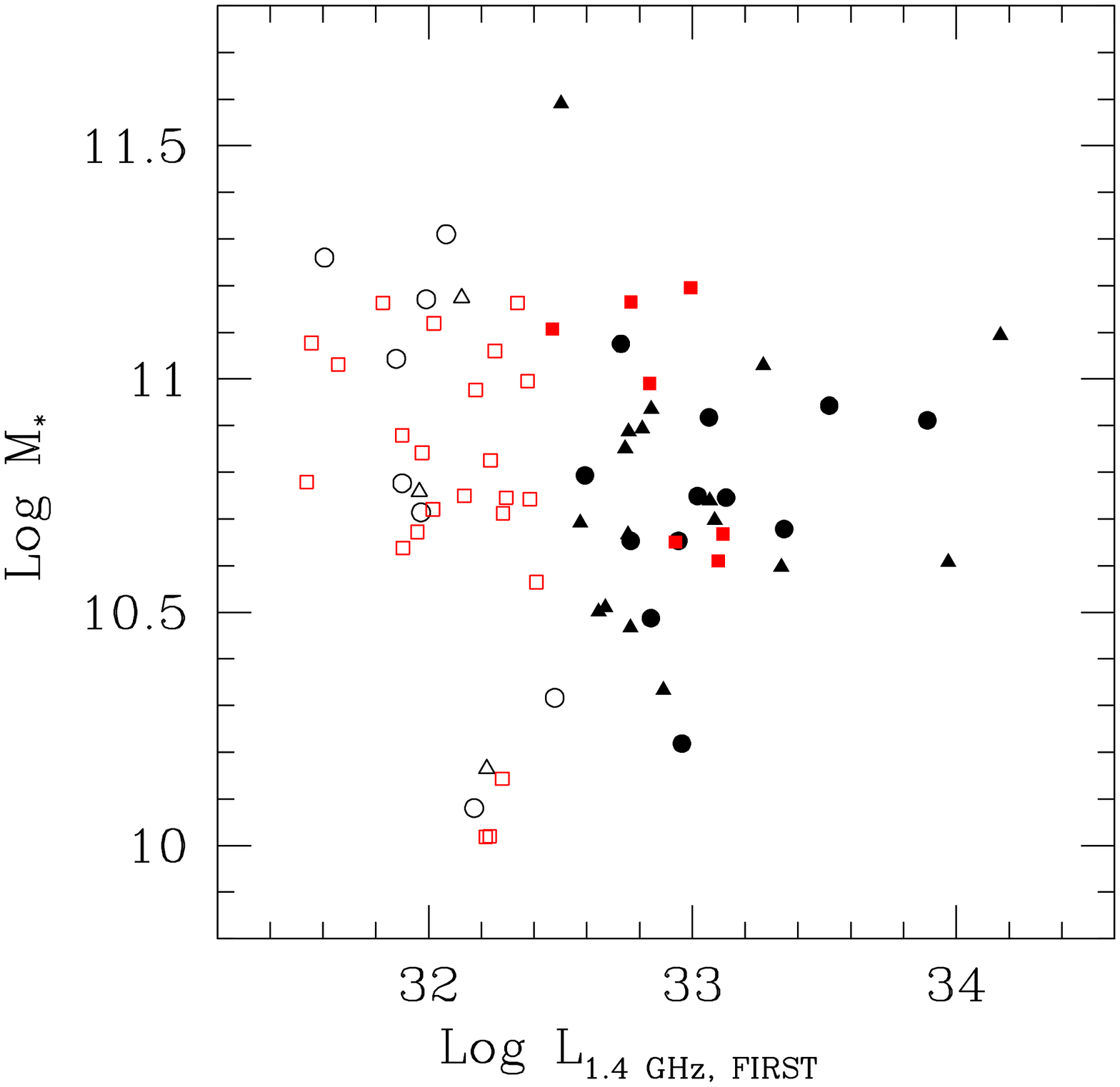}
\includegraphics[scale=0.45]{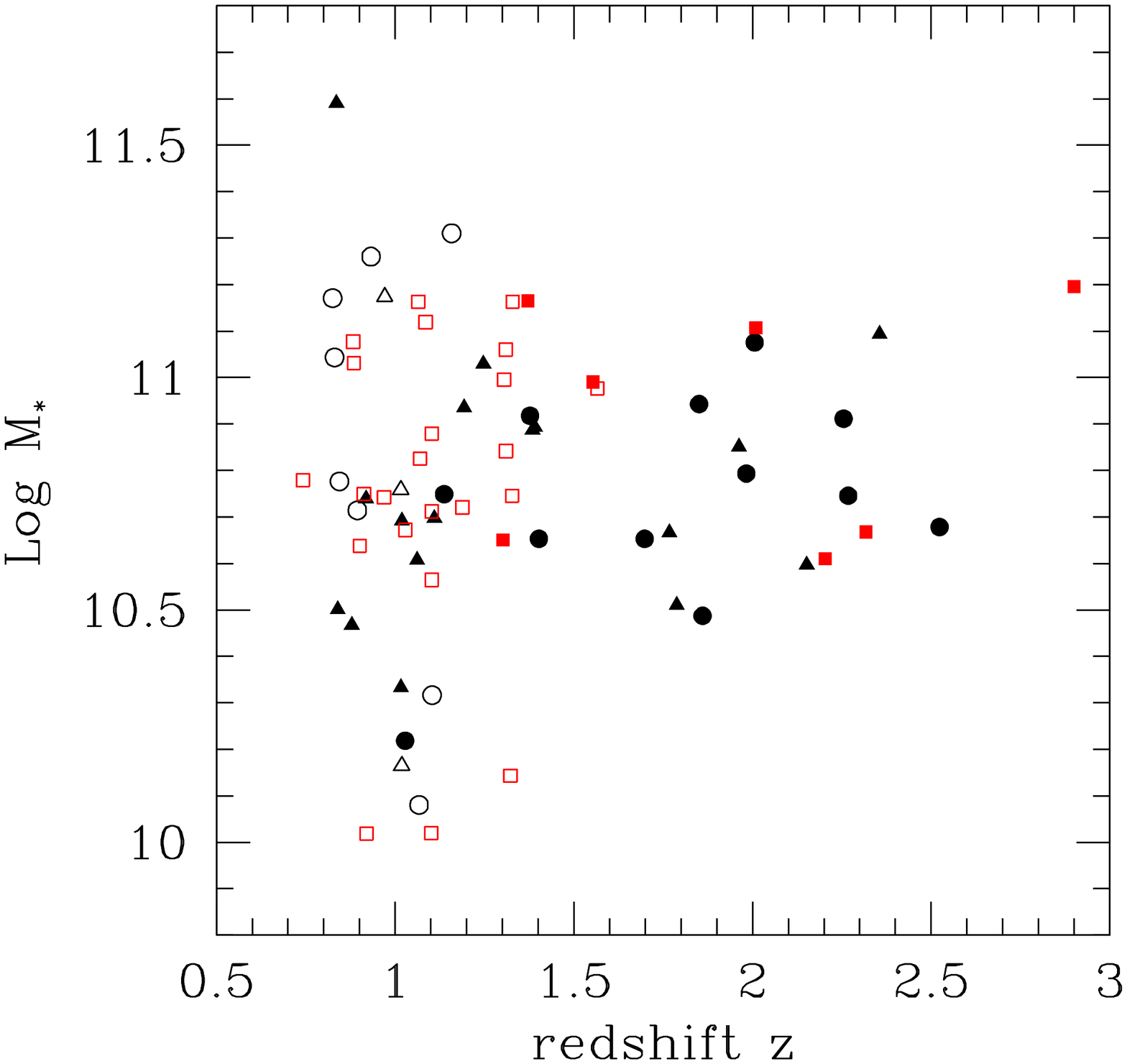}
\caption{Stellar masses (in M$_{\odot}$) measured with {\it 2SPD} in relation
  with the rest-frame FIRST radio powers (in erg s$^{-1}$ Hz$^{-1}$) (left
  panel) and redshifts (right panel) of the sample. Black empty points are
  LPs, while filled black points are HPs. The  FR~IIs are the black
  triangles. The red squared points represent the sample studied by B13: LPs
  are the empty points, and HPs the filled ones.}
\label{Ms}
\end{figure*}

\begin{figure*}
\includegraphics[scale=0.45]{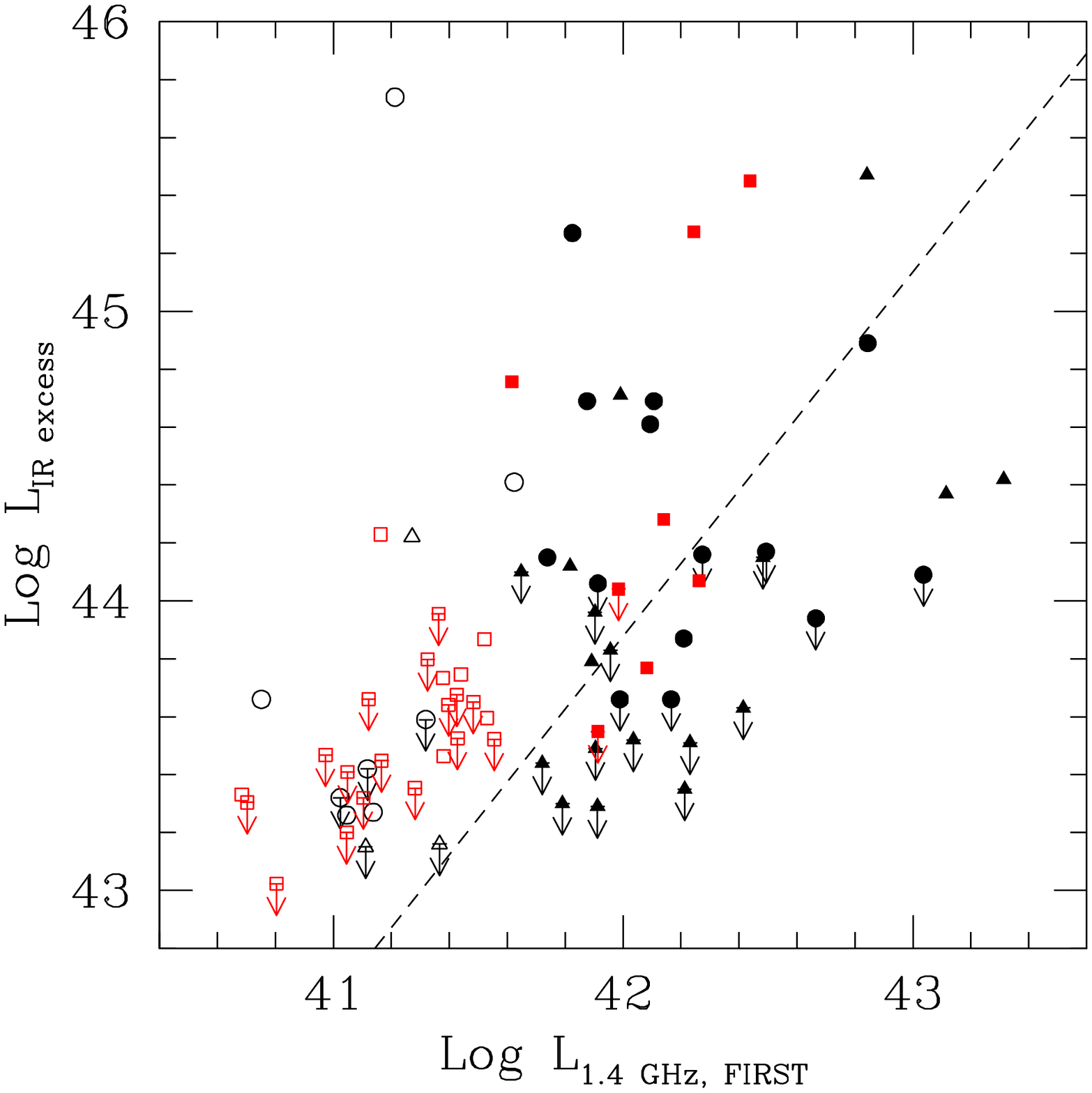}  
\includegraphics[scale=0.45]{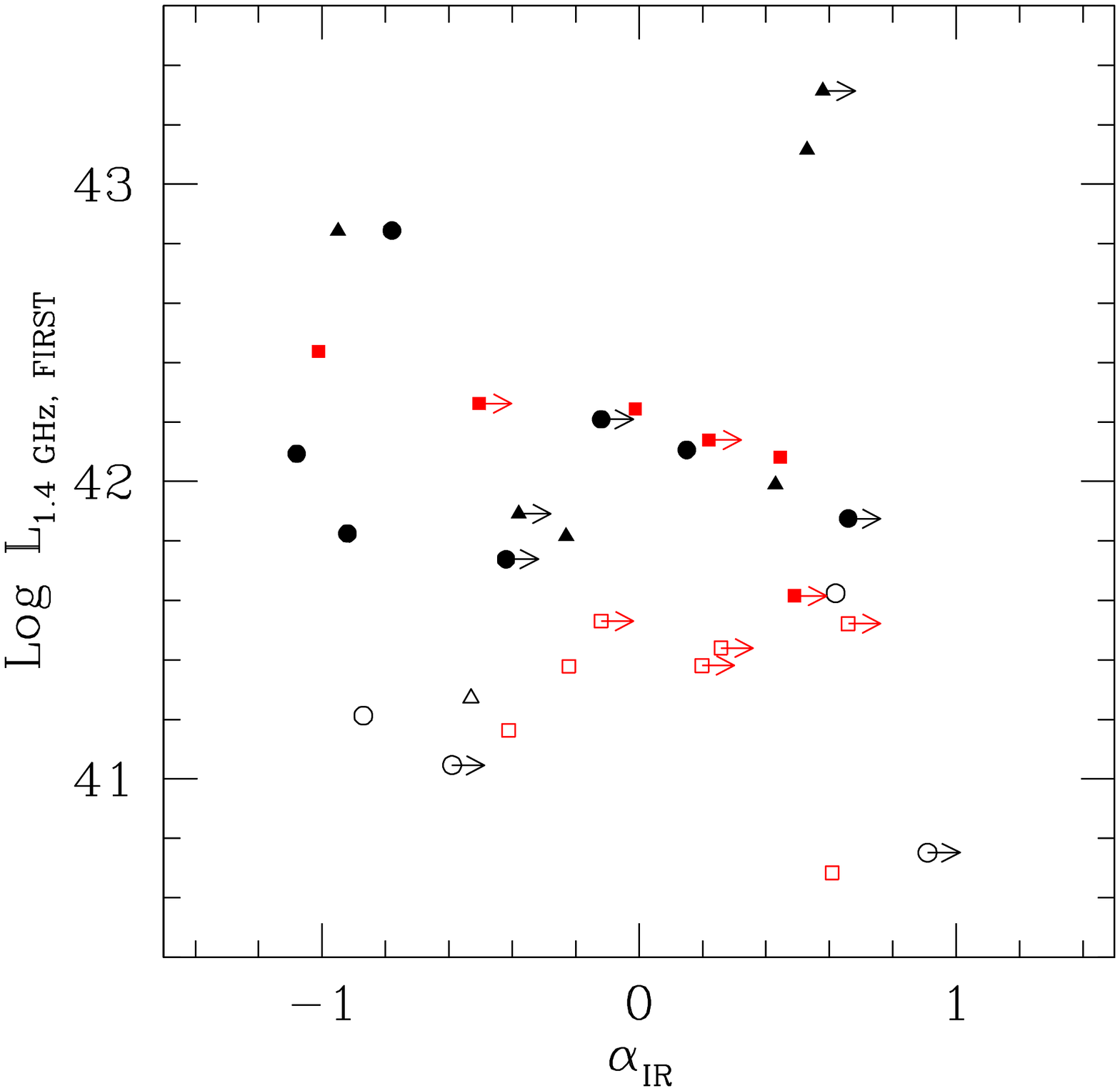}  
\caption{Infrared excess luminosity ($\ergs$) versus (left panel) rest-frame
  FIRST radio luminosity ($\ergs$) and (right panel) spectral index from 8 to
  24 \mum\ estimated from the IR excess in the SED above the stellar emission
  (right panel). Black empty points are LPs, while filled black points are
  HPs. The FR~IIs are the black triangles. The red squared points represent
  the sample studied by B13: LPs are the empty points, and HPs the filled
  ones. The radio-IR correlation is the dashed line and its parameters are
  shown in Table~\ref{statistics}.}
\label{dustfigs}
\end{figure*}

The radio galaxies at z$\gtrsim$1 selected in the COSMOS field are associated
with hosts which are in first approximation early-type galaxies. However, a
further study is necessary to investigate quantitatively the morphology of the
hosts to classify them more precisely. This will be addressed in a forthcoming
paper.

Another important result is that the galaxies which harbor our sample of
distant RL AGN are massive and old. In fact the median stellar mass of the
hosts of these RL AGN is 5.8 $\times$ 10$^{10}$ M$_{\odot}$. Figure~\ref{Ms}
does not show any significantly evident relation between the stellar content and
the radio power and the redshifts.  In fact, if we divide the sample in LPs
and HPs or FR~IIs and no-FR~IIs, no clear differences in stellar masses are
present. Furthermore, the typical age of the dominant stellar population in
those galaxies is a few Gyr-yr old.

The dust luminosities of the sample cover the range between 10$^{43}$
to 10$^{46}$ $\ergs$, with a median value of 7 $\times$ 10$^{43}$ $\ergs$
(Fig.~\ref{dustfigs}, left panel). Considering the radio classes, the LP and
HP difference in dust luminosity is a factor $\sim$3.3. FR~IIs cover the
entire range of IR excess luminosities, showing also non detections at 24
$\mu$m.

Figure~\ref{dustfigs} (right panel) shows a broad overlapping between the radio classes in
dust luminosities and in IR spectral indices. This indicates that an increase
of the AGN power (L$_{radio}$) is not necessarily associated with an increase
of the dust luminosity and/or temperature. This denotes different possible contribution
for the IR emission, star-formation and/or AGN.  However, we will focus on the
association between the dust and the AGN in Sect.~\ref{multi}.

\begin{figure*}
\includegraphics[scale=0.45]{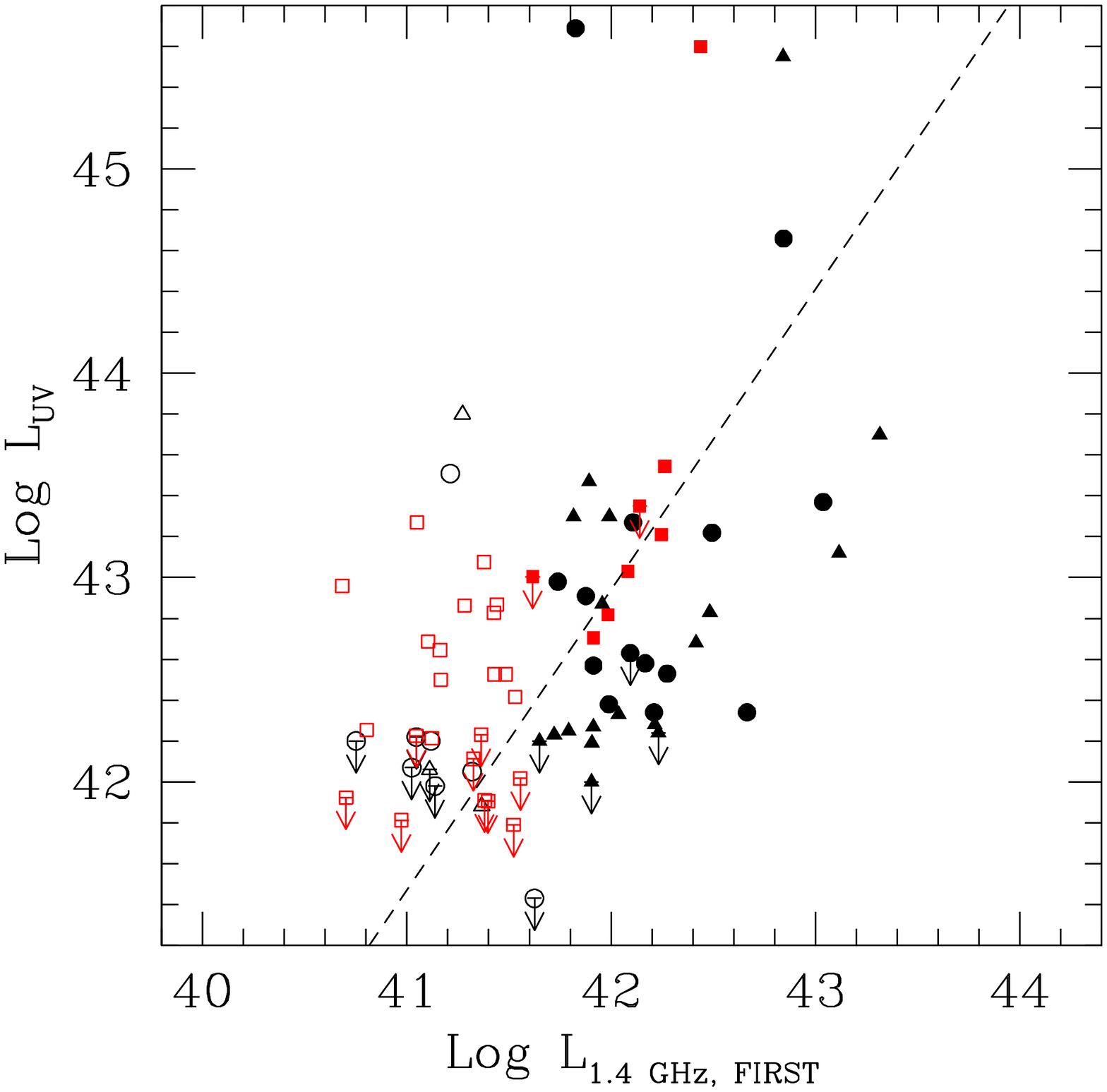}  
\includegraphics[scale=0.45]{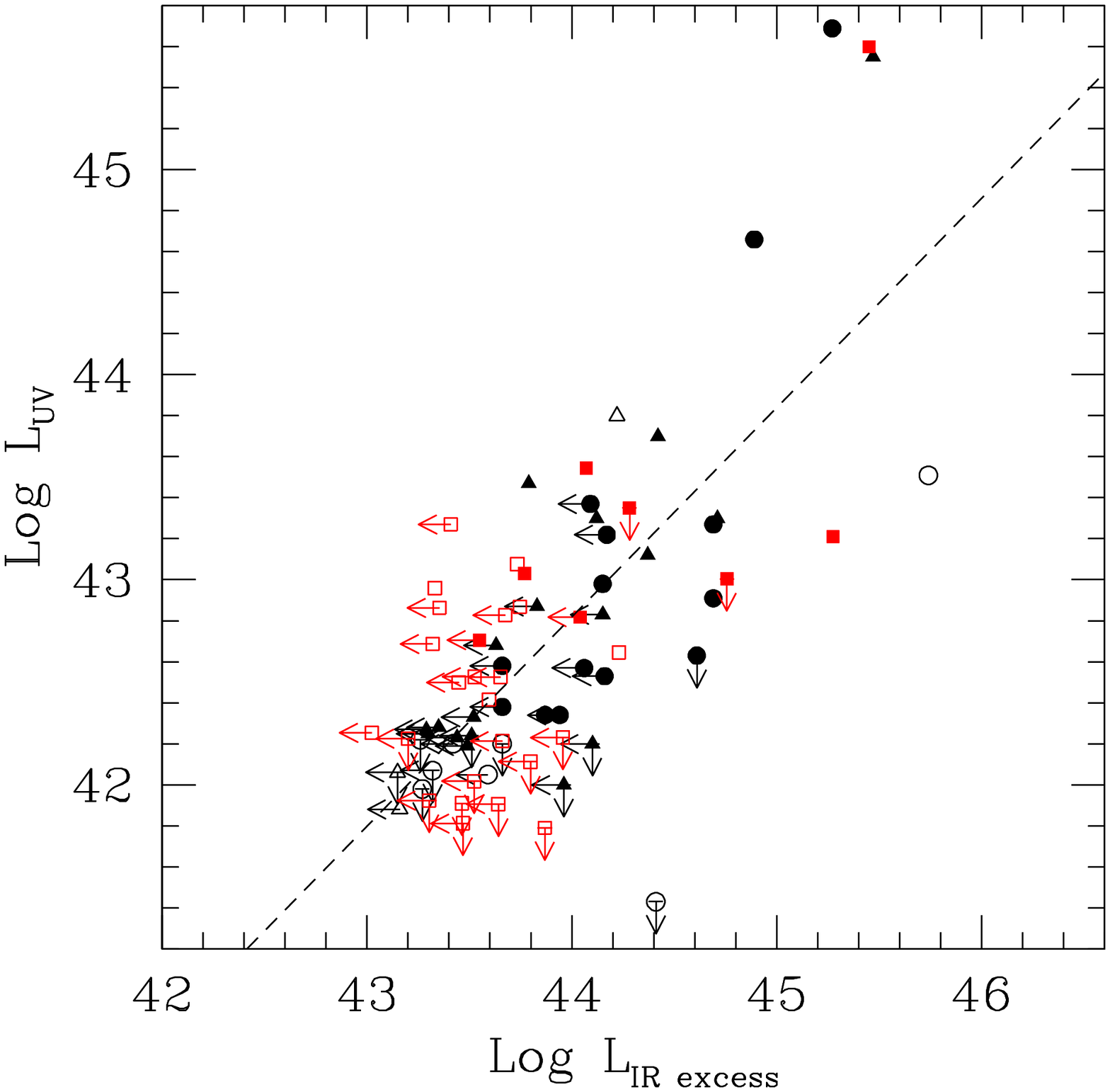}  
\caption{UV luminosity ($\ergs$) measured at 2000 \AA\, versus: (left panel)
  rest-frame FIRST radio luminosity ($\ergs$), and (right panel) infrared excess
  luminosity ($\ergs$). Black empty points are LPs, while filled black points are HPs. The
   FR~IIs are the black triangles. The red squared points represent
  the sample studied by B13: LPs are the empty points, and HPs the
  filled ones. The dashed lines represent the linear correlations whose
  parameters are reported in Table~\ref{statistics}.}
\label{UVplot}
\end{figure*}

The UV luminosities of the radio population range between $L_{\rm UV} \sim
10^{41.5}$ and $10^{45.5}$ erg s$^{-1}$ (Fig.~\ref{UVplot}), with a median
value of 6 $\times$ 10$^{42}$ $\ergs$. Most of LPs are
faint in UV, while HPs show UV luminosities larger than LPs by a factor
$\sim$6. HPs include most of the significant UV detections. FR~II sources
cover the entire range of UV luminosities, showing also upper
limits. 

The global picture emerging from the high-z RL AGN population is substantially
a mild bimodality. The LP are typically UV- and IR-faint, while the HPs are on
the opposite side of the luminosity range.  This behavior is not clearly a
one-to-one relation and needs to be statistically tested (see next section).

We prefer to address the comparisons of our results with other samples of
distant and local radio galaxies in a forthcoming paper.

\begin{table*}
\begin{center}
\caption{Statistics} 
\begin{tabular}{ll|cc|cc|cc} 
\hline
\hline
    $X$  & $Y$ & $\rho_{XY}$ & $P_{\rho_{XY}}$ &  $\rho_{XY,z}$ & $P_{\rho_{XY,z}}$ &  Slope  & Intercept  \\
\hline
    Log L$_{1.4 \,\,\rm GHz \,\ FIRST}$  & Log L$_{\rm IR ,\ excess}$  & 0.396  & 0.0007 &  0.151 & 0.193 & 1.3$\pm$0.3  &  -9$\pm$19  \\
    Log L$_{1.4 \,\,\rm GHz \rm \,\ FIRST}$  & Log L$_{\rm UV}$  &   0.456  &  0.0001 &  0.273  &   0.0188  &  1.5$\pm$0.4  &  -19$\pm$24 \\
    Log L$_{\rm IR ,\ excess}$     & Log L$_{\rm UV}$  &  0.509   &  $<$0.0001 &  0.397 & 0.0007 & 1.0$\pm$0.3  &  -2$\pm$17 \\
\hline 
\end{tabular} 
\label{statistics} 
\end{center} 
Column description: (1)-(2) the two variables of the considered
  relation; (3)-(4) the generalized Spearman correlation coefficient
  (computed including upper limits) and the probability that there is no
  correlation between the variables; (5)-(6) the partial rank coefficient
  after excluding the common dependence of redshift and the probability
  that there is no correlation between the variables; (7)-(8) the slope and
  the intercept with their errors of the possible linear correlation.
\end{table*}

\subsection{The connection between radio, dust, and UV emission}
\label{multi}

In this section we focus on the possible link between the nuclear and host
properties for the entire sample of high-z RL AGN in the COSMOS field.  The
key point is to understand the origin of the IR and UV emission which might be
stellar or associated with the AGN. The radio luminosity is a fundamental
indicator of the output energetics for a RL AGN. The IR and UV luminosities
can provide clues on the amount of dust, of star formation and AGN
contribution.

Figure~\ref{dustfigs} (left panel) and Fig.~\ref{UVplot} compare the radio,
IR, and UV luminosities, showing broad relations between them, although the
dataset includes a large number of upper limits. A statistical approach which
takes into account the censored data is necessary to analyze the significance
of such trends. We performed a statistical analysis using the Astronomy
Survival Analysis (ASURV) package \citep{lavalley92}. We used the {\it
  schmittbin} task \citep{schmitt85} to calculate the associated linear
regression coefficients for two sets of variables.  Operatively, we carried
out this procedure twice, obtaining two linear regressions: first, considering
the former quantity as the independent variable and the latter as the
dependent one and second switching the roles of the variables. The best fit is
represented by the bisector of these two regression lines. This followed the
suggestion of \citet{isobe90} that considers such a method preferable for
problems that require a symmetrical treatment of the two variables. In order
to estimate the quality of the linear regression, we used the generalized
Spearman's rank order correlation coefficient, using the {\it spearman} task
\citep{akritas89}. Furthermore, since the sample covers a large range of
redshifts, we tested the effects of such a quantity in driving these correlations (both
luminosities depend on $z^2$) estimating the Spearman Partial Rank correlation
coefficient.\footnote{The Spearman Partial Rank correlation coefficient
  estimates the linear correlation coefficient between two variables taking
  the presence of a third into account. If X and Y are both related to the
  variable z, the Spearman partial correlation coefficient is
  $\rho_{XY,z}=\frac{\rho_{XY}-\rho_{Xz}\rho_{Yz}}{[(1-\rho_{Xz}^2)(1-\rho_{Yz}^2)]^{1/2}}$}
All the statistical parameters obtained are reported in
Table~\ref{statistics}.

Firstly, let us consider the relation between IR excess luminosity and 1.4-GHz
FIRST power (Fig.~\ref{dustfigs}, left panel). Using the generalized
Spearman's $\rho$ test, the probability that a fortuitous correlation is
$P=0.0007$. However, considering the common dependence of the two luminosities
on redshift, such a probability increases to $P=0.193$.

Secondly, we focus on the UV excess and its possible link with radio power and
dust luminosity (Fig.~\ref{UVplot}, left and right panels). For the radio-UV
relation, the generalized rank coefficient $\rho$ returns a probability of no
correlation of $P=0.0001$. The exclusion of the common redshift dependence
with the partial rank coefficient yields a probability, $P=0.019$ that a
fortuitous correlation appears. Instead, for the UV-IR relation, according to
the generalized Spearman's rank coefficient, the probability that there is no
correlation between the variables is $P<0.0001$. The effect of the redshift
which might stretch the relation is negligible as the value of the partial
$\rho$ slightly changes and the associated probability is still very small
($P=0.0007$).

Since the four QSO present in the sample are among the brightest sources in
the UV and IR bands, they potentially might drive the relations we find. We
then re-measure the censored statistics parameters by excluding them. We
obtain that the relations improve, apart from the radio-IR trend that slightly
worsens. This indicates that the QSOs are not responsible to drive the
relations in the radio-IR-UV planes.

Summarizing, these tests enable us to conclude that the UV emission is
significantly correlated with both the IR and the radio luminosities, the
former being the stronger link. Conversely, the radio-IR relation might be
real, but it has a not negligible probability of being just driven by the
common redshift dependence.

\section{Summary and conclusions}
\label{discussion}

We select a sample of radio sources in the COSMOS field looking for objects at
high redshifts (z$\gtrsim$1), and extending a previous analysis focused on
low-luminosity radio galaxies selected by C09 and studied by B13.  While C09
selected the sample with the aim of searching for FR~I candidates, we relaxed
the C09 selection criteria in order to include all radio sources likely to be
associated with galaxies at $z\gtrsim 1$, independently of the radio
morphology. In particular we include in this analysis also objects with a
FR~II radio morphology and we do not set a high limit to the radio flux in
order to obtain a complete view of the RL AGN phenomenon in this cosmological
era.  We then consider all radio sources with a flux $>$ 1 mJy looking for
faint optical counterparts ($I > 21$), typical of radio galaxies in the
redshift region of interest. We take advantage of the COSMOS multiband survey
to select and identify their host galaxies with a careful visual inspection of
the multiwavelength counterparts. The wide spectral coverage from the FUV to
the MIR enables us to derive their SEDs and model them with our own code,
{\it 2SPD}, which includes two stellar population of different ages and dust
component(s). We obtained a sample formed by 74 members, most of them indeed
having $z\gtrsim1$.

The sample displays a large variety of radio and photometric properties, such
as, compact radio sources, FR~Is, FR~IIs and QSOs. We analyzed the properties
of the SEDs of the sample analogously to what done by B13. Here we summarize
the main properties of the entire high-z COSMOS RL AGN population and briefly
discuss them.

\begin{itemize}

\item The photometric redshifts of the sample range between $\sim0.7$ and 3,
  with a median of $z = 1.2$. Most of the sources have already a photo-z
    derivation from the COSMOS collaboration, but some do not mainly because
    their I-band magnitudes are beyond the $I=25$ limit of the COSMOS
    photometric redshift catalog. Our photo-z measurements are in agreement
  with those present in literature, but more robust because of the careful
  visual inspections of the multi-band counterparts. For those source not
  present in the COSMOS catalog, we provide new photo-z estimates.

\item Once we obtain the photo-z, we infer their radio luminosities starting
  from their FIRST and NVSS radio fluxes. The rest-frame 1.4-GHz radio
  power distribution of the sample covers the range from 10$^{31.5}$ to
  10$^{34.3}$ erg s$^{-1}$ Hz$^{-1}$, straddling the local FR~I/FR~II break
  ($L_{\rm 1.4 \,\,\, GHz} $= 10$^{32.6}$ erg s$^{-1}$ Hz$^{-1}$). Based on such a
  separation we divide the sample in low- and high-power (LP and HP) sources.
  The radio sources are mostly compact (or slightly resolved) with sizes
  smaller than a few tens of kpc. Nevertheless, 21 objects show FR~II radio
  morphology and are mostly HPs. Three QSOs show compact radio structure,
  while one is associated with the center of a FR~II.

\item The most robust result of the SED modeling is the derivation of the host
  stellar mass. The stellar content distribution covers values from
  $\sim$10$^{10}$ to 10$^{11.5}$ M$_{\odot}$ with a median value of 6 $\times$
  10$^{10}$ M$_{\odot}$. LP and HP sources have similar stellar mass
  distributions.

\item Most sources show SED dominated by an old stellar population with an age
  of $\sim$ 1-3 Gyr. However, significant excesses above the old stellar
  population are often observed in the IR and/or UV part of the SEDs.

\item A dust component is necessary to account for the 24$\mu$m emission in 32
  objects in the sample and significant infrared excesses even at shorter
  wavelengths are observed in 13 of these galaxies (not considering the four
  QSOs). The dust luminosities derived from these infrared excesses range from
  $\sim$10$^{43}$ to 10$^{45.5}$ erg s$^{-1}$. The dust temperatures estimated
  for these 13 objects are 350-1200 K, values dependent on the MIR spectral
  shape and redshifts.

\item The UV excess are present significantly in 30 sources and marginally in
  19 sources. Estimates of the UV luminosities measured at 2000 \AA\ (at rest
  frame) yield values in the range of 10$^{41.5}$-10$^{45.5}$ erg s$^{-1}$.

\item We test the statistical significance of the radio-IR-UV luminosity
  relations, taking into account censored statistics and the influence of the
  common dependence of luminosities on redshift. The UV emission is
  significantly correlated with both the IR and the radio luminosities, the
  former being the stronger link. Conversely, more doubts are present for the
  radio-IR relation.

\end{itemize}

It is important to further address at this stage the sample completeness and
purity, an essential issue for any future statistical study.

A potential worry concerns the completeness at low fluxes. Indeed, a radio
source with a total flux exceeding the adopted flux threshold of 1 mJy, might
be split into two separate components both below the flux limit. Such objects
would not be included causing incompleteness for radio fluxes between 1 and 2
mJy. We looked for sources in the COSMOS field between 0.6 (the FIRST catalog
flux limit) and 1 mJy, and then focused on the 20 objects detected by the
COSMOS-VLA images.  13 of them have clear optical/IR counterparts and thus are
unlikely to be one component of a double radio source. None of remaining 7
shows evident FIRST double morphology on a scale of 180 arcsec. This suggests
that the incompleteness caused by this effect is negligible.

A further problem appears to exist at higher radio fluxes. 20 FIRST radio
sources, including extended radio sources, are not part of the final sample:
10 of them were discarded because they are not detected in the VLA-COSMOS
survey, while in another 10 cases the association with an optical/IR
counterpart is not possible or univocal. On the other hand, the identification
of further 10 FR~II appears to be much more convincing. We do not expect more
than a few false positive among them. With this approach we favor the sample
purity over its completeness.  Another 6 objects were instead excluded,
despite a successful counterpart identification, because the obtained modeling
of their SEDs is less convincing than the others. We then conclude that while
no more than a few per cent of interlopers might be present (and all at high
radio fluxes), a fraction as high as $\sim$25\% of incompleteness may affect
the sample.

The total radio power distribution of the sample is rather broad and straddle
the luminosity separation between FR~Is and FR~IIs in the local Universe.
Note that the selected sample is, as planned, of much lower luminosity than
classical samples of radio sources at similar redshifts and comparable to
those of the local population of radio galaxies.  However, the radio
luminosities of our sample are far larger than the level at which starburst
activity may significantly contribute and this indicates that they are genuine
RL AGN. Unfortunately, we cannot establish the radio morphological
classification for most of the sources; furthermore, the separation between
the two FR types is less sharp at higher radio frequencies, considering that
the rest-frame of these observations is $\sim$ 3 GHz (at
z=1.2). Nonetheless, for the objects for which this is possible the radio
power generally predicts correctly their FR classification. In particular all
FR~IIs are HPs, with only 3 exceptions that still have radio luminosities
larger than $10^{32}$ erg s$^{-1}$ Hz$^{-1}$. Also local FR~II radio galaxies
exist at this lower radio power (e.g., \citealt{zirbel95,best05a}).

The hosts of these high-z low-luminosity radio sources are massive (likely,
early-type) galaxies, $M_* \sim$10$^{10}$ to 10$^{11.5}$ M$_{\odot}$. These
values are similar with those derived for ellipticals in the COSMOS field in a
similar range of redshifts \citep{ilbert10}. However, they display a wide
range of properties. On the one hand, roughly half of the sample appears to be
similar to those of local FR~Is which live in red massive early-type galaxies
(e.g., \citealt{zirbel96,best05b,smolcic09,baldi10a}).  Indeed, among the
least radio luminous objects, 28 sources are faint in both the UV and MIR
bands. On the other hand, significant excesses in the UV and/or in the MIR
band are often present. These excesses are observed in LP and HP sources, in
contrast to low-z FR~I which are typically UV faint and poor in dust (e.g.,
\citealt{chiaberge02b,baldi08,hardcastle09}).  These photometric properties
are, instead, more familiar to local FR~IIs, which typically show bluer color
(e.g., \citealt{baldi08,smolcic09}) and larger amount of dust (e.g.,
\citealt{dekoff00,dicken10}) than FR~Is. These two behaviors correspond, but
only approximatively, to the two radio classes, LPs and HPs. In fact, overall,
LPs (HPs) seem to better conform with the local FR~Is (FR~IIs) than HPs (LPs).

A key issue is to establish the origin of the MIR and UV emission,
i.e. whether they are produced by the AGN and/or by star formation within the
host. The statistical analysis indicates a connection between these two
processes. However, this does not directly imply that for each source the IR
and UV emission have the same origin.  If we focus on the 13 sources which
show the significant MIR excess (to which we add the 4 QSOs) the dust
temperatures crudely derived from the spectral indices indicate values that
largely exceed those measured in star forming galaxies (where $T_{\rm dust}
\lesssim 200$ K, e.g., \citealt{dunne01,hwang10}). This warrants that in these
objects we deal with a dust heating mechanism from the AGN, where $T_{\rm
  dust} \gtrsim 300$ K (e.g., \citealt{ogle06}). The UV properties of these
sources are complex: two of them are undetected, while the remaining galaxies
show a large scatter in the MIR/UV ratio. Furthermore, for our sample of radio
galaxies this ratio is $L_{\rm IR \,\, excess} / L_{\rm UV} \sim 1 - 100$, much
higher than typical for QSO \citep{elvis94}. This suggests that nuclear
obscuration also plays an important role, an indication also supported by
their resolved optical appearance. The observed UV emission might still be
nuclear in origin (e.g., due to scattering) but we cannot exclude a
contribution (even dominant) from star formation.

For the remaining 57 objects, the situation is even more complex, with the
only common feature of an absence of an excess in the IRAC bands. In fact,
some of them show a 24 $\mu$m detection but not always accompanied with an UV
excess, and viceversa. The upper limits on the dust temperature for the
galaxies detected in Spitzer are not sufficiently stringent to exclude an AGN
origin. However, for the 28 objects which are faint in UV and IR band, the
emission in such bands is likely ascribed to the galaxy and/or
synchrotron radiation from the jet, as seen in local FR~Is (e.g.,
\citealt{chiaberge02b,baldi08,baldi10b}).

The correlation between radio and UV indicates that a more powerful AGN has a
larger probability to be associated with bluer host, but it remains unclear
whether this is due to a larger star formation rate in the host or to a
brighter AGN. The connection between radio and IR correlation has a
significantly larger dispersion. A similar scatter is found also in the local
RL AGN population \citep{dicken09}; here, this is clearly related to the
different properties of the various classes of RL AGN.

In summary, the sample of the high-z RL AGN population in the COSMOS field
includes active galaxies in which the contribution from AGN and stellar
emission may largely different from one object to the other. Weak AGN
dominated by an old stellar population, bright quasar-like AGN, and star
formation are the fundamental ingredients for the 'recipe' to account for the
range of properties of the sample, as broad as their radio morphology variety.

A further study is clearly necessary to investigate the properties of this
sample which also includes information on the X-ray and radio core data,
already available from the COSMOS survey. Furthermore, a more detailed and
quantitative comparison of this sample of RL AGN at high redshifts with
samples of local and distant radio galaxies as well as with population of
quiescent galaxies at high $z$ is required to explore their differences and
the similarities in order to shed further light on the galaxy-BH evolution
across the cosmic time.

\begin{acknowledgements}

  R.D.B. acknowledges the financial support from SISSA (Trieste) and from the
  Technion Institute (Haifa). We are grateful to the anonymous referee for the
  extremely useful comments to improve the paper and Sara Calabrese for the
  helpful support. This work is primarily based on the COSMOS data.

\end{acknowledgements}

\bibliography{my}

\begin{thebibliography}{63}
\expandafter\ifx\csname natexlab\endcsname\relax\def\natexlab#1{#1}\fi

\bibitem[{{Akritas}(1989)}]{akritas89}
{Akritas}, M. 1989, {Aligned Rank Tests for Regression With Censored Data}
  (Penn State Dept. of Statistics Technical Report, 1989)

\bibitem[{{Baldi} \& {Capetti}(2008)}]{baldi08}
{Baldi}, R.~D. \& {Capetti}, A. 2008, \aap, 489, 989

\bibitem[{{Baldi} \& {Capetti}(2010)}]{baldi10a}
{Baldi}, R.~D. \& {Capetti}, A. 2010, \aap, 519, A48+

\bibitem[{{Baldi} {et~al.}(2013){Baldi}, {Chiaberge}, {Capetti},
  {Rodriguez-Zaurin}, {Deustua}, \& {Sparks}}]{baldi13}
{Baldi}, R.~D., {Chiaberge}, M., {Capetti}, A., {et~al.} 2013, \apj, 762, 30

\bibitem[{{Baldi} {et~al.}(2010){Baldi}, {Chiaberge}, {Capetti}, {Sparks},
  {Macchetto}, {O'Dea}, {Axon}, {Baum}, \& {Quillen}}]{baldi10b}
{Baldi}, R.~D., {Chiaberge}, M., {Capetti}, A., {et~al.} 2010, \apj, 725, 2426

\bibitem[{{Bardelli} {et~al.}(2010){Bardelli}, {Schinnerer}, {Smol{\v c}ic},
  {Zamorani}, {Zucca}, {Mignoli}, {Halliday}, {Kova{\v c}}, {Ciliegi},
  {Caputi}, {Koekemoer}, {Bongiorno}, {Bondi}, {Bolzonella}, {Vergani},
  {Pozzetti}, {Carollo}, {Contini}, {Kneib}, {Le F{\`e}vre}, {Lilly},
  {Mainieri}, {Renzini}, {Scodeggio}, {Coppa}, {Cucciati}, {de la Torre}, {de
  Ravel}, {Franzetti}, {Garilli}, {Iovino}, {Kampczyk}, {Knobel}, {Lamareille},
  {Le Borgne}, {Le Brun}, {Maier}, {Pell{\`o}}, {Peng}, {Perez-Montero},
  {Ricciardelli}, {Silverman}, {Tanaka}, {Tasca}, {Tresse}, {Abbas}, {Bottini},
  {Cappi}, {Cassata}, {Cimatti}, {Guzzo}, {Leauthaud}, {Maccagni}, {Marinoni},
  {McCracken}, {Memeo}, {Meneux}, {Oesch}, {Porciani}, {Scaramella}, {Capak},
  {Sanders}, {Scoville}, {Taniguchi}, \& {Jahnke}}]{bardelli10}
{Bardelli}, S., {Schinnerer}, E., {Smol{\v c}ic}, V., {et~al.} 2010, \aap, 511,
  A1

\bibitem[{{Becker} {et~al.}(1995){Becker}, {White}, \& {Helfand}}]{FIRST}
{Becker}, R.~H., {White}, R.~L., \& {Helfand}, D.~J. 1995, \apj, 450, 559

\bibitem[{{Bertin} \& {Arnouts}(1996)}]{bertin96}
{Bertin}, E. \& {Arnouts}, S. 1996, \aaps, 117, 393

\bibitem[{{Best} {et~al.}(2005{\natexlab{a}}){Best}, {Kauffmann}, {Heckman},
  {Brinchmann}, {Charlot}, {Ivezi{\'c}}, \& {White}}]{best05b}
{Best}, P.~N., {Kauffmann}, G., {Heckman}, T.~M., {et~al.} 2005{\natexlab{a}},
  \mnras, 362, 25

\bibitem[{{Best} {et~al.}(2005{\natexlab{b}}){Best}, {Kauffmann}, {Heckman}, \&
  {Ivezi{\'c}}}]{best05a}
{Best}, P.~N., {Kauffmann}, G., {Heckman}, T.~M., \& {Ivezi{\'c}}, {\v Z}.
  2005{\natexlab{b}}, \mnras, 362, 9

\bibitem[{{Bolzonella} {et~al.}(2000){Bolzonella}, {Miralles}, \&
  {Pell{\'o}}}]{bolzonella00}
{Bolzonella}, M., {Miralles}, J., \& {Pell{\'o}}, R. 2000, \aap, 363, 476

\bibitem[{{Bruzual} \& {Charlot}(2009)}]{bc09}
{Bruzual}, G. \& {Charlot}, S. 2009, private comunication

\bibitem[{{Calzetti} {et~al.}(2000){Calzetti}, {Armus}, {Bohlin}, {Kinney},
  {Koornneef}, \& {Storchi-Bergmann}}]{calzetti00}
{Calzetti}, D., {Armus}, L., {Bohlin}, R.~C., {et~al.} 2000, \apj, 533, 682

\bibitem[{{Capak} {et~al.}(2007){Capak}, {Aussel}, {Ajiki}, {McCracken},
  {Mobasher}, {Scoville}, {Shopbell}, {Taniguchi}, {Thompson}, {Tribiano},
  {Sasaki}, {Blain}, {Brusa}, {Carilli}, {Comastri}, {Carollo}, {Cassata},
  {Colbert}, {Ellis}, {Elvis}, {Giavalisco}, {Green}, {Guzzo}, {Hasinger},
  {Ilbert}, {Impey}, {Jahnke}, {Kartaltepe}, {Kneib}, {Koda}, {Koekemoer},
  {Komiyama}, {Leauthaud}, {Le Fevre}, {Lilly}, {Liu}, {Massey}, {Miyazaki},
  {Murayama}, {Nagao}, {Peacock}, {Pickles}, {Porciani}, {Renzini}, {Rhodes},
  {Rich}, {Salvato}, {Sanders}, {Scarlata}, {Schiminovich}, {Schinnerer},
  {Scodeggio}, {Sheth}, {Shioya}, {Tasca}, {Taylor}, {Yan}, \&
  {Zamorani}}]{capak07}
{Capak}, P., {Aussel}, H., {Ajiki}, M., {et~al.} 2007, \apjs, 172, 99

\bibitem[{{Capak} {et~al.}(2008){Capak}, {Aussel}, {Ajiki}, {McCracken},
  {Mobasher}, {Scoville}, {Shopbell}, {Taniguchi}, {Thompson}, {Tribiano},
  {Sasaki}, {Blain}, {Brusa}, {Carilli}, {Comastri}, {Carollo}, {Cassata},
  {Colbert}, {Ellis}, {Elvis}, {Giavalisco}, {Green}, {Guzzo}, {Hasinger},
  {Ilbert}, {Impey}, {Jahnke}, {Kartaltepe}, {Kneib}, {Koda}, {Koekemoer},
  {Komiyama}, {Leauthaud}, {Lefevre}, {Lilly}, {Liu}, {Massey}, {Miyazaki},
  {Murayama}, {Nagao}, {Peacock}, {Pickles}, {Porciani}, {Renzini}, {Rhodes},
  {Rich}, {Salvato}, {Sanders}, {Scarlata}, {Schiminovich}, {Schinnerer},
  {Scodeggio}, {Sheth}, {Shioya}, {Tasca}, {Taylor}, {Yan}, \&
  {Zamorani}}]{capak08}
{Capak}, P., {Aussel}, H., {Ajiki}, M., {et~al.} 2008, VizieR Online Data
  Catalog, 2284, 0

\bibitem[{{Chabrier}(2003)}]{chabrier03}
{Chabrier}, G. 2003, \pasp, 115, 763

\bibitem[{{Chiaberge} {et~al.}(2002){Chiaberge}, {Macchetto}, {Sparks},
  {Capetti}, {Allen}, \& {Martel}}]{chiaberge02b}
{Chiaberge}, M., {Macchetto}, F.~D., {Sparks}, W.~B., {et~al.} 2002, \apj, 571,
  247

\bibitem[{{Chiaberge} {et~al.}(2009){Chiaberge}, {Tremblay}, {Capetti},
  {Macchetto}, {Tozzi}, \& {Sparks}}]{chiaberge09}
{Chiaberge}, M., {Tremblay}, G., {Capetti}, A., {et~al.} 2009, \apj, 696, 1103

\bibitem[{{Condon} {et~al.}(1998){Condon}, {Cotton}, {Greisen}, {Yin},
  {Perley}, {Taylor}, \& {Broderick}}]{NVSS}
{Condon}, J.~J., {Cotton}, W.~D., {Greisen}, E.~W., {et~al.} 1998, \aj, 115,
  1693

\bibitem[{{de Koff} {et~al.}(2000){de Koff}, {Best}, {Baum}, {Sparks},
  {R{\"o}ttgering}, {Miley}, {Golombek}, {Macchetto}, \& {Martel}}]{dekoff00}
{de Koff}, S., {Best}, P., {Baum}, S.~A., {et~al.} 2000, \apjs, 129, 33

\bibitem[{{Dicken} {et~al.}(2009){Dicken}, {Tadhunter}, {Axon}, {Morganti},
  {Inskip}, {Holt}, {Gonz{\'a}lez Delgado}, \& {Groves}}]{dicken09}
{Dicken}, D., {Tadhunter}, C., {Axon}, D., {et~al.} 2009, \apj, 694, 268

\bibitem[{{Dicken} {et~al.}(2010){Dicken}, {Tadhunter}, {Axon}, {Robinson},
  {Morganti}, \& {Kharb}}]{dicken10}
{Dicken}, D., {Tadhunter}, C., {Axon}, D., {et~al.} 2010, \apj, 722, 1333

\bibitem[{{Dunne} \& {Eales}(2001)}]{dunne01}
{Dunne}, L. \& {Eales}, S.~A. 2001, \mnras, 327, 697

\bibitem[{{Elvis} {et~al.}(1994){Elvis}, {Wilkes}, {McDowell}, {Green},
  {Bechtold}, {Willner}, {Oey}, {Polomski}, \& {Cutri}}]{elvis94}
{Elvis}, M., {Wilkes}, B.~J., {McDowell}, J.~C., {et~al.} 1994, \apjs, 95, 1

\bibitem[{{Fabian} {et~al.}(2006){Fabian}, {Sanders}, {Taylor}, {Allen},
  {Crawford}, {Johnstone}, \& {Iwasawa}}]{fabian06}
{Fabian}, A.~C., {Sanders}, J.~S., {Taylor}, G.~B., {et~al.} 2006, \mnras, 366,
  417

\bibitem[{{Fanaroff} \& {Riley}(1974)}]{fanaroff74}
{Fanaroff}, B.~L. \& {Riley}, J.~M. 1974, \mnras, 167, 31P

\bibitem[{{Hardcastle} {et~al.}(2009){Hardcastle}, {Evans}, \&
  {Croston}}]{hardcastle09}
{Hardcastle}, M.~J., {Evans}, D.~A., \& {Croston}, J.~H. 2009, \mnras, 396,
  1929

\bibitem[{{Hewett} \& {Wild}(2010)}]{hewett10}
{Hewett}, P.~C. \& {Wild}, V. 2010, \mnras, 405, 2302

\bibitem[{{Hopkins} {et~al.}(2006){Hopkins}, {Hernquist}, {Cox}, {Di Matteo},
  {Robertson}, \& {Springel}}]{hopkins06a}
{Hopkins}, P.~F., {Hernquist}, L., {Cox}, T.~J., {et~al.} 2006, \apjs, 163, 1

\bibitem[{{Huertas-Company} {et~al.}(2007){Huertas-Company}, {Rouan},
  {Soucail}, {Le F{\`e}vre}, {Tasca}, \& {Contini}}]{huertas07}
{Huertas-Company}, M., {Rouan}, D., {Soucail}, G., {et~al.} 2007, \aap, 468,
  937

\bibitem[{{Hwang} {et~al.}(2010){Hwang}, {Elbaz}, {Magdis}, {Daddi},
  {Symeonidis}, {Altieri}, {Amblard}, {Andreani}, {Arumugam}, {Auld}, {Aussel},
  {Babbedge}, {Berta}, {Blain}, {Bock}, {Bongiovanni}, {Boselli}, {Buat},
  {Burgarella}, {Castro-Rodr{\'{\i}}guez}, {Cava}, {Cepa}, {Chanial}, {Chapin},
  {Chary}, {Cimatti}, {Clements}, {Conley}, {Conversi}, {Cooray},
  {Dannerbauer}, {Dickinson}, {Dominguez}, {Dowell}, {Dunlop}, {Dwek}, {Eales},
  {Farrah}, {Schreiber}, {Fox}, {Franceschini}, {Gear}, {Genzel}, {Glenn},
  {Griffin}, {Gruppioni}, {Halpern}, {Hatziminaoglou}, {Ibar}, {Isaak},
  {Ivison}, {Jeong}, {Lagache}, {Le Borgne}, {Le Floc'h}, {Lee}, {Lee}, {Lee},
  {Levenson}, {Lu}, {Lutz}, {Madden}, {Maffei}, {Magnelli}, {Mainetti},
  {Maiolino}, {Marchetti}, {Mortier}, {Nguyen}, {Nordon}, {O'Halloran},
  {Okumura}, {Oliver}, {Omont}, {Page}, {Panuzzo}, {Papageorgiou}, {Pearson},
  {P{\'e}rez-Fournon}, {Garc{\'{\i}}a}, {Poglitsch}, {Pohlen}, {Popesso},
  {Pozzi}, {Rawlings}, {Rigopoulou}, {Riguccini}, {Rizzo}, {Rodighiero},
  {Roseboom}, {Rowan-Robinson}, {Saintonge}, {Portal}, {Santini}, {Sauvage},
  {Schulz}, {Scott}, {Seymour}, {Shao}, {Shupe}, {Smith}, {Stevens}, {Sturm},
  {Tacconi}, {Trichas}, {Tugwell}, {Vaccari}, {Valtchanov}, {Vieira},
  {Vigroux}, {Wang}, {Ward}, {Wright}, {Xu}, \& {Zemcov}}]{hwang10}
{Hwang}, H.~S., {Elbaz}, D., {Magdis}, G., {et~al.} 2010, \mnras, 409, 75

\bibitem[{{Ilbert} {et~al.}(2009){Ilbert}, {Capak}, {Salvato}, {Aussel},
  {McCracken}, {Sanders}, {Scoville}, {Kartaltepe}, {Arnouts}, {Le Floc'h},
  {Mobasher}, {Taniguchi}, {Lamareille}, {Leauthaud}, {Sasaki}, {Thompson},
  {Zamojski}, {Zamorani}, {Bardelli}, {Bolzonella}, {Bongiorno}, {Brusa},
  {Caputi}, {Carollo}, {Contini}, {Cook}, {Coppa}, {Cucciati}, {de la Torre},
  {de Ravel}, {Franzetti}, {Garilli}, {Hasinger}, {Iovino}, {Kampczyk},
  {Kneib}, {Knobel}, {Kovac}, {Le Borgne}, {Le Brun}, {F{\`e}vre}, {Lilly},
  {Looper}, {Maier}, {Mainieri}, {Mellier}, {Mignoli}, {Murayama}, {Pell{\`o}},
  {Peng}, {P{\'e}rez-Montero}, {Renzini}, {Ricciardelli}, {Schiminovich},
  {Scodeggio}, {Shioya}, {Silverman}, {Surace}, {Tanaka}, {Tasca}, {Tresse},
  {Vergani}, \& {Zucca}}]{ilbert09}
{Ilbert}, O., {Capak}, P., {Salvato}, M., {et~al.} 2009, \apj, 690, 1236

\bibitem[{{Ilbert} {et~al.}(2010){Ilbert}, {Salvato}, {Le Floc'h}, {Aussel},
  {Capak}, {McCracken}, {Mobasher}, {Kartaltepe}, {Scoville}, {Sanders},
  {Arnouts}, {Bundy}, {Cassata}, {Kneib}, {Koekemoer}, {Le F{\`e}vre}, {Lilly},
  {Surace}, {Taniguchi}, {Tasca}, {Thompson}, {Tresse}, {Zamojski}, {Zamorani},
  \& {Zucca}}]{ilbert10}
{Ilbert}, O., {Salvato}, M., {Le Floc'h}, E., {et~al.} 2010, \apj, 709, 644

\bibitem[{{Isobe} {et~al.}(1990){Isobe}, {Feigelson}, {Akritas}, \&
  {Babu}}]{isobe90}
{Isobe}, T., {Feigelson}, E.~D., {Akritas}, M.~G., \& {Babu}, G.~J. 1990, \apj,
  364, 104

\bibitem[{{Jarosik} {et~al.}(2011){Jarosik}, {Bennett}, {Dunkley}, {Gold},
  {Greason}, {Halpern}, {Hill}, {Hinshaw}, {Kogut}, {Komatsu}, {Larson},
  {Limon}, {Meyer}, {Nolta}, {Odegard}, {Page}, {Smith}, {Spergel}, {Tucker},
  {Weiland}, {Wollack}, \& {Wright}}]{jarosik11}
{Jarosik}, N., {Bennett}, C.~L., {Dunkley}, J., {et~al.} 2011, \apjs, 192, 14

\bibitem[{{Koekemoer} {et~al.}(2007){Koekemoer}, {Aussel}, {Calzetti}, {Capak},
  {Giavalisco}, {Kneib}, {Leauthaud}, {Le F{\`e}vre}, {McCracken}, {Massey},
  {Mobasher}, {Rhodes}, {Scoville}, \& {Shopbell}}]{koekemoer07}
{Koekemoer}, A.~M., {Aussel}, H., {Calzetti}, D., {et~al.} 2007, \apjs, 172,
  196

\bibitem[{{Kroupa}(2001)}]{kroupa01}
{Kroupa}, P. 2001, \mnras, 322, 231

\bibitem[{{Lavalley} {et~al.}(1992){Lavalley}, {Isobe}, \&
  {Feigelson}}]{lavalley92}
{Lavalley}, M., {Isobe}, T., \& {Feigelson}, E. 1992, in Astronomical Society
  of the Pacific Conference Series, Vol.~25, Astronomical Data Analysis
  Software and Systems I, ed. D.~M. {Worrall}, C.~{Biemesderfer}, \&
  J.~{Barnes}, 245--+

\bibitem[{{Lilly} {et~al.}(2007){Lilly}, {Le F{\`e}vre}, {Renzini}, {Zamorani},
  {Scodeggio}, {Contini}, {Carollo}, {Hasinger}, {Kneib}, {Iovino}, {Le Brun},
  {Maier}, {Mainieri}, {Mignoli}, {Silverman}, {Tasca}, {Bolzonella},
  {Bongiorno}, {Bottini}, {Capak}, {Caputi}, {Cimatti}, {Cucciati}, {Daddi},
  {Feldmann}, {Franzetti}, {Garilli}, {Guzzo}, {Ilbert}, {Kampczyk}, {Kovac},
  {Lamareille}, {Leauthaud}, {Borgne}, {McCracken}, {Marinoni}, {Pello},
  {Ricciardelli}, {Scarlata}, {Vergani}, {Sanders}, {Schinnerer}, {Scoville},
  {Taniguchi}, {Arnouts}, {Aussel}, {Bardelli}, {Brusa}, {Cappi}, {Ciliegi},
  {Finoguenov}, {Foucaud}, {Franceschini}, {Halliday}, {Impey}, {Knobel},
  {Koekemoer}, {Kurk}, {Maccagni}, {Maddox}, {Marano}, {Marconi}, {Meneux},
  {Mobasher}, {Moreau}, {Peacock}, {Porciani}, {Pozzetti}, {Scaramella},
  {Schiminovich}, {Shopbell}, {Smail}, {Thompson}, {Tresse}, {Vettolani},
  {Zanichelli}, \& {Zucca}}]{lilly07}
{Lilly}, S.~J., {Le F{\`e}vre}, O., {Renzini}, A., {et~al.} 2007, \apjs, 172,
  70

\bibitem[{{Maraston}(2005)}]{ma05}
{Maraston}, C. 2005, \mnras, 362, 799

\bibitem[{{Mauch} \& {Sadler}(2007)}]{mauch07}
{Mauch}, T. \& {Sadler}, E.~M. 2007, \mnras, 375, 931

\bibitem[{{Ogle} {et~al.}(2006){Ogle}, {Whysong}, \& {Antonucci}}]{ogle06}
{Ogle}, P., {Whysong}, D., \& {Antonucci}, R. 2006, \apj, 647, 161

\bibitem[{{Richards} {et~al.}(2001{\natexlab{a}}){Richards}, {Fan},
  {Schneider}, {Vanden Berk}, {Strauss}, {York}, {Anderson}, {Anderson},
  {Annis}, {Bahcall}, {Bernardi}, {Briggs}, {Brinkmann}, {Brunner}, {Burles},
  {Carey}, {Castander}, {Connolly}, {Crocker}, {Csabai}, {Doi}, {Finkbeiner},
  {Friedman}, {Frieman}, {Fukugita}, {Gunn}, {Hindsley}, {Ivezi{\'c}}, {Kent},
  {Knapp}, {Lamb}, {Leger}, {Long}, {Loveday}, {Lupton}, {McKay}, {Meiksin},
  {Merrelli}, {Munn}, {Newberg}, {Newcomb}, {Nichol}, {Owen}, {Pier}, {Pope},
  {Richmond}, {Rockosi}, {Schlegel}, {Siegmund}, {Smee}, {Snir}, {Stoughton},
  {Stubbs}, {SubbaRao}, {Szalay}, {Szokoly}, {Tremonti}, {Uomoto}, {Waddell},
  {Yanny}, \& {Zheng}}]{richards01b}
{Richards}, G.~T., {Fan}, X., {Schneider}, D.~P., {et~al.} 2001{\natexlab{a}},
  \aj, 121, 2308

\bibitem[{{Richards} {et~al.}(2001{\natexlab{b}}){Richards}, {Weinstein},
  {Schneider}, {Fan}, {Strauss}, {Vanden Berk}, {Annis}, {Burles}, {Laubacher},
  {York}, {Frieman}, {Johnston}, {Scranton}, {Gunn}, {Ivezi{\'c}}, {Nichol},
  {Budav{\'a}ri}, {Csabai}, {Szalay}, {Connolly}, {Szokoly}, {Bahcall},
  {Ben{\'{\i}}tez}, {Brinkmann}, {Brunner}, {Fukugita}, {Hall}, {Hennessy},
  {Knapp}, {Kunszt}, {Lamb}, {Munn}, {Newberg}, \& {Stoughton}}]{richards01a}
{Richards}, G.~T., {Weinstein}, M.~A., {Schneider}, D.~P., {et~al.}
  2001{\natexlab{b}}, \aj, 122, 1151

\bibitem[{{Salpeter}(1955)}]{salpeter55}
{Salpeter}, E.~E. 1955, \apj, 121, 161

\bibitem[{{Salvato} {et~al.}(2009){Salvato}, {Hasinger}, {Ilbert}, {Zamorani},
  {Brusa}, {Scoville}, {Rau}, {Capak}, {Arnouts}, {Aussel}, {Bolzonella},
  {Buongiorno}, {Cappelluti}, {Caputi}, {Civano}, {Cook}, {Elvis}, {Gilli},
  {Jahnke}, {Kartaltepe}, {Impey}, {Lamareille}, {Le Floc'h}, {Lilly},
  {Mainieri}, {McCarthy}, {McCracken}, {Mignoli}, {Mobasher}, {Murayama},
  {Sasaki}, {Sanders}, {Schiminovich}, {Shioya}, {Shopbell}, {Silverman},
  {Smol{\v c}i{\'c}}, {Surace}, {Taniguchi}, {Thompson}, {Trump}, {Urry}, \&
  {Zamojski}}]{salvato09}
{Salvato}, M., {Hasinger}, G., {Ilbert}, O., {et~al.} 2009, \apj, 690, 1250

\bibitem[{{Sanders} {et~al.}(2007){Sanders}, {Salvato}, {Aussel}, {Ilbert},
  {Scoville}, {Surace}, {Frayer}, {Sheth}, {Helou}, {Brooke}, {Bhattacharya},
  {Yan}, {Kartaltepe}, {Barnes}, {Blain}, {Calzetti}, {Capak}, {Carilli},
  {Carollo}, {Comastri}, {Daddi}, {Ellis}, {Elvis}, {Fall}, {Franceschini},
  {Giavalisco}, {Hasinger}, {Impey}, {Koekemoer}, {Le F{\`e}vre}, {Lilly},
  {Liu}, {McCracken}, {Mobasher}, {Renzini}, {Rich}, {Schinnerer}, {Shopbell},
  {Taniguchi}, {Thompson}, {Urry}, \& {Williams}}]{sanders07}
{Sanders}, D.~B., {Salvato}, M., {Aussel}, H., {et~al.} 2007, \apjs, 172, 86

\bibitem[{{Schinnerer} {et~al.}(2004){Schinnerer}, {Carilli}, {Scoville},
  {Bondi}, {Ciliegi}, {Vettolani}, {Le F{\`e}vre}, {Koekemoer}, {Bertoldi}, \&
  {Impey}}]{schinnerer04}
{Schinnerer}, E., {Carilli}, C.~L., {Scoville}, N.~Z., {et~al.} 2004, \aj, 128,
  1974

\bibitem[{{Schinnerer} {et~al.}(2007){Schinnerer}, {Smol{\v c}i{\'c}},
  {Carilli}, {Bondi}, {Ciliegi}, {Jahnke}, {Scoville}, {Aussel}, {Bertoldi},
  {Blain}, {Impey}, {Koekemoer}, {Le Fevre}, \& {Urry}}]{schinnerer07}
{Schinnerer}, E., {Smol{\v c}i{\'c}}, V., {Carilli}, C.~L., {et~al.} 2007,
  \apjs, 172, 46

\bibitem[{{Schmitt}(1985)}]{schmitt85}
{Schmitt}, J.~H.~M.~M. 1985, \apj, 293, 178

\bibitem[{{Scoville} {et~al.}(2007){Scoville}, {Aussel}, {Brusa}, {Capak},
  {Carollo}, {Elvis}, {Giavalisco}, {Guzzo}, {Hasinger}, {Impey}, {Kneib},
  {LeFevre}, {Lilly}, {Mobasher}, {Renzini}, {Rich}, {Sanders}, {Schinnerer},
  {Schminovich}, {Shopbell}, {Taniguchi}, \& {Tyson}}]{scoville07}
{Scoville}, N., {Aussel}, H., {Brusa}, M., {et~al.} 2007, \apjs, 172, 1

\bibitem[{{Smol{\v c}i{\'c}}(2009)}]{smolcic09}
{Smol{\v c}i{\'c}}, V. 2009, \apjl, 699, L43

\bibitem[{{Smol{\v c}i{\'c}} {et~al.}(2008){Smol{\v c}i{\'c}}, {Schinnerer},
  {Scodeggio}, {Franzetti}, {Aussel}, {Bondi}, {Brusa}, {Carilli}, {Capak},
  {Charlot}, {Ciliegi}, {Ilbert}, {Ivezi{\'c}}, {Jahnke}, {McCracken},
  {Obri{\'c}}, {Salvato}, {Sanders}, {Scoville}, {Trump}, {Tremonti}, {Tasca},
  {Walcher}, \& {Zamorani}}]{smolcic08}
{Smol{\v c}i{\'c}}, V., {Schinnerer}, E., {Scodeggio}, M., {et~al.} 2008,
  \apjs, 177, 14

\bibitem[{{Spergel} {et~al.}(2003){Spergel}, {Verde}, {Peiris}, {Komatsu},
  {Nolta}, {Bennett}, {Halpern}, {Hinshaw}, {Jarosik}, {Kogut}, {Limon},
  {Meyer}, {Page}, {Tucker}, {Weiland}, {Wollack}, \& {Wright}}]{spergel03}
{Spergel}, D.~N., {Verde}, L., {Peiris}, H.~V., {et~al.} 2003, \apjs, 148, 175

\bibitem[{{Taniguchi} {et~al.}(2007){Taniguchi}, {Scoville}, {Murayama},
  {Sanders}, {Mobasher}, {Aussel}, {Capak}, {Ajiki}, {Miyazaki}, {Komiyama},
  {Shioya}, {Nagao}, {Sasaki}, {Koda}, {Carilli}, {Giavalisco}, {Guzzo},
  {Hasinger}, {Impey}, {LeFevre}, {Lilly}, {Renzini}, {Rich}, {Schinnerer},
  {Shopbell}, {Kaifu}, {Karoji}, {Arimoto}, {Okamura}, \& {Ohta}}]{taniguchi07}
{Taniguchi}, Y., {Scoville}, N., {Murayama}, T., {et~al.} 2007, \apjs, 172, 9

\bibitem[{{Trump} {et~al.}(2007){Trump}, {Impey}, {McCarthy}, {Elvis},
  {Huchra}, {Brusa}, {Hasinger}, {Schinnerer}, {Capak}, {Lilly}, \&
  {Scoville}}]{trump07}
{Trump}, J.~R., {Impey}, C.~D., {McCarthy}, P.~J., {et~al.} 2007, \apjs, 172,
  383

\bibitem[{{White} {et~al.}(2007){White}, {Helfand}, {Becker}, {Glikman}, \& {de
  Vries}}]{white07}
{White}, R.~L., {Helfand}, D.~J., {Becker}, R.~H., {Glikman}, E., \& {de
  Vries}, W. 2007, \apj, 654, 99

\bibitem[{{Willott} {et~al.}(2001){Willott}, {Rawlings}, {Blundell}, {Lacy}, \&
  {Eales}}]{willott01}
{Willott}, C.~J., {Rawlings}, S., {Blundell}, K.~M., {Lacy}, M., \& {Eales},
  S.~A. 2001, \mnras, 322, 536

\bibitem[{{Willott} {et~al.}(2003){Willott}, {Rawlings}, {Jarvis}, \&
  {Blundell}}]{willott03}
{Willott}, C.~J., {Rawlings}, S., {Jarvis}, M.~J., \& {Blundell}, K.~M. 2003,
  \mnras, 339, 173

\bibitem[{{Wilman} {et~al.}(2008){Wilman}, {Miller}, {Jarvis}, {Mauch},
  {Levrier}, {Abdalla}, {Rawlings}, {Kl{\"o}ckner}, {Obreschkow}, {Olteanu}, \&
  {Young}}]{wilman08}
{Wilman}, R.~J., {Miller}, L., {Jarvis}, M.~J., {et~al.} 2008, \mnras, 388,
  1335

\bibitem[{{York} {et~al.}(2000){York}, {Adelman}, {Anderson}, {Anderson},
  {Annis}, {Bahcall}, {Bakken}, {Barkhouser}, {Bastian}, {Berman}, {Boroski},
  {Bracker}, {Briegel}, {Briggs}, {Brinkmann}, {Brunner}, {Burles}, {Carey},
  {Carr}, {Castander}, {Chen}, {Colestock}, {Connolly}, {Crocker}, {Csabai},
  {Czarapata}, {Davis}, {Doi}, {Dombeck}, {Eisenstein}, {Ellman}, {Elms},
  {Evans}, {Fan}, {Federwitz}, {Fiscelli}, {Friedman}, {Frieman}, {Fukugita},
  {Gillespie}, {Gunn}, {Gurbani}, {de Haas}, {Haldeman}, {Harris}, {Hayes},
  {Heckman}, {Hennessy}, {Hindsley}, {Holm}, {Holmgren}, {Huang}, {Hull},
  {Husby}, {Ichikawa}, {Ichikawa}, {Ivezi{\'c}}, {Kent}, {Kim}, {Kinney},
  {Klaene}, {Kleinman}, {Kleinman}, {Knapp}, {Korienek}, {Kron}, {Kunszt},
  {Lamb}, {Lee}, {Leger}, {Limmongkol}, {Lindenmeyer}, {Long}, {Loomis},
  {Loveday}, {Lucinio}, {Lupton}, {MacKinnon}, {Mannery}, {Mantsch}, {Margon},
  {McGehee}, {McKay}, {Meiksin}, {Merelli}, {Monet}, {Munn}, {Narayanan},
  {Nash}, {Neilsen}, {Neswold}, {Newberg}, {Nichol}, {Nicinski}, {Nonino},
  {Okada}, {Okamura}, {Ostriker}, {Owen}, {Pauls}, {Peoples}, {Peterson},
  {Petravick}, {Pier}, {Pope}, {Pordes}, {Prosapio}, {Rechenmacher}, {Quinn},
  {Richards}, {Richmond}, {Rivetta}, {Rockosi}, {Ruthmansdorfer}, {Sandford},
  {Schlegel}, {Schneider}, {Sekiguchi}, {Sergey}, {Shimasaku}, {Siegmund},
  {Smee}, {Smith}, {Snedden}, {Stone}, {Stoughton}, {Strauss}, {Stubbs},
  {SubbaRao}, {Szalay}, {Szapudi}, {Szokoly}, {Thakar}, {Tremonti}, {Tucker},
  {Uomoto}, {Vanden Berk}, {Vogeley}, {Waddell}, {Wang}, {Watanabe},
  {Weinberg}, {Yanny}, \& {Yasuda}}]{york00}
{York}, D.~G., {Adelman}, J., {Anderson}, Jr., J.~E., {et~al.} 2000, \aj, 120,
  1579

\bibitem[{{Zirbel}(1996)}]{zirbel96}
{Zirbel}, E.~L. 1996, \apj, 473, 713

\bibitem[{{Zirbel} \& {Baum}(1995)}]{zirbel95}
{Zirbel}, E.~L. \& {Baum}, S.~A. 1995, \apj, 448, 521

\end{thebibliography}

\begin{appendix}
\section{Radio maps, host identifications and SEDs}

In this Appendix we include the Tables~A.1 and A.2 which present the final
magnitudes and fluxes of the multi-band counterparts of the radio sources of
the sample.  Figures~\ref{radio1}, \ref{hard}, and \ref{noident} show the
VLA-COSMOS radio maps and the optical/IR images of the sources, for which we
easily identify the host, for the FR~IIs lacking of a clear point-like
emission from the radio core, and for the sources we fail in the host
identification, respectively. Figure~\ref{sed1} collects the modelled SEDs of
the radio galaxies selected in this work.

\begin{table*}
\addtolength{\tabcolsep}{-2pt}
  \begin{center}
  \caption{COSMOS multiwavelength counterparts of the sample}
  \begin{tabular}{l|c|c|c|c|c|c|c|c|c|c|c|c}
\hline\hline
{\tiny ID} & {\tiny $u^{*}$} & {\tiny $B_{J}$} & {\tiny $g^{+}$}& {\tiny $V_{J}$}& {\tiny $r^{+}$}& {\tiny $i^{*}$}& {\tiny $i^{+}$}& {\tiny $F814W$}&
{\tiny $z^{+}$}& {\tiny $J$}& {\tiny $K_{S}$}& {\tiny $K$} \\
\hline
{\tiny  (1)} & {\tiny  (2)}  & {\tiny  (3)}  & {\tiny (4)} & {\tiny (5)}  & {\tiny (6)}  & {\tiny (7)}  & {\tiny (8)} & {\tiny (9)} &  {\tiny (10)} & {\tiny (11)} &
{\tiny (12)} & {\tiny (13)}  \\ 
\hline\hline
{\tiny 1}  &  {\tiny 23.88$\pm$0.04} & {\tiny 23.50$\pm$0.03$^{*}$} & {\tiny 23.54$\pm$0.03$^{*}$} & {\tiny 23.20$\pm$0.03} & {\tiny 22.91$\pm$0.02} & {\tiny 22.28$\pm$0.04} & {\tiny 22.28$\pm$0.02} & {\tiny  22.10$\pm$0.07} & {\tiny 21.50$\pm$0.01} & {\tiny 20.71$\pm$0.11$^{*}$} & {\tiny 19.21 $\pm$0.01} &  \\  
{\tiny 2}  &  {\tiny $<$27.40} & {\tiny $<$28.30} & {\tiny $<$28.00} & {\tiny $<$27.60} & {\tiny $<$27.45} & {\tiny $<$25.90} & {\tiny $<$26.72} & {\tiny $<$26.22} & {\tiny $<$26.20} & {\tiny $<$24.25} & {\tiny 22.85$\pm$0.10} &  \\ 
{\tiny 3}  &  {\tiny $<$27.10$^{*}$} & {\tiny 26.47$\pm$0.37$^{*}$} & {\tiny $<$26.02$^{*}$} & {\tiny  25.83$\pm$0.35$^{*}$} & {\tiny 25.81$\pm$0.30$^{*}$} & {\tiny 25.20$\pm$0.75$^{*}$} & {\tiny 24.97$\pm$0.20$^{*}$} & {\tiny 25.34$\pm$0.85} & {\tiny 24.53$\pm$0.40$^{*}$} & {\tiny 23.81$\pm$0.50$^{*}$}  & {\tiny 22.43$\pm$0.10$^{*}$}  &   \\ 
{\tiny 4}  &  {\tiny $<$27.40} & {\tiny $<$28.30} & {\tiny $<$28.00} & {\tiny  $<$27.60} & {\tiny $<$27.14$^{*}$} & {\tiny $<$25.55$^{*}$} & {\tiny 25.90$\pm$0.30$^{*}$} & {\tiny $<$25.34} & {\tiny $<$24.70$^{*}$} & {\tiny 23.90$\pm$0.40$^{*}$}  & {\tiny 22.06$\pm$0.20$^{*}$}  &   \\ 
{\tiny 5}  &  {\tiny 22.14$\pm$0.03$^{*}$} & {\tiny 22.04$\pm$0.02} & {\tiny 22.016$\pm$0.01} & {\tiny  21.59$\pm$0.01} & {\tiny 21.62$\pm$0.01} & {\tiny  21.19$\pm$0.03} & {\tiny 21.21$\pm$0.01} & {\tiny   21.01$\pm$0.04} & {\tiny 20.49$\pm$0.01} & {\tiny 20.01$\pm$0.10} & {\tiny 19.01$\pm$0.01} &  \\ 
{\tiny 6}  &  {\tiny 24.27$\pm$0.04} & {\tiny 24.27$\pm$0.05} & {\tiny  24.50$\pm$0.12$^{*}$} & {\tiny   24.02$\pm$0.05$^{*}$} & {\tiny 23.69$\pm$0.03$^{*}$} & {\tiny 23.31$\pm$0.22} & {\tiny 23.30$\pm$0.03} & {\tiny  23.37$\pm$0.18} & {\tiny 23.06$\pm$0.04} & {\tiny 22.33$\pm$0.23$^{*}$} & {\tiny 21.14$\pm$0.10$^{*}$} &  {\tiny 21.36$\pm$0.23} \\ 
{\tiny 7}  &  {\tiny $<$27.40} & {\tiny 26.89$\pm$0.24} & {\tiny $<$26.25$^{*}$} & {\tiny 25.97$\pm$0.14} & {\tiny 25.70$\pm$0.11} & {\tiny $<$25.03$^{*}$} & {\tiny 25.24$\pm$0.12$^{*}$} & {\tiny 25.23$\pm$0.96} & {\tiny 24.25$\pm$0.09} & {\tiny 23.13$\pm$0.20$^{*}$}  & {\tiny 21.51$\pm$0.03$^{*}$}  &    {\tiny 21.43$\pm$0.27} \\ 
{\tiny 8}  &  {\tiny 26.00$\pm$0.20$^{*}$} & {\tiny 25.08$\pm$0.08} & {\tiny 24.81$\pm$0.06} & {\tiny 23.63$\pm$0.03} & {\tiny  22.72$\pm$0.02} & {\tiny  21.44$\pm$0.03} & {\tiny 21.45$\pm$0.01} & {\tiny  21.18$\pm$0.04} & {\tiny 20.64$\pm$0.01} & {\tiny 20.16$\pm$0.20$^{*}$} & {\tiny 19.15$\pm$0.01} &  {\tiny 19.22$\pm$0.04} \\ 
{\tiny 9}  &  {\tiny $<$27.40} & {\tiny $<$26.70$^{*}$} & {\tiny $<$26.48$^{*}$} & {\tiny 26.43$\pm$0.35$^{*}$} & {\tiny 26.05$\pm$0.25$^{*}$} & {\tiny $<$25.40$^{*}$} & {\tiny 25.80$\pm$0.15$^{*}$} & {\tiny 24.85$\pm$0.80} & {\tiny 25.05$\pm$0.25$^{*}$} & {\tiny 23.80$\pm$0.35$^{*}$}  & {\tiny 21.17$\pm$0.11$^{*}$}  &   \\ 
{\tiny 10}  &  {\tiny $<$27.40} & {\tiny $<$28.30} & {\tiny $<$28.00} & {\tiny 27.60$\pm$0.50$^{*}$} & {\tiny 27.21$\pm$0.36} & {\tiny $<$25.90$^{*}$} & {\tiny 26.82$\pm$0.35} & {\tiny $<$26.24} & {\tiny $<$25.91$^{*}$} & {\tiny $<$24.70$^{*}$}  & {\tiny 23.53$\pm$0.30$^{*}$}  &   \\ 
{\tiny 11}  &  {\tiny 26.57$\pm$0.19} & {\tiny 26.04$\pm$0.15} & {\tiny 26.27$\pm$0.20} & {\tiny 25.48$\pm$0.10} & {\tiny  24.52$\pm$0.05} & {\tiny 23.22$\pm$0.20} & {\tiny 23.22$\pm$0.03} & {\tiny 22.86$\pm$0.13} & {\tiny  22.23$\pm$0.02} & {\tiny 21.43$\pm$0.15$^{*}$}  & {\tiny 20.30$\pm$0.01}  &  {\tiny 20.25$\pm$0.09} \\ 
{\tiny 12}  &  {\tiny $<$27.40} & {\tiny $<$28.30} & {\tiny $<$28.00} & {\tiny $<$27.19$^{*}$} & {\tiny  26.30$\pm$0.55} & {\tiny $<$25.50$^{*}$} & {\tiny 25.95$\pm$0.45$^{*}$} & {\tiny $<$25.20$^{*}$} & {\tiny  $<$25.21$^{*}$} & {\tiny $<$24.03$^{*}$}  & {\tiny 22.65$\pm$0.25}  & \\ 
{\tiny 13}  &  {\tiny $<$27.40} & {\tiny 26.93$\pm$0.26} & {\tiny 26.49$\pm$0.19} & {\tiny 26.22$\pm$0.17} & {\tiny  25.36$\pm$0.09} & {\tiny 24.14$\pm$0.37} & {\tiny 23.99$\pm$0.04} & {\tiny 23.96$\pm$0.24} & {\tiny  23.12$\pm$0.04} & {\tiny 22.40$\pm$0.15$^{*}$}  & {\tiny 21.41$\pm$0.03}  & {\tiny 21.63$\pm$0.32} \\ 
{\tiny 14}  &  {\tiny $<$27.40} & {\tiny 26.86$\pm$0.25} & {\tiny 26.73$\pm$0.23} & {\tiny 26.55$\pm$0.22} & {\tiny  25.81$\pm$0.12} & {\tiny 24.74$\pm$0.66} & {\tiny 24.58$\pm$0.06} & {\tiny 24.42$\pm$0.50} & {\tiny 23.73$\pm$0.06} & {\tiny 22.28$\pm$0.15$^{*}$}  & {\tiny 21.05$\pm$0.02}  & {\tiny 20.92$\pm$0.14} \\ 
{\tiny 15}  &  {\tiny $<$27.40} & {\tiny 26.33$\pm$0.17} & {\tiny 26.23$\pm$0.16} & {\tiny 25.07$\pm$0.07} & {\tiny   23.93$\pm$0.04} & {\tiny 22.66$\pm$0.12}  & {\tiny  22.66$\pm$0.02} & {\tiny 22.23$\pm$0.08} & {\tiny  21.67$\pm$0.02}  & {\tiny  20.98$\pm$0.15}  & {\tiny 19.91$\pm$0.01} & {\tiny 19.95$\pm$0.07}  \\ 
{\tiny 16}  &  {\tiny 26.24$\pm$0.40$^{*}$} & {\tiny  25.50$\pm$0.30$^{*}$} & {\tiny 24.92$\pm$0.35$^{*}$} & {\tiny   23.99$\pm$0.30$^{*}$} & {\tiny 23.13$\pm$0.25$^{*}$} & {\tiny  21.94$\pm$0.25$^{*}$} & {\tiny 21.87$\pm$0.20$^{*}$} & {\tiny 21.68$\pm$0.25$^{*}$} & {\tiny 21.12$\pm$0.15$^{*}$} & {\tiny 20.60$\pm$0.25$^{*}$} &  {\tiny 19.59$\pm$0.15$^{*}$} & {\tiny 19.48$\pm$0.20$^{*}$} \\ 
{\tiny 17}  &  {\tiny 26.05$\pm$0.13} & {\tiny 25.66$\pm$0.10$^{*}$} & {\tiny 25.50$\pm$0.11} & {\tiny  25.31$\pm$0.09} & {\tiny 25.03$\pm$0.08} & {\tiny 24.08$\pm$0.38} & {\tiny 24.08$\pm$0.05} & {\tiny  24.01$\pm$0.34} & {\tiny 23.38$\pm$0.05} & {\tiny  22.17$\pm$0.15$^{*}$} & {\tiny 20.94$\pm$0.02} &  \\ 
{\tiny 18}  &  {\tiny 24.72$\pm$0.06} & {\tiny 24.02$\pm$0.10$^{*}$} & {\tiny 23.84$\pm$0.04} & {\tiny  23.34$\pm$0.03} & {\tiny 23.02$\pm$0.02} & {\tiny 22.39$\pm$0.07} & {\tiny 22.38$\pm$0.02} & {\tiny  22.11$\pm$0.07} & {\tiny 21.69$\pm$0.02} & {\tiny 21.00$\pm$0.15$^{*}$} & {\tiny 19.74$\pm$0.01} &  {\tiny 19.70$\pm$0.06} \\ 
{\tiny 19}  &  {\tiny 26.79$\pm$0.35$^{*}$} & {\tiny 25.10$\pm$0.30$^{*}$} & {\tiny 24.40$\pm$0.40$^{*}$} & {\tiny  23.40$\pm$0.30$^{*}$} & {\tiny 22.54$\pm$0.25$^{*}$} & {\tiny 21.42$\pm$0.25$^{*}$} & {\tiny 21.52$\pm$0.20$^{*}$} & {\tiny  21.20$\pm$0.20$^{*}$} & {\tiny 20.87$\pm$0.20$^{*}$} & {\tiny 20.12$\pm$0.40$^{*}$} & {\tiny 19.03$\pm$0.20$^{*}$} &  \\ 
{\tiny 20}  &  {\tiny 26.29$\pm$0.16} & {\tiny 26.23$\pm$0.20$^{*}$} & {\tiny 26.18$\pm$0.17} & {\tiny 25.52$\pm$0.15$^{*}$} & {\tiny 24.91$\pm$0.15$^{*}$} & {\tiny 23.85$\pm$0.35} & {\tiny 23.81$\pm$0.15$^{*}$} & {\tiny  23.49$\pm$0.22} & {\tiny  23.00$\pm$0.15$^{*}$} & {\tiny 22.24$\pm$0.30$^{*}$} & {\tiny  21.21$\pm$0.12$^{*}$} &  {\tiny 21.09$\pm$0.17} \\ 
{\tiny 21}  &  {\tiny 26.480$\pm$0.35$^{*}$} & {\tiny 26.12$\pm$0.15$^{*}$} & {\tiny 25.50$\pm$0.30$^{*}$} & {\tiny  24.39$\pm$0.20$^{*}$} & {\tiny 23.58$\pm$0.15$^{*}$} & {\tiny 22.42$\pm$0.13$^{*}$} & {\tiny 22.40$\pm$0.10$^{*}$} & {\tiny  22.28$\pm$0.11$^{*}$} & {\tiny  21.79$\pm$0.016} & {\tiny 21.24$\pm$0.15$^{*}$} & {\tiny 20.18$\pm$0.01} & {\tiny 20.28$\pm$0.07} \\ 
{\tiny 22}  &  {\tiny 26.42$\pm$0.42$^{*}$} & {\tiny 25.55$\pm$0.20$^{*}$} & {\tiny 25.75$\pm$0.20$^{*}$} & {\tiny 25.64$\pm$0.20$^{*}$} & {\tiny   25.38$\pm$0.15$^{*}$} & {\tiny 25.13$\pm$0.40$^{*}$}  & {\tiny  24.91$\pm$0.10} & {\tiny 24.72$\pm$0.66} & {\tiny  25.02$\pm$0.20}  & {\tiny  24.08$\pm$0.50$^{*}$}  & {\tiny 23.12$\pm$0.15} &  \\ 
{\tiny 23}  &  {\tiny 25.84$\pm$0.12} & {\tiny 25.70$\pm$0.14} & {\tiny 25.69$\pm$0.21} & {\tiny  24.44$\pm$0.06} & {\tiny 23.67$\pm$0.04} & {\tiny 22.33$\pm$0.09} & {\tiny 22.42$\pm$0.02} & {\tiny 22.14$\pm$0.08} & {\tiny 21.68$\pm$0.02} & {\tiny 21.13$\pm$0.15$^{*}$} & {\tiny 20.21$\pm$0.01} &  \\ 
{\tiny 24}  &  {\tiny $<$26.90$^{*}$} & {\tiny 26.73$\pm$0.24} & {\tiny 26.40$\pm$0.18} & {\tiny 26.11$\pm$0.16} & {\tiny   25.77$\pm$0.11} & {\tiny $<$25.10$^{*}$}  & {\tiny  25.63$\pm$0.15} & {\tiny 25.40$\pm$0.40$^{*}$} & {\tiny  24.66$\pm$0.14}  & {\tiny  23.27$\pm$0.20$^{*}$}  & {\tiny 21.67$\pm$0.04} &   {\tiny 21.60$\pm$0.30$^{*}$}\\ 
{\tiny 25}  &  {\tiny 26.69$\pm$0.35$^{*}$} & {\tiny 24.95$\pm$0.25$^{*}$} & {\tiny 24.65$\pm$0.30$^{*}$} & {\tiny  23.44$\pm$0.25$^{*}$} & {\tiny 22.50$\pm$0.23$^{*}$} & {\tiny 21.20$\pm$0.25$^{*}$} & {\tiny 21.21$\pm$0.20$^{*}$} & {\tiny  20.98$\pm$0.15$^{*}$} & {\tiny  20.46$\pm$0.10$^{*}$} & {\tiny 19.90$\pm$0.17$^{*}$} & {\tiny 18.96$\pm$0.10$^{*}$} &  {\tiny 18.90$\pm$0.07} \\ 
{\tiny 26}  &  {\tiny 25.82$\pm$0.13} & {\tiny 24.86$\pm$0.07} & {\tiny  24.47$\pm$0.05} & {\tiny 23.24$\pm$0.03} & {\tiny 22.46$\pm$0.02} & {\tiny 21.12$\pm$0.03} & {\tiny  21.14$\pm$0.01} & {\tiny  20.97$\pm$0.03} & {\tiny 20.55$\pm$0.01} & {\tiny  19.99$\pm$0.15} & {\tiny 19.04$\pm$0.01} &  {\tiny 19.09$\pm$0.03} \\ 
{\tiny 27}  &  {\tiny 26.69$\pm$0.28} & {\tiny 26.32$\pm$0.16} & {\tiny 26.52$\pm$0.20} & {\tiny 26.41$\pm$0.19} & {\tiny 25.77$\pm$0.11} & {\tiny 24.70$^{*}$} & {\tiny 25.22$\pm$0.10} & {\tiny  24.74$\pm$0.59} & {\tiny 24.79$\pm$0.14} & {\tiny 24.20$\pm$0.45} & {\tiny 22.30$\pm$0.20} & {\tiny 22.31$\pm$0.41} \\ 
{\tiny 28}  &  {\tiny 26.20$\pm$0.35$^{*}$} & {\tiny 25.22$\pm$0.20$^{*}$} & {\tiny 25.15$\pm$0.25$^{*}$} & {\tiny  24.70$\pm$0.15$^{*}$} & {\tiny 24.59$\pm$0.15$^{*}$} & {\tiny 24.51$\pm$0.79} & {\tiny 24.27$\pm$0.10$^{*}$} &  & {\tiny  23.85$\pm$0.07} & {\tiny 22.83$\pm$0.15$^{*}$} & {\tiny 21.92$\pm$0.10$^{*}$} & {\tiny 21.78$\pm$0.33} \\ 
{\tiny 29}  &  {\tiny 19.89$\pm$0.01} & {\tiny 19.73$\pm$0.01} & {\tiny 19.67$\pm$0.01} & {\tiny 19.60$\pm$0.01} & {\tiny 19.48$\pm$0.01} & {\tiny 19.35$\pm$0.01} & & {\tiny  19.27$\pm$0.01} & {\tiny 19.42$\pm$0.01} & {\tiny  19.47$\pm$0.10$^{*}$} & {\tiny 19.05$\pm$0.01} &  \\ 
{\tiny 30}  &  {\tiny $<$27.40} & {\tiny $<$27.92$^{*}$} & {\tiny $<$27.45$^{*}$} & {\tiny $<$27.28$^{*}$} & {\tiny   $<$26.72$^{*}$} & {\tiny  $<$25.90} & {\tiny $<$26.10$^{*}$}  &  & {\tiny 25.33$\pm$0.60$^{*}$} & {\tiny  $<$24.00$^{*}$}  & {\tiny 22.86$\pm$0.25}  & \\ 
{\tiny 31}  &  {\tiny 27.29$\pm$0.43} & {\tiny 26.95$\pm$0.28} & {\tiny 26.84$\pm$0.26} & {\tiny 26.68$\pm$0.25} & {\tiny 26.17$\pm$0.15} & {\tiny 25.65$\pm$0.30$^{*}$} & {\tiny 25.84$\pm$0.17} &   & {\tiny 25.48$\pm$0.28} & {\tiny $<$23.50$^{*}$} & {\tiny 22.46$\pm$0.09} &  \\ 
{\tiny 32}  &  {\tiny $<$27.20} & {\tiny 26.85$\pm$0.65$^{*}$} & {\tiny $<$25.38$^{*}$} & {\tiny $<$25.22$^{*}$} & {\tiny  25.40$\pm$0.40$^{*}$} & {\tiny 24.12$\pm$0.43} & {\tiny 24.35$\pm$0.30$^{*}$} &  & {\tiny $<$23.11$^{*}$}  & {\tiny $<$22.90$^{*}$} & {\tiny 22.20$\pm$0.20$^{*}$}  & \\ 
{\tiny 33}  &  {\tiny $<$27.40} & {\tiny $<$28.30} & {\tiny $<$28.00} & {\tiny $<$27.60} & {\tiny  $<$26.49$^{*}$} & {\tiny $<$25.40$^{*}$}  &  &   & {\tiny $<$23.70$^{*}$} & {\tiny $<$22.90$^{*}$}  &  &\\ 
{\tiny 34}  &  {\tiny $<$27.03$^{*}$} & {\tiny 26.20$\pm$0.50$^{*}$} & {\tiny $<$25.62$^{*}$} & {\tiny 25.64$\pm$0.35$^{*}$} & {\tiny  25.40$\pm$0.25$^{*}$} & {\tiny 24.19$\pm$0.48} & {\tiny 24.07$\pm$0.15$^{*}$} &  & {\tiny 23.13$\pm$0.04}  & {\tiny 21.81$\pm$0.15$^{*}$} & {\tiny 20.56$\pm$0.02}  & {\tiny 20.63$\pm$0.14} \\ 
{\tiny 35}  &  {\tiny 22.72$\pm$0.020} & {\tiny 22.30$\pm$0.02$^{*}$} & {\tiny  22.60$\pm$0.02$^{*}$} & {\tiny 22.19$\pm$0.02} & {\tiny 22.05$\pm$0.02} & {\tiny 21.63$\pm$0.05} & & {\tiny  21.56$\pm$0.05} & {\tiny 21.49$\pm$0.01} & {\tiny 21.00$\pm$0.10$^{*}$} & {\tiny  20.33$\pm$0.01} &  {\tiny 20.37$\pm$0.11} \\ 
{\tiny 36}  &  {\tiny 26.84$\pm$0.22} & {\tiny 25.38$\pm$0.09} & {\tiny 25.44$\pm$0.09} & {\tiny  25.18$\pm$0.08} & {\tiny 25.08$\pm$0.07} & {\tiny 24.78$\pm$0.63} & {\tiny 24.89$\pm$0.08} & {\tiny  24.83$\pm$0.62} & {\tiny  24.38$\pm$0.11} & {\tiny 23.53$\pm$0.36$^{*}$} & {\tiny 22.14$\pm$0.06} &  \\ 
{\tiny 37}  &  {\tiny 19.65$\pm$0.01} & {\tiny 19.86$\pm$0.01} & {\tiny 19.55$\pm$0.01} & {\tiny 19.39$\pm$0.01} & {\tiny  19.30$\pm$0.01} & {\tiny 19.20$\pm$0.01} &  & {\tiny  19.16$\pm$0.01} & {\tiny 19.26$\pm$0.01} & {\tiny 18.88$\pm$0.10$^{*}$} & {\tiny 18.46$\pm$0.01} &  \\ 
{\tiny 38}  &  {\tiny 25.41$\pm$0.20$^{*}$} & {\tiny 24.60$\pm$0.20$^{*}$} & {\tiny 24.55$\pm$0.20$^{*}$} & {\tiny  24.29$\pm$0.15$^{*}$} & {\tiny 24.18$\pm$0.15$^{*}$} & {\tiny  23.86$\pm$0.32} & {\tiny 23.89$\pm$0.04} & {\tiny   24.09$\pm$0.36} & {\tiny 23.69$\pm$0.06} & {\tiny 23.80$\pm$0.50} & {\tiny 21.43$\pm$0.03} &  {\tiny 21.67$\pm$0.32} \\ 
{\tiny 39}  &  {\tiny $<$27.40} & {\tiny $<$28.30} & {\tiny $<$28.00} & {\tiny 27.40$\pm$0.35$^{*}$} & {\tiny 26.42$\pm$0.22} & {\tiny $<$25.30$^{*}$} & {\tiny 25.72$\pm$0.17} &  {\tiny 25.83$\pm$0.35$^{*}$}  &  {\tiny  25.36$\pm$0.29} & {\tiny 23.55$\pm$0.40$^{*}$}  & {\tiny 21.53$\pm$0.06} & {\tiny 21.57$\pm$0.29}\\
{\tiny 40}  &  {\tiny $<$27.40} & {\tiny 26.07$\pm$0.15} & {\tiny 26.11$\pm$0.30} & {\tiny 25.09$\pm$0.12} & {\tiny  24.17$\pm$0.18} & {\tiny 23.12$\pm$0.24} & {\tiny 23.06$\pm$0.03} &  {\tiny 22.70$\pm$0.12}  &  {\tiny  21.96$\pm$0.02} & {\tiny 21.13$\pm$0.15$^{*}$}  & {\tiny 20.23$\pm$0.01} & {\tiny 20.42$\pm$0.18}\\ 
{\tiny 41}  &  {\tiny  24.18$\pm$0.04} & {\tiny 23.81$\pm$0.04} & {\tiny 23.97$\pm$0.04} & {\tiny  23.59$\pm$0.03} & {\tiny 23.14$\pm$0.03} & {\tiny  22.41$\pm$0.07} & {\tiny 22.39$\pm$0.02} &  & {\tiny 21.84$\pm$0.02} & {\tiny 21.30$\pm$0.20$^{*}$} & {\tiny 20.62$\pm$0.01} & {\tiny 20.70$\pm$0.17} \\ 
{\tiny 42}  &  {\tiny 26.45$\pm$0.25$^{*}$} & {\tiny 25.70$\pm$0.15$^{*}$} & {\tiny  25.35$\pm$0.30$^{*}$} & {\tiny  24.85$\pm$0.11$^{*}$} & {\tiny 24.47$\pm$0.10$^{*}$} & {\tiny 23.25$\pm$0.20} & {\tiny 23.29$\pm$0.10$^{*}$} & {\tiny 22.94$\pm$0.14} & {\tiny 22.28$\pm$0.10$^{*}$} & {\tiny 21.30$\pm$0.20$^{*}$} & {\tiny 20.13$\pm$0.11$^{*}$} &  {\tiny 20.25$\pm$0.14} \\ 
{\tiny 43}  &  {\tiny 26.61$\pm$0.35$^{*}$} & {\tiny 25.89$\pm$0.13} & {\tiny 25.55$\pm$0.20$^{*}$} & {\tiny 25.21$\pm$0.08} & {\tiny  24.24$\pm$0.05} & {\tiny 23.03$\pm$0.17} & {\tiny 22.97$\pm$0.02} &  {\tiny 22.64$\pm$0.11}  &  {\tiny  21.954$\pm$0.02} & {\tiny 20.89$\pm$0.04}  & {\tiny 19.76$\pm$0.01} & {\tiny 19.91$\pm$0.08}\\
{\tiny 44}  &  {\tiny 24.73$\pm$0.15$^{*}$} & {\tiny 24.44$\pm$0.12$^{*}$} & {\tiny 24.61$\pm$0.12$^{*}$} & {\tiny  24.20$\pm$0.11$^{*}$} & {\tiny 23.80$\pm$0.10$^{*}$} & {\tiny 22.97$\pm$0.16} & {\tiny  22.95$\pm$0.02} & {\tiny  22.74$\pm$0.12} & {\tiny 2.25$\pm$0.02} & {\tiny 21.33$\pm$0.15$^{*}$} & {\tiny 20.31$\pm$0.01} &  {\tiny  20.42$\pm$0.09} \\ 
{\tiny 45}  &  {\tiny 26.05$\pm$0.15} & {\tiny 25.61$\pm$0.11} & {\tiny 25.62$\pm$0.11} & {\tiny 24.79$\pm$0.06} & {\tiny  23.83$\pm$0.04} & {\tiny 22.62$\pm$0.11} & {\tiny 22.59$\pm$0.02} &  & {\tiny 21.63$\pm$0.01} & {\tiny 20.95$\pm$0.06$^{*}$} & {\tiny 19.90$\pm$0.01} &  {\tiny 19.91$\pm$0.09} \\ 
{\tiny 46}  &  {\tiny 26.27$\pm$0.19} & {\tiny  25.41$\pm$0.10} & {\tiny 25.39$\pm$0.09} & {\tiny  24.49$\pm$0.05} & {\tiny 23.74$\pm$0.03} & {\tiny 22.62$\pm$0.11} & {\tiny 22.57$\pm$0.02} & {\tiny  22.40$\pm$0.09} & {\tiny 21.81$\pm$0.02} & {\tiny 21.25$\pm$0.11$^{*}$} & {\tiny 20.34$\pm$0.01} &  {\tiny 20.42$\pm$0.12} \\ 
%{\tiny COSMOS-FR~I~7}  &  {\tiny $<$27.40} & {\tiny $<$28.30} & {\tiny $<$28.30} & {\tiny $<$27.60} & {\tiny $<$27.30} & {\tiny $<$25.90} & {\tiny $<$27.20} & {\tiny $<$26.24} & {\tiny $<$25.86} & {\tiny $<$24.41} & {\tiny 23.15$\pm$0.18} &  \\ 
\hline                                                                    
\end{tabular}
\end{center}
\label{tabmag}

\small{Column description: (1) ID number of the object; (2) CFHT
  $u^{*}$ magnitude with its error; (3)-(4)-(5)-(6) Subaru $B_{J}$, $g^{+}$,
  $V_{J}$, $r^{+}$ magnitudes with their errors; (7) CFHT $i^{*}$ magnitude
  with its error; (8) Subaru $i^{+}$ magnitude with its error; (9) HST/ACS
  $F814W$ magnitude with its error; (10) Subaru $z^{+}$ magnitude with its
  error; (11) UKIRT $J$ magnitude with its error; (12) CFHT $K$ magnitude with
  its error; (13) NOAO$K_{S}$ with its error\footnote{Since the CFHT telescope
    is more sensitive and has a higher resolution than the images from the
    NOAO telescopes, this usually results in far smaller errors for the K
    -band magnitudes. In such cases, we prefer to use only the CFHT K -band
    data.}. The values marked by $*$ are measured by our 3\arcsec-aperture
  photometry on the images. }
\end{table*}

\begin{table*}
  \begin{center}
  \caption{COSMOS GALEX and Spitzer counterparts of the sample}
  \begin{tabular}{r|cc|cccc|c}
    \hline \hline
 ID & $FUV$ & $NUV$ & $IRAC1$ & $IRAC2$  & $IRAC3$ & $IRAC4$ & $MIPS$ \\
\hline
1  &                 &  25.17$\pm$0.60$^{*}$  &  415.89 $\pm$ 0.46 &  835.39$\pm$0.73 & 1431.50$\pm$1.92 &  2070.05$\pm$2.51 & 7.20$\pm$0.06      \\	 
2  &                 &                                       &  7.48$\pm$0.15 & 9.79$\pm$0.27 &  7.14$\pm$0.98 & 10.32$\pm$2.22  &  $<$0.15         \\	 
3  &                 &                                       & 11.96$\pm$0.15 & 14.94$\pm$0.26 & 13.26$\pm$0.91 &  9.02$\pm$2.13$^{*}$ &  0.22$\pm$0.02        \\	 
4  &                 &                                       &  11.37 $\pm$0.15 & 15.10$\pm$0.26 &  24.18$\pm$0.96 & 50.98$\pm$2.09 & 0.89$\pm$0.01     \\	 
5  &                 & 23.44$\pm$0.40$^{*}$  &   179.79$\pm$0.29 & 149.72$\pm$0.35 & 121.56$\pm$1.05 & 122.63$\pm$2.12 & 0.48$\pm$0.02          \\	 
6  &                 &                                      &  25.97$\pm$0.16 & 29.95$\pm$0.25 &  29.85$\pm$1.00 &  23.57$\pm$1.93 &  0.16$\pm$0.04$^{*}$   \\	 
7  &                 &                                      &  24.54$\pm$0.16 & 28.28$\pm$0.25 &  28.45$\pm$1.00 & 13.74$\pm$1.92 &  $<$0.15          \\	 
8  &                 &                                     &  120.53$\pm$1.25$^{*}$ &  90.79$\pm$1.32$^{*}$ & 70.25$\pm$1.94$^{*}$ & 49.12$\pm$3.50$^{*}$ & $<$0.15        \\	 
9  &                 &                                     &  10.18$\pm$0.15 & 12.21$\pm$0.25 & 13.58$\pm$0.91 & 16.94$\pm$2.01 & $<$0.15     \\	 
10&                 &                                     &  6.66$\pm$0.14 & 10.09$\pm$0.24 &  13.51$\pm$0.89 & 33.11$\pm$2.04 & 0.23$\pm$0.02        \\	 
11&                 &                                     & 55.75$\pm$0.19 & 42.22$\pm$0.28 & 24.53$\pm$0.94 &  12.45$\pm$2.06 &  $<$0.15        \\	 
12&                 &                                     &   9.46$\pm$0.15 &  11.29$\pm$0.24 & 10.95$\pm$0.94 &  8.76$\pm$1.98 &  $<$0.08        \\	 
13&                 &                                     &  17.10$\pm$0.15 &  14.62$\pm$0.27 & 7.18$\pm$0.92 &  7.88$\pm$2.22 & $<$0.08       \\	 
14&                 &                                      &  31.60$\pm$0.16 &  39.58$\pm$0.26 & 27.60$\pm$0.93 &  22.73$\pm$1.97 & 0.19$\pm$0.03$^{*}$       \\	 
15 &                 &                                    &  62.48$\pm$0.18 & 46.91$\pm$0.30 & 28.36$\pm$0.89 & 22.13$\pm$2.21 & $<$0.08        \\	 
16 &                 &                                   & 73.50$\pm$0.40$^{*}$   & 50.67$\pm$0.50$^{*}$   & 40.40$\pm$1.60$^{*}$   & 25.54$\pm$3.16$^{*}$   &  0.15$\pm$0.02          \\	 
17  &                 &                                  & 30.72$\pm$1.75$^{*}$ &  35.95$\pm$1.50$^{*}$ &  28.05$\pm$1.90$^{*}$ & 24.00$\pm$2.20$^{*}$ & $<$0.15         \\	 
18  &                 &                                  & 93.82$\pm$0.35$^{*}$ &  89.19$\pm$0.43$^{*}$ & 83.76$\pm$1.13 &  133.58$\pm$2.46 & 1.43$\pm$0.01    \\	 
19  &                 &                                  & 144.75$\pm$0.26 & 115.41$\pm$0.30 & 75.15$\pm$1.04 &  48.2$\pm$1.98 & 0.30$\pm$0.02          \\	 
20  &                 &                                  & 22.54$\pm$0.40$^{*}$  & 20.05$\pm$0.50$^{*}$  & 16.15$\pm$2.00$^{*}$  & 15.00$\pm$2.50$^{*}$  & $<$0.15      \\	 
21  &                 &                                  & 44.56$\pm$0.18 & 30.88$\pm$0.25 & 17.57$\pm$0.99 & 22.25$\pm$2.06 & $<$0.15        \\	 
22  &                 &                                  & 7.28$\pm$0.14 &  7.95$\pm$0.26 & $<$9.81$^{*}$ & 14.23$\pm$2.15 &  0.16$\pm$0.03$^{*}$         \\	 
23  &                 &                                  & 41.87$\pm$0.16 & 31.54$\pm$0.27 & 26.96$\pm$0.79 &13.40$\pm$2.05 & $<$0.15       \\	 
24  &                 &                                 &  22.21$\pm$0.16 & 27.75$\pm$0.29 & 37.36$\pm$0.84 &  27.45$\pm$2.40 &  0.67$\pm$0.05$^{*}$        \\	 
25  &                 &                                 &  144.73$\pm$0.26 &  96.56$\pm$0.33 & 75.37$\pm$1.03 & 34.72$\pm$2.45 & 0.17$\pm$0.05$^{*}$   \\	 
26  &                 &                                 &  109.67$\pm$0.22 & 66.80$\pm$0.27 & 46.35$\pm$0.93 &  23.93$\pm$1.94 & $<$0.15      \\	 
27  &                 &                                 &  11.29$\pm$0.13 & 13.09$\pm$0.22 &  15.46$\pm$0.77 &  5.99$\pm$1.68 &  $<$0.15        \\	 
28  &                 &                                 & 10.48$\pm$0.14 &  13.85$\pm$0.21 & 16.98$\pm$0.88 & 22.97$\pm$1.66 & 0.13$\pm$0.03$^{*}$    \\
29  & 24.07$\pm$0.09 & 20.62$\pm$0.02 & 140.90$\pm$0.23 & 226.38$\pm$0.37 & 316.10$\pm$1.02 & 446.18$\pm$2.18 & 1.49$\pm$0.02    \\	 
30  &                 &                                &  7.85$\pm$0.16 & 9.91$\pm$0.23 &  15.63$\pm$1.08 & 14.24$\pm$1.78 & $<$0.15         \\	 
31  &                 &                                &  10.59$\pm$0.16 & 15.30$\pm$0.26 & 14.69$\pm$0.97 & 24.92$\pm$2.23 & $<$0.15    \\	 
32  &                 &                                &  11.63$\pm$0.14 &  9.35$\pm$0.23 & 8.76$\pm$0.84 & 13.57$\pm$1.87 &  $<$0.15       \\	 
33  &                 &                               & 19.68$\pm$0.19 &   &  17.11$\pm$1.17 &  & $<$0.30   \\	 
34  &                 &                               & 62.26$\pm$0.46 & 88.70$\pm$0.58 & 96.12$\pm$1.16 &  108.93$\pm$2.14 &  0.27$\pm$0.02    \\	 
35  &                 &  23.42$\pm$0.06  & 52.47$\pm$0.20 & 71.41$\pm$0.30 & 81.98$\pm$1.10 & 100.70$\pm$2.17 &  0.40$\pm$0.02        \\	 
36  &                 &                               & 14.83$\pm$0.14 & 16.54$\pm$0.26 & 19.27$\pm$0.88 & 13.52$\pm$2.05 &  $<$0.15   \\	 
37  &                 &  21.15$\pm$0.02  &  258.63$\pm$0.39 & 405.22$\pm$0.55 & 594.43$\pm$1.54 & 794.89$\pm$2.69 & 2.59$\pm$0.02    \\	 
38  &                 &                              & 23.74$\pm$0.18 & 27.74$\pm$0.27 & 33.27$\pm$1.04 & 22.81$\pm$2.06 & 0.32$\pm$0.02       \\	 
39  &                 &                              & 16.22$\pm$0.16 & 17.71$\pm$0.27 & 17.13$\pm$0.99 & 10.73$\pm$2.15 & $<$0.15         \\	 
40  &                 &                              &  53.68$\pm$0.19 & 42.13$\pm$0.29 & 22.46$\pm$1.07 & 16.74$\pm$2.31 & $<$0.15      \\	 
41  &                 &  24.05$\pm$0.09  & 38.92$\pm$0.17 & 38.52$\pm$0.26 & 52.85$\pm$0.93 & 167.18$\pm$2.02 & 1.73$\pm$0.14    \\	 
42  &                 &                              & 55.38$\pm$4.00$^{*}$  & 49.06$\pm$5.00$^{*}$  & 30.02$\pm$4.00$^{*}$  & 27.03$\pm$5.00$^{*}$  & $<$0.15         \\	 
43  &                 &                              & 79.54$\pm$2.00$^{*}$   & 69.96$\pm$3.00$^{*}$  & 46.10$\pm$3.00$^{*}$ &  36.23$\pm$4.00$^{*}$  & $<$0.15      \\	 
44  &                 &  24.50$\pm$0.25$^{*}$ & 49.14$\pm$0.18 & 44.33$\pm$0.28 & 36.46$\pm$0.97 & 65.43$\pm$2.26 & 0.84$\pm$0.10           \\	 
45  &                 &                             & 66.08$\pm$0.19 &  51.05$\pm$0.25 & 34.95$\pm$0.94 & 24.60$\pm$1.78 &  $<$0.15        \\	 
46  &                 &                             & 36.56$\pm$0.32$^{*}$  & 25.53$\pm$0.48$^{*}$ & 17.10$\pm$1.00$^{*}$ & 12.70$\pm$2.00$^{*}$ &  $<$0.15      \\	 
%COSMOS-FR~7  &    &                         &  7.72$\pm$0.18$^{*}$  &  9.29$\pm$0.32$^{*}$   & 10.95$\pm$1.18$^{*}$  & 9.36$\pm$1.44$^{*}$  & 0.20$\pm$0.02         \\
\hline
\end{tabular}
\end{center}
  \label{tabmag2}
Column description: (1) ID number of the object; (2)-(3)
GALEX FUV and NUV magnitudes with their errors; (4)-(5)-(6)-(7) Spitzer/IRAC
4-channel (3.6, 4.5, 5.8, and 8.0 $\mu$m) fluxes with their
errors; (8) Spitzer/MIPS flux at 24$\mu$m with its error. The values marked by
$*$ are measured by our 3\arcsec-aperture photometry on the images.
\end{table*}

\begin{figure*}
\includegraphics[scale=0.80]{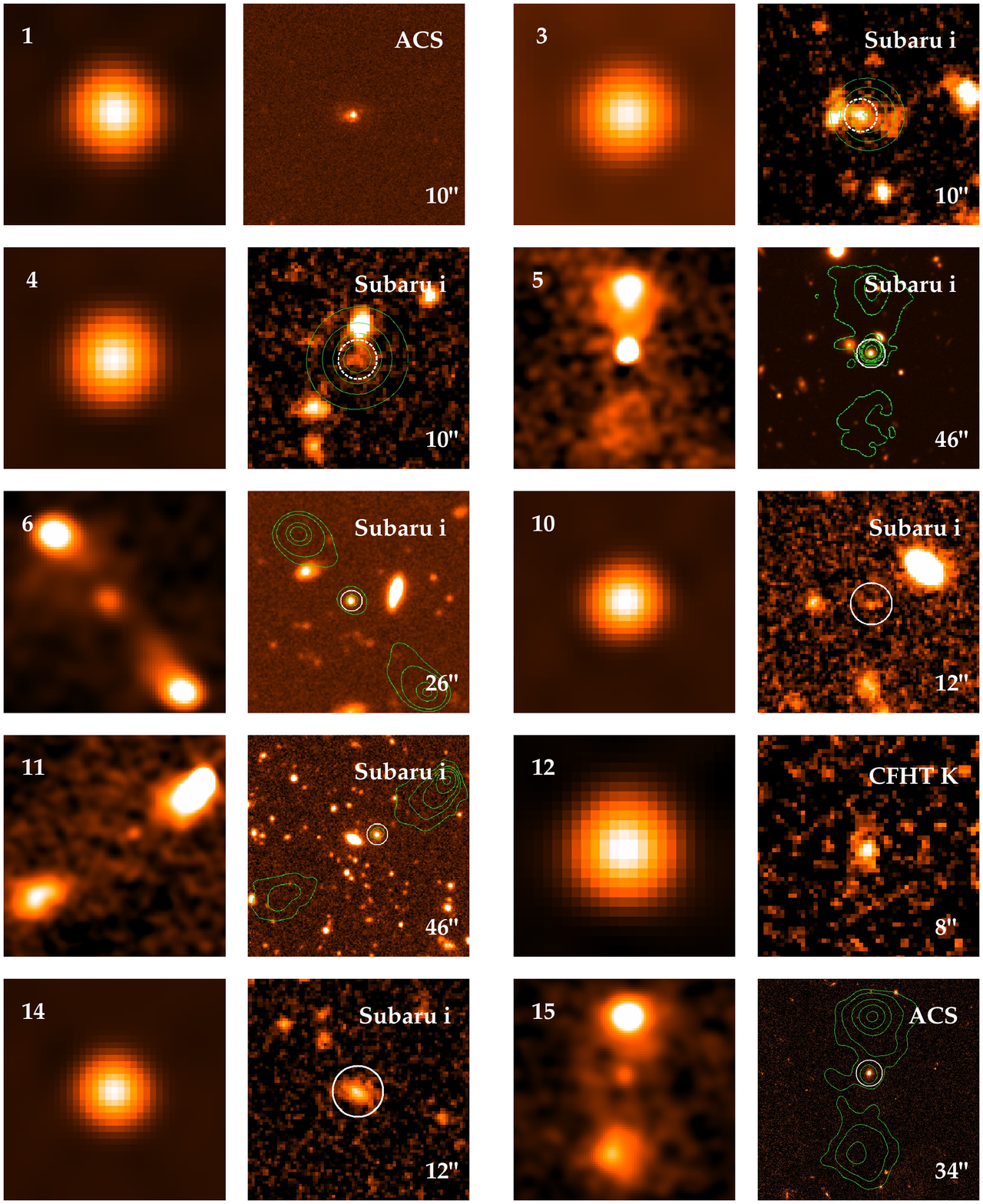}
\caption{Images of the radio sources selected in the COSMOS field associated
  with a host galaxy with $I >$ 21 (see Table~\ref{1table}). The left image
  for each source is from the VLA-COSMOS survey; the right image shows radio contours and
  the host identification in the band labelled on top. The image size in
  arcsec is marked on the bottom of each panel. When necessary we mark the
  identified host galaxy with a white circle.}
\label{radio1}
\end{figure*}

\begin{figure*}
\addtocounter{figure}{-1}
\includegraphics[scale=0.80]{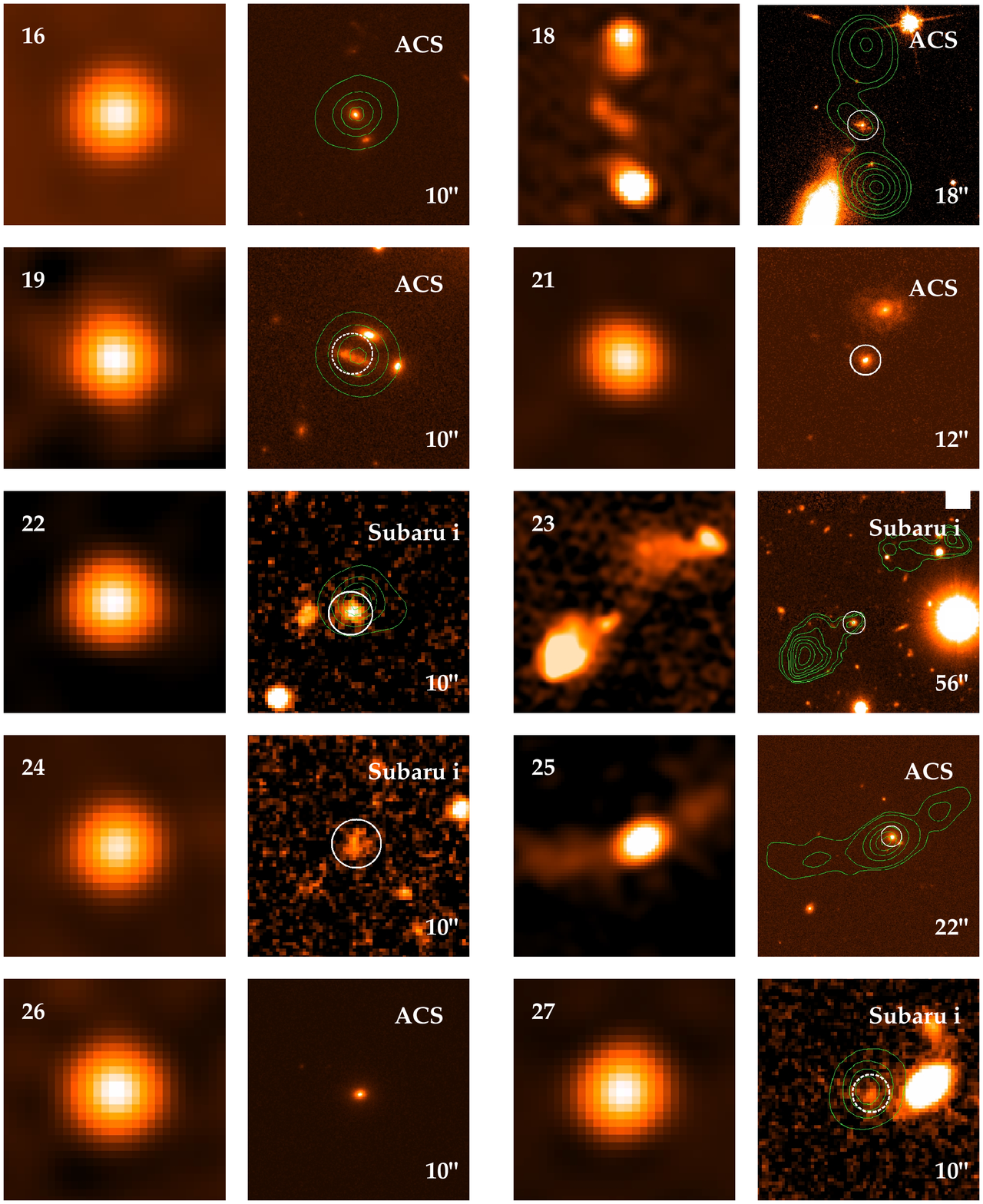}
\caption{CONTINUED}
\end{figure*}

\begin{figure*}
\addtocounter{figure}{-1}
\includegraphics[scale=0.80]{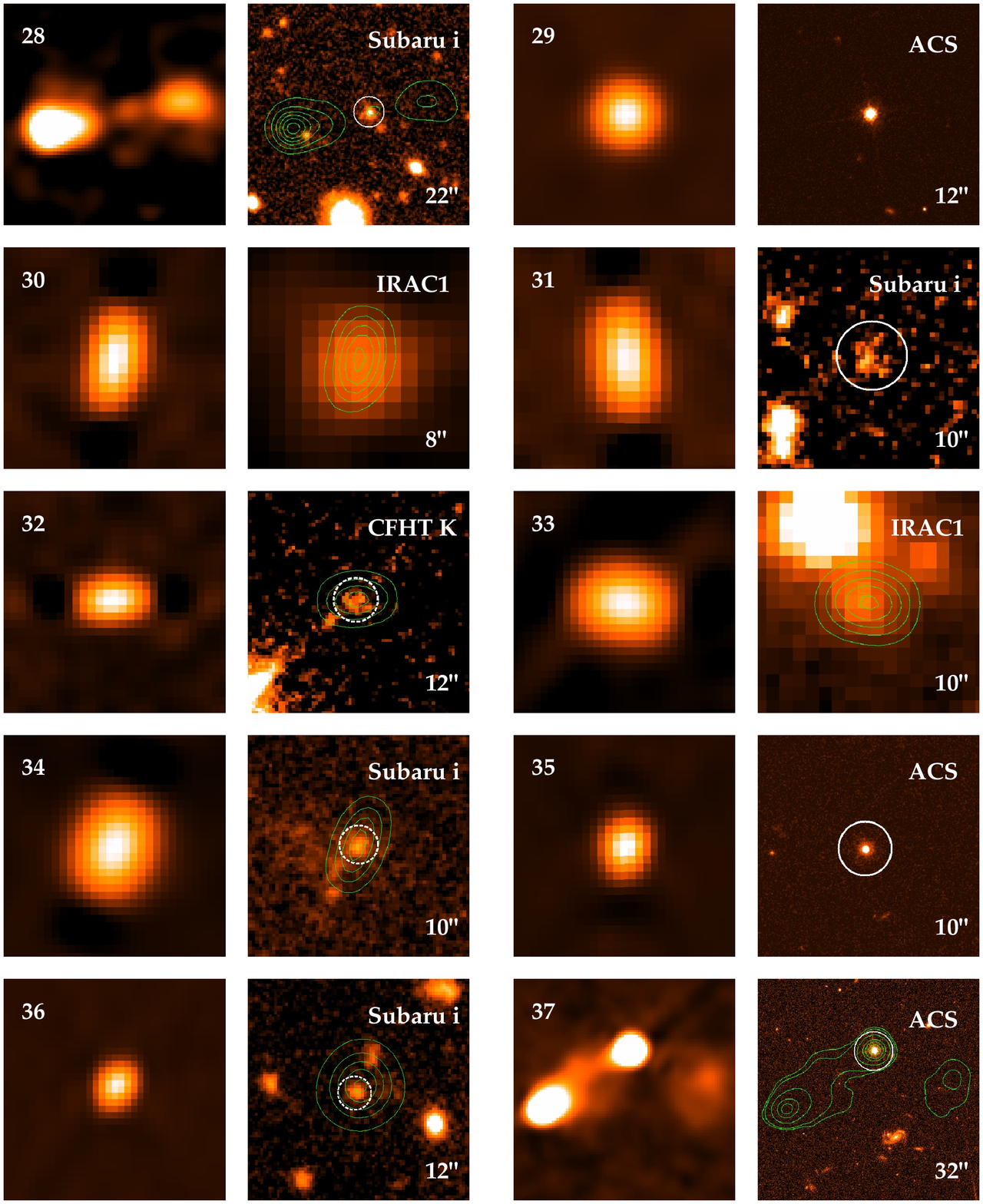}
\caption{CONTINUED}
\end{figure*}

\begin{figure*}
\addtocounter{figure}{-1}
\includegraphics[scale=0.80]{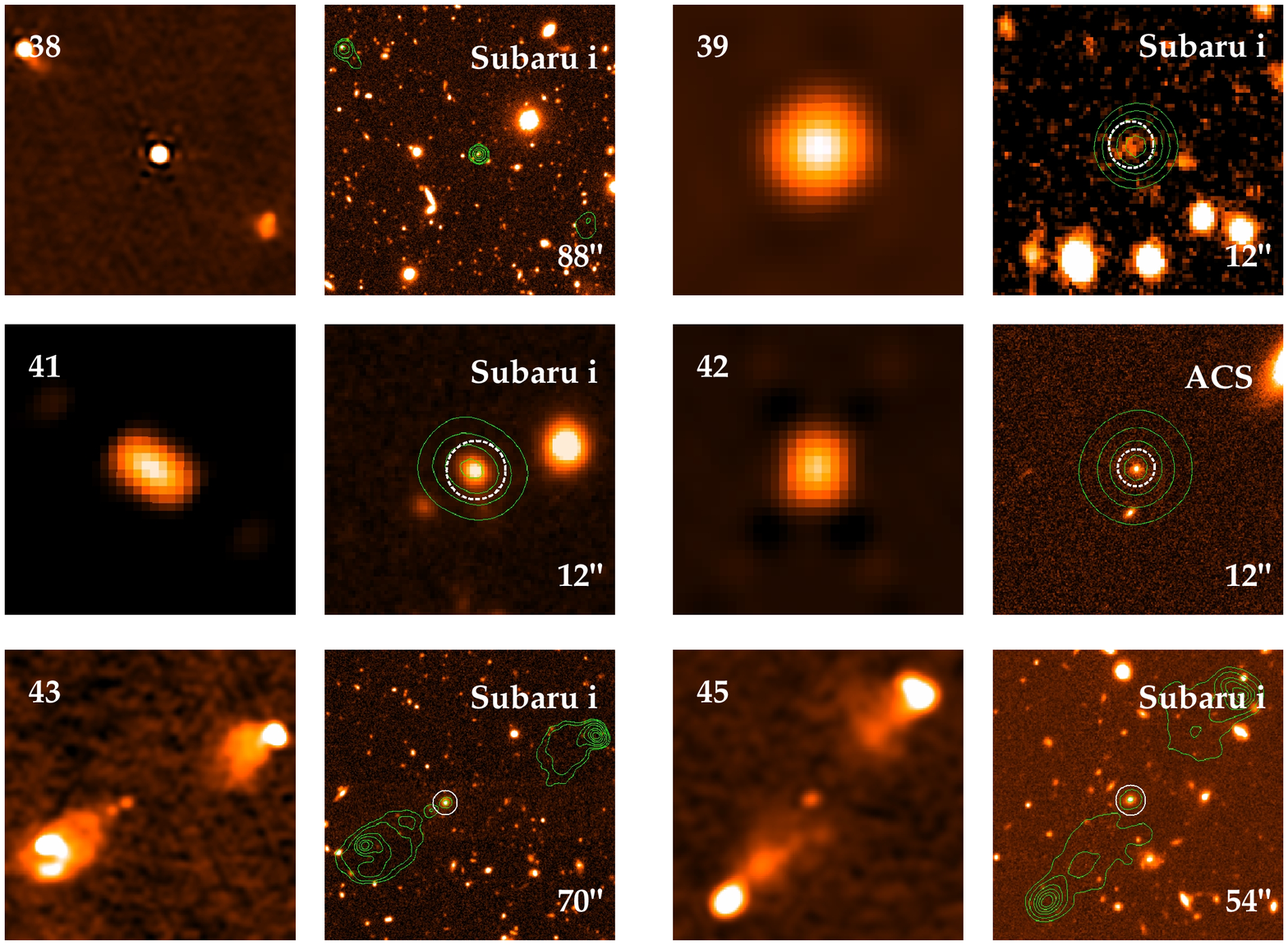}
\caption{CONTINUED}
\end{figure*}

\begin{figure*}
\includegraphics[scale=0.80]{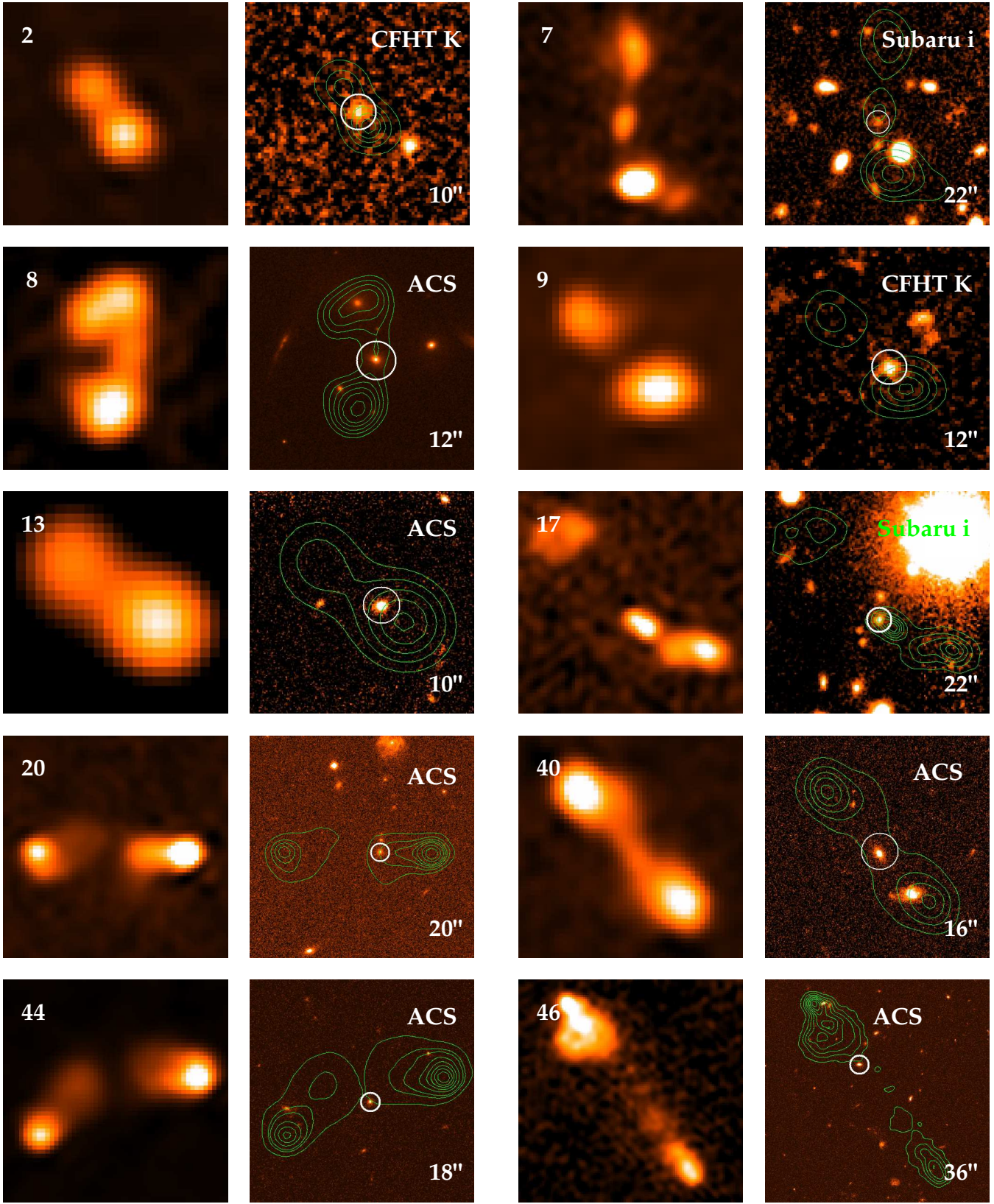}
\caption{Host identification for FR~IIs lacking of a clear point-like emission
  from the radio core. The labels and marks are as in Fig. \ref{radio1}.}
\label{hard}
\end{figure*}

\begin{figure*}
\includegraphics[scale=0.80]{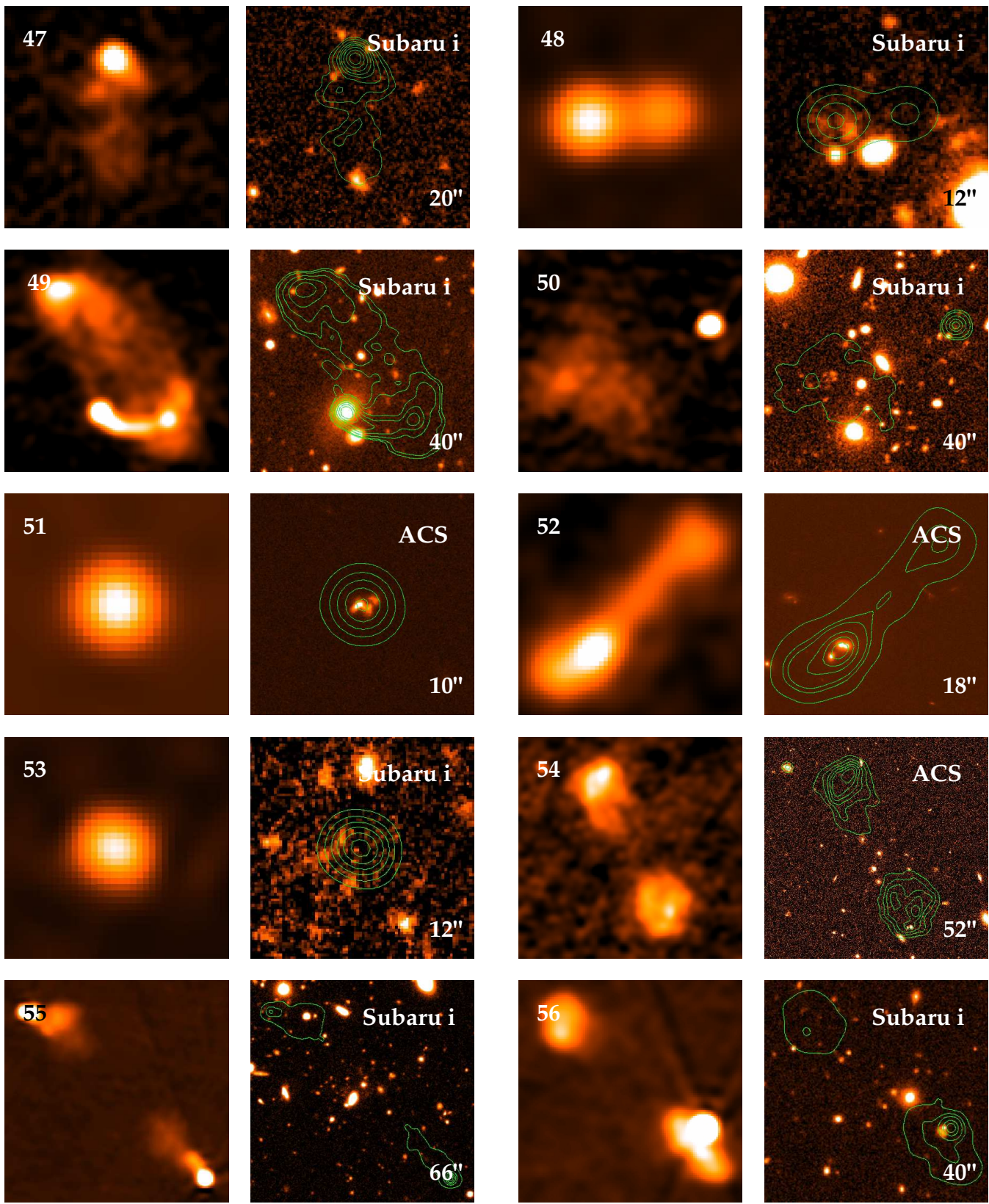}
\caption{Images of the failed host identification. The labels and marks are as
  in Fig. \ref{radio1}.}
\label{noident}
\end{figure*}

\begin{figure*}[h]
\begin{center}$
\begin{array}{ccc}
\vspace{2em}
\includegraphics[scale=0.31,angle=90]{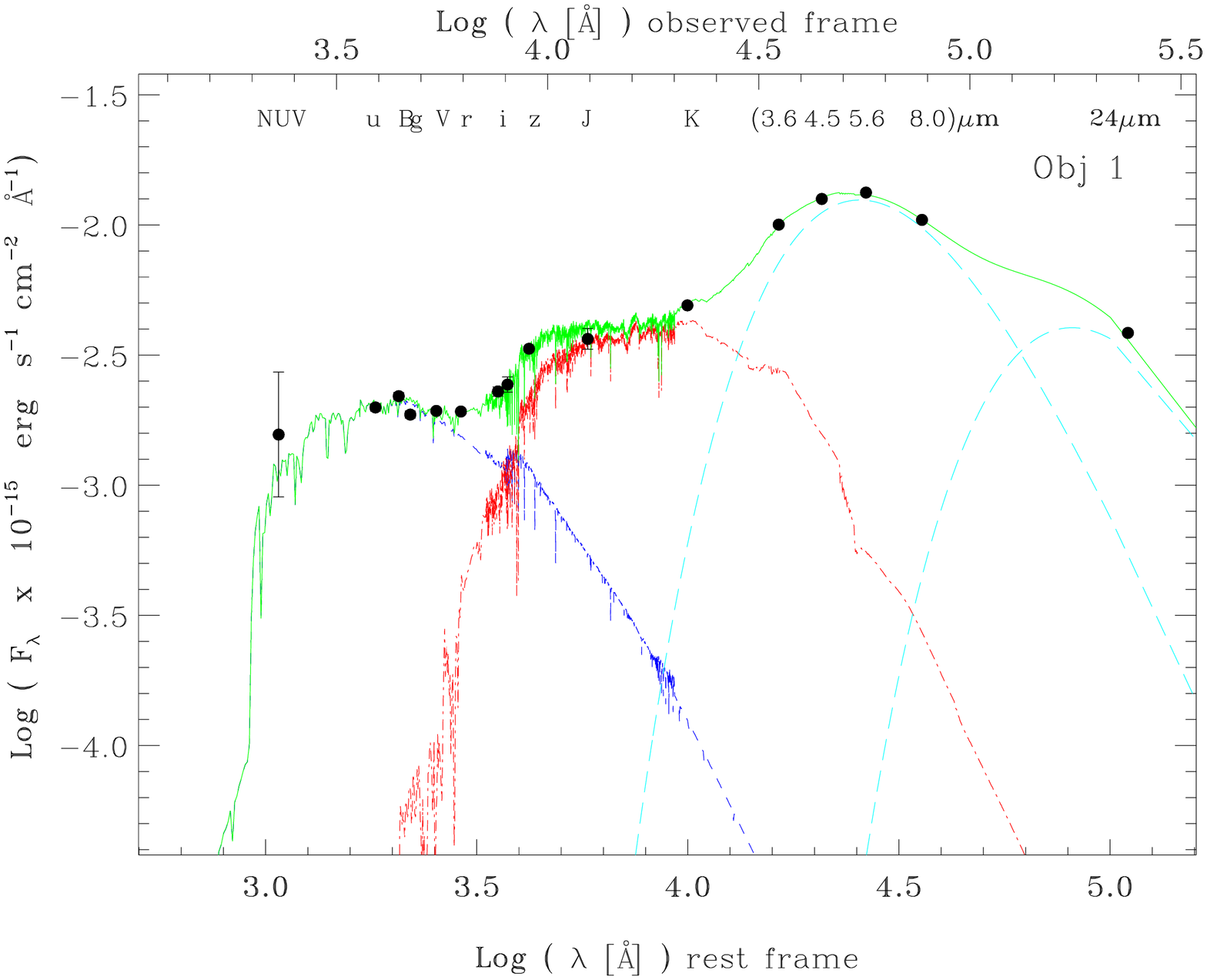} &
\includegraphics[scale=0.31,angle=90]{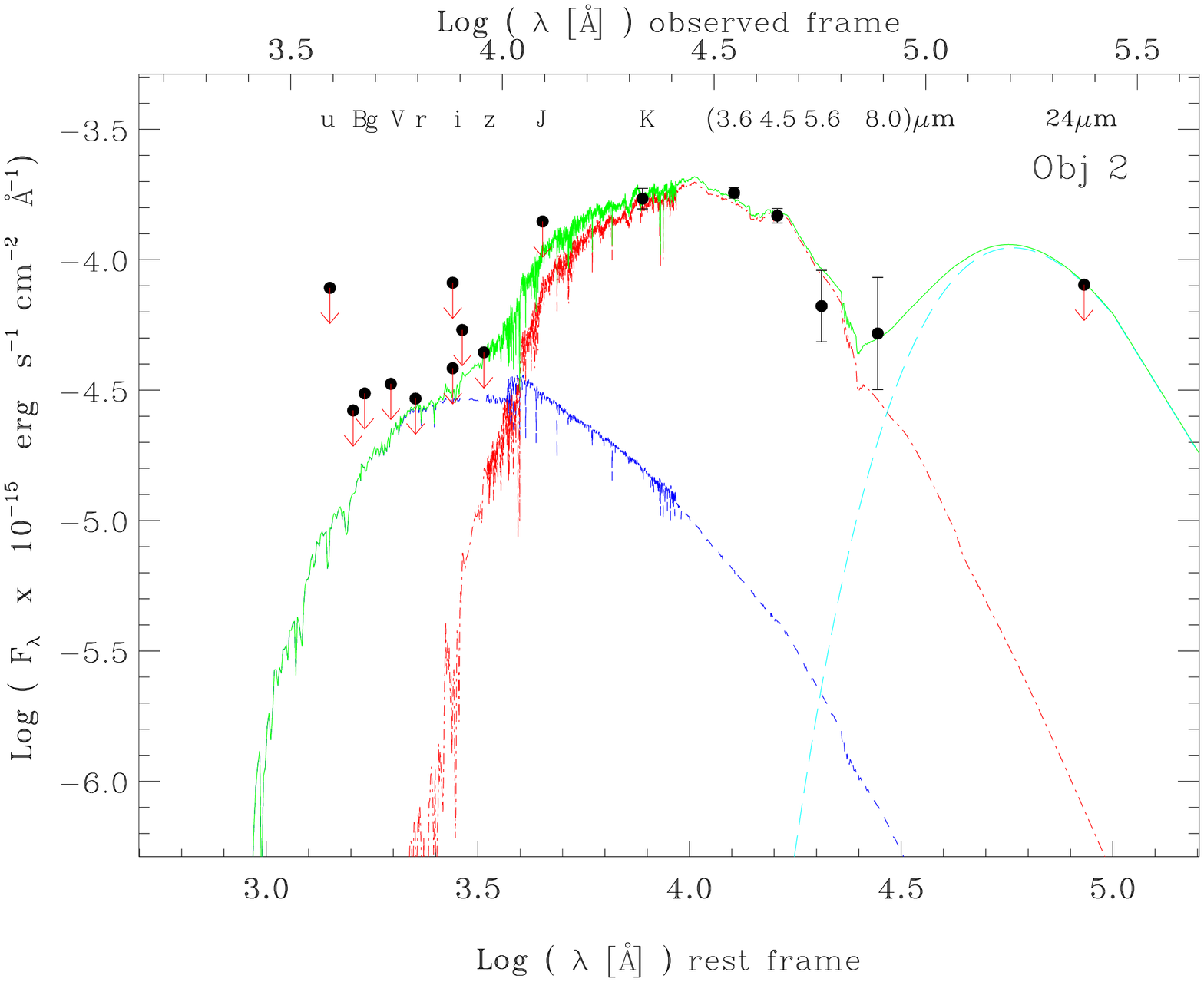} \\
\vspace{1em}
\includegraphics[scale=0.31,angle=90]{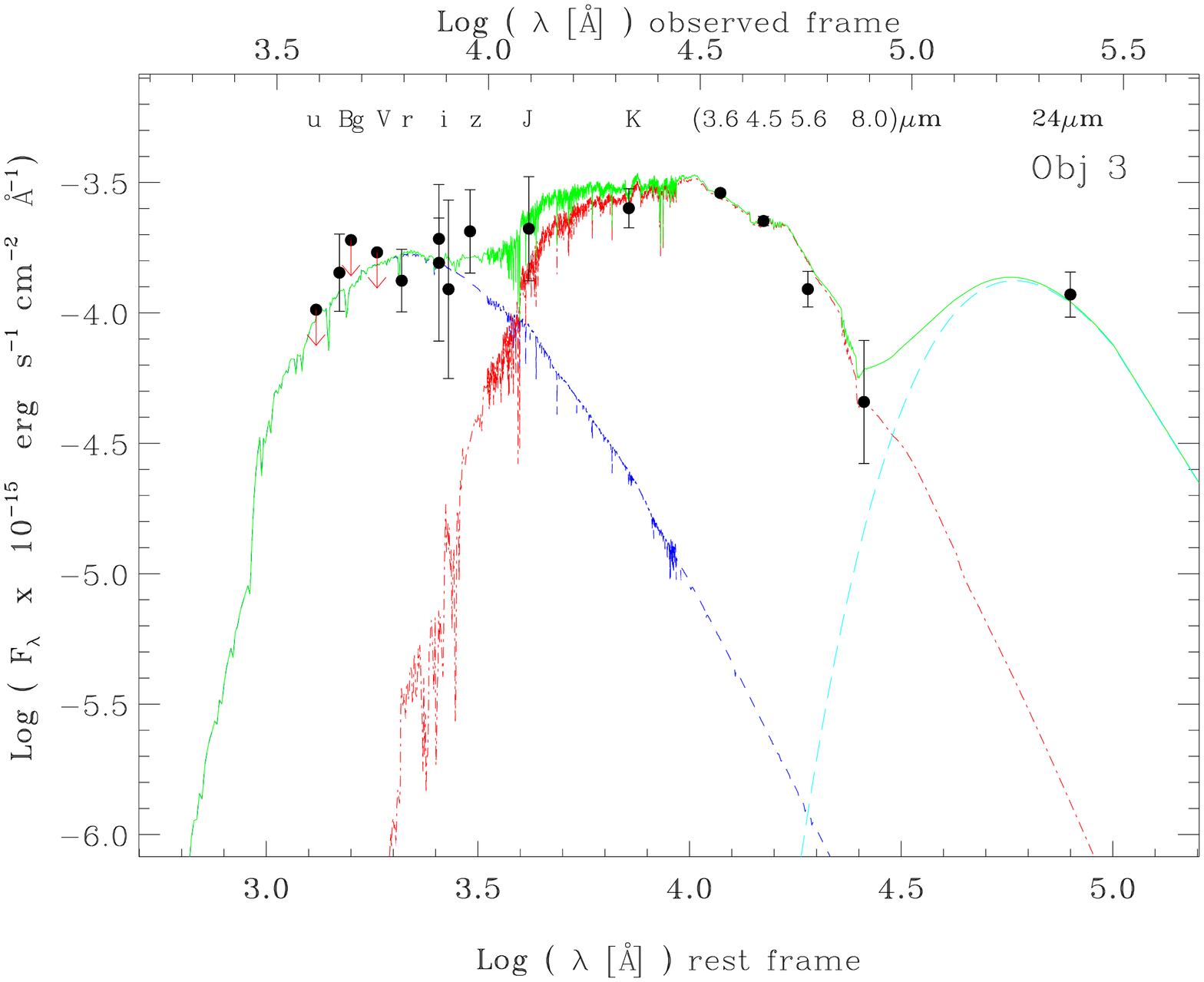} &
\includegraphics[scale=0.31,angle=90]{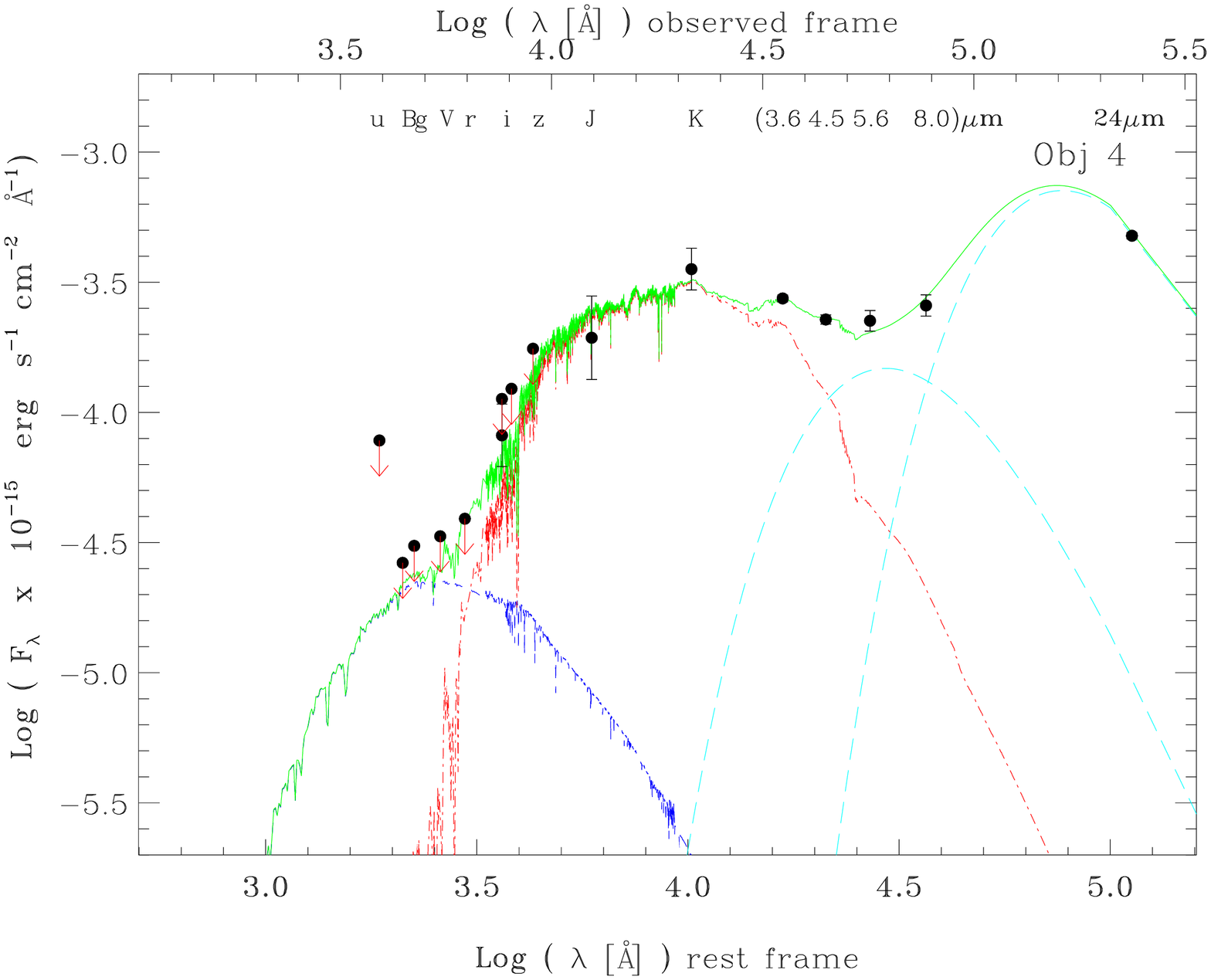} \\
\vspace{1em}
\includegraphics[scale=0.31,angle=90]{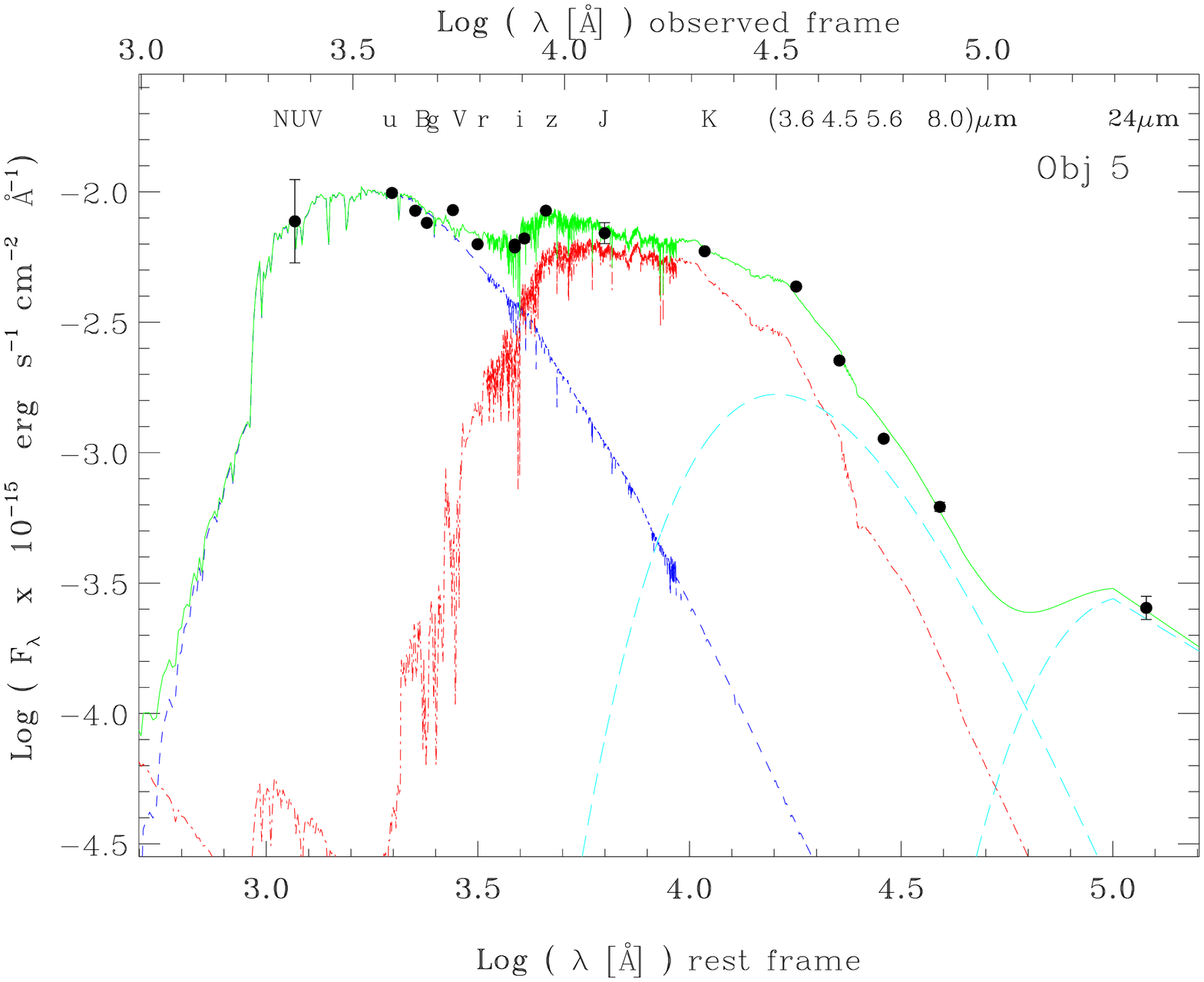}  &
\includegraphics[scale=0.31,angle=90]{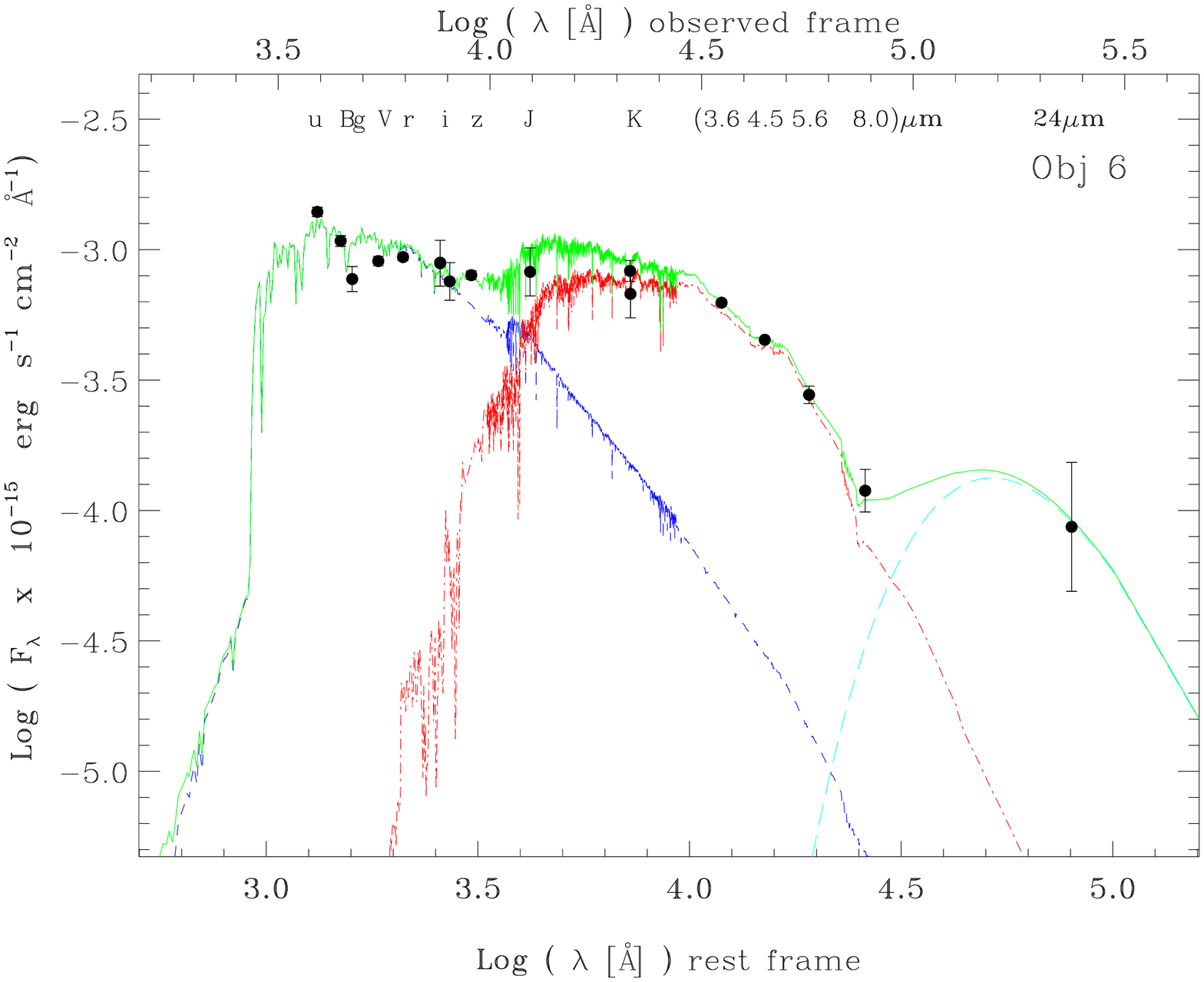}  \\
\vspace{1em}
\includegraphics[scale=0.31,angle=90]{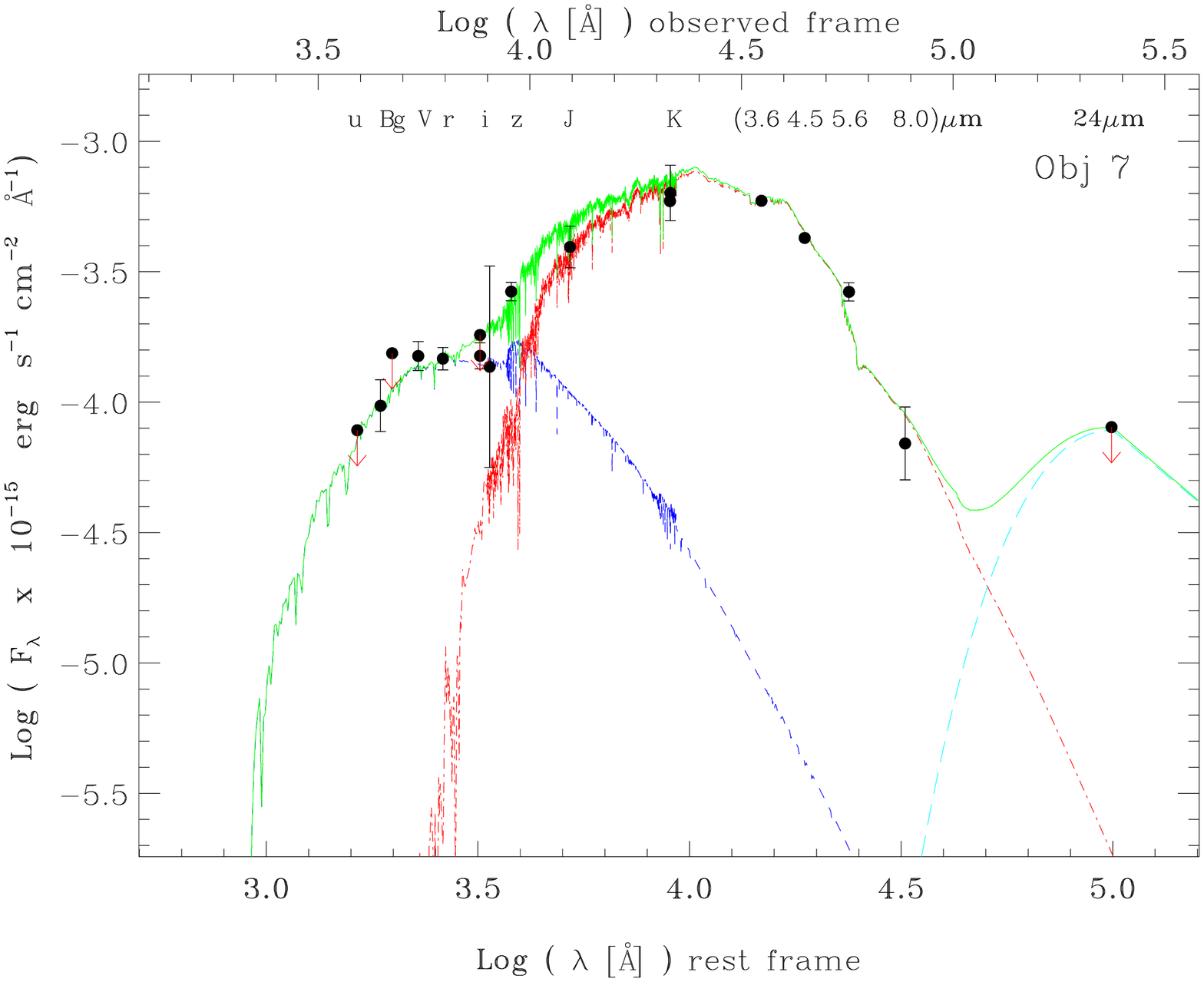} &
\includegraphics[scale=0.31,angle=90]{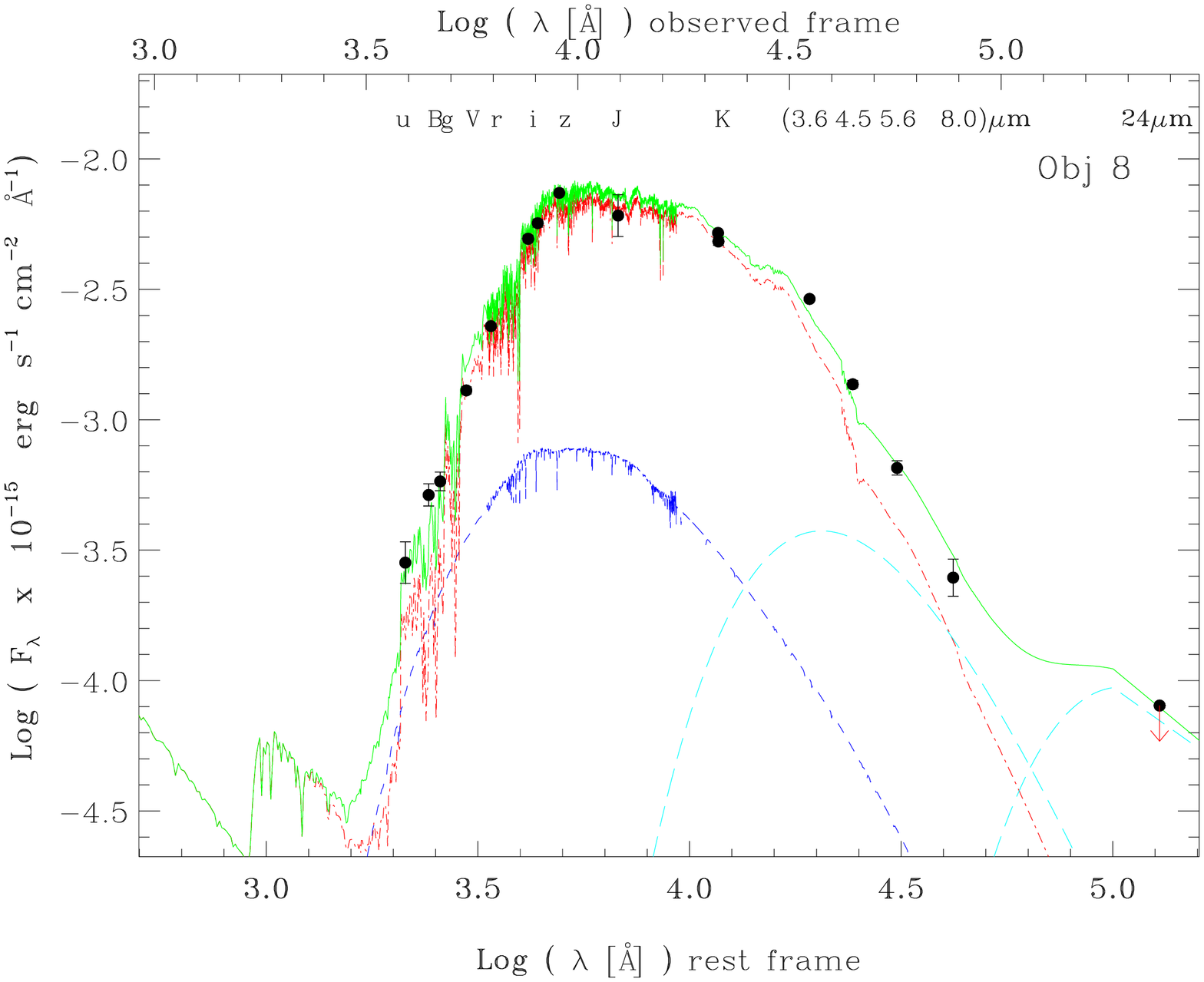} 
\end{array}$
\end{center}
\caption{SEDs of the objects of the sample. The model is the green line, which
  includes the YSP (blue), the OSP (red), and the dust component(s)
  (cyan). The wavelengths on top of the plots correspond to observed
  wavelengths, while those on bottom are at rest frame. For the 2
  spectroscopically confirmed QSOs (objects 29 and 37) we only show the
  photometric points.}
\label{sed1}
\end{figure*}

\begin{figure*}[h]
\addtocounter{figure}{-1}
\begin{center}$
\begin{array}{ccc}
\vspace{2em}
\includegraphics[scale=0.31,angle=90]{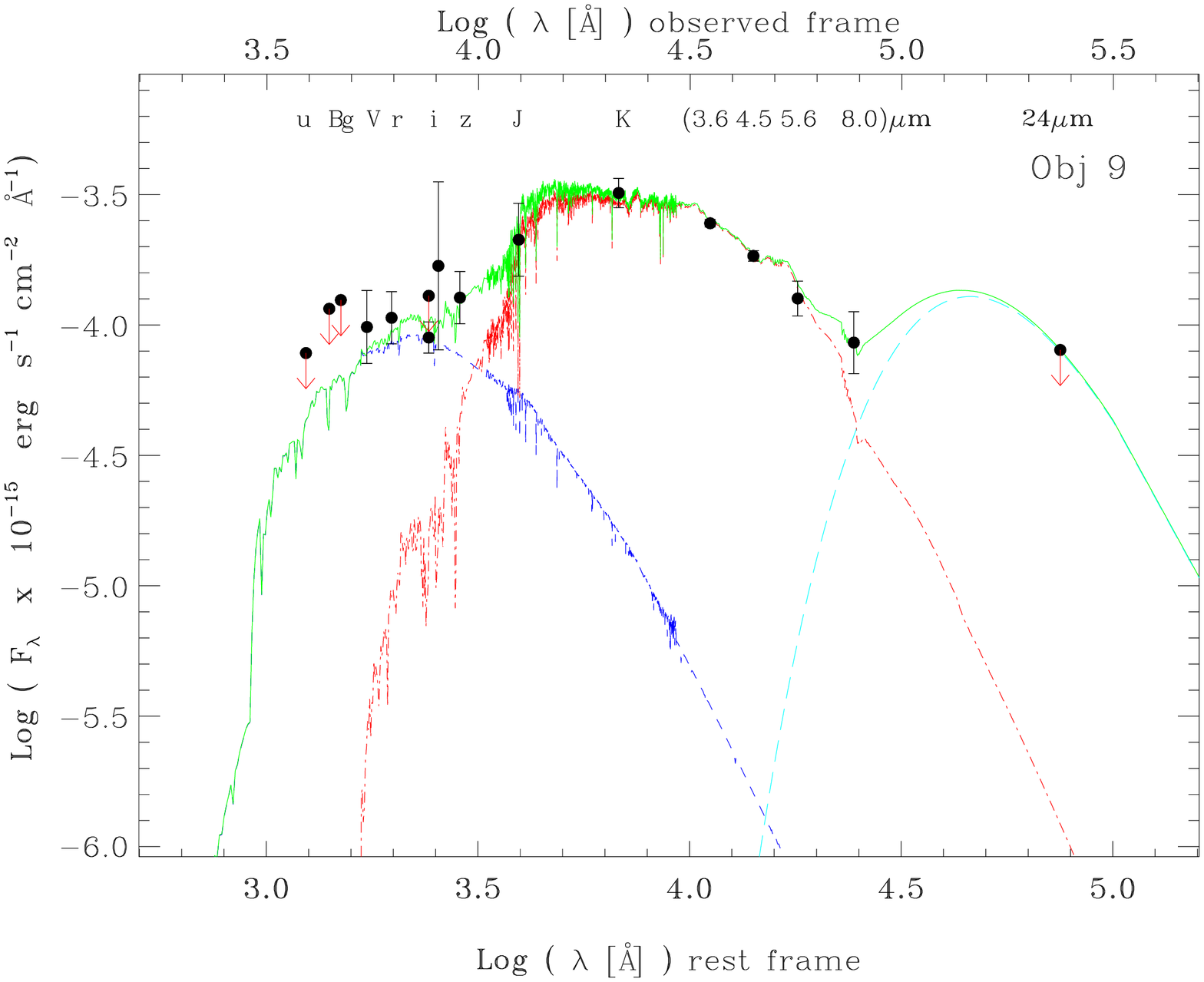} &
\includegraphics[scale=0.31,angle=90]{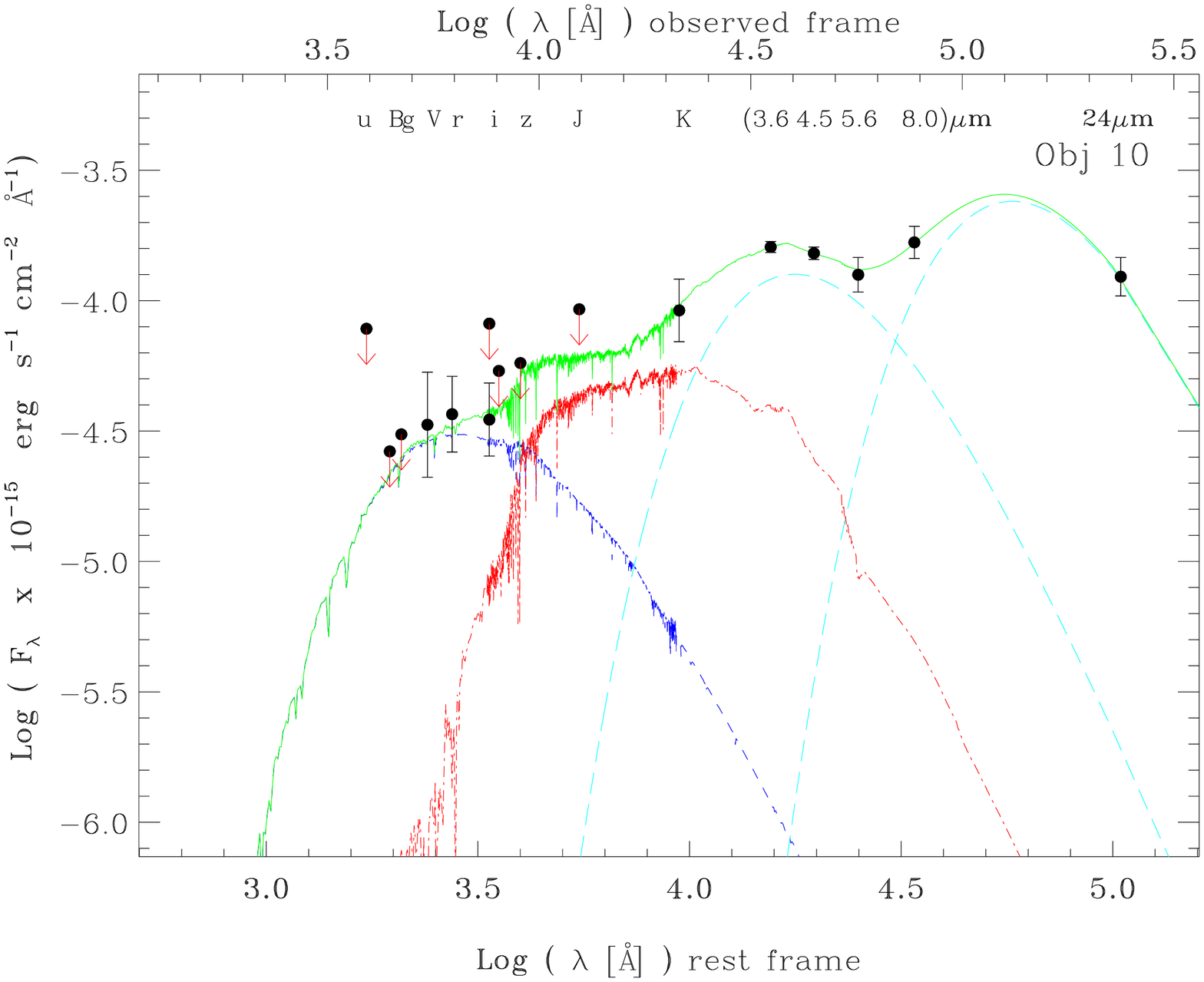} \\
\vspace{1em}
\includegraphics[scale=0.31,angle=90]{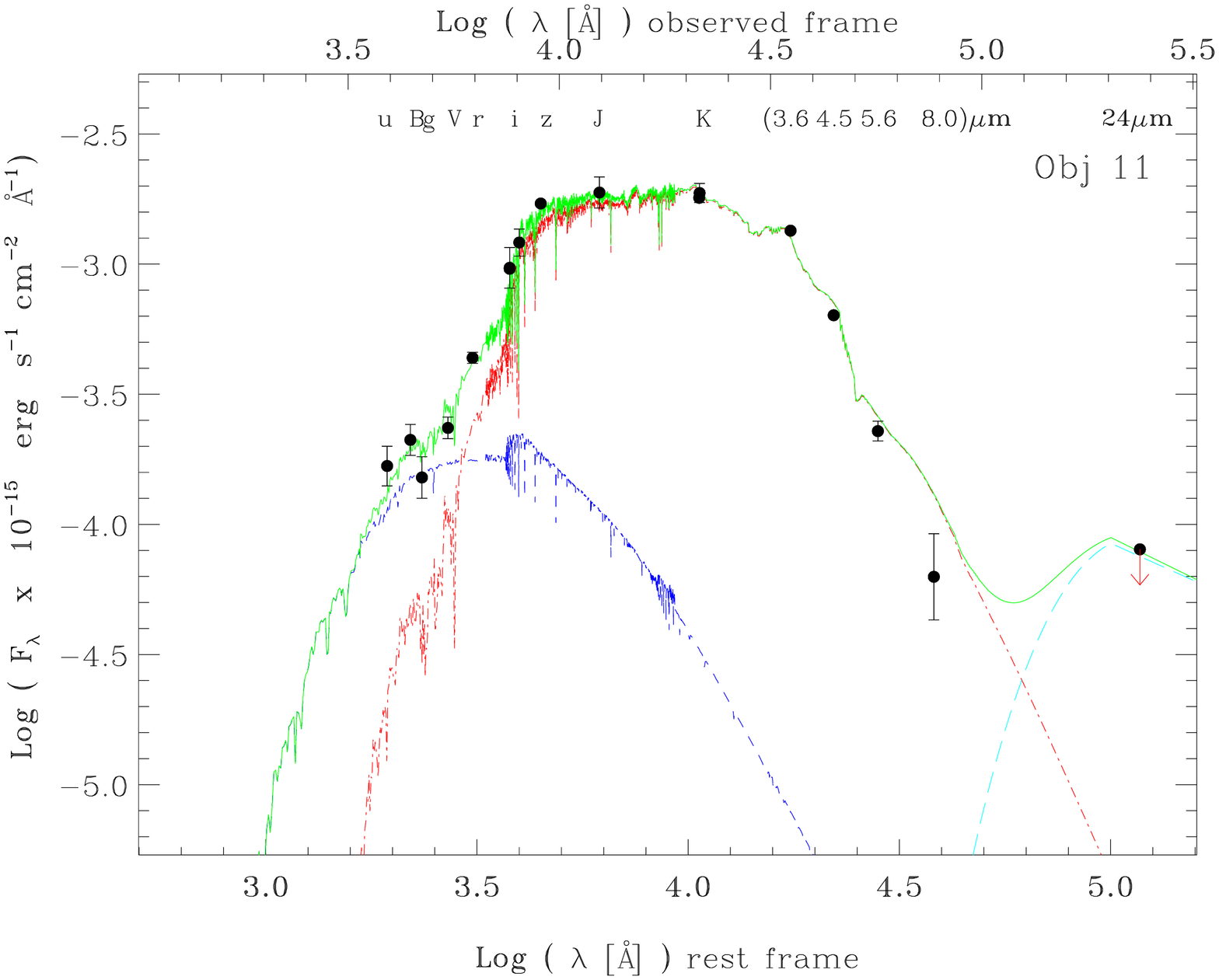} &
\includegraphics[scale=0.31,angle=90]{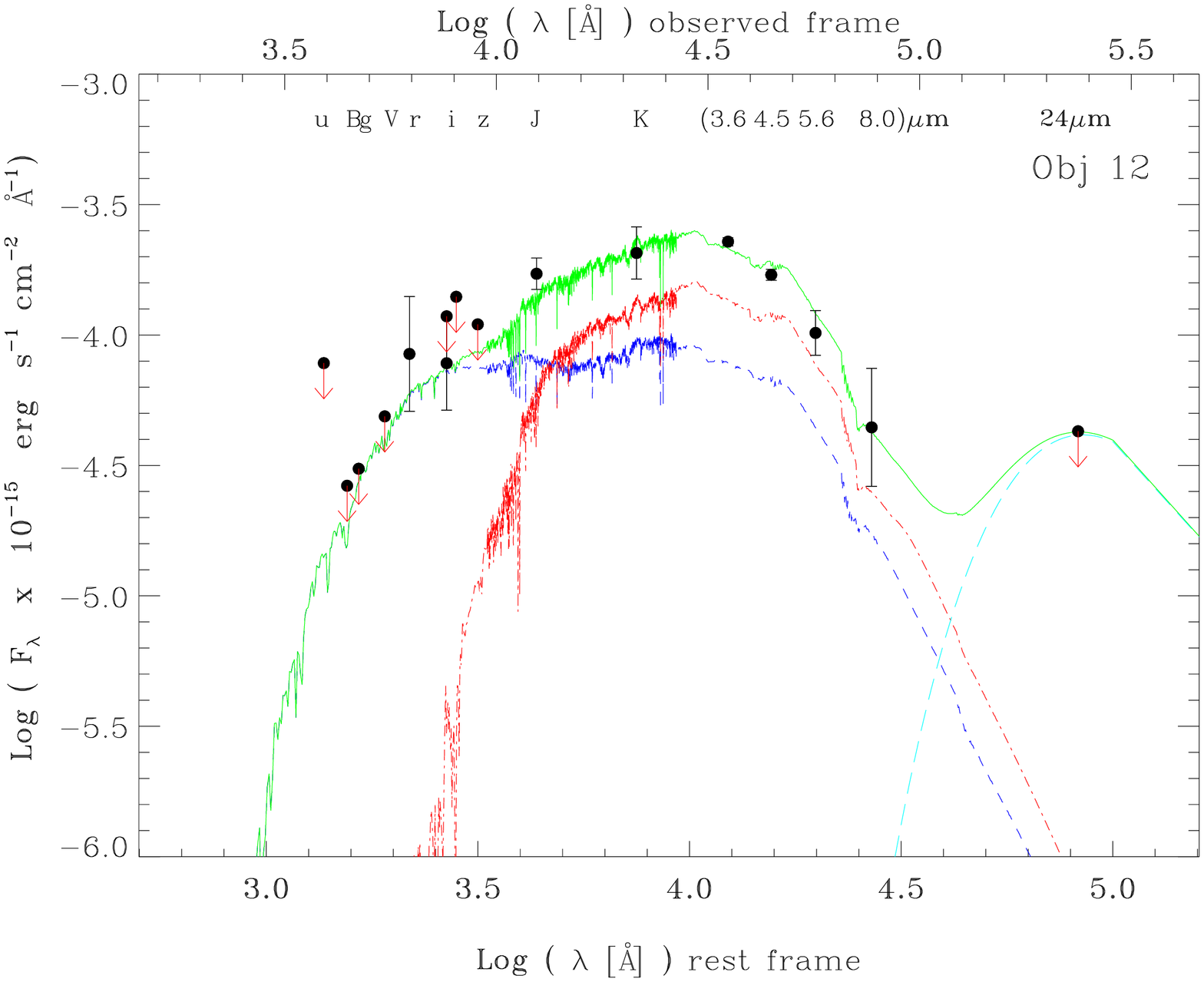} \\
\vspace{1em}
\includegraphics[scale=0.31,angle=90]{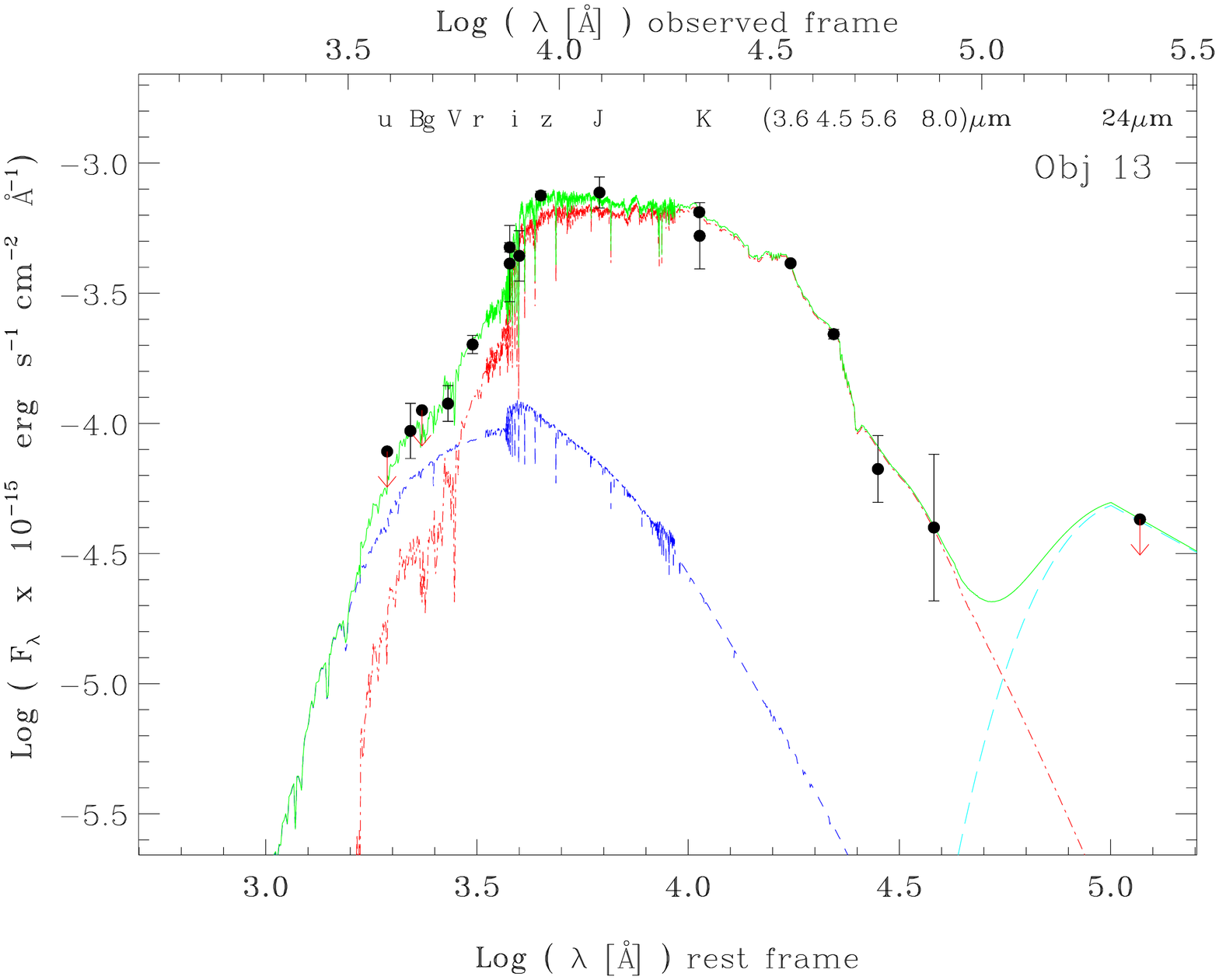}  &
\includegraphics[scale=0.31,angle=90]{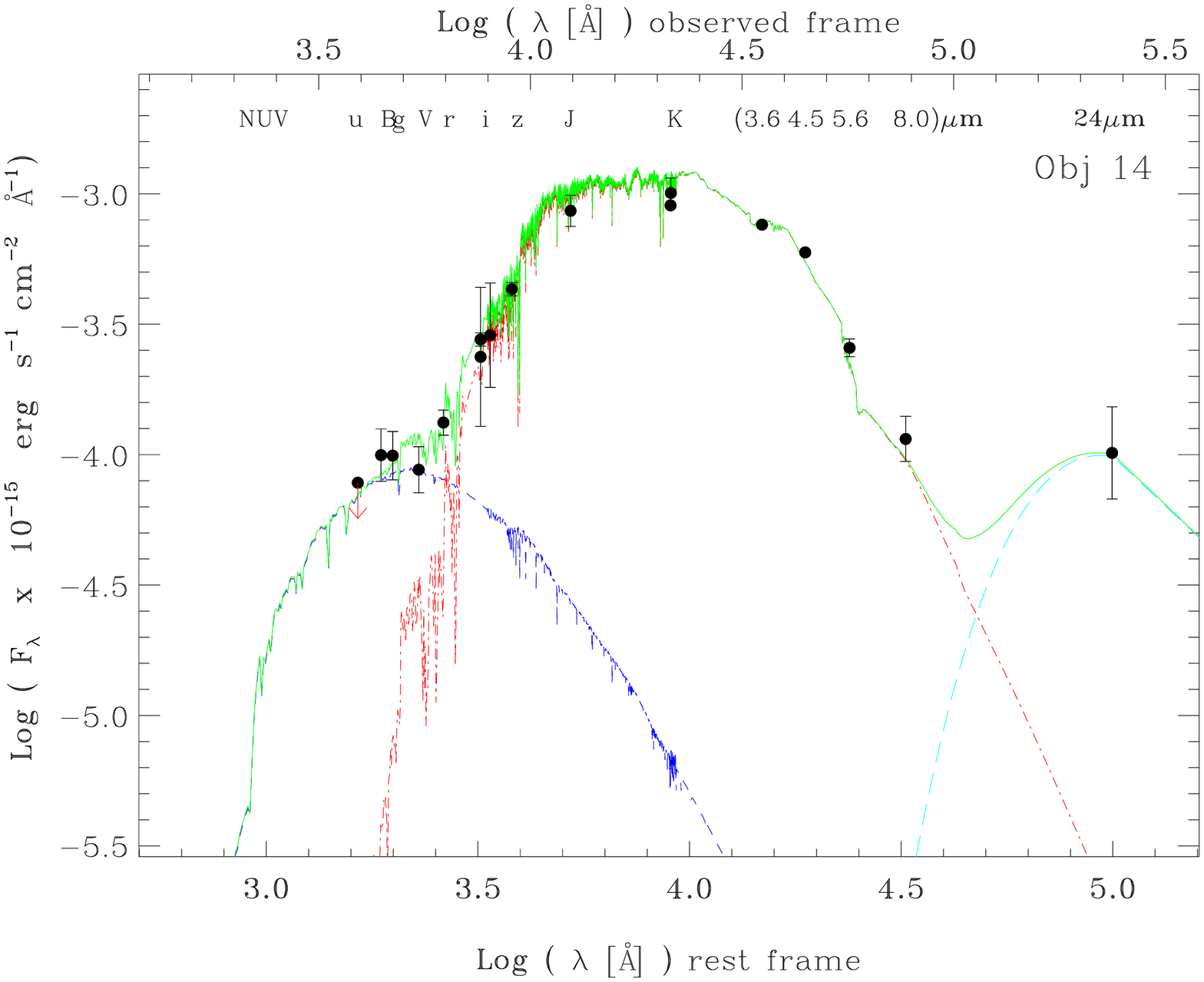}  \\
\vspace{1em}
\includegraphics[scale=0.31,angle=90]{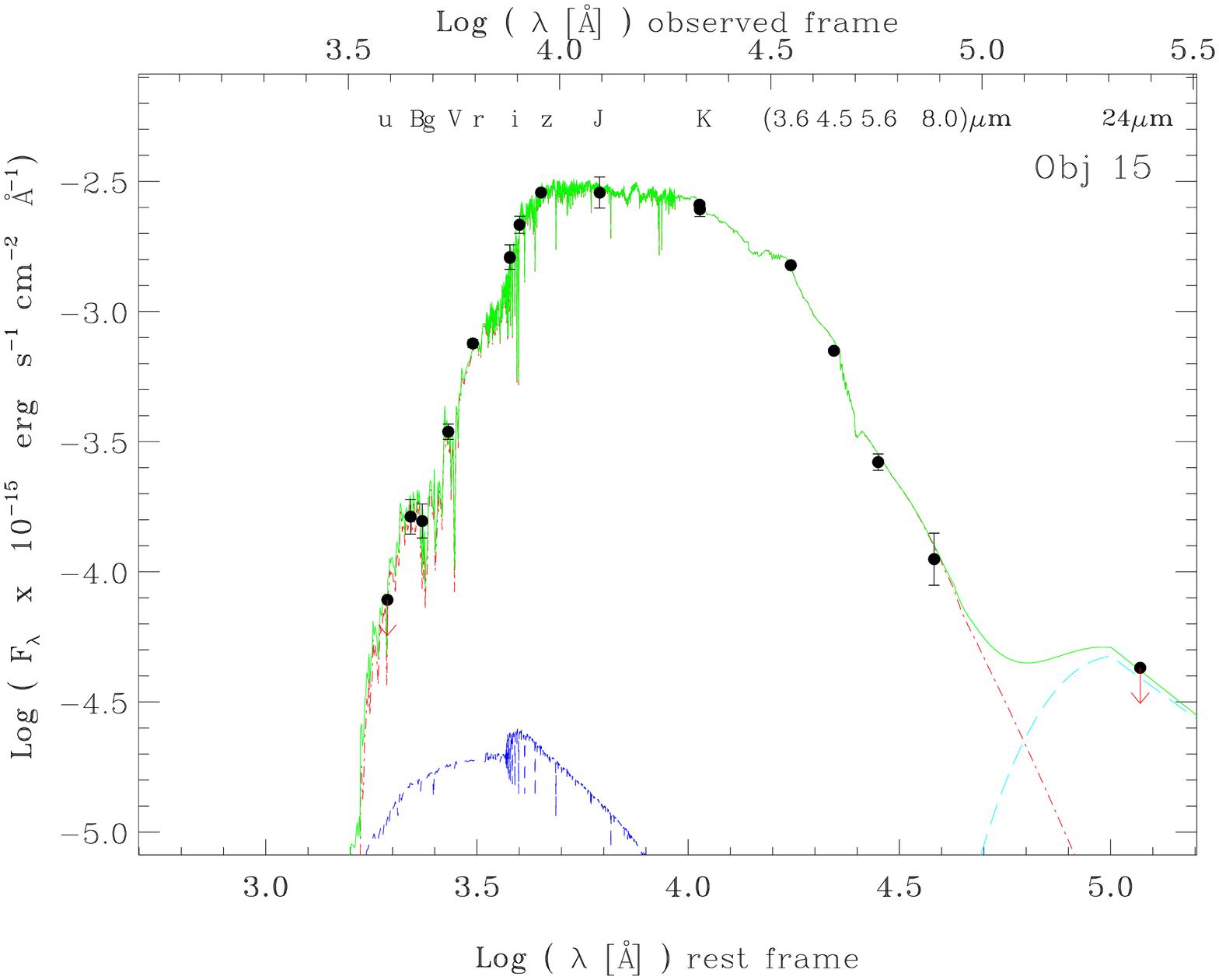} &
\includegraphics[scale=0.31,angle=90]{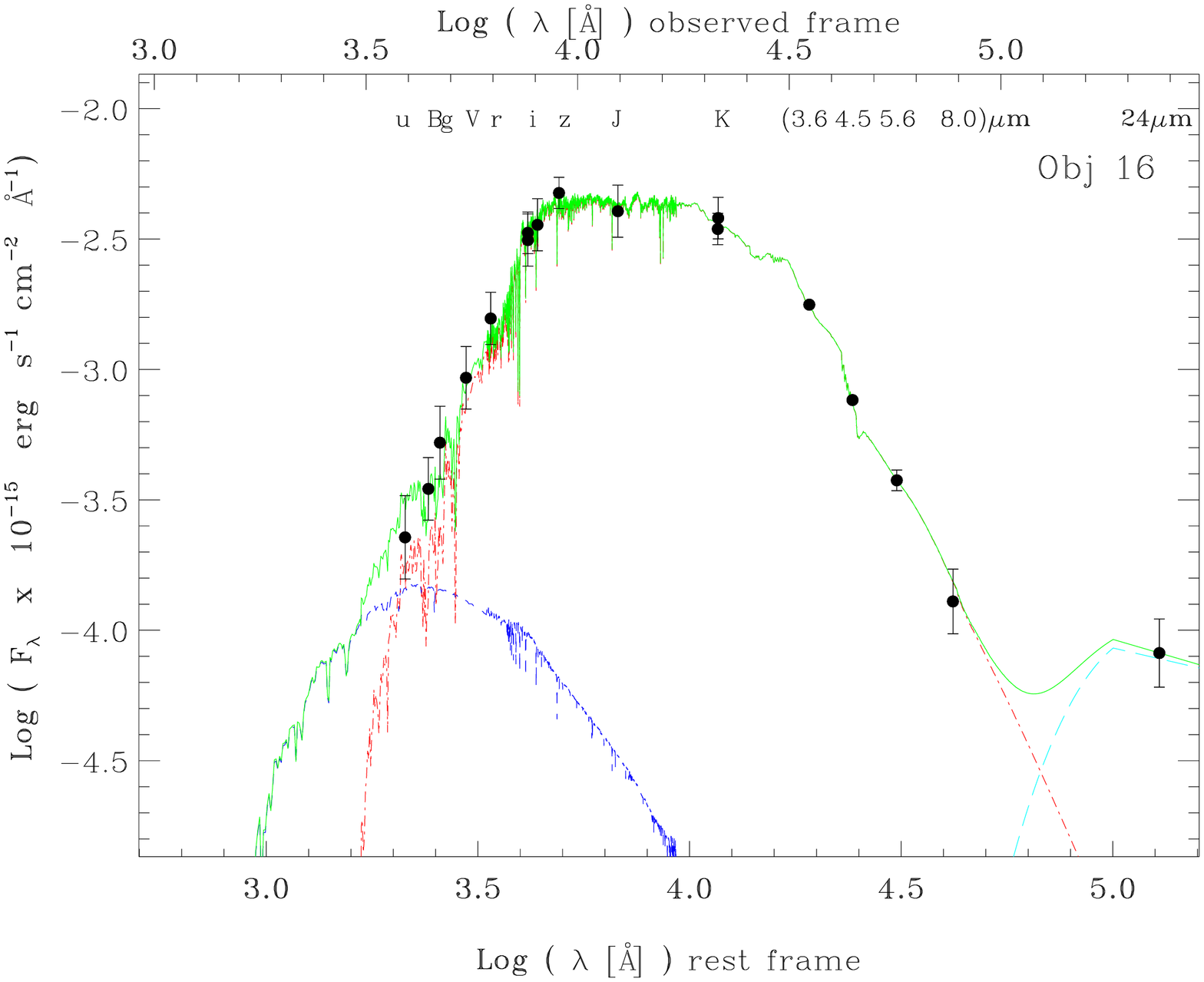} 
\end{array}$
\end{center}
\caption{CONTINUED}
\end{figure*}

\begin{figure*}[h]
\addtocounter{figure}{-1}
\begin{center}$
\begin{array}{ccc}
\vspace{2em}
\includegraphics[scale=0.31,angle=90]{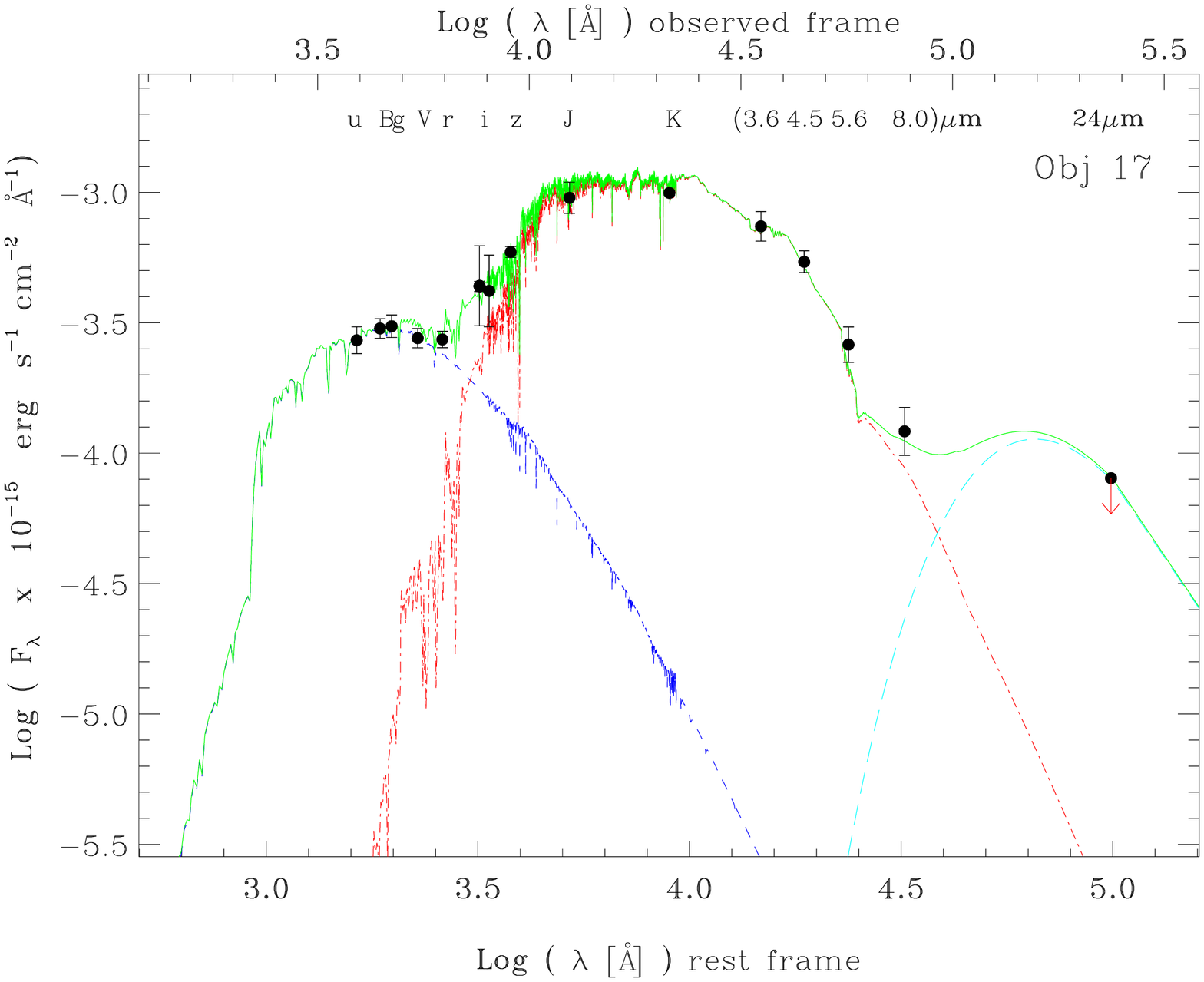} &
\includegraphics[scale=0.31,angle=90]{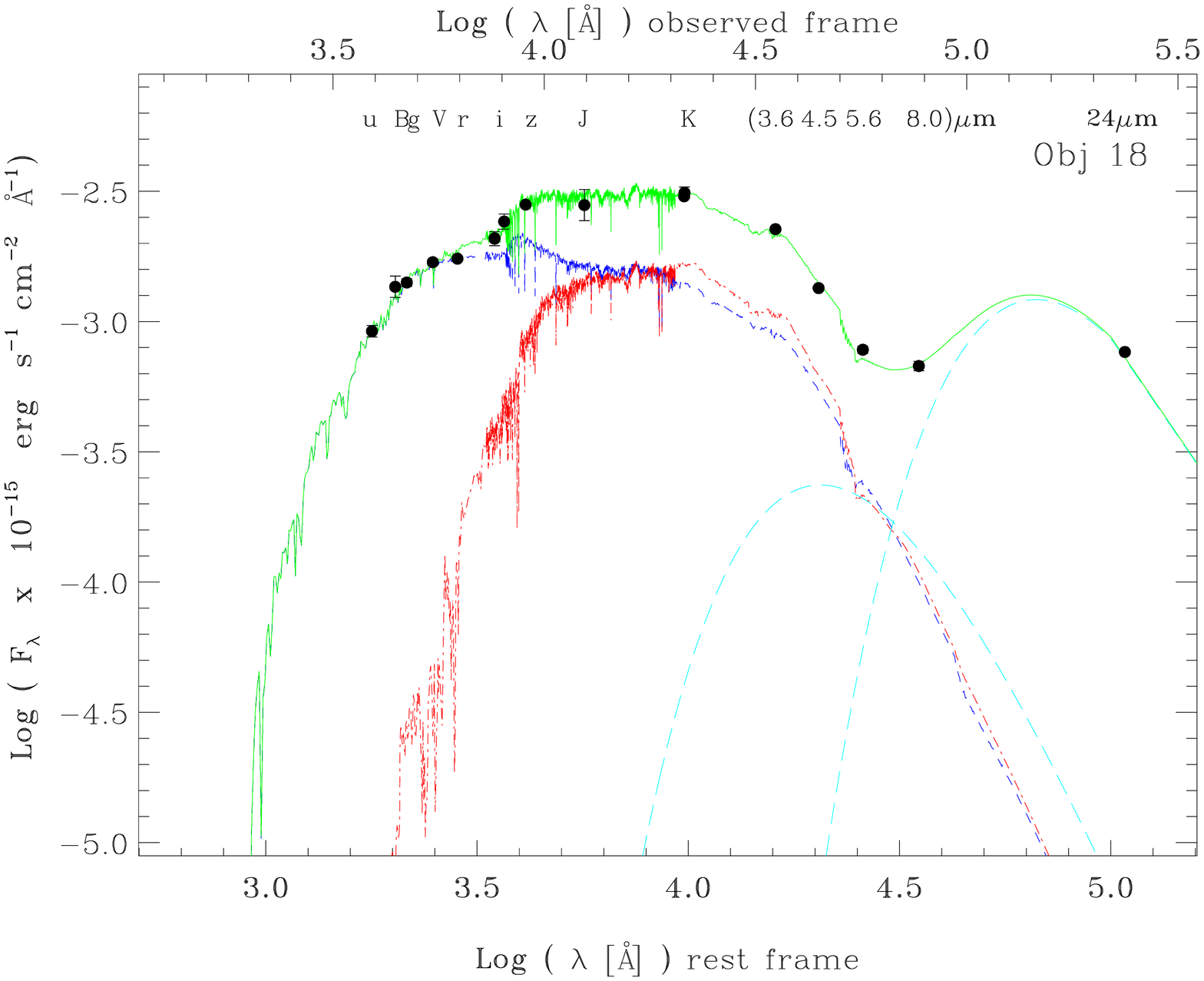} \\
\vspace{1em}
\includegraphics[scale=0.31,angle=90]{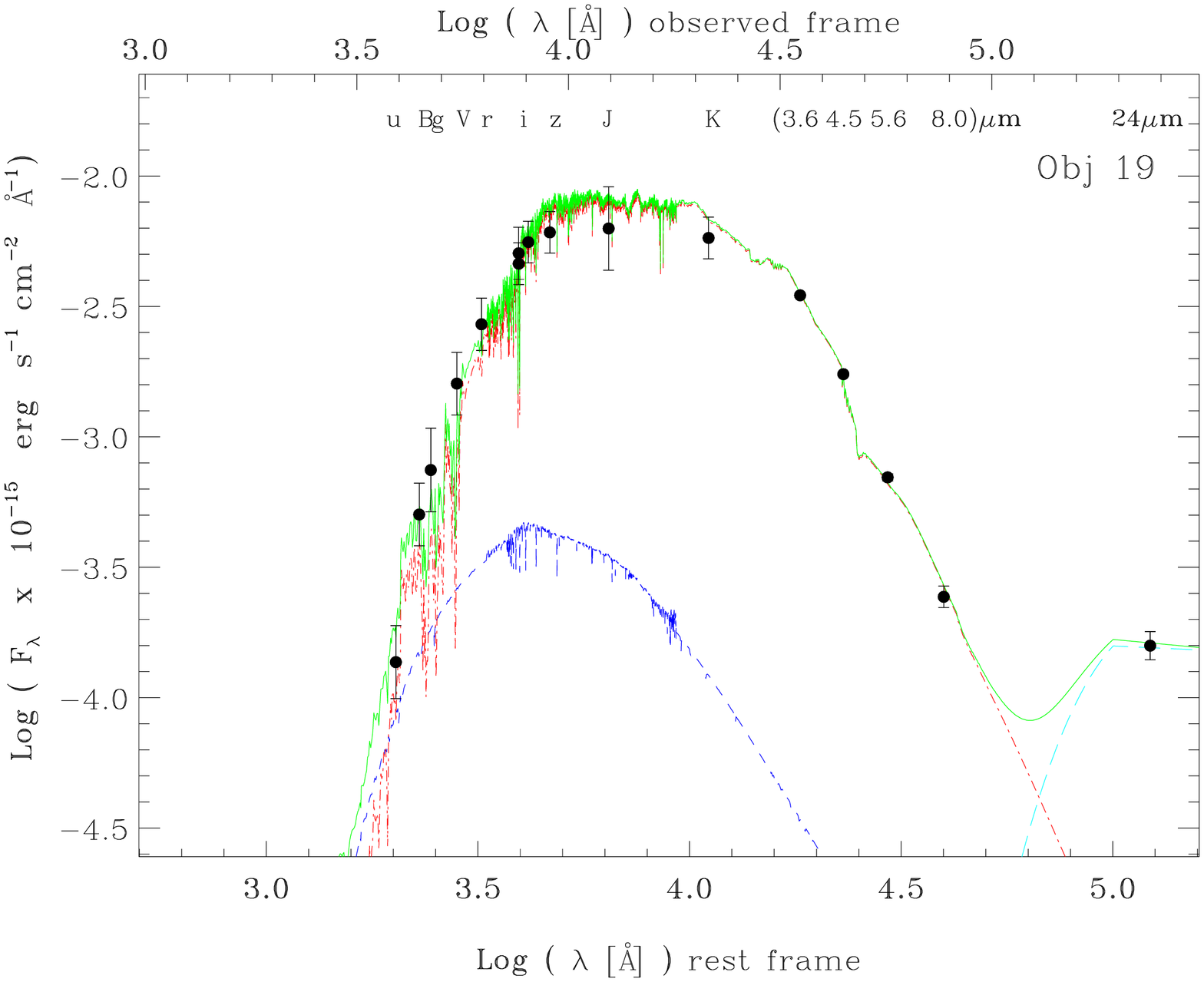} &
\includegraphics[scale=0.31,angle=90]{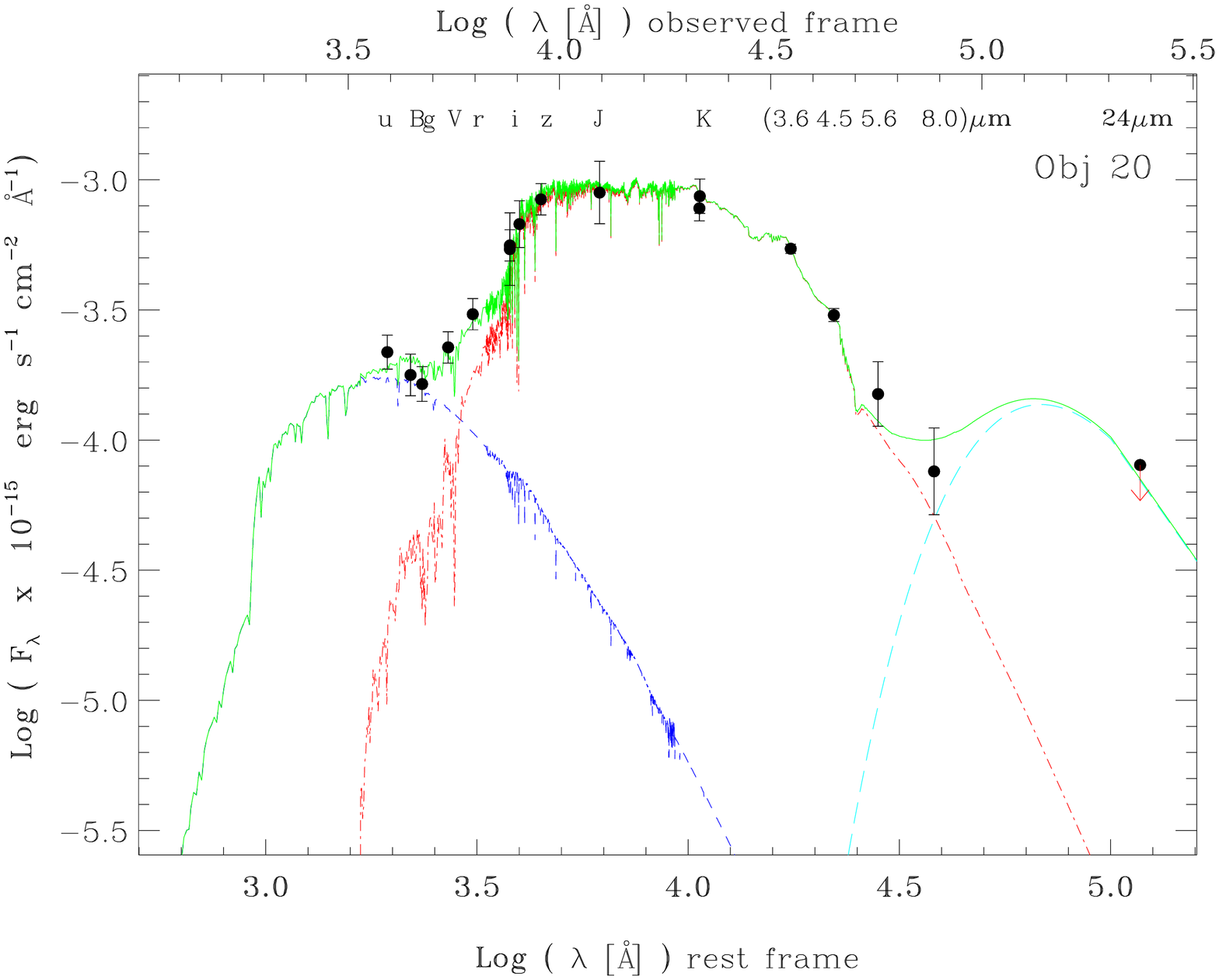} \\
\vspace{1em}
\includegraphics[scale=0.31,angle=90]{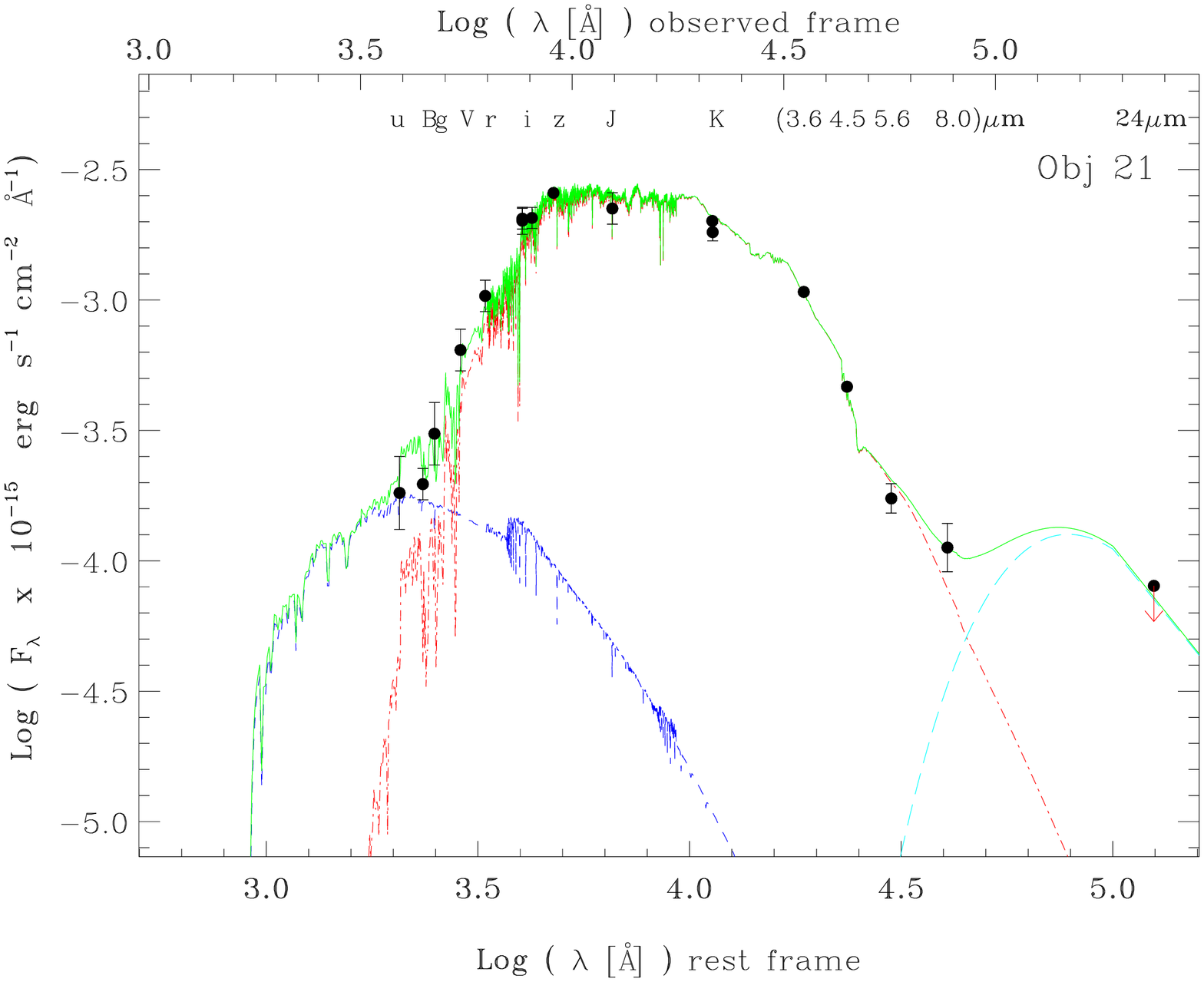} &
\includegraphics[scale=0.31,angle=90]{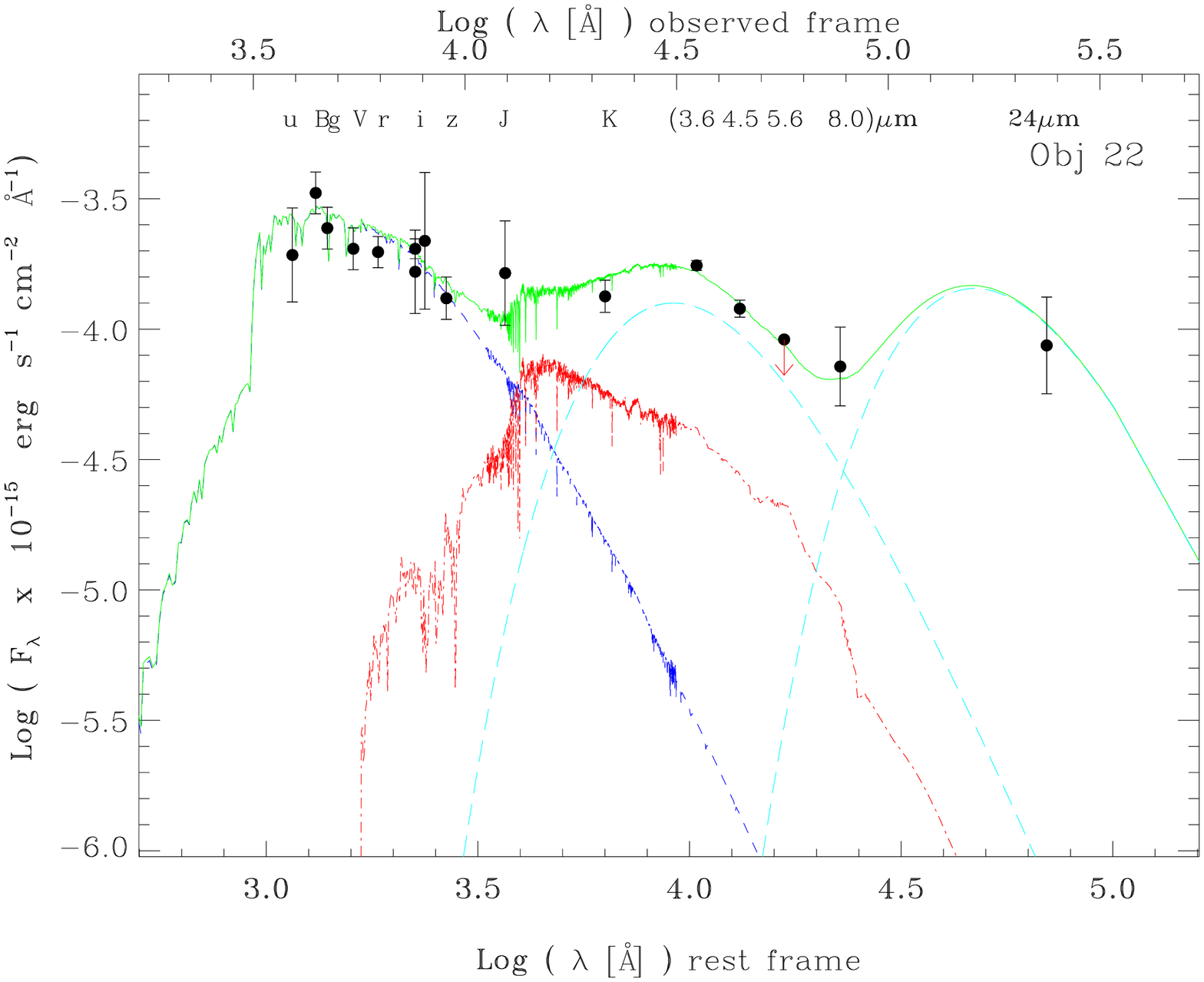} \\
\vspace{1em}
\includegraphics[scale=0.31,angle=90]{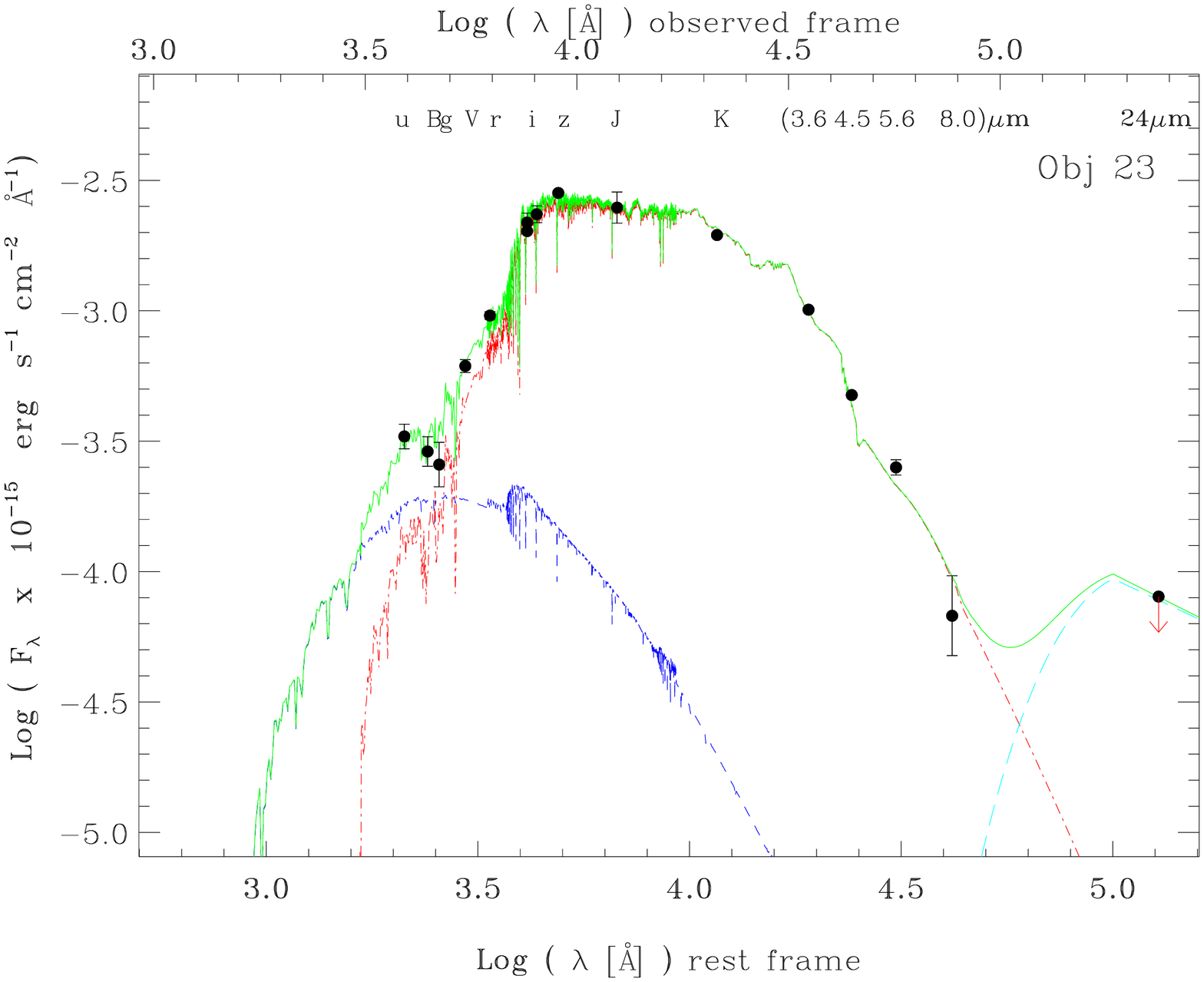} &
\includegraphics[scale=0.31,angle=90]{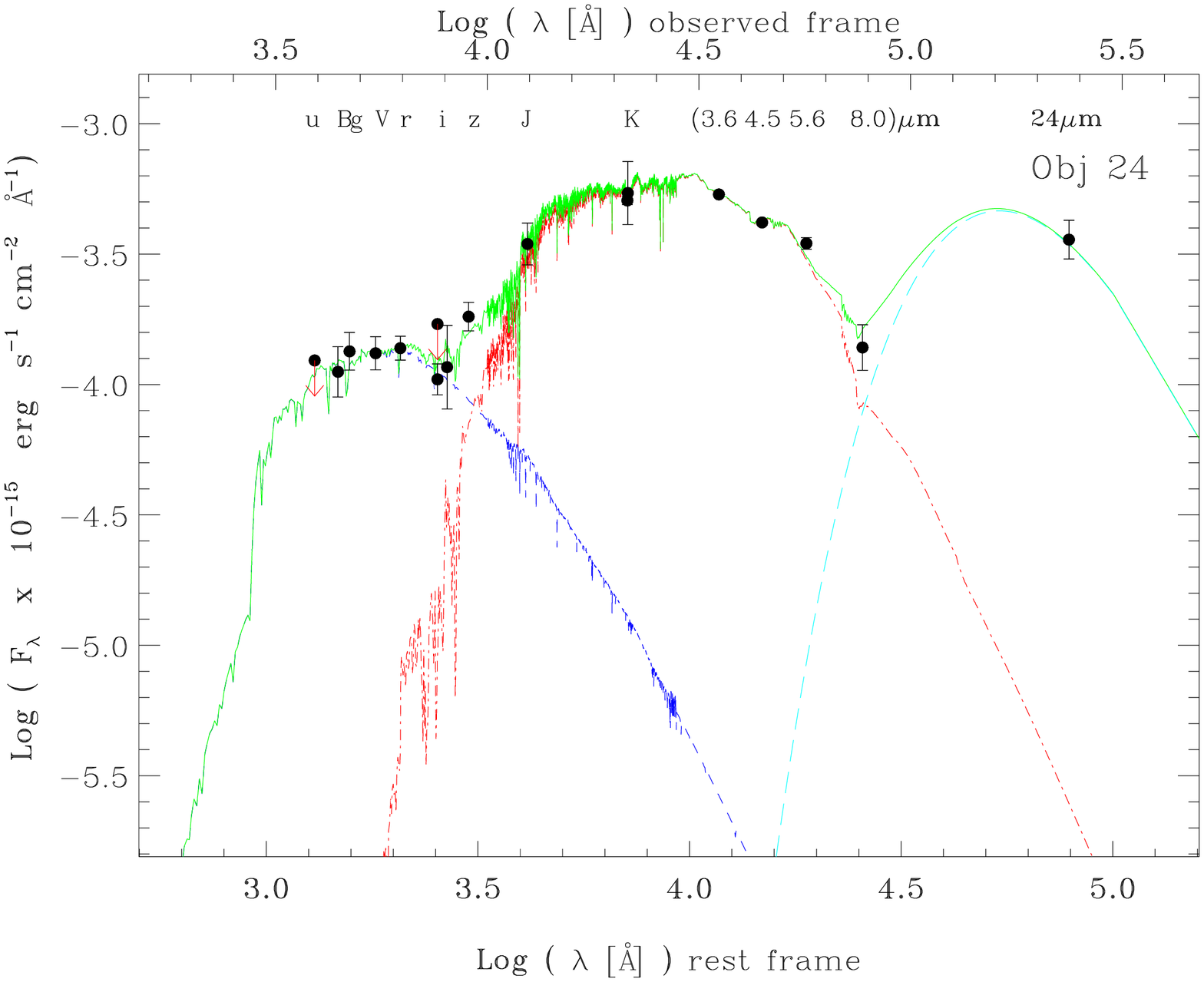} 
\end{array}$
\end{center}
\caption{CONTINUED}
\end{figure*}

\begin{figure*}[h]
\addtocounter{figure}{-1}
\begin{center}$
\begin{array}{ccc}
\vspace{2em}
\includegraphics[scale=0.31,angle=90]{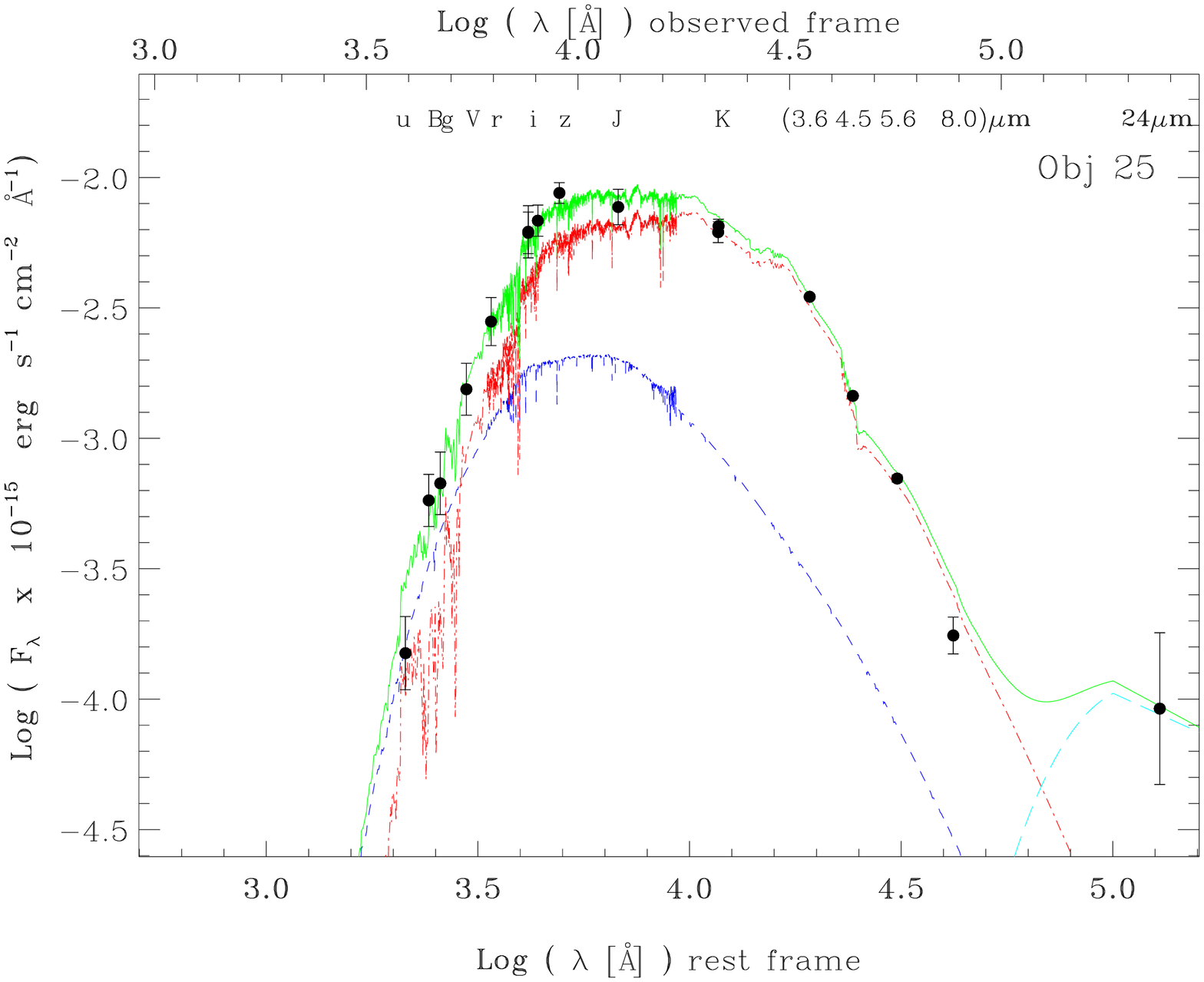} &
\includegraphics[scale=0.31,angle=90]{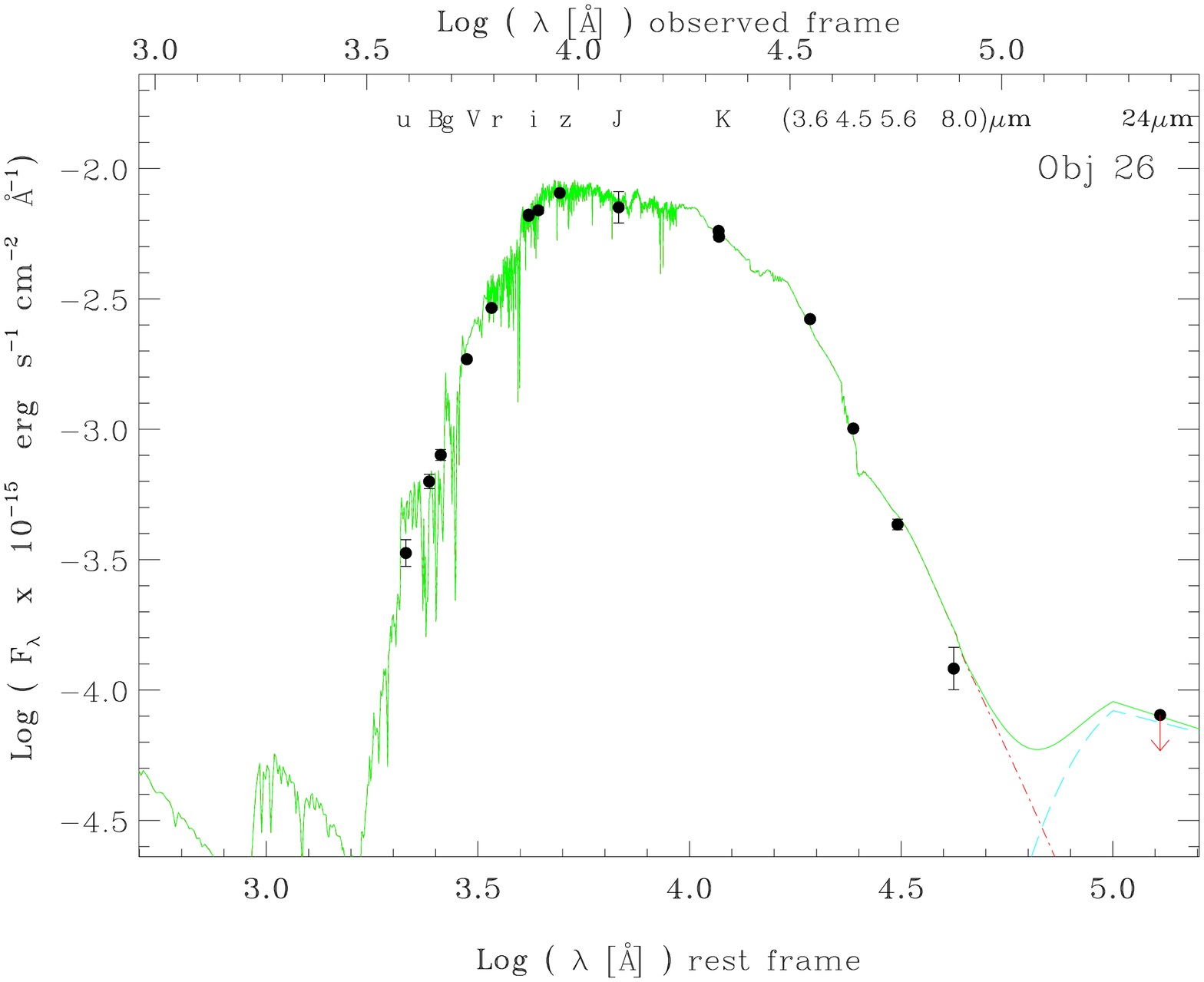} \\
\vspace{1em}
\includegraphics[scale=0.31,angle=90]{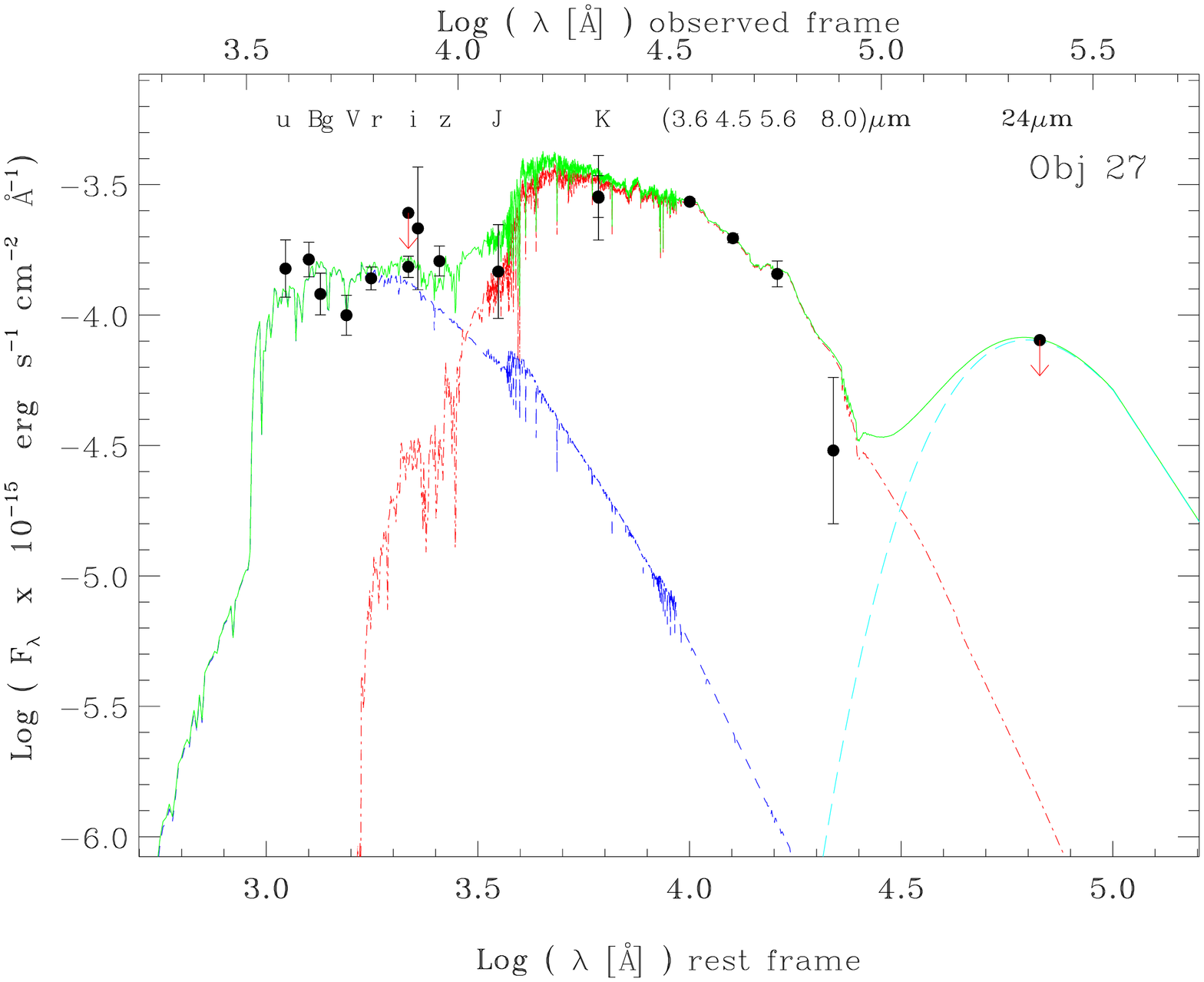} &
\includegraphics[scale=0.31,angle=90]{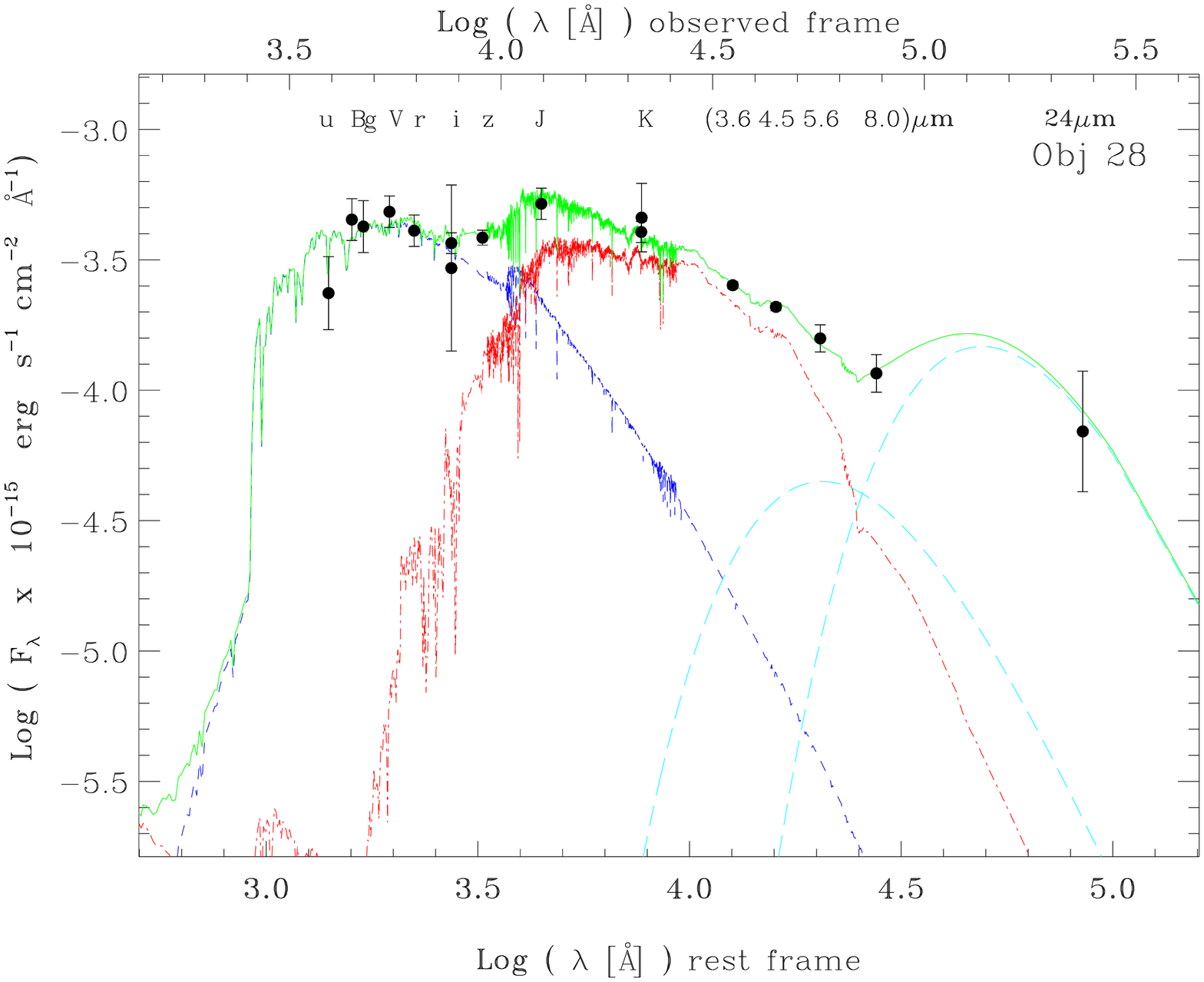} \\
\vspace{1em}
\includegraphics[scale=0.31,angle=90]{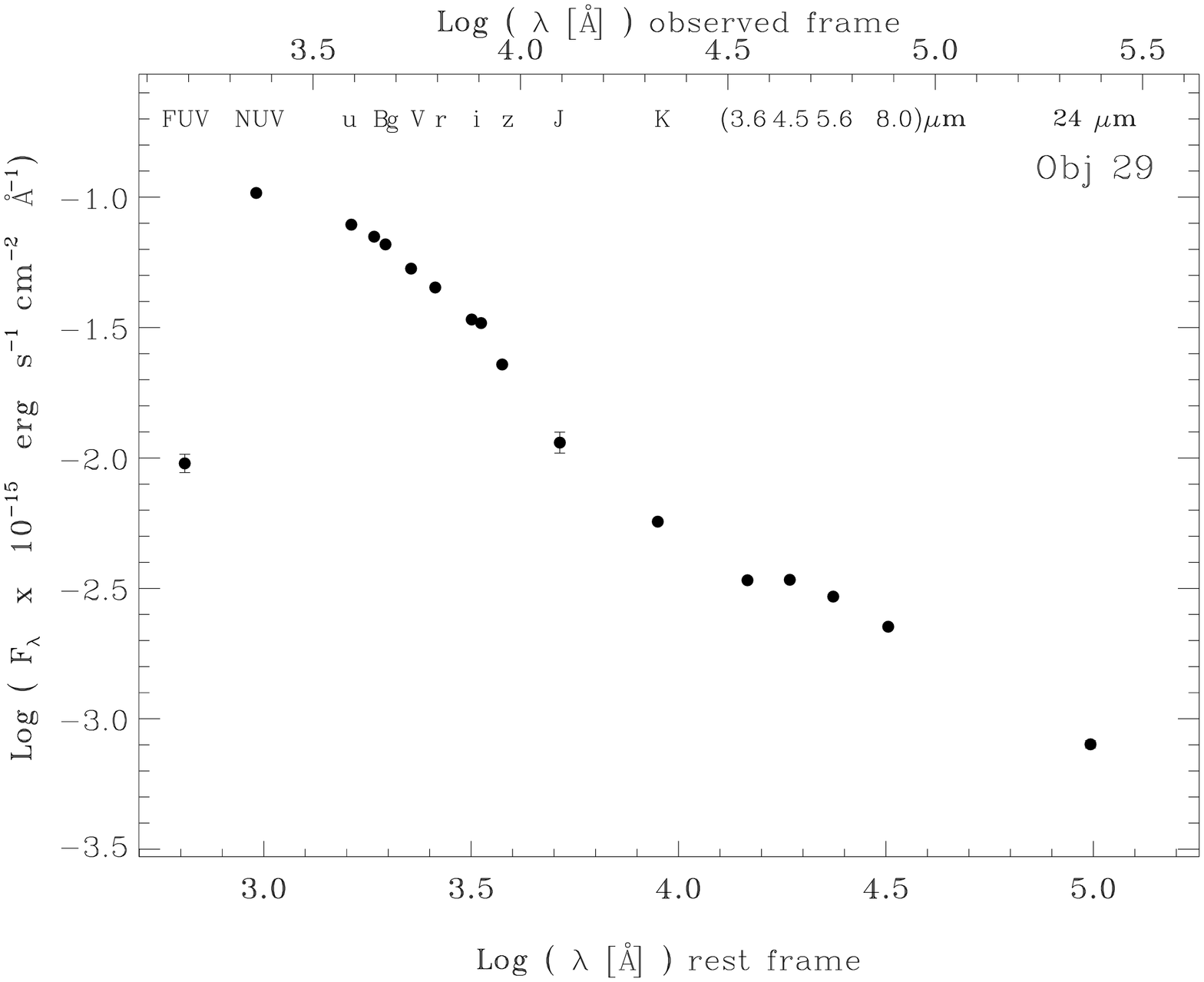} &
\includegraphics[scale=0.31,angle=90]{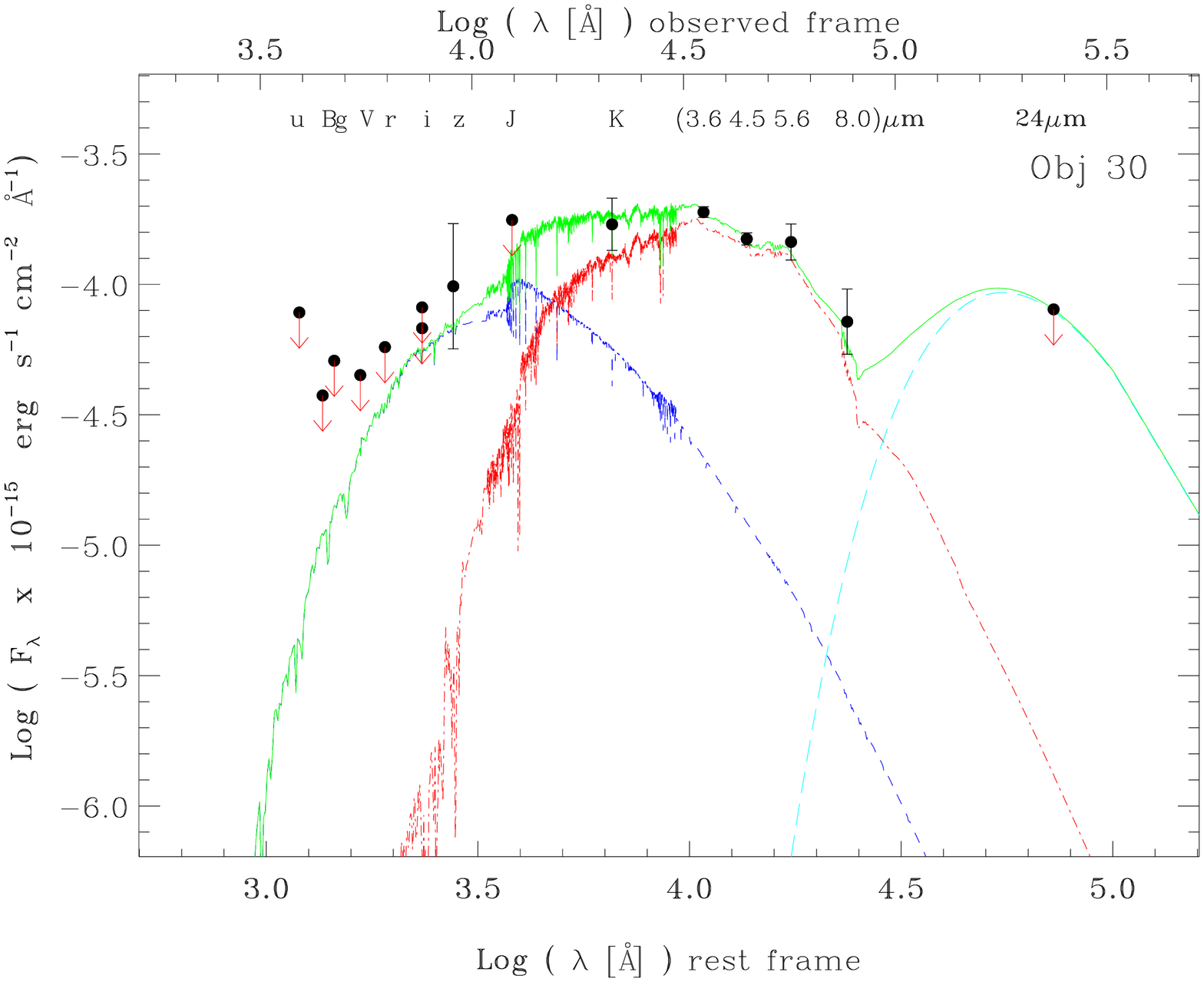} \\
\vspace{1em}
\includegraphics[scale=0.31,angle=90]{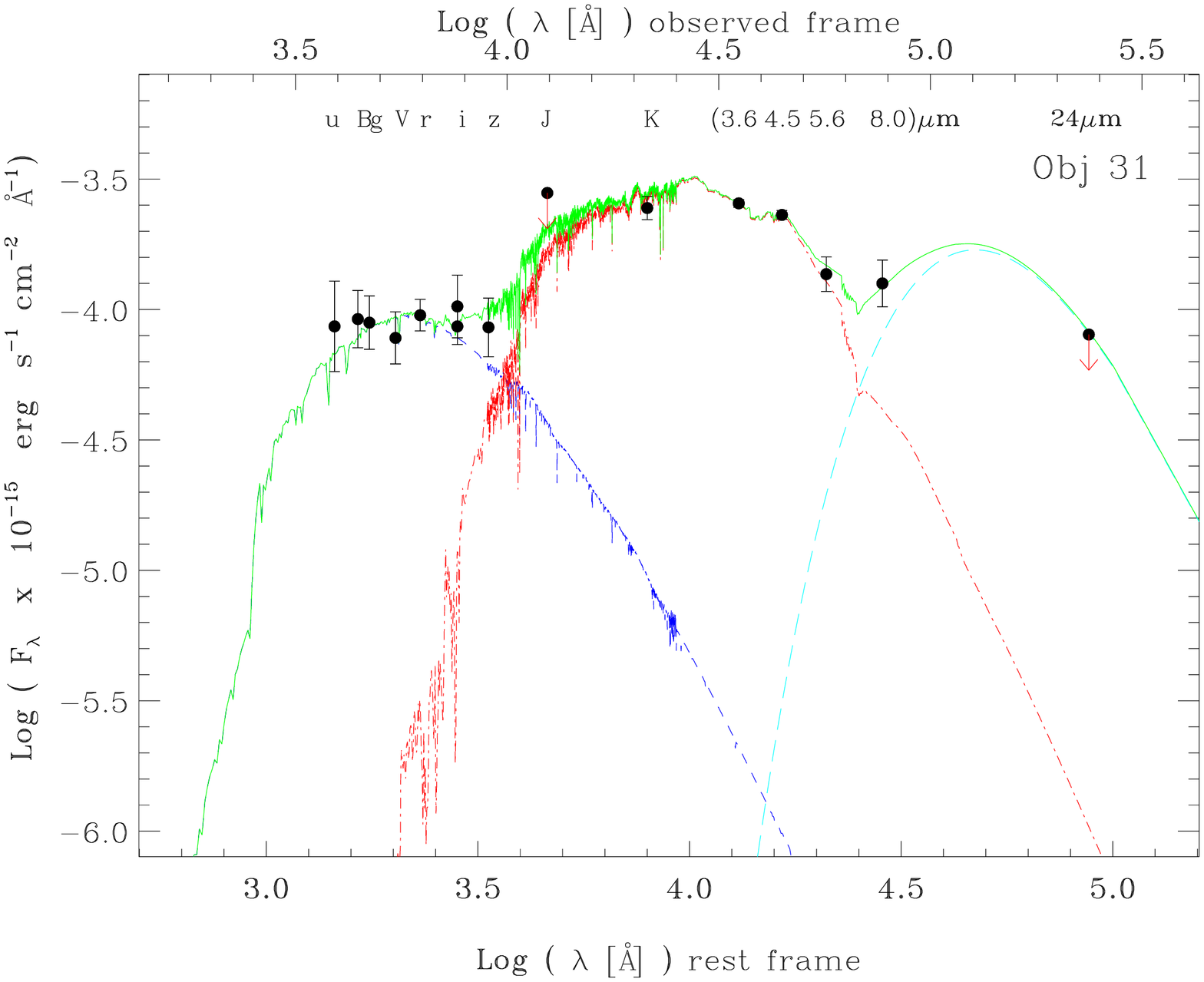}  &
\includegraphics[scale=0.31,angle=90]{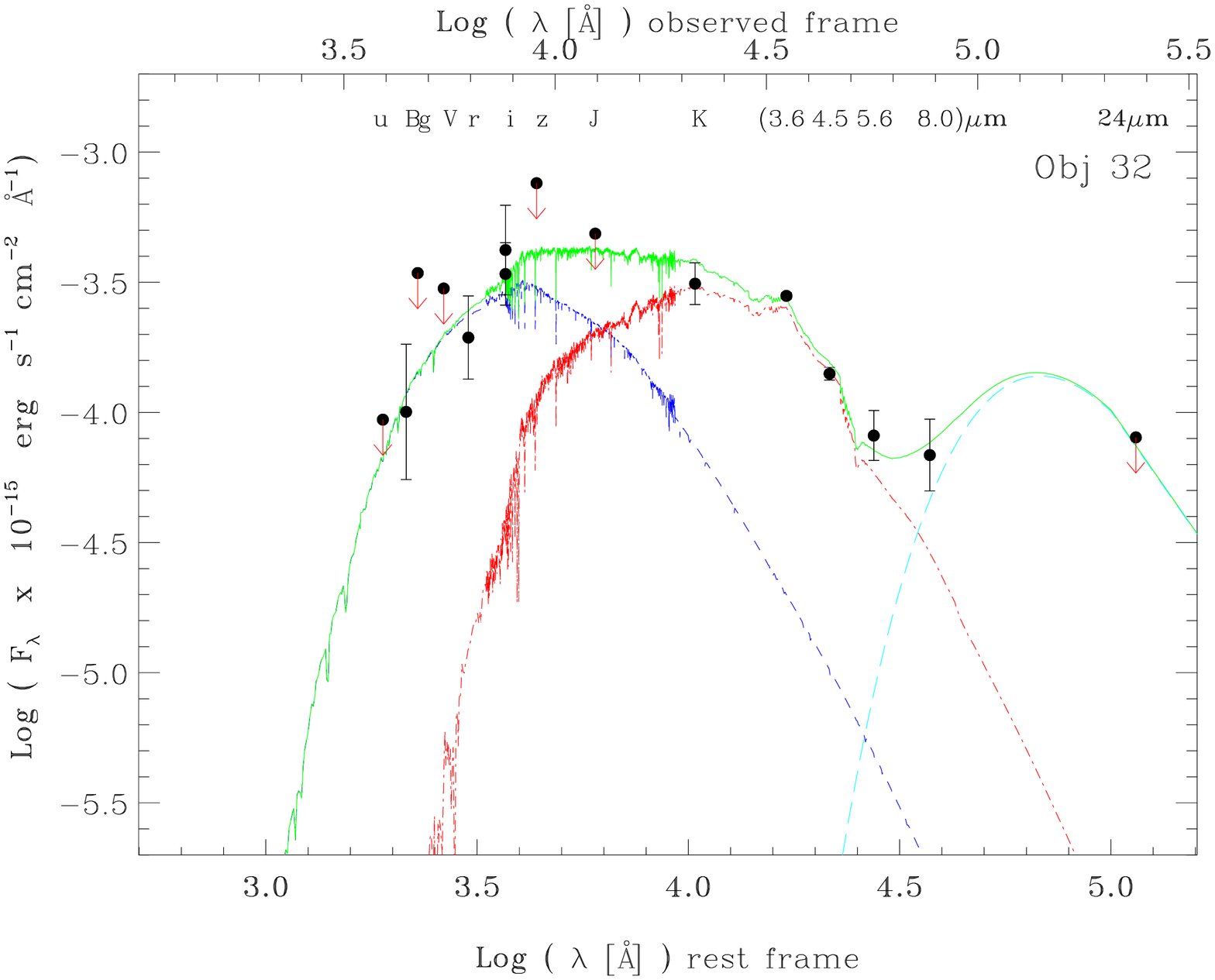} 
\end{array}$
\end{center}
\caption{CONTINUED}
\end{figure*}

\begin{figure*}[h]
\addtocounter{figure}{-1}
\begin{center}$
\begin{array}{ccc}
\vspace{2em}
\includegraphics[scale=0.31,angle=90]{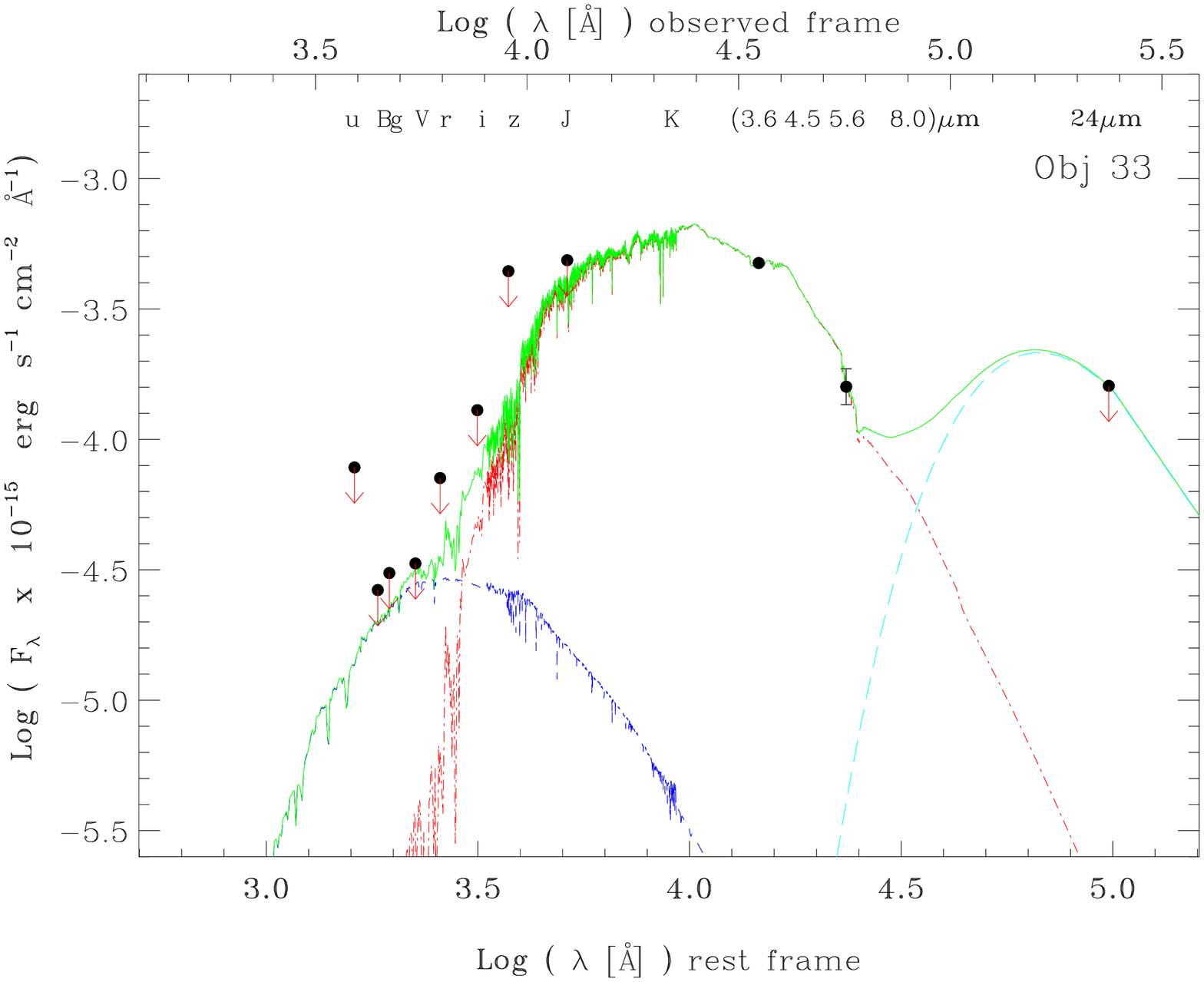} &
\includegraphics[scale=0.31,angle=90]{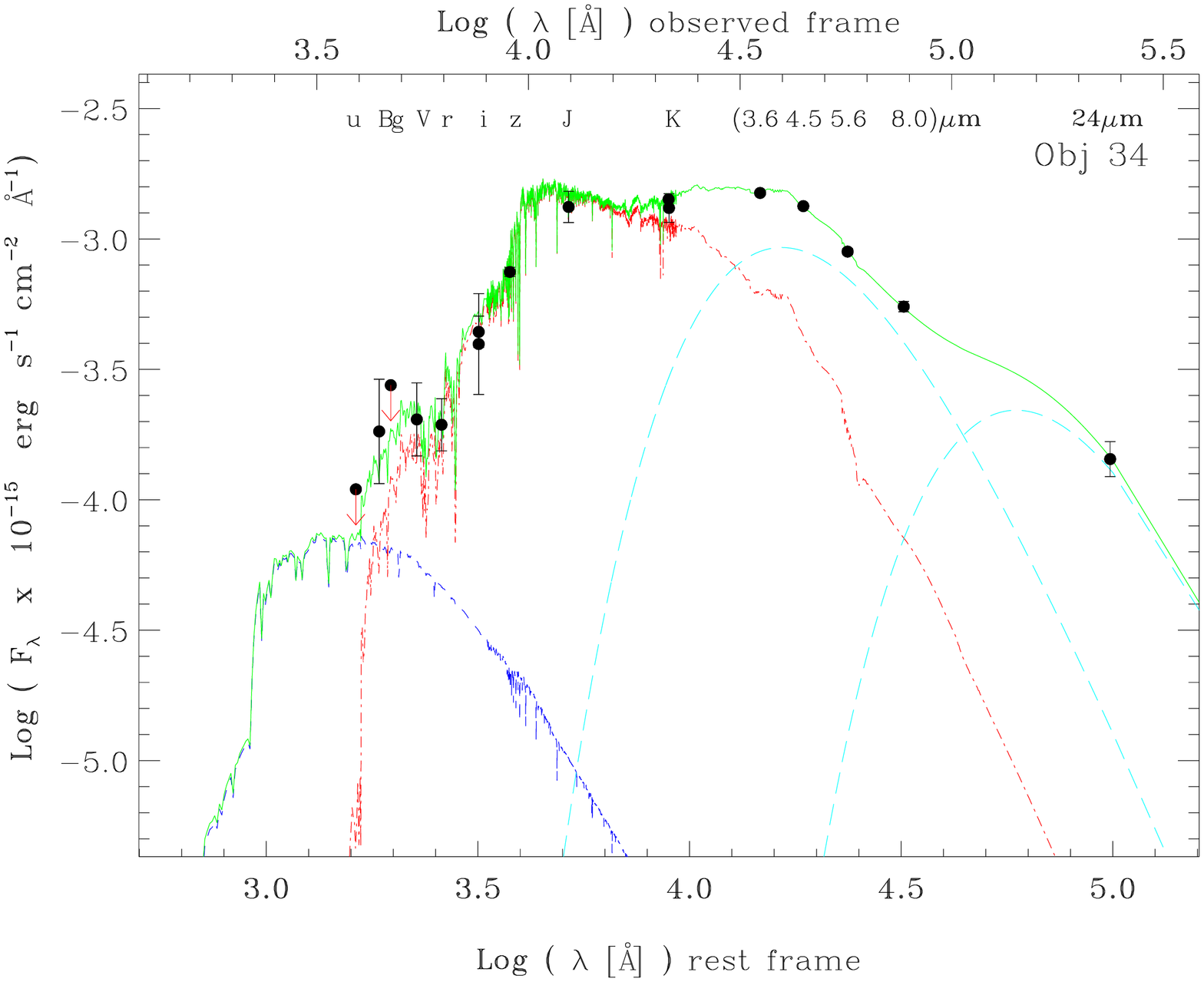} \\
\vspace{1em} 
\includegraphics[scale=0.31,angle=90]{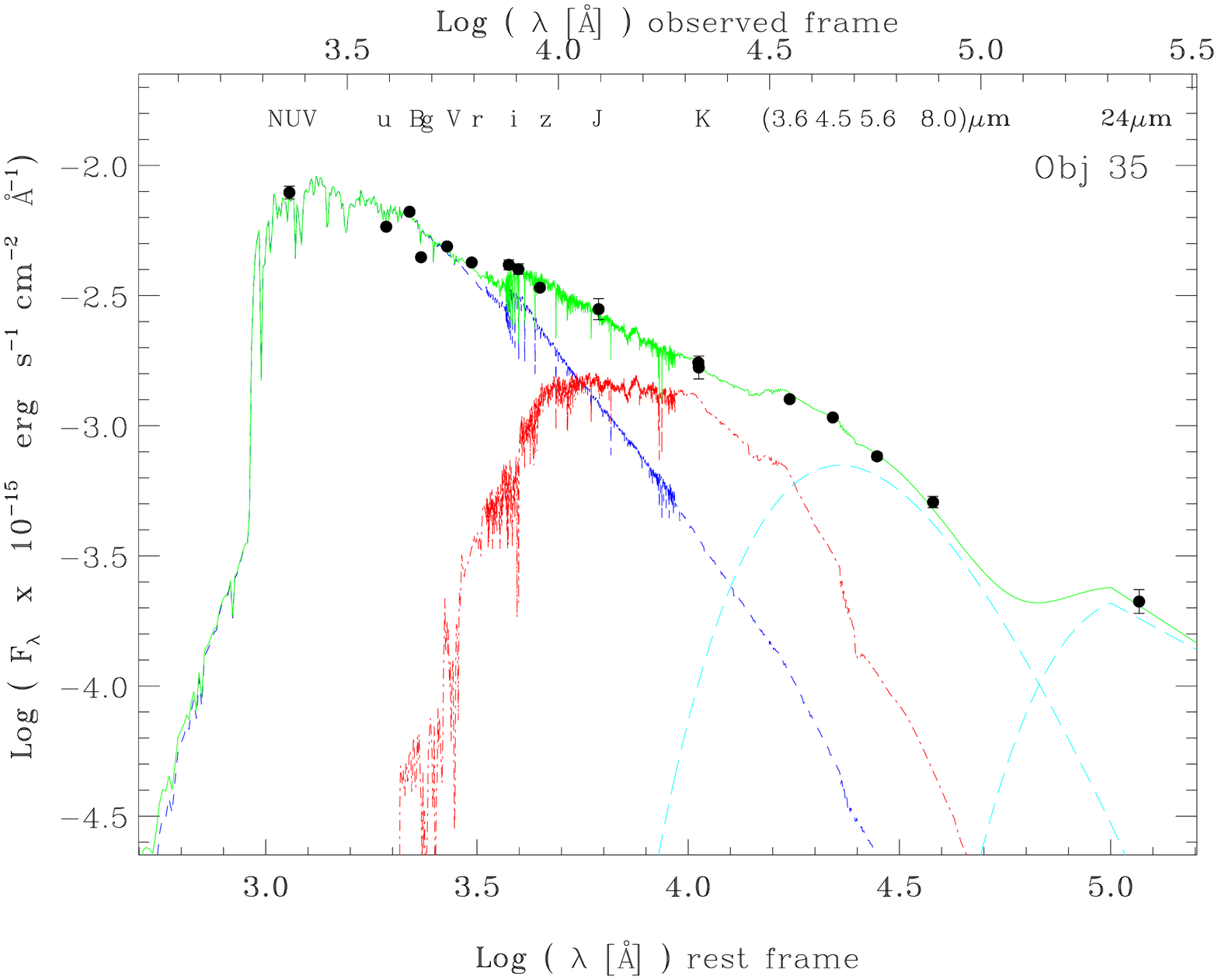}  &
\includegraphics[scale=0.31,angle=90]{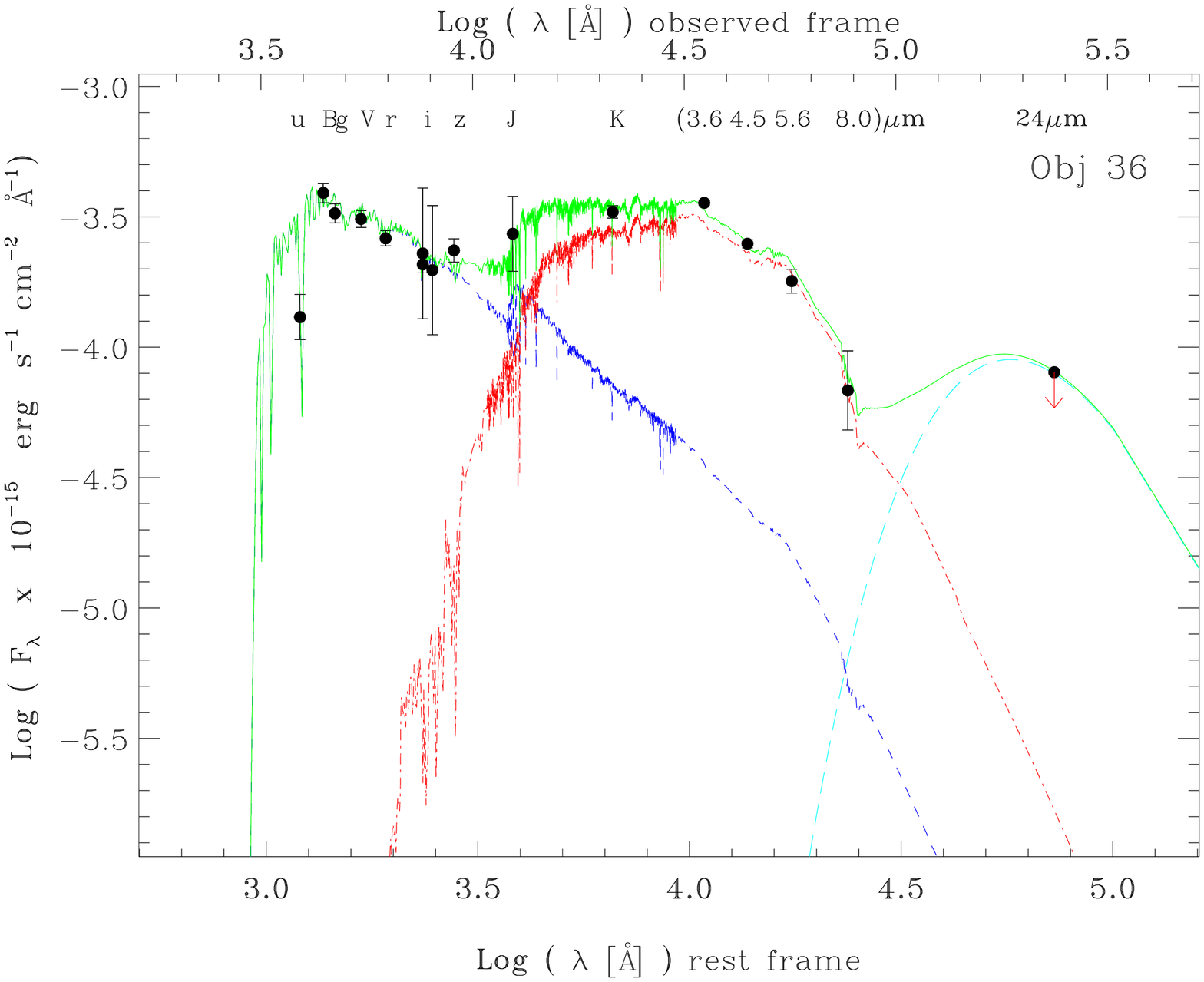}  \\
\vspace{1em} 
\includegraphics[scale=0.31,angle=90]{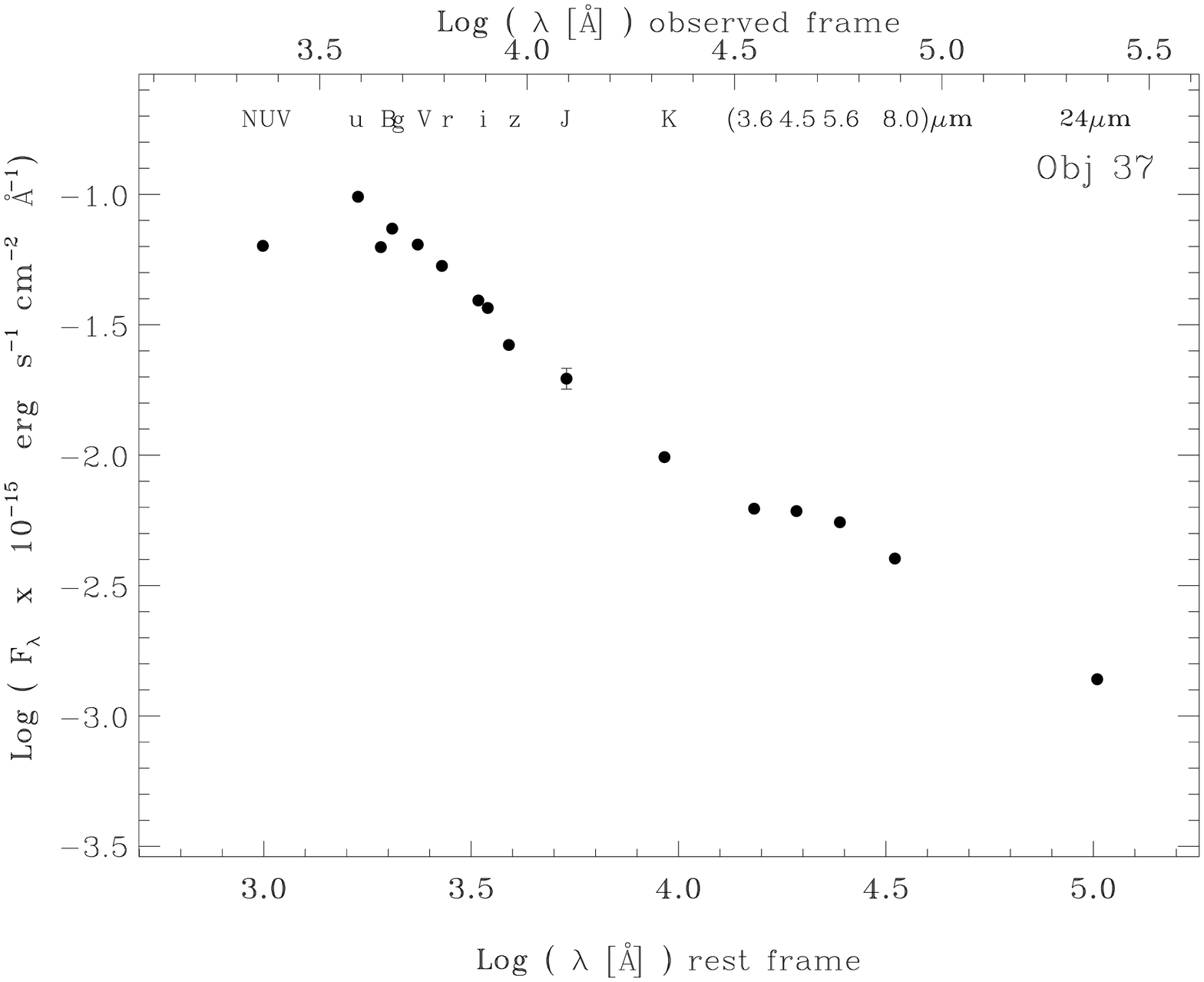} &
\includegraphics[scale=0.31,angle=90]{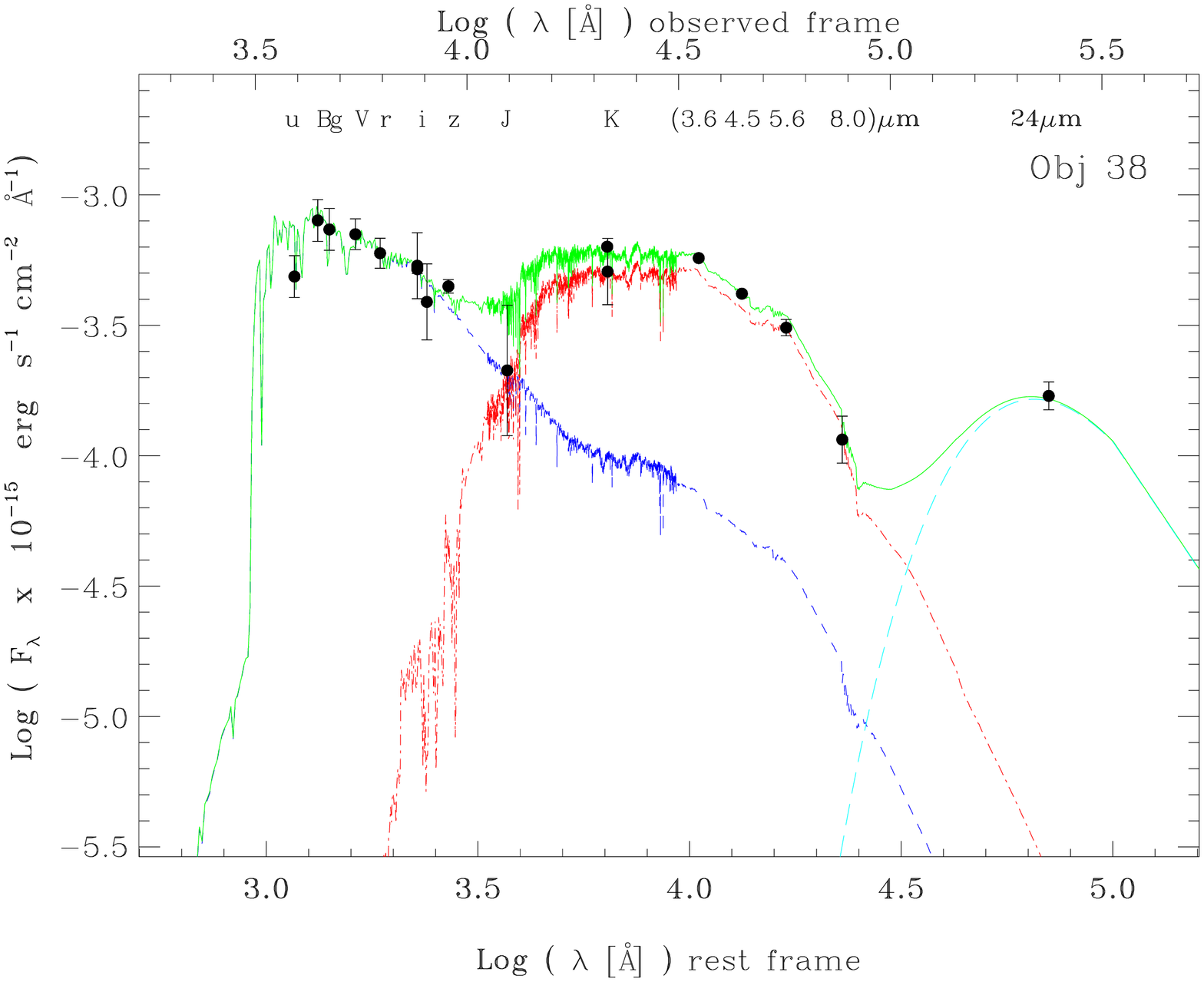} \\
\vspace{1em}
\includegraphics[scale=0.31,angle=90]{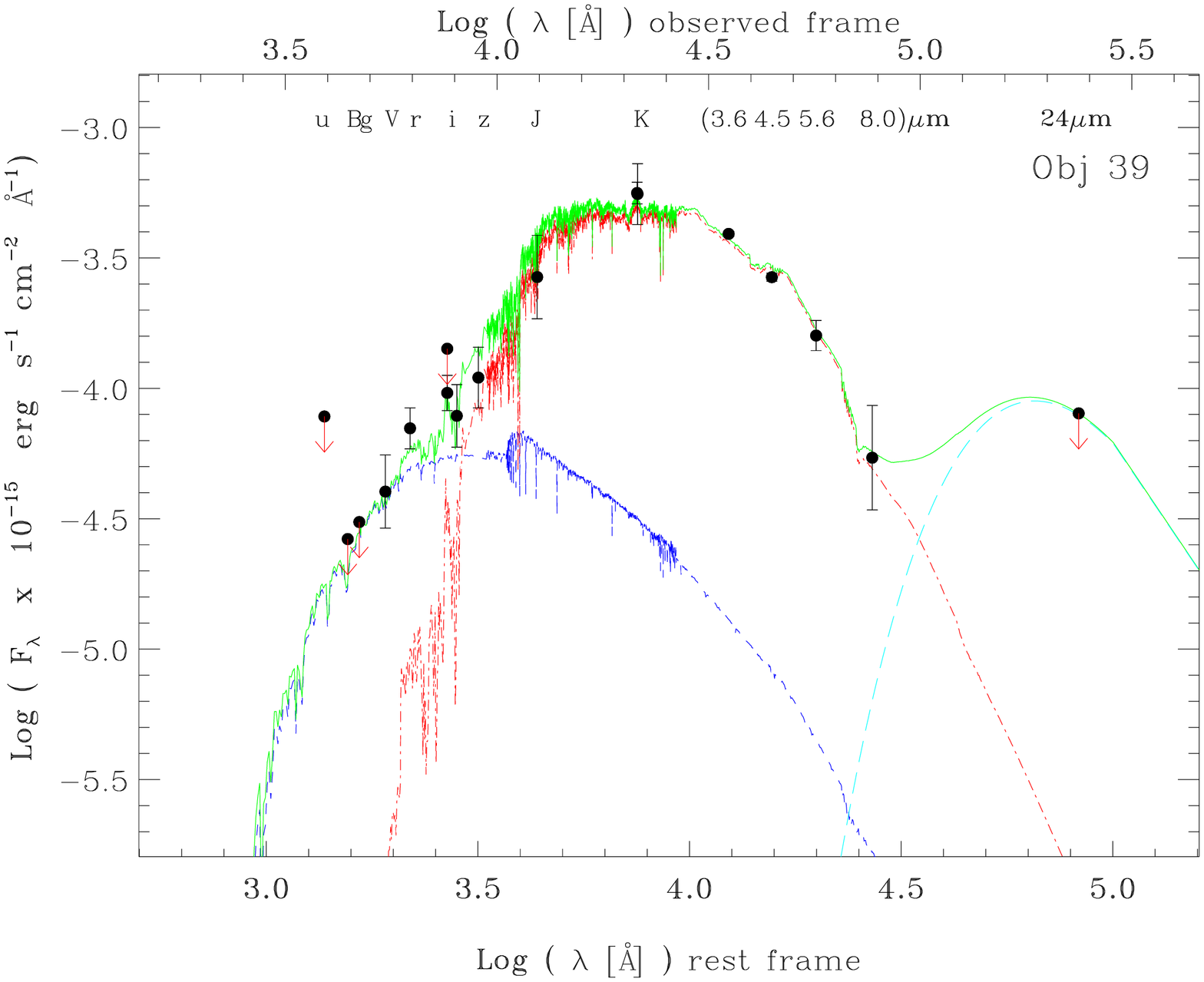} &
\includegraphics[scale=0.31,angle=90]{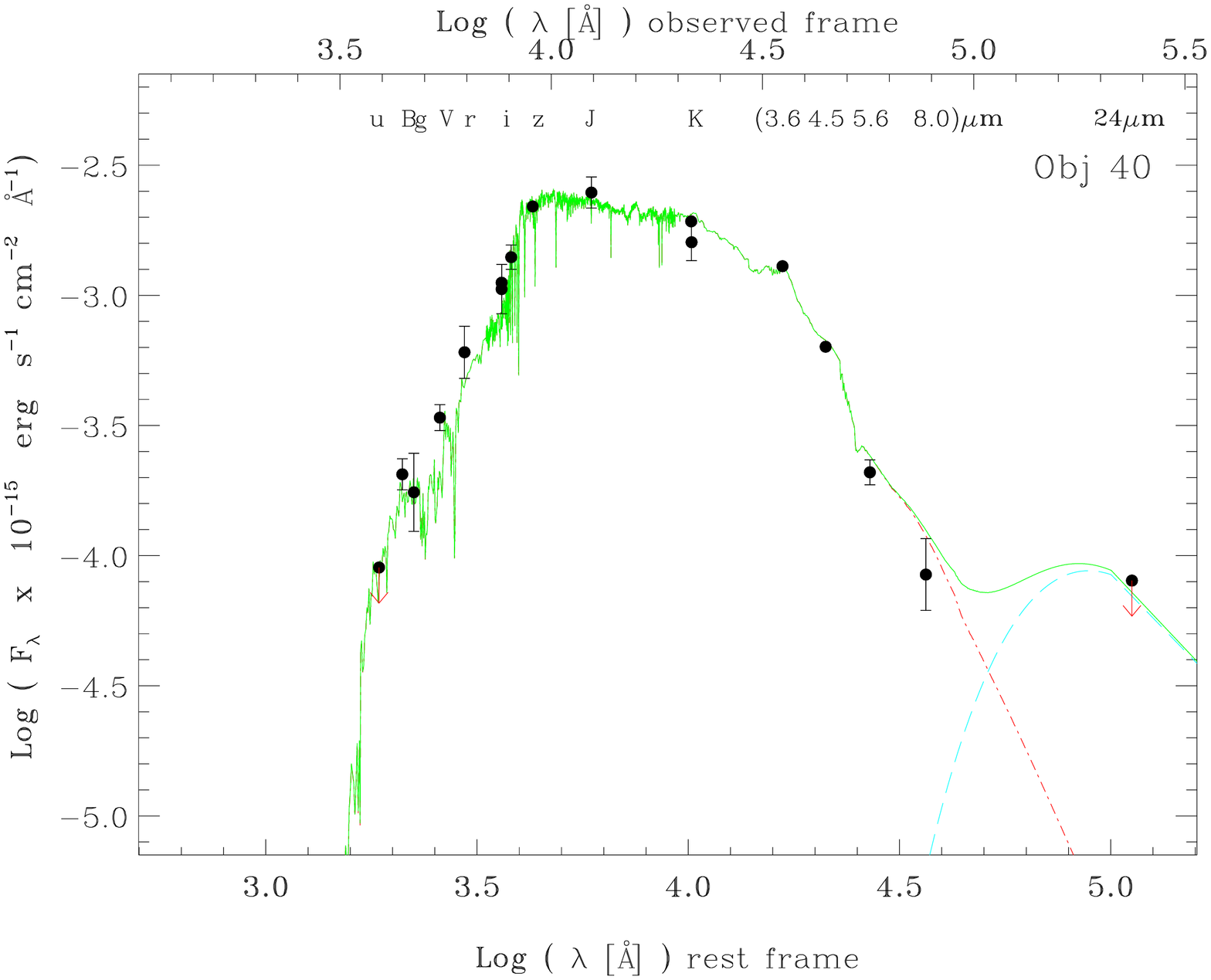}  
\end{array}$
\end{center}
\caption{CONTINUED}
\end{figure*}

\begin{figure*}[h]
\addtocounter{figure}{-1}
\begin{center}$
\begin{array}{ccc}
\vspace{2em}
\includegraphics[scale=0.31,angle=90]{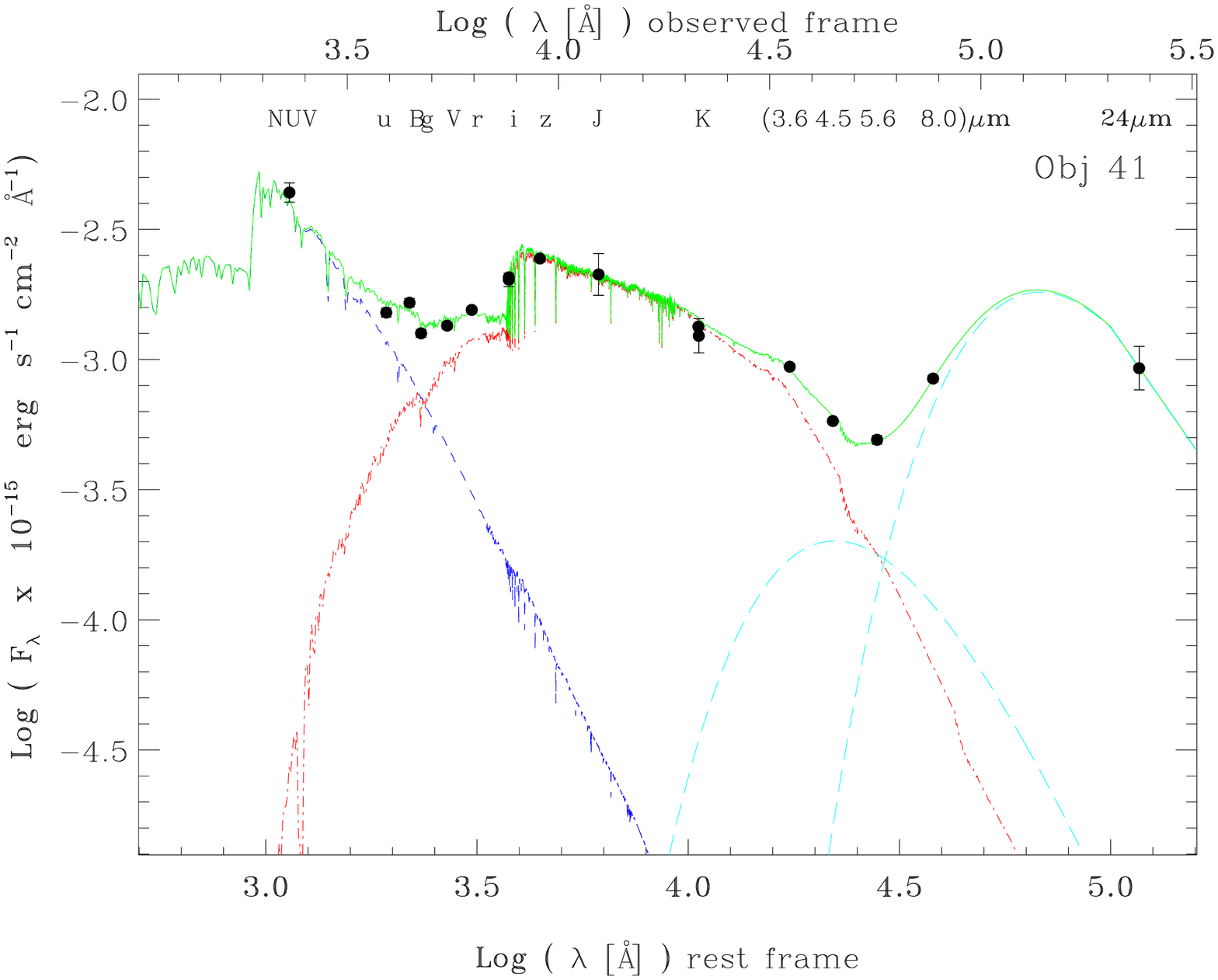} &
\includegraphics[scale=0.31,angle=90]{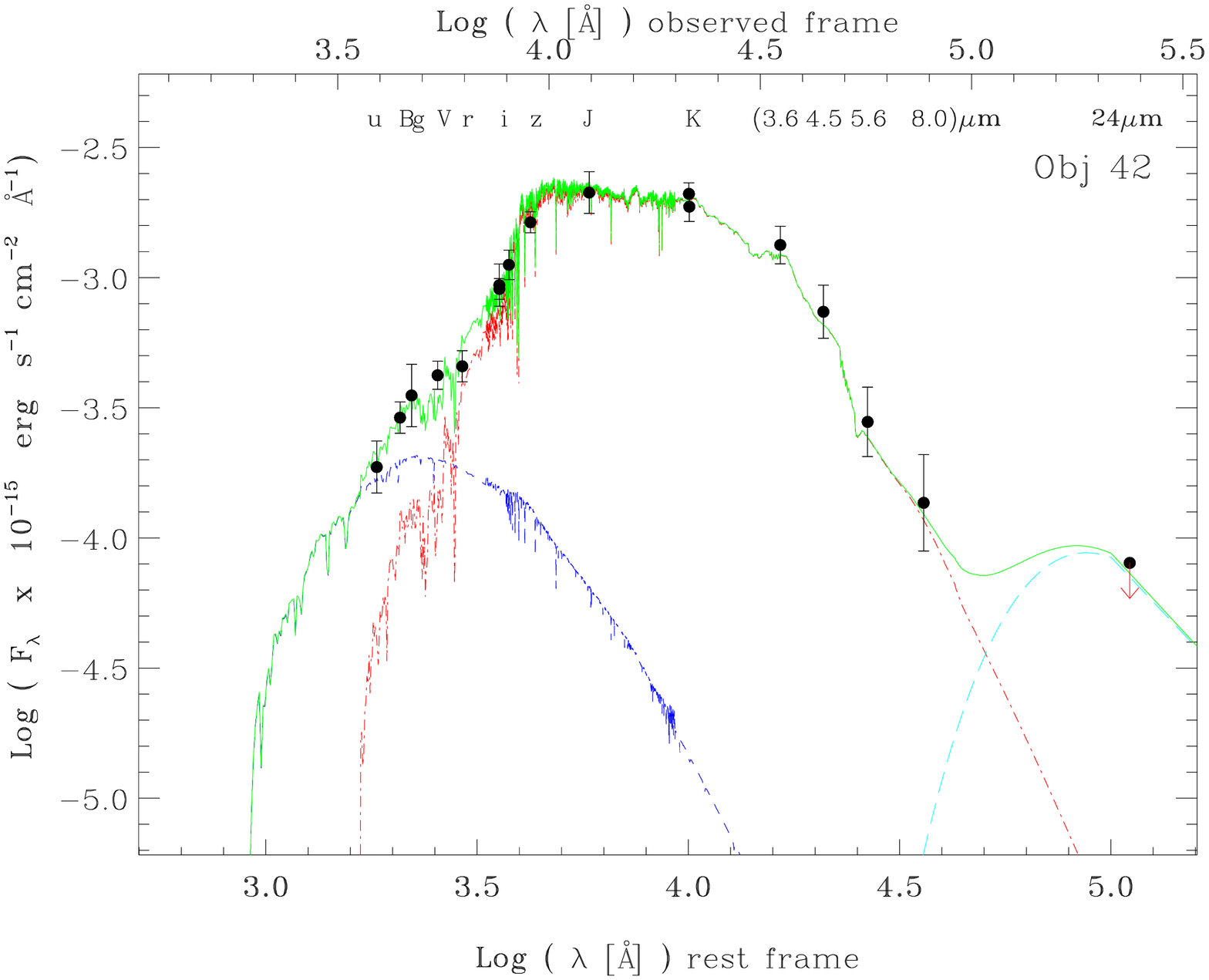}  \\
\vspace{1em}
\includegraphics[scale=0.31,angle=90]{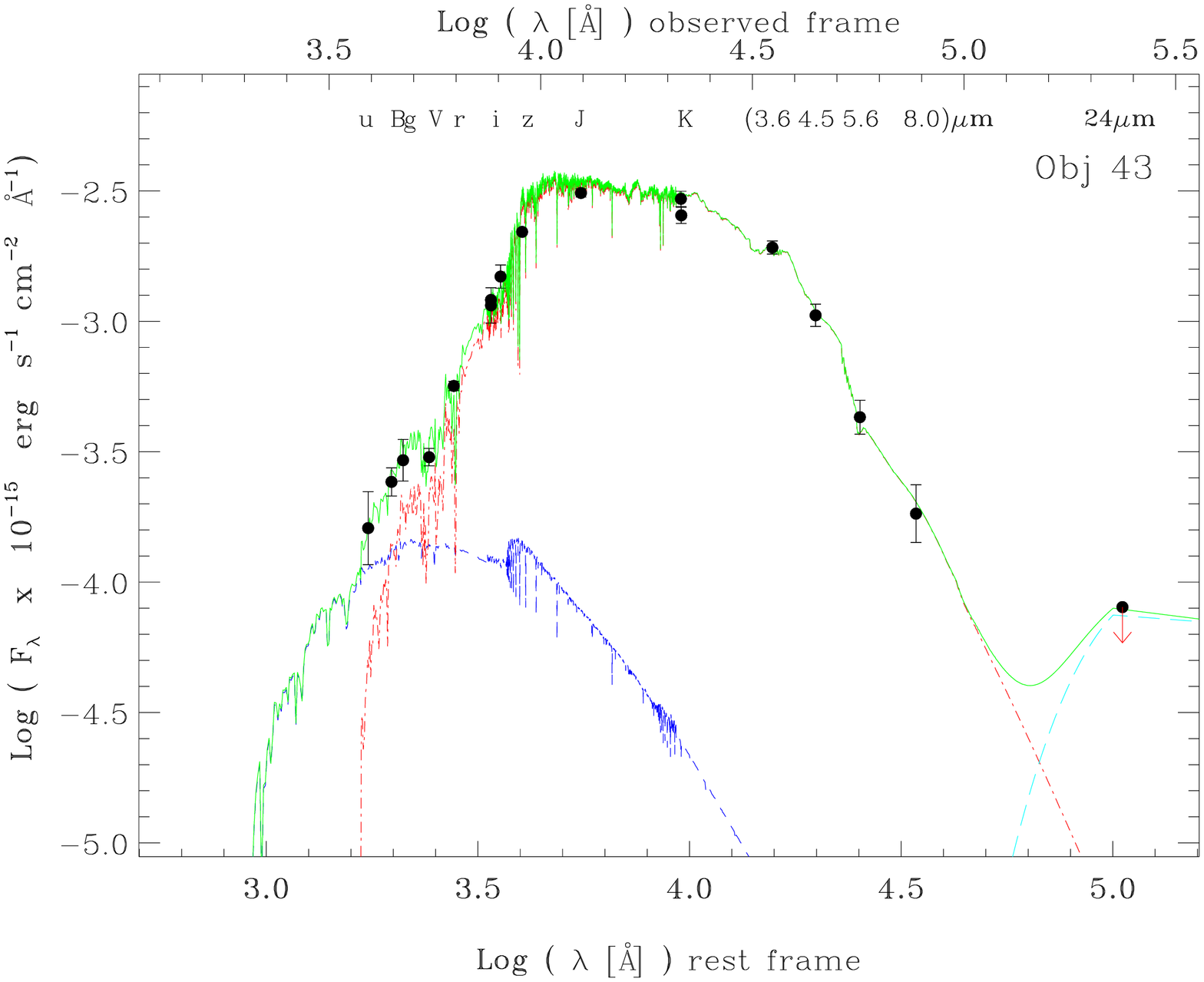}  &
\includegraphics[scale=0.31,angle=90]{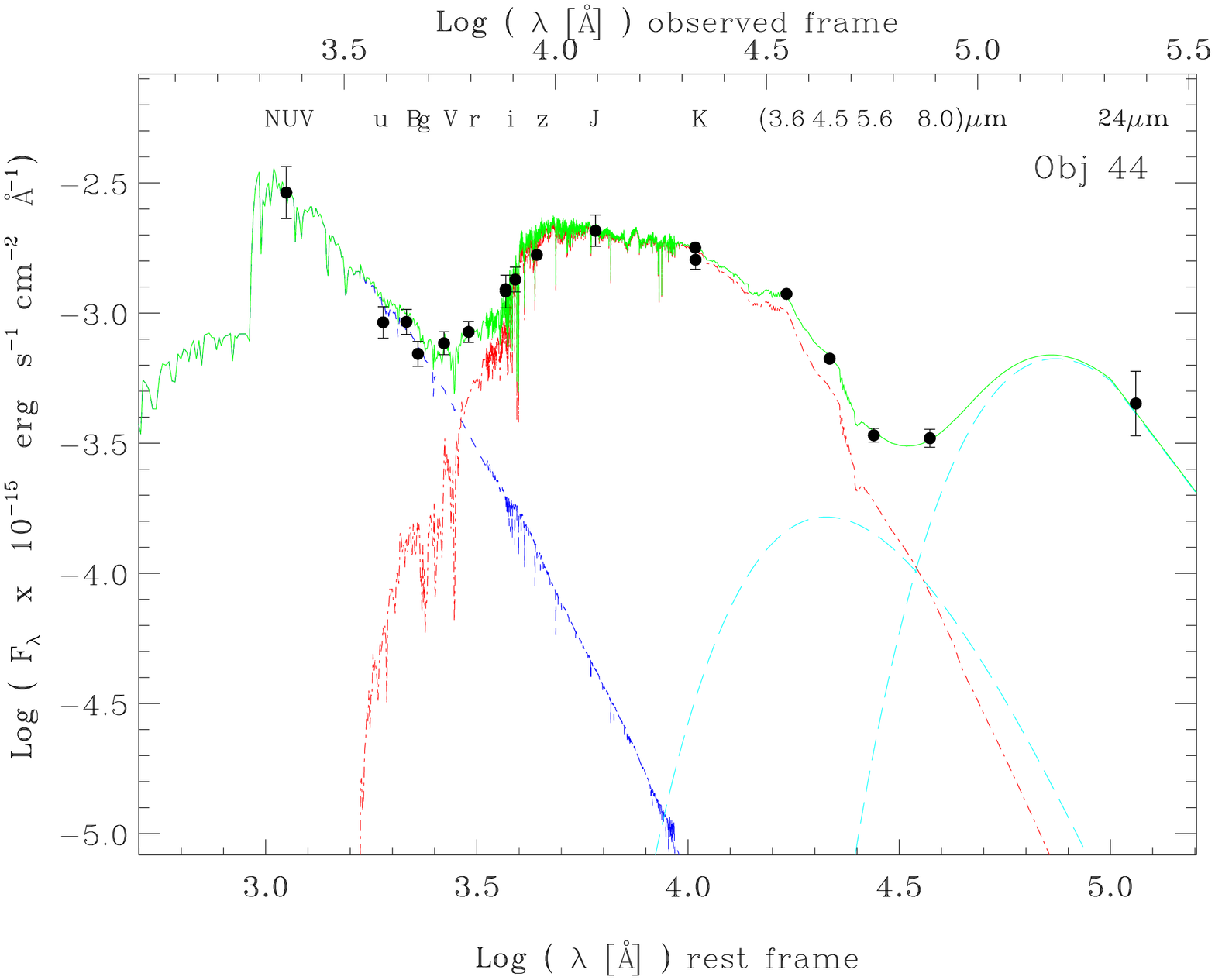} \\
\vspace{1em}
\includegraphics[scale=0.31,angle=90]{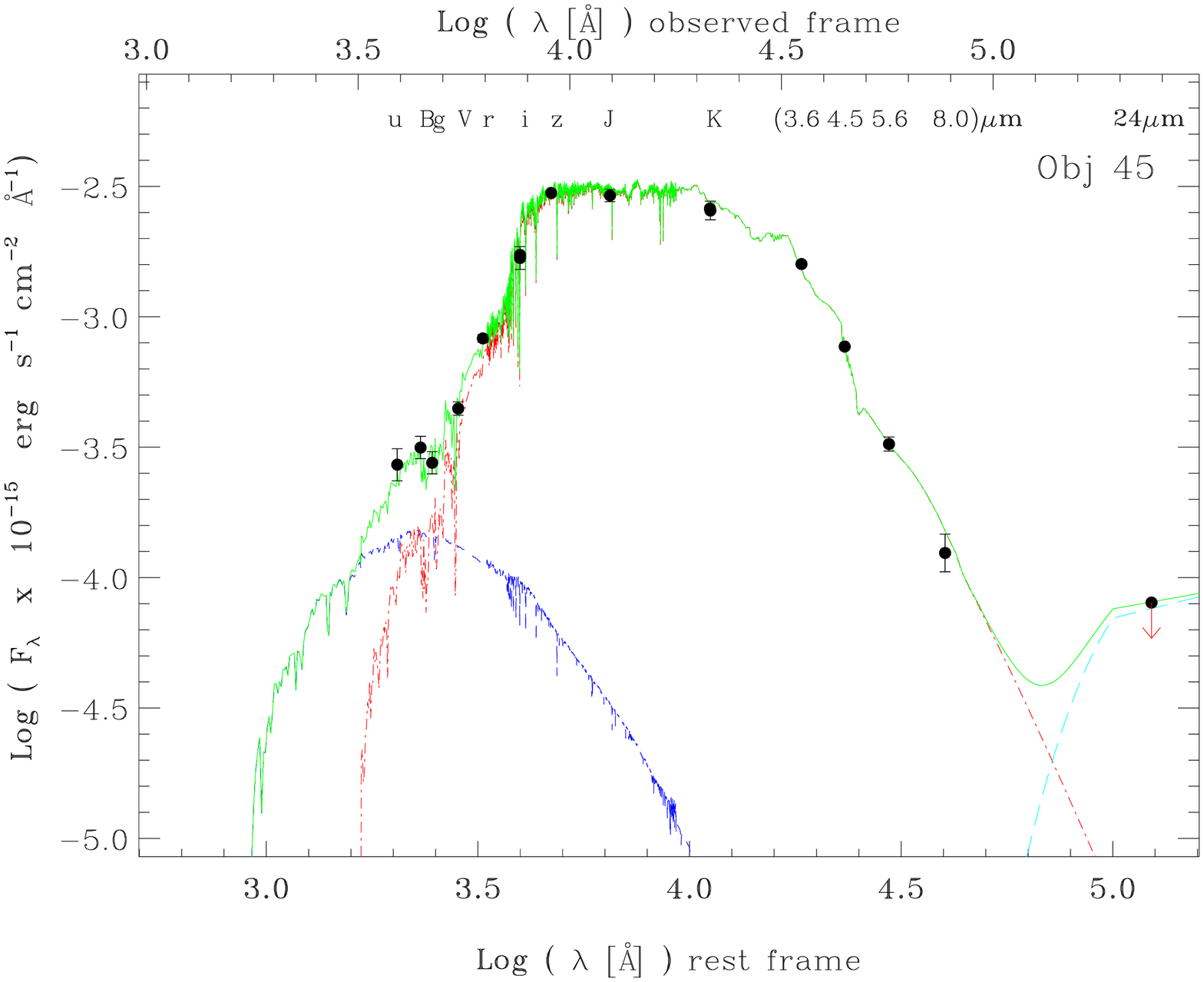} &
\includegraphics[scale=0.31,angle=90]{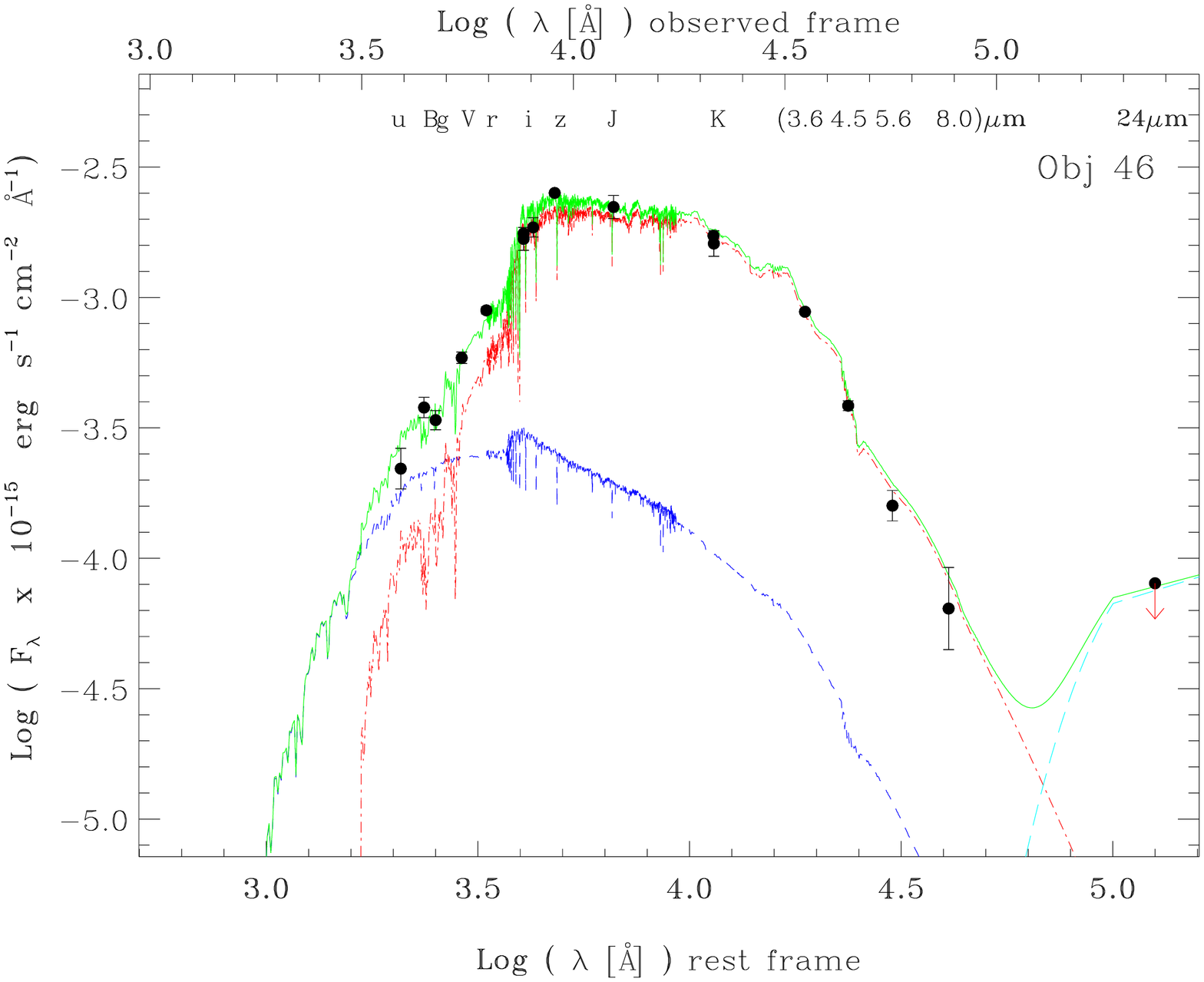} \\
\end{array}$
\end{center}
\caption{CONTINUED}
\end{figure*}

\end{appendix}

\end{document}